\documentclass[12pt,nofootinbib,tightenlines,preprintnumbers,superscriptaddress,floatfix]{revtex4}

\makeatletter
\g@addto@macro\bfseries{\boldmath}
\makeatother

\usepackage{hyperref}

\usepackage{graphicx}
\usepackage{amssymb}
\usepackage{amsmath}

\usepackage{mathtools,siunitx,booktabs,braket,slashed}
\usepackage{caption}
\usepackage{tikz}

\usepackage{color}

\newcommand{\be}{\begin{equation}}
\newcommand{\ee}{\end{equation}}
\newcommand{\ba}{\begin{eqnarray}}
\newcommand{\ea}{\end{eqnarray}}
\newcommand{\la}{\label}

\newcommand{\<}{\langle}
\renewcommand{\>}{\rangle}

\DeclareMathOperator\trace{Tr}

\newcommand\mcI{\mathcal{I}}

\newcommand{\bea}{\begin{eqnarray}}
\newcommand{\eea}{\end{eqnarray}}

\newcommand{\amu}{a_\mu}
\newcommand{\psib}{\overline{\psi}}	
	
\newcommand{\Lb}{\overline{\mathcal{L}}}

\begin{document}

\title{Hadronic light-by-light scattering contribution to the muon
  $g-2$ from lattice QCD: semi-analytical calculation of
  the QED kernel} 

\preprint{MITP-22-083~~~~~~~DESY-22-163}

\author{Nils Asmussen}
\affiliation{Department of Physics and Astronomy,
University of Southampton, Southampton SO17 1BJ, UK}

\author{En-Hung Chao}
 \affiliation{PRISMA$^+$ Cluster of Excellence  \& Institut f\"ur Kernphysik,
Johannes Gutenberg-Universit\"at Mainz, D-55099 Mainz, Germany}

\author{Antoine G\'erardin}
\affiliation{Aix-Marseille-Universit\'e, Universit\'e de Toulon, CNRS, CPT, Marseille, France}

\author{Jeremy R.\ Green}
\affiliation{Deutsches Elektronen-Synchrotron DESY, Platanenallee 6, 15738 Zeuthen, Germany}

\author{Renwick J.\ Hudspith}
\affiliation{GSI Helmholtzzentrum f\"ur Schwerionenforschung, 64291 Darmstadt, Germany}

\author{Harvey~B.~Meyer} 
\affiliation{Helmholtz-Institut Mainz, Johannes Gutenberg-Universit\"at Mainz,
D-55099 Mainz, Germany}
 \affiliation{PRISMA$^+$ Cluster of Excellence  \& Institut f\"ur Kernphysik,
Johannes Gutenberg-Universit\"at Mainz, D-55099 Mainz, Germany}

\author{Andreas Nyffeler\footnote{Currently no affiliation.}}
 \affiliation{PRISMA$^+$ Cluster of Excellence  \& Institut f\"ur Kernphysik,
Johannes Gutenberg-Universit\"at Mainz, D-55099 Mainz, Germany}

 \begin{abstract}
   \medskip
   
\noindent Hadronic light-by-light scattering is one of the virtual processes
 that causes the gyromagnetic factor $g$ of the muon to deviate from the value of two
 predicted by Dirac's theory. This process makes one of the largest contributions to the
 uncertainty of the Standard Model prediction for the muon $(g-2)$.
Lattice QCD allows for a first-principles approach to
computing this non-perturbative effect. In order to avoid power-law
finite-size artifacts generated by virtual photons in lattice simulations, we
follow a coordinate-space approach involving
a weighted integral over the vertices of the QCD four-point function
of the electromagnetic current carried by the quarks.
Here we present in detail the semi-analytical calculation of the QED part of the amplitude,
employing position-space perturbation theory in continuous, infinite four-dimensional Euclidean space.
We also provide some useful information about a computer code for the numerical implementation
of our approach that has been made public at {\tt https://github.com/RJHudspith/KQED}.
\end{abstract}

\maketitle

\centerline{\today}

\newpage

\tableofcontents

\section{Introduction}

The anomalous magnetic moment of the muon, $a_\mu \equiv (g-2)_\mu/2$,
characterizes its response to a magnetic field, and is one of the most
precisely known quantities in fundamental physics. Currently, the
experimental world average~\cite{Muong-2:2021ojo,Muong-2:2006rrc} is
in tension with the theoretical evaluation based on the Standard Model
(SM) of particle physics.  On the basis of the Muon $g-2$ Theory Initiative's 2020 White Paper
(WP)~\cite{Aoyama:2020ynm} with input from Refs.\!~\cite{Aoyama:2012wk,Aoyama:2019ryr,Czarnecki:2002nt,Gnendiger:2013pva,Davier:2017zfy,Keshavarzi:2018mgv,Colangelo:2018mtw,Hoferichter:2019mqg,Davier:2019can,Keshavarzi:2019abf,Kurz:2014wya,Melnikov:2003xd,Masjuan:2017tvw,Colangelo:2017fiz,Hoferichter:2018kwz,Gerardin:2019vio,Bijnens:2019ghy,Colangelo:2019uex,Blum:2019ugy,Colangelo:2014qya},
the tension is at the $4.2\sigma$ level. Theoretical and experimental uncertainties are practically
equal and just under the level of 0.4\,ppm. While a tension between
theory and experiment has persisted for about twenty years,
the 2021 result of the Fermilab Muon $(g-2)$ experiment~\cite{Muong-2:2021ojo} has increased this tension 
and thereby revived the general interest in possible explanations involving
beyond-the-Standard-Model physics, see e.g.~\!\cite{Athron:2021iuf}.

The leading prediction for $a_\mu$ in QED is
$\frac{\alpha}{2\pi}$~\cite{Schwinger:1948iu}, where $\alpha$ is the
fine-structure constant. Effects of the strong interaction enter
at O($\alpha^2$).  Due to the low mass scale of the muon,
strong-interaction effects in the muon $(g-2)$ must be treated in
their full, non-perturbative complexity. As a result, the theory
uncertainty of this precision observable is entirely dominated by the
hadronic contributions.

The leading hadronic contribution goes under the name of hadronic
vacuum polarization (HVP).  The situation around the muon $(g-2)$ has
become more intricate with the publication of a lattice-QCD based
calculation~\cite{Borsanyi:2020mff} of the HVP contribution, which
finds a larger value than the dispersion-theory based estimate of the
WP and would bring the overall theory prediction into far better
agreement with the experimental value of $a_\mu$. Thus it will be
crucial to resolve the tension between the different determinations of
the HVP contribution in order to capitalize on the
expected improvements in the experimental determinations of $a_\mu$: a
reduction by more than a factor of two is expected from the Fermilab
Muon $g-2$ experiment~\cite{Muong-2:2015xgu}, and further measurements
are planned at J-PARC~\cite{Abe:2019thb} and considered at
PSI~\cite{Aiba:2021bxe}.

An O($\alpha^3$) hadronic contribution to $a_\mu$, known as the
hadronic light-by-light (HLbL) contribution, also adds
significantly to the error budget of the SM prediction.
It can be represented as the Feynman diagram depicted in Fig.~\ref{fig:muonhlbl}.
In the WP error budget for $a_\mu$, its assigned uncertainty is
0.15\,ppm. Therefore, anticipating error reductions in the HVP
contribution and in the experimental measurements, it is crucial to
further reduce the uncertainty on the HLbL contribution by at least a factor of two.

The HLbL contribution is conceptually more complex than the HVP
contribution. 
On the other hand, being suppressed by an additional
power of the fine-structure constant $\alpha$, the requirements on its
relative precision are far less stringent: the uncertainty quoted in
the WP corresponds to 20\%. In recent years, the HLbL contribution has been evaluated
using either dispersive methods, 
for which a full result can be found in the WP~\cite{Aoyama:2020ynm}
based on Refs.\!~\cite{Melnikov:2003xd,Masjuan:2017tvw,Colangelo:2017fiz,Hoferichter:2018kwz,Gerardin:2019vio,Bijnens:2019ghy,Colangelo:2019uex,Pauk:2014rta,Danilkin:2016hnh,Jegerlehner:2017gek,Knecht:2018sci,Eichmann:2019bqf,Roig:2019reh,Colangelo:2014qya},
or lattice QCD (\cite{Blum:2019ugy} and \cite{Chao:2021tvp}--\cite{Chao:2022xzg}).  Good agreement is found among
the three evaluations within the quoted uncertainties.

The purpose of the present paper is to provide a detailed account of
the computational strategy underlying our recent
calculation~\cite{Chao:2021tvp,Chao:2022xzg}.  Its full development spanned several
years, with progress reported in a number of conferences since
2015~\cite{Talk_Asmussen_DPG_2015,Green:2015mva,Asmussen:2016lse,Asmussen:2017bup,Asmussen:2018ovy,Asmussen:2019act}.
The basic idea is to treat the muon and photon propagators of Fig.~\ref{fig:muonhlbl}
in position-space perturbation theory, in the continuum and
in infinite-volume, while the `hadronic blob' is to be treated in
lattice QCD on a spatial torus.  Thus much of this paper is concerned
with the semi-analytical calculation of the QED part of the amplitude.

The idea to compute the HLbL contribution to $a_\mu$ was first
proposed in 2005~\cite{Hayakawa:2005eq}, with a follow-up three years
later~\cite{Chowdhury:2008zz}.  These initial methods finally led to
the 2014 publication~\cite{Blum:2014oka}.  In parallel to the
development of our strategy, the RBC/UKQCD collaboration then also
worked on improving its computational methods~\cite{Blum:2015gfa},
with a first exploratory calculation at physical quark masses
published in~\cite{Blum:2016lnc}.  These methods are based on treating
the QED parts of Fig.~\ref{fig:muonhlbl} within the lattice field
theory set up on a finite torus.  Starting with
Ref.~\!\cite{Blum:2017cer}, the RBC/UKQCD collaboration also developed
its own tools to treat the muon and photon propagators in infinite
volume. We will return in section~\ref{sec:KernelSubtractions} to some aspects
of the cross-fertilization that occurred between the two groups.

It is also worth pointing out other, less direct approaches that have
been pursued towards better determining the HLbL 
contribution to the muon $(g-2)$ using lattice QCD. Of all meson
exchanges, the neutral-pion pole contribution is by far the largest,
and we have published two lattice calculations of its transition form
factor describing its coupling to two (in general) virtual
photons~\cite{Gerardin:2016cqj,Gerardin:2019vio}. Since the $\pi^0$
contribution is the numerically dominant one at long distances, having
a dedicated determination thereof also helps control systematic errors
at long distances in the direct lattice calculation~\cite{Chao:2020kwq}
based on the formalism presented in this paper.  As a separate
line of study, we have investigated the HLbL 
scattering amplitude at Euclidean kinematics~\cite{Green:2015sra},
particularly its eight independent forward-scattering components,
which depend on three invariant kinematic variables. Knowing these
amplitudes allows one to constrain the contributions of various meson
exchanges~\cite{Green:2015sra,Gerardin:2017ryf} by parametrizing their transition
form factors, information which may subsequently be used to estimate
the HLbL contribution to the muon $(g-2)$.

This manuscript is organized as follows.  Section \ref{sec:master}
presents the general features of our position-space approach and the
master-formula for $a_\mu^{\rm HLbL}$.  The ingredients necessary for
the evaluation of the `QED kernel' describing all purely QED elements
of the amplitude depicted in Fig.~\ref{fig:muonhlbl} are collected
in section~\ref{sec:starting}, at the end of which the averaging over the direction
of the muon momentum is performed. A relatively straightforward method of
evaluating the final convolution integral yielding
the weight functions parametrizing the QED kernel is
described in section~\ref{sec:direct}. An alternative, ultimately
favored method based on the multipole expansion of the photon
propagator in Gegenbauer polynomials is presented in
section~\ref{sec:alternative_evaluation}. Some technical aspects of
the numerical implementation are given in
section~\ref{sec:discussion_master_formula}.  Then several models are
used in section~\ref{sec:ipihats} to compute various contributions to
the four-point function of the electromagnetic current in QED and
QCD. Since these contributions to the muon $(g-2)$ have been computed
previously (using analytical methods in momentum-space), we use them
to perform tests of our position-space QED kernel in section
\ref{sec:numtests}. Published results obtained 
in lattice QCD by the present methods for the quark-connected contribution 
are also reviewed in that section. Section~\ref{sec:concl} collects
our concluding remarks.  The appendices contain additional material
useful for numerical implementations, providing in particular the
kernel asymptotics for various special kinematic regimes.
The final appendix (\ref{app:code}) provides some
information about a computer code available for the numerical implementation
of our approach based on the results of section~\ref{sec:alternative_evaluation}.

\section{Master formula for $a_\mu^{\rm HLbL}$ in position space\la{sec:master}}

\begin{figure}
\centering
\includegraphics[width=0.5\textwidth]{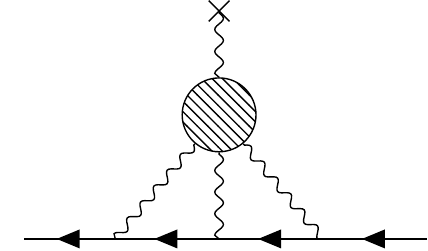}
\caption{Hadronic light-by-light scattering diagram in the muon
  $(g-2)$. 
} 
\label{fig:muonhlbl}
\end{figure}

We are interested in the hadronic light-by-light (HLbL) scattering
contribution to the anomalous magnetic moment of the muon, see
Fig.~\ref{fig:muonhlbl}.
The basic idea of our approach is to treat the four-point function of
hadronic electromagnetic currents, represented by the blob in
Fig.~\ref{fig:muonhlbl}, in lattice QCD regularization, while for the
remaining QED part with photons and muons, we use continuum, Euclidean
position-space perturbation theory in infinite
volume~\cite{Talk_Asmussen_DPG_2015, Green:2015mva,
  Asmussen:2016lse}. As we will show, this allows a Lorentz covariant,
semi-analytical calculation of the QED part which avoids power-law
finite-volume corrections $1/L^2$ in $a_\mu^{\rm HLbL}$ due to the
massless photons. An approach in position space is most natural, since
in lattice QCD one obtains the four-point correlation function
directly in position space. Furthermore, it will be possible to get
directly the HLbL contribution $a_\mu^{\rm HLbL}$ as a spatial moment
of the four-point correlation function, i.e.\ no extrapolation of the
Pauli form factor $F_2(k^2)$ for $k^2 \to 0$ is needed as tried in
earlier attempts in Ref.~\cite{Blum:2014oka}.

The HLbL contribution to the muon $(g-2)$ from the light quarks can be
obtained from the matrix element of the electromagnetic current 
\be
j_\rho(x) = \frac{2}{3} (\bar u\gamma_\rho u)(x) - \frac{1}{3} (\bar
d\gamma_\rho d)(x) - \frac{1}{3} (\bar s\gamma_\rho s)(x),  
\ee
between muon states, which can be parametrized by two form factors
(assuming Lorentz symmetry, current conservation as well as parity
and charge conjugation invariance)
\be
(ie) \<\mu^-(p')|j_\rho(0)|\mu^-(p)\> 
= - (ie) \bar u(p') \Big[\gamma_\rho F_1(k^2) +
\frac{\sigma_{\rho\sigma}k_\sigma }{2m} F_2(k^2)\Big] u(p), 
\ee
where $e$ is the electric charge of the electron, $m$ is the muon
mass, $\sigma_{\rho\sigma} \equiv \frac{i}{2}
[\gamma_\rho,\gamma_\sigma]$ and we use $\gamma$-matrices in Euclidean
space with $\{\gamma_\mu , \gamma_\nu\} = 2 \delta_{\mu\nu}$
that are Hermitian, $\gamma_\mu^\dagger = \gamma_\mu$. The
on-shell momenta in Euclidean space fulfill $p^2 = p'^2 = -m^2$ and
the momentum transfer from the external photon is denoted by $k_\mu =
p_\mu' - p_\mu$. The anomalous magnetic moment is then given by the
Pauli form factor at vanishing momentum transfer $a_\mu = F_2(0)$.

From the expression for the HLbL diagram in Fig.~\ref{fig:muonhlbl} in
Minkowski space given in Ref.~\cite{KN_02}, we obtain the
corresponding result in Euclidean space by performing a Wick rotation
($\int_{q}\equiv \int \frac{d^4q}{(2\pi)^4}$)
\ba
 (ie)\<\mu^-(p')|j_\rho(0)|\mu^-(p)\> &=&  (-ie)^3  \,(ie)^4
\int_{q_1, q_2} \frac{1}{q_1^2 q_2^2 (q_1+q_2-k)^2} 
\nonumber \\
&& 
\qquad \times \; \frac{-1}{(p'-q_1)^2+m^2}\,  \frac{-1}{(p'-q_1-q_2)^2+m^2} 
\nonumber 
\phantom{\frac{1}{1}} \\ &&  
\qquad \times \; \bar u(p') \gamma_\mu (i\slashed{p}'-i\slashed{q}_1 -
m) \gamma_\nu (i\slashed{p}'-i\slashed{q}_1 -i\slashed{q}_2 - m)
\gamma_\lambda u(p) 
\nonumber \\
&& 
\qquad \times \; \Pi_{\mu\nu\lambda\rho}(q_1,q_2,k-q_1-q_2),
\label{eq:vertex1_momentum_space}
\ea
with the QCD four-point correlation function ($\int_x \equiv \int d^4x$) 
\be \label{eq:4_point_function}
\Pi_{\mu\nu\lambda\rho}(q_1,q_2,q_3) = \int_{x,y,z} \;
e^{-i(q_1 \cdot x + q_2 \cdot y + q_3 \cdot z)} 
\Big\< j_\mu(x) j_\nu(y) j_\lambda(z) j_\rho(0)\Big\>_{\rm QCD}. 
\ee
The issue of the Wick rotation in general requires some care.
Starting from the time-ordered correlation function with interpolating
operators for the muon initial and final states,
the standard recipe requires one to Wick rotate the loop momenta, here
$q_1$ and $q_2$ in Eq.~(\ref{eq:vertex1_momentum_space}), and the
external momenta, here $p$ and $p'$; see the derivation for a general Feynman
diagram in Ref.~\cite{Sterman_QFT}. The expression for the loop
integral is initially valid for  real Euclidean vectors $p,p'$
and one needs, in principle, to perform an analytic continuation of
the final result after all loop integrations have been performed from
$p'{}^2,p^2 > 0$ to $p^2 =p'{}^2 = -m^2$ to recover the result for an on-shell
momentum. In Eq.\ (\ref{eq:vertex1_momentum_space}), we have declared the Euclidean
norm of $p$ and $p'$ to be $-m^2$ from the outset, and return to the issue around
Eq.\ (\ref{eq:hat_epsilon}) below.

Since the electromagnetic current is conserved, the tensor
$\Pi_{\mu\nu\lambda\rho}(q_1,q_2,q_3)$ satisfies the Ward identities
(momentum conservation entails $q_1 + q_2 + q_3 + q_4 = 0$) 
\be 
\left\{q_{1\mu}; q_{2\nu}; q_{3\lambda}; (q_1 + q_2 + q_3)_\rho
\right\} \Pi_{\mu\nu\lambda\rho}(q_1,q_2,q_3) = 0.   
\ee
This implies the relation~\cite{Aldins:1970id}  
\be 
 \Pi_{\mu\nu\lambda\rho}(q_1,q_2,k-q_1-q_2) = 
- k_\sigma \frac{\partial}{\partial
  k_\rho}\Pi_{\mu\nu\lambda\sigma}(q_1,q_2,k-q_1-q_2),  
\ee 
which allows one to pull out the factor $k_\sigma$ from the vertex 
function in Eq.~(\ref{eq:vertex1_momentum_space}) to obtain the needed
form factor $F_2(0)$ with the projection operator~\cite{Aldins:1970id}
\be    \label{eq:KinoshitaTrace}
   \left. a_\mu^{{\rm HLbL}} =
   F_2(0)=\frac{-i}{48m}\trace\{[\gamma_\rho,\gamma_\sigma]
   (-i\slashed{p}+m)\Gamma_{\rho\sigma}(p,p) (-i\slashed{p}+m)\}
 \right|_{p^2 = -m^2} \,. 
\ee

The HLbL contribution to the vertex function reads for non-vanishing
momentum transfer 
\bea 
\Gamma_{\rho\sigma}(p',p) & = & 
   -e^6\int_{q_1,q_2} 
   \frac{1}{q_1^2q_2^2(q_1+q_2-k)^2}
   \frac{1}{(p'-q_1)^2+m^2}\frac{1}{(p'-q_1-q_2)^2+m^2}
   \nonumber \\
&& \qquad  
   \times \Big(\gamma_\mu(i\slashed{p}'-i\slashed{q}_1-m)
   \gamma_\nu(i\slashed{p}'-i\slashed{q}_1-i\slashed{q}_2-m)
   \gamma_\lambda\Big)
   \nonumber \\ 
&& \qquad 
   \times \frac{\partial}{\partial
     k_\rho}\Pi_{\mu\nu\lambda\sigma}(q_1,q_2,k-q_1-q_2).  
\label{eq:vertex_2_momentum_space}
\eea 

In order go over to a position-space representation for $a_\mu^{{\rm
    HLbL}}$, we insert into Eq.~(\ref{eq:vertex_2_momentum_space}) the
expression for the four-point function from
Eq.~(\ref{eq:4_point_function}) and interchange the integrals over
momenta and positions. One can then write the momenta $\slashed{q}_1$
and $\slashed{q}_2$ in the numerator as derivatives with respect to
$x$ and $y$ of the exponential function in
Eq.~(\ref{eq:4_point_function}) and also perform the derivative with
respect to $k_\rho$ to obtain a factor $z_\rho$.

In this way one gets the following expression for the vertex function
at vanishing momentum transfer that enters in the projector in
Eq.~(\ref{eq:KinoshitaTrace}) in terms of position-space functions
\be \label{eq:gammarhosigmakernel}
\Gamma_{\rho\sigma}(p,p)\;=\; -e^6\int_{x,y} 
K_{\mu\nu\lambda}(p,x,y) \, 
\widehat\Pi_{\rho;\mu\nu\lambda\sigma}(x,y),
\ee  
with the QED 
kernel\footnote{Now that the momentum transfer has been set to zero,
  we use the letter $k$ to denote an integration variable in
  Eq.~(\ref{eq:I_function}).}
\bea  
K_{\mu\nu\lambda}(p,x,y) &=&
\;\gamma_\mu(i\slashed{p}+\slashed{\partial}^{(x)}-m)
\gamma_\nu(i\slashed{p}+\slashed{\partial}^{(x)}+\slashed{\partial}^{(y)}-m)
\gamma_\lambda \, \mathcal{I}(p,x,y)_{\rm IR~reg.}\,,
\label{eq:K_function}
\\
\mathcal{I}(p,x,y)_{\rm IR~reg.}
&=&\;\int_{q,k}\frac{1}{q^2\,k^2\,(q+k)^2}\, 
\frac{1}{(p-q)^2+m^2}\,\frac{1}{(p-q-k)^2+m^2}\; 
e^{-i(q \cdot x + k \cdot y)}\,.  
\label{eq:I_function}
\eea 
The kernel $K_{\mu\nu\lambda}(p,x,y)$ in
Eqs.~(\ref{eq:gammarhosigmakernel}) and (\ref{eq:K_function}) is
understood to be a function, not a differential operator.

The function 
\be 
\widehat\Pi_{\rho;\mu\nu\lambda\sigma}(x,y)\;=\;\int_{z}
iz_\rho\;
\Big\langle{j_\mu(x)j_\nu(y)j_\sigma(z)j_\lambda(0)}\Big\rangle_{\rm
  QCD} 
\ee 
in Eq.~(\ref{eq:gammarhosigmakernel}) is a spatial moment of the
four-point function in QCD. Note the order of the indices
$\lambda\sigma$ on left-hand side and the order of the currents
$j_\sigma(z) j_\lambda(0)$ on the right-hand side. We have frequently
used the translation invariance of the four-point function to shift
the integration variables $x,y$ or to reverse their direction $z \to
-z$. The most important properties of $\widehat\Pi$
(Bose and reflection symmetries, transversality from current conservation)
are reviewed in subsection~\ref{sec:iPihatgenprop}.

Note that the function $\mathcal{I}(p,x,y)_{\rm IR~reg.}$ in
Eq.~(\ref{eq:I_function}) has a logarithmic infrared divergence for
on-shell muon momentum $p^2 = - m^2$ inside the loop integration,
i.e.\ for small $q,k$ with the three massless photon propagators and
the two on-shell massive muon propagators.  The IR divergence
disappears in the kernel $K_{\mu\nu\lambda}(p,x,y)$, after the
projection on $a_\mu^{\rm HLbL}$ in Eq.~(\ref{eq:KinoshitaTrace}), as
it should be, since the latter is well defined. After the projection,
only terms with derivatives with respect to $x$ and / or $y$ remain,
which bring down additional factors of $k$ and / or $q$ from the
exponential in Eq.~(\ref{eq:I_function}). However, since it is
convenient to first compute the scalar function ${\cal I}(p,x,y)_{\rm
  IR~reg.}$, we will regulate the infrared divergence, see details
below. After projecting on $a_\mu^{\rm HLbL}$ the regulator can be
removed.

We insert Eq.~(\ref{eq:gammarhosigmakernel}) into
Eq.~(\ref{eq:KinoshitaTrace}), evaluate the trace of the Dirac
matrices and obtain the expression 
\be \label{eq:amuHLbLprelim} 
a_\mu^{\rm HLbL} = \frac{m e^6}{3}\int_{x,y}  
{\cal L}_{[\rho,\sigma];\mu\nu\lambda}(p,x,y) \; 
i \widehat\Pi_{\rho;\mu\nu\lambda\sigma}(x,y), 
\ee
where the QED kernel is given by  
\ba 
&& {\cal L}_{[\rho,\sigma];\mu\nu\lambda}(p,x,y)
 \nonumber \\ 
&=& \frac{1}{16 m^2} {\rm Tr} \Big\{ (-i \slashed{p} + m)
[\gamma_\rho,\gamma_\sigma] (-i \slashed{p} + m) K_{\mu\nu\lambda}(p,x,y)  \Big\}   \\ 
&=& - \frac{i}{8m} {\rm Tr}\Big\{ \Big( \slashed{p}
[\gamma_\rho,\gamma_\sigma] + 
2 (p_\sigma\gamma_\rho - p_\rho\gamma_\sigma) \Big)
\gamma_\mu \gamma_\alpha \gamma_\nu \gamma_\beta \gamma_\lambda
\Big\}\, \partial^{(x)}_\alpha 
(\partial^{(x)}_\beta + \partial^{(y)}_\beta) \, {\cal I}
\nonumber \\ 
&& + \frac{1}{4m}  {\rm Tr}\Big\{
\Big( \slashed{p} [\gamma_\rho,\gamma_\sigma] +
2 (p_\sigma\gamma_\rho - p_\rho\gamma_\sigma)\Big)
\gamma_\mu \gamma_\alpha \gamma_\nu \Big\}\; p_\lambda
\;\partial^{(x)}_\alpha \,  {\cal I} 
\nonumber \\ 
&& + \frac{1}{4m}  {\rm Tr}\Big\{
\Big(\slashed{p} [\gamma_\rho,\gamma_\sigma] +
2(p_\sigma\gamma_\rho - p_\rho\gamma_\sigma)
\Big) 
 \gamma_\mu  (p_\lambda\gamma_\nu\gamma_\beta -
 p_\beta \gamma_\nu\gamma_\lambda 
+ p_\nu \gamma_\beta\gamma_\lambda) \Big\}\;
(\partial^{(x)}_\beta + \partial^{(y)}_\beta) \, {\cal I},  
\nonumber \\ 
&& \label{eq:traces}
\ea
and where we used $\slashed{p} \slashed{p} = p^2 = -m^2$ and
$(-i\slashed{p} + m) (i \slashed{p} + m) = 0$. The use of an IR
regulator in the function $\mcI(p,x,y)_{\rm IR~reg.}$ is always
understood. As noted above, only terms with derivatives acting on
$\mcI(p,x,y)_{\rm IR~reg.}$ survive after projecting on $a_\mu^{\rm
  HLbL}$.

It may be worth pointing out some discrete symmetries of the quantities introduced above.
First, we note that, for a general vector $p$, 
\be\la{eq:Ifullrefl}
{\cal I}(p,x,y)_{\rm IR~reg.} = {\cal I}(-p,-x,-y)_{\rm IR~reg.}.
\ee
Second, it is easy to show that
\ba\la{eq:Ireality}
{\cal I}(p,x,y)_{\rm IR~reg.} &=& ({\cal I}(-p^*,x,y)_{\rm IR~reg.})^*,
\\
  {\cal I}(p,x,y)_{\rm IR~reg.} &=& {\cal I}(p,x,x-y)_{\rm IR~reg.} ,
\la{eq:Iyswapxmy}
\ea
whence it follows that 
\be
K_{\lambda\nu\mu}(p,x,x-y) = K_{\mu\nu\lambda}(-p^*,x,y)^\dagger.
\ee
Finally, the latter equation entails the following property for the full kernel,
\be\la{eq:Lxswapxmy}
{\cal L}_{[\rho,\sigma];\lambda\nu\mu}(p,x,x-y) = -{\cal L}_{[\rho,\sigma];\mu\nu\lambda}(-p^*,x,y)^*. 
\ee

Our goal is to perform as many integrations of the 8-dimensional
integral over $x,y$ in Eq.~(\ref{eq:amuHLbLprelim}) as possible
(semi-) analytically to have full control over the QED kernel function
${\cal L}_{[\rho,\sigma];\mu\nu\lambda}(p,x,y)$. To achieve this, we
will rewrite the function $\mcI(p,x,y)_{\rm IR~reg.}$ in
Eq.~(\ref{eq:I_function}) in terms of position-space
propagators~\cite{Position_space_methods_old,
  Position_space_methods_new} and use the method of Gegenbauer
polynomials~\cite{Gegenbauer_momentum_space_early,
  Gegenbauer_momentum_space, Gegenbauer_momentum_space_dim_reg,
  Roskies_et_al_90, Gegenbauer_position_space} to perform the angular
integrals and average over the direction of the muon
momentum~\cite{Jegerlehner:2009ry} (see also Ref.~\cite{Roskies_et_al_90,Barbieri:1974nc}).  We will show the details
of this calculation in the next sections, but present here first the
structure of the final result, our master formula for $a_\mu^{\rm
  HLbL}$ in position space.

As mentioned earlier, we adopt an approach where the Euclidean vector $p$ obeys $p^2=-m^2$
from the outset, exploiting the fact that the muon is the ground state in the channel of its symmetry.
In the context of the use of Gegenbauer polynomials for loop integrals in
momentum space in Ref.~\cite{Gegenbauer_momentum_space} this procedure
only works in a straightforward way, if the integrand is a meromorphic
function of all the integration variables and all the external
invariants, like $p^2$. Fortunately, in our case one can show that the relevant $p$-dependent part of the
integrand  is a meromorphic function and the analytical
continuation to $p^2 = -m^2$ can be performed without
problems.
Aiming at keeping the absolute size of the imaginary components as small as possible,
we thus parametrize the on-shell momentum as follows,
\be \label{eq:hat_epsilon}
p = i m \hat\epsilon, \quad \hat\epsilon^2 = 1, \quad p^2 = -m^2, 
\ee
where the unit vector $\hat\epsilon$ parametrizes the direction of the muon
momentum.  
From Eq.\ (\ref{eq:Ireality}), 
for a vector $p$ with purely imaginary components, ${\cal I}(p,x,y)_{\rm IR~reg.}$ is real, 
and so is\footnote{Indeed, the trace of a product of 
linear combinations of Euclidean Dirac matrices with real coefficients is real.}
 ${\cal L}_{[\rho,\sigma];\mu\nu\lambda}(p,x,y)$.
Furthermore, the general property Eq.\ (\ref{eq:Lxswapxmy}) becomes
\be\la{eq:Lx,x-y}
{\cal L}_{[\rho,\sigma];\lambda\nu\mu}(p,x,x-y) = -{\cal L}_{[\rho,\sigma];\mu\nu\lambda}(p,x,y), 
\qquad p = im\hat\epsilon.
\ee

Since $a_\mu^{\rm HLbL}$ is a Lorentz scalar, the expression in
Eq.~(\ref{eq:amuHLbLprelim}) can be averaged over the direction
$\hat\epsilon$ of the muon momentum ($\int d\Omega_{\hat\epsilon} = 2
\pi^2$), 
\be
\bar {\cal L}_{[\rho,\sigma];\mu\nu\lambda}(x,y) = 
\frac{1}{2\pi^2} \int d\Omega_{\hat\epsilon} \, {\cal
  L}_{[\rho,\sigma];\mu\nu\lambda}(p,x,y) 
\equiv \Big\< {\cal
  L}_{[\rho,\sigma];\mu\nu\lambda}(p,x,y)\Big\>_{\hat \epsilon}.   
\ee

In this way we obtain 
\ba
\bar{\cal L}_{[\rho,\sigma];\mu\nu\lambda}(x,y) 
& = & {\cal G}^{\rm I}_{\delta[\rho,\sigma]\mu\alpha\nu\beta\lambda}\,
 \<\hat\epsilon_\delta \partial^{(x)}_\alpha
(\partial^{(x)}_\beta + \partial^{(y)}_\beta)   {\cal I}
\>_{\hat\epsilon} \nonumber \\ 
&& 
+ m \, {\cal G}^{\rm II}_{\delta[\rho,\sigma]\mu\alpha\nu\beta\lambda}
\< \hat\epsilon_\delta\hat\epsilon_\beta \;\partial^{(x)}_\alpha
{\cal I}\>_{\hat\epsilon} 
\nonumber \\ 
&& 
+ m \, {\cal G}^{\rm III}_{\delta[\rho,\sigma]\mu\alpha\nu\beta\lambda}\;  
\<\hat\epsilon_\alpha \hat\epsilon_\delta (\partial^{(x)}_\beta
+ \partial^{(y)}_\beta)   {\cal I} \>_{\hat\epsilon}, 
\la{eq:Lbar2} 
\ea
where we have defined 
\ba \la{eq:GI}
{\cal G}^{\rm I}_{\delta[\rho,\sigma]\mu\alpha\nu\beta\lambda} &\equiv &
{\frac{1}{8}} {\rm Tr}\Big\{\Big( \gamma_\delta [\gamma_\rho,\gamma_\sigma] +
2 (\delta_{\delta\sigma}\gamma_\rho - \delta_{\delta\rho}\gamma_\sigma)\Big)
\gamma_\mu \gamma_\alpha \gamma_\nu \gamma_\beta \gamma_\lambda \Big\},
\\ \la{eq:GII}
{\cal G}^{\rm II}_{\delta[\rho,\sigma]\mu\alpha\nu\beta\lambda} &\equiv &
 -\frac{1}{4} {\rm Tr}\Big\{
\Big( \gamma_\delta [\gamma_\rho,\gamma_\sigma] +
2 (\delta_{\delta\sigma}\gamma_\rho - \delta_{\delta\rho}\gamma_\sigma)\Big)
\gamma_\mu \gamma_\alpha \gamma_\nu \Big\}\; \delta_{\beta\lambda}, 
\\ \la{eq:GIII}
{\cal G}^{\rm III}_{\delta[\rho,\sigma]\mu\alpha\nu\beta\lambda} &\equiv&
 -\frac{1}{4} {\rm Tr}\Big\{
\Big(\gamma_\delta [\gamma_\rho,\gamma_\sigma] +
2(\delta_{\delta\sigma}\gamma_\rho - \delta_{\delta\rho}\gamma_\sigma)
\Big)  \gamma_\mu  (\delta_{\alpha\lambda}\gamma_\nu\gamma_\beta -
 \delta_{\alpha\beta} \gamma_\nu\gamma_\lambda 
+ \delta_{\alpha\nu} \gamma_\beta\gamma_\lambda) \Big\}.
\qquad 
\ea
The tensors ${\cal G}^{\rm A}_{\delta[\rho,\sigma]\mu\alpha\nu\beta\lambda}$ are sums of
products of Kronecker deltas from the traces of the Dirac matrices in
Euclidean space.
The QED kernel $\bar {\cal L}_{[\rho,\sigma];\mu\nu\lambda}(x,y)$ inherits
from the kernel ${\cal L}_{[\rho,\sigma];\mu\nu\lambda}(p,x,y)$ the antisymmetry property
\be\la{eq:barLx,x-y}
\bar {\cal L}_{[\rho,\sigma];\lambda\nu\mu}(x,x-y) = -\bar {\cal L}_{[\rho,\sigma];\mu\nu\lambda}(x,y)
\ee
under the transformation ($\mu\leftrightarrow\lambda$, $y\to x-y$)
upon averaging both sides of Eq.~(\ref{eq:Lx,x-y}) over the direction of the muon momentum.

\begin{figure}
\centering 
\includegraphics[width=0.5\textwidth]{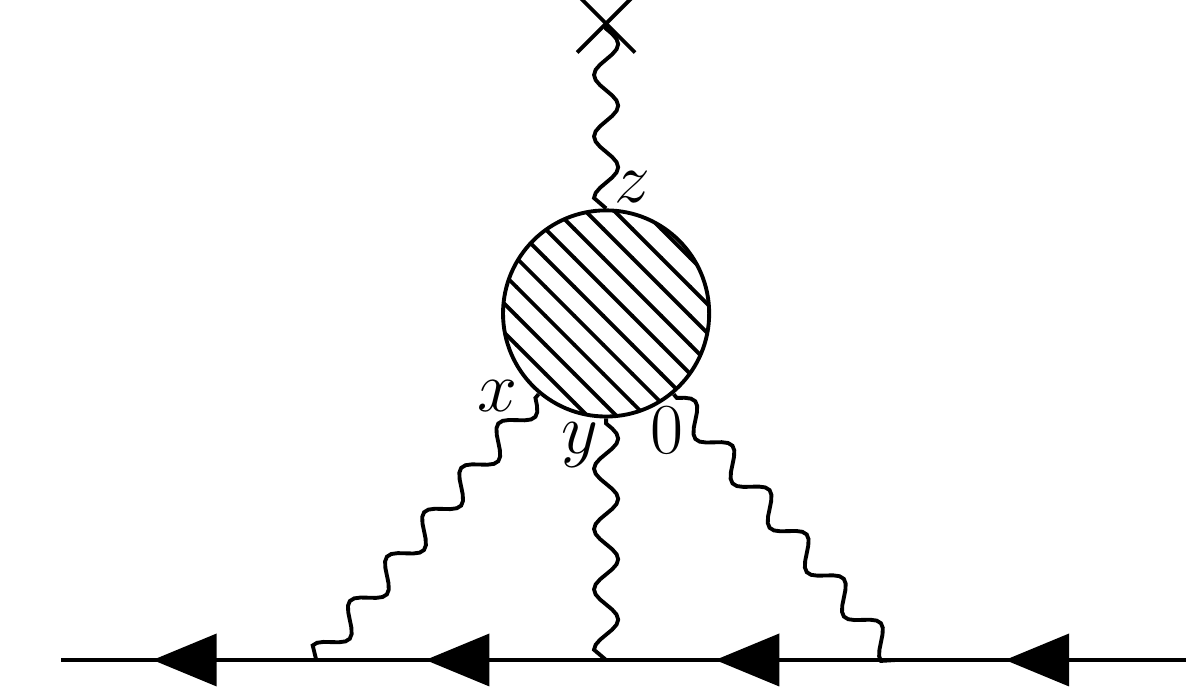}
\caption{Hadronic light-by-light scattering diagram in the muon $(g-2)$
  with the locations of the vertices $x,y,z,0$ in the master formula
  in Eq.~(\ref{eq:master_formula}).} 
\label{fig:muonhlbl_points}
\end{figure}

We thus arrive at our master formula for $a_\mu^{\rm HLbL}$ in
position space (which we have already presented previously in
Refs.~\cite{Talk_Asmussen_DPG_2015, Green:2015mva, Asmussen:2016lse})
\ba
a_\mu^{\rm HLbL} &=& \frac{m e^6}{3} \int_{x,y}
\bar {\cal L}_{[\rho,\sigma];\mu\nu\lambda}(x,y)\;
i\widehat\Pi_{\rho;\mu\nu\lambda\sigma}(x,y),  
\label{eq:master_formula} \\ 
\bar {\cal L}_{[\rho,\sigma];\mu\nu\lambda}(x,y) 
&=& \sum_{{\rm A}={\rm I,II,III}} 
{\cal G}^{\rm A}_{\delta[\rho,\sigma]\mu\alpha\nu\beta\lambda} 
T^{\rm A}_{\alpha\beta\delta}(x,y), \label{eq:Lbar}
\\
i\widehat \Pi_{\rho;\mu\nu\lambda\sigma}( x, y)  &=& 
-\int_{z} z_\rho\, \Big\<\,j_\mu(x)\,j_\nu(y)\,j_\sigma(z)\,
j_\lambda(0)\Big\>_{\rm QCD}. 
\ea
After contracting the Lorentz indices in
Eq.~(\ref{eq:master_formula}), the integration reduces to a
3-dimensional integral over $|x|, |y|$ and $x \cdot y$.
For illustration we depict in
Fig.~\ref{fig:muonhlbl_points} the HLbL diagram in the muon $(g-2)$
indicating the positions $x,y,z,0$ of the four vector currents
(attached to the photons) in the master formula in
Eq.~(\ref{eq:master_formula}).
We emphasize at this point that the kernel $\bar{\cal L}$ is far from being
unique.  A significant amount of freedom remains to adjust the kernel
to the needs of practical calculations, wihtout modifying the final
value of $a_\mu^{\rm HLbL}$. We return to this aspect in subsection
\ref{sec:KernelSubtractions}.

The tensors $T^{\rm A}_{\alpha\beta\delta}(x,y)$ in
Eq.~(\ref{eq:Lbar}) can be decomposed into a scalar $S$, a
vector $V_\delta$ and a tensor part $T_{\beta\delta}$:  
\ba
T^{\rm I}_{\alpha\beta\delta}(x,y) &=&   \partial^{(x)}_\alpha
(\partial^{(x)}_\beta + \partial^{(y)}_\beta) V_\delta(x,y),
\label{eq:T_I}
\\
T^{\rm II}_{\alpha\beta\delta}(x,y) &=& 
m \partial^{(x)}_\alpha 
\Big( T_{\beta\delta}(x,y) + \frac{1}{4}\delta_{\beta\delta}
S(x,y)\Big), \label{eq:T_II}
\\
T^{\rm III}_{\alpha\beta\delta}(x,y) &=&  m (\partial^{(x)}_\beta
+ \partial^{(y)}_\beta) 
\Big( T_{\alpha\delta}(x,y) + \frac{1}{4}\delta_{\alpha\delta}
S(x,y)\Big).
\label{eq:T_III}
\ea

These parts are given in terms of the function
$\mcI(p,x,y)_{\rm IR~reg.}$ from Eq.~(\ref{eq:I_function}) as follows
(see Eq.~(\ref{eq:Lbar2})):   
\ba
S(x,y) &=& \Big\<
\mcI(p,x,y)_{\rm{IR~reg.}}\Big\>_{\hat\epsilon},  
\label{eq:scalar_average}  \\
V_\delta(x,y) &=& \Big\<\hat\epsilon_\delta \; \mcI(p,x,y)_{\rm{IR~reg.}}
\Big\>_{\hat\epsilon}, \label{eq:vector_average}
\\
T_{\beta\delta}(x,y) &=& 
\Big\<
\Big(\hat\epsilon_\beta\hat\epsilon_\delta-\frac{1}{4}\delta_{\beta\delta}\Big)
\; \mcI(p,x,y)_{\rm{IR~reg.}} \Big\>_{\hat\epsilon}. \label{eq:tensor_average}
\ea
From the property (\ref{eq:Ifullrefl}) of the scalar function ${\cal I}(p,x,y)_{\rm IR~reg.}$,
it follows that  $S(x,y)$ and $T_{\beta\delta}(x,y)$ are even under the simultaneous sign
reflection of both their arguments, while $V_\delta(x,y)$ is odd. From here, it follows that
the $T_{\alpha\beta\delta}^{\rm A}(x,y)$ are all odd under $(x,y)\to (-x,-y)$, so that the QED kernel
is too,
\be\la{eq:LbarIsOdd}
\bar {\cal L}_{[\rho,\sigma];\mu\nu\lambda}(x,y) = - \bar {\cal L}_{[\rho,\sigma];\mu\nu\lambda}(-x,-y) .
\ee
Since the function ${\cal I}(p,x,y)_{\rm IR~reg.}$ is ultraviolet finite by power-counting (including at $x=y=0$),
we conclude that so is the QED kernel, and Eq.\ (\ref{eq:LbarIsOdd}) then implies the property
\be\la{eq:LbarEq0at0}
\bar {\cal L}_{[\rho,\sigma];\mu\nu\lambda}(0,0) = 0.
\ee

The quantity $S(x,y)$ can only be a scalar function $\bar{\mathfrak{g}}^{(0)}$
of the three invariants $|x|,x \cdot y, |y|$; we call this a weight function. The vector and
tensor functions will be parametrized by respectively two and three weight functions: 
\ba\label{eq:define_g_0}
S(x,y) &=& \bar{\mathfrak{g}}^{(0)}(|x|, x \cdot y, |y|),
\phantom{\frac{1}{1}} \\
V_\delta(x,y)
&=& x_\delta  \bar{\mathfrak{g}}^{(1)}(|x|,x \cdot y,|y|)
+ y_\delta  \bar{\mathfrak{g}}^{(2)}(|x|,x \cdot y,|y|),
\label{eq:define_g_1_2}
\\
 T_{\alpha\beta}(x,y) 
 &=& (x_\alpha x_\beta - \frac{x^2}{4}\delta_{\alpha\beta})\;
 \bar{\mathfrak{l}}^{(1)}(|x|, x \cdot y, |y|)  
+ (y_\alpha y_\beta - \frac{y^2}{4}\delta_{\alpha\beta})\;
\bar{\mathfrak{l}}^{(2)}(|x|, x \cdot y, |y|) \nonumber \\ 
& &  
+ (x_\alpha y_\beta + y_\alpha x_\beta  - \frac{x\cdot
  y}{2}\delta_{\alpha\beta})\; \bar{\mathfrak{l}}^{(3)}(|x|, x \cdot
y, |y|).    
\label{eq:define_l_1_2_3}
\ea
In total, the QED kernel $\bar {\cal  L}_{[\rho,\sigma];\mu\nu\lambda}(x,y)$ in Eq.~(\ref{eq:Lbar}) is
thus parametrized by six weight functions and their derivatives.
See Appendix~\ref{sec:chainrT} for the explicit expressions of the tensors $T^{\rm A}_{\alpha\beta\delta}(x,y)$. 
As we will see in the explicit calculation later, the IR divergence of
$\mcI(p,x,y)_{\rm IR~reg.}$ will only be important in the scalar
weight function $\bar{\mathfrak{g}}^{(0)}(|x|, x \cdot y, |y|)$,
before performing the derivatives in Eqs.~(\ref{eq:T_II}) and
(\ref{eq:T_III}).

It is clear that the tensors $S(x,y)$, $V_\delta(x,y)$ and $T_{\beta\delta}(x,y)$
inherit from ${\cal I}(p,x,y)_{\rm IR~reg.}$ the invariance under $y\to (x-y)$.
In turn, their invariance implies the following symmetry properties for the weight functions,
\ba\la{eq:g0symmetry}
\bar{\mathfrak{g}}^{(0)}_* &=& \bar{\mathfrak{g}}^{(0)},
\\
\bar{\mathfrak{g}}^{(1)}_* &=& \bar {\mathfrak{g}}^{(1)} +\bar {\mathfrak{g}}^{(2)},
\\
\bar {\mathfrak{g}}^{(2)}_* &=&  - \bar {\mathfrak{g}}^{(2)},
\\
\bar{\mathfrak{l}}^{(1)}_* &=& \bar{\mathfrak{l}}^{(1)} + \bar{\mathfrak{l}}^{(2)} +2 \bar{\mathfrak{l}}^{(3)} ,
\\
\bar{\mathfrak{l}}^{(2)}_* &=& \bar{\mathfrak{l}}^{(2)},
\\
\bar{\mathfrak{l}}^{(3)}_* &=&  - \bar{\mathfrak{l}}^{(2)}  - \bar{\mathfrak{l}}^{(3)}.
\la{eq:l3symmetry}
\ea
Unstarred functions have as argument $(|x|,x\cdot y,|y|)$, while starred functions refer to the same weight functions but
with argument $(|x|,x\cdot(x-y),|x-y|)$.
Furthermore, the rank-three tensors contributing to the QED kernel
satisfy\footnote{In fact, the contributions $S(x,y)$ and $T_{\beta\delta}(x,y)$
to the rank-three tensors separately satisfy Eq.\ (\ref{eq:TIIandTIII}).}
\ba\la{eq:TIinv}
T^{\rm I}_{\beta\alpha\delta}(x,x-y) &=& T^{\rm I}_{\alpha\beta\delta}(x,y),
\\
T^{\rm III}_{\beta\alpha\delta}(x,x-y) &=& T^{\rm II}_{\alpha\beta\delta}(x,y).
\la{eq:TIIandTIII}
\ea
We already know that the QED kernel as a whole obeys the antisymmetry property Eq.\ (\ref{eq:barLx,x-y})
under the transformation ($\mu\leftrightarrow\lambda$, $y\to x-y$).
Eq.\ (\ref{eq:TIinv}), combined with the fact that 
 ${\cal G}^{\rm I}$ is antisymmetric under the simultaneous index exchanges
 $\mu\leftrightarrow \lambda$ and $\alpha\leftrightarrow\beta$, implies that 
the contribution to the QED kernel  ${\cal G}^{\rm I}T^{\rm I}$ 
by itself is antisymmetric under the transformation ($\mu\leftrightarrow\lambda$, $y\to x-y$).

\section{Preparatory steps for the calculation of the QED weight functions
  in position space\la{sec:starting}}

In this section, we present some (partly known) results on propagators
and their expansion in Gegenbauer polynomials. These preliminaries
will allow us to provide the expansion in Gegenbauer polynomials of
the function $J(\hat \epsilon, u)$, defined in Eq.\ (\ref{eq:J_function}),
which plays a crucial role in the entire calculation.
Then, starting in subsection \ref{sec:muonmom_average},
we perform the average over the direction of the muon momentum analytically.
The final convolution integrals yielding the tensors
$S$, $V_\delta$ and $T_{\beta\delta}$ are treated in sections
\ref{sec:direct} and \ref{sec:alternative_evaluation}.

\subsection{Starting point for the calculation of
  $\mcI(p,x,y)_{\rm IR~reg.}$}

\begin{figure}
\centering 
\includegraphics[width=0.75\textwidth]{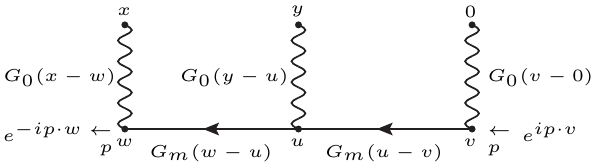}
\caption{Feynman diagram of the QED part of the HLbL contribution in
  the muon $g-2$ in position space, see
  Fig.~\ref{fig:muonhlbl_points}, corresponding to the scalar function
  $\mcI(p,x,y)_{\rm IR~reg.}$ in Eq.~(\ref{eq:I_function}). Note that
  the arrows on the muon line are only meant for illustration; all
  propagators are scalar functions.
In the derivation, the Euclidean momentum $p$ flowing through the diagram was assumed to be real.}
\label{fig:diagram_I_position_space}
\end{figure}

To obtain a convenient expression for the scalar function
$\mcI(p,x,y)_{\rm IR~reg.}$ in Eq.~(\ref{eq:I_function}), we
translate the momentum integrals into position-space perturbation
theory integrals, where the integration variables correspond to the
positions of the vertices, see
Fig.~\ref{fig:diagram_I_position_space}.

The relevant Fourier transforms can be performed by using the
well-known expressions for the massless and massive propagators in
position space~\cite{Bogoliubov_Shirkov}:\footnote{The positions $x,y$
  in the definitions of the propagators are generic Euclidean
  four-vectors and do not correspond to the vertices in the Feynman
  diagrams in Figs.~\ref{fig:muonhlbl_points} and
  \ref{fig:diagram_I_position_space}.} 
\bea \la{eq:massless_scalar_prop}
G_0(x-y) & = & \int_k \frac{e^{i k \cdot (x-y)}}{k^2} = \frac{1}{4\pi^2
  (x-y)^2}\,, 
\\ 
G_m(x-y) & = & \int_k \frac{e^{i k \cdot (x-y)}}{k^2+m^2} = \frac{m}{4\pi^2
  |x-y|} K_1(m|x-y|)\,,
\la{eq:massive_scalar_prop}
\eea 
where $K_1$ is a modified Bessel function and we use the conventions
from Ref.~\cite{Abramowitz_Stegun}.  The propagators in position space
are Green's functions of the `Euclidean Klein-Gordon' equation, 
\be \label{eq:Klein_Gordon}
(-\triangle^{(x)} + m^2) G_m(x-y) = \delta^{(4)}(x-y),
\ee 
and analogously for $m=0$. Here $\triangle^{(x)} = \sum_{\mu=0}^3 \partial_\mu^{(x)} \partial_\mu^{(x)} $
is the four-dimensional Laplacian.

These position-space representations of propagators have been used for
a long time to evaluate Feynman diagrams, see for instance
Refs.~\cite{Position_space_methods_old, Gegenbauer_position_space,
  Position_space_methods_new}. However, these calculations were mostly
dealing with loop integrals, not their Fourier transform as in
Eq.~(\ref{eq:I_function}); they involved massless particles (relevant for QCD)
and were either aimed at evaluating renormalization constants at higher loop order, or
treated Feynman diagrams with a special topology, e.g.\ of the
sun-rise type, where $L+1$ propagators connect the two points $x$ and
$y$ in a $L$-loop diagram. Thus our goal required the development of
additional computational methods.

We then obtain as the starting point for the evaluation of the
weight-functions of the QED kernel in position space the following
representation of the scalar function $\mcI(p, x,y)_{\rm IR~reg.}$ for
$p = i m \hat\epsilon$:
\bea 
\mcI(p,x,y)_{\rm IR~reg.} & = & \int_u \, 
G_0(y-u) \, J(\hat\epsilon,u) \, J(\hat\epsilon,x-u)
\,, \label{eq:I_with_J_J} \\ 
J(\hat\epsilon,u) & = & \int_{\tilde u} \, G_0(u-\tilde u) e^{m \hat\epsilon \cdot \tilde u}
\, G_m(\tilde u) \,. \label{eq:J_function} 
\eea 
Recall again the need to regulate the IR divergence of the function
$\mcI(p,x,y)_{\rm IR~reg.}$, which is related in position space to the
behavior of the integrand in the final integration in
Eq.~(\ref{eq:I_with_J_J}) for large $|u|$, i.e.\ at long
distances. Below we will show that $J(\hat\epsilon,u) \sim 1/|u|$ for
large $|u|$ and therefore the integral for $\mcI(p,x,y)_{\rm IR~reg.}$
is logarithmically divergent for large $|u|$, since $\int_u \equiv
\int_0^\infty d|u| |u|^3 d\Omega_{\hat u}$ (with unit-vector $\hat u =
u/|u|$) and $G_0(y-u) \sim 1/u^2$.

In the derivation of Eq.~(\ref{eq:I_with_J_J}) we encounter the
integral  
\be \label{eq:H_x} 
\int_k \frac{e^{i k \cdot x}}{(k-p)^2 + m^2} =
e^{i x \cdot p} \, G_m(x) = e^{-m \hat\epsilon \cdot x} \, G_m(x) \,. 
\ee
The first equality follows, formally, by shifting the integration
variable $k \to k-p$. The derivation is formal because with our
parametrization of the on-shell momentum $p = i m \hat\epsilon$ in the
integrand, $p$ is a complex Euclidean vector, while the shift in the
integration variable assumes that $p$ is a real Euclidean vector. The
final result, also used in Eq.~(\ref{eq:J_function}), follows by
analytical continuation of the expression from a real off-shell vector
$p$ to a complex on-shell vector $p$. The second equality with the
on-shell momentum $p$ can be derived directly by using the Schwinger
representation of the propagator $((k-p)^2 + m^2)^{-1} = \int_0^\infty
d\alpha \, \exp(-\alpha ((k-p)^2 + m^2))$ and then, after the
dependence on the components of $k_\mu$ has been factorized,
performing four simple Gaussian integrals.

Note that although the Feynman diagram from
Fig.~\ref{fig:diagram_I_position_space} is only a tree-level diagram
and thus trivial in momentum space, the expression in position space
in Eq.~(\ref{eq:I_with_J_J}) is a non-trivial convolution of
propagators and exponentials.

\subsection{Gegenbauer method for angular integrals in position space
  in four dimensions} 

We summarize first some basic properties of Gegenbauer polynomials
which have been used since a long time ago to perform the angular
integrations of Feynman loop integrals (hyperspherical approach) in
momentum space~\cite{Gegenbauer_momentum_space} and in position
space~\cite{Gegenbauer_position_space}.  Since we work in $d=4$ dimensions,
we only need the special case of the Gegenbauer polynomials $C_n(\xi)
\equiv C_n^{(\alpha=1)}(\xi) \equiv U_n(\xi)$, which are actually
equal to the Chebyshev polynomials of the second kind, see
Ref.~\cite{Abramowitz_Stegun}.\footnote{Note that in some references
$C_n(\xi)$ denotes a dilated Chebyshev polynomial of the first kind.}
The corresponding generating function is given by 
\be \label{eq:C_n_generating_function} 
\frac{1}{\tau^2 - 2 \xi \tau + 1} = \sum_{n=0}^{\infty} \tau^n \,
C_n(\xi), \quad -1 \leq \xi \leq 1, \ \ |\tau| < 1. 
\ee 
Some low-order Gegenbauer polynomials are given by $C_0(\xi) = 1$,
$C_1(\xi) = 2\xi$, $C_2(\xi) = 4\xi^2 -1$ and therefore $\xi =
C_1(\xi)/2$, $\xi^2 = \left( C_2(\xi) + C_0(\xi) \right)/4$.

Some simple properties of the Gegenbauer polynomials (Chebyshev
polynomials of the second kind) are~\cite{Abramowitz_Stegun}
\ba
C_n(\cos\theta) & = & \frac{\sin((n+1)\theta)}{\sin(\theta)},
\la{eq:C_n_cos_theta} \\ 
C_n(-\xi) & = & (-1)^n C_n(\xi), \la{eq:C_n_parity} \\
C_n(1) & = & n+1, \la{eq:C_n_normalization} \\ 
\xi\,C_n(\xi) & = & \frac{1}{2} \Big[ C_{n+1}(\xi) + (1- \delta_{n0}) 
C_{n-1}(\xi)\Big], \qquad n \geq 0, \la{eq:xiCn} \\ 
\xi^2 C_n(\xi) & = & \frac{1}{4} \Big[ C_{n+2}(\xi) + (2-\delta_{n0})
C_n(\xi) + C_{n-2}(\xi) \Big],\qquad n\geq 0,  \la{eq:xisqCn}
\ea
where we adopt in Eqs.~(\ref{eq:xiCn}) and (\ref{eq:xisqCn}) the
convention that $C_{n}(\xi) \equiv 0$ for $n \leq -1$. The property
(\ref{eq:C_n_parity}) under parity transformations and the
normalization (\ref{eq:C_n_normalization}) follow easily from the
generating function in Eq.~(\ref{eq:C_n_generating_function}).

For the evaluation of angular integrals in Feynman diagrams one makes
use of the fact that the Gegenbauer polynomials (hyperspherical
polynomials) are the polynomials that obey the orthogonality relations 
\ba 
\Big\<C_n(\hat\epsilon\cdot \hat x)\; C_m(\hat\epsilon\cdot \hat
y)\Big\>_{\hat\epsilon}  
&=& \frac{\delta_{nm}}{n+1} \,C_n(\hat x\cdot\hat y),
\nonumber \\ 
\Big\<C_n(\hat\epsilon\cdot \hat x)\; C_m(\hat\epsilon\cdot \hat
x)\Big\>_{\hat\epsilon} 
&=& {\delta_{nm}}, 
\label{eq:orthogonality_relations}
\ea
on the unit sphere,
where we denote unit four-vectors as $\hat x = x/|x|, \hat y =
y/|y|$.

\subsection{Expansion in Gegenbauer polynomials of propagators in
  position space, of 
  the exponential function and of the function $J(\hat\epsilon,y)$}

For our goal to perform the angular integrations in the function
$\mcI(p,x,y)_{\rm IR~reg.}$ in Eq.~(\ref{eq:I_with_J_J}) and the
averages in Eqs~(\ref{eq:scalar_average}), (\ref{eq:vector_average})
and (\ref{eq:tensor_average}), it is important to note that if we have
a function $f=f(\hat\epsilon \cdot x,x^2)$, its dependence on the
angle between the vectors $\hat\epsilon$ and $x$ can be expanded in
Gegenbauer polynomials as follows 
\ba
f(\hat\epsilon\cdot x,x^2) & = & \sum_{n=0}^\infty \zeta_n(x^2) \,
C_n(\hat\epsilon\cdot \hat x), \label{eq:f_expansion} \\
\zeta_n(x^2) & = & \Big\< f(\hat\epsilon\cdot x,x^2) \,
C_n(\hat\epsilon\cdot \hat x) \Big\>_{\hat \epsilon}, \label{eq:zeta_n}
\ea
where one uses the orthogonality relations in
Eq.~(\ref{eq:orthogonality_relations}) to derive the expression for
$\zeta_n(x^2)$.

For later reference we write down the expansions into Gegenbauer
polynomials for the massless and massive propagators in position
space, the exponential function and the function $J(p,y)$ from
Eq.~(\ref{eq:J_function}) (again, the vectors $x,y$ in the propagators
and in the exponential function are meant to be generic Euclidean
four-vectors)
\bea
4\pi^2 G_0(x-y) & \equiv & \frac{1}{(x-y)^2} = \sum_{n=0}^{\infty}
d_n(x^2,y^2) \, C_n(\hat x\cdot\hat y), \label{eq:G0_expansion} \\
d_n(x^2,y^2) & = & \theta(x^2-y^2) \frac{|y|^n}{|x|^{n+2}} +
\theta(y^2-x^2) \frac{|x|^n}{|y|^{n+2}}, \label{eq:G0_d_n} \\
G_m(x-y) & = & \sum_{n=0}^{\infty} \gamma_n(x^2,y^2)\, C_n(\hat x
\cdot \hat y), \label{eq:Gm_expansion} \\ 
\gamma_n(x^2,y^2) & = & \frac{n + 1}{2\pi^2|x||y|} \Big(
\theta(x^2-y^2) \, I_{n+1}(m|y|) \, K_{n+1}(m|x|) \nonumber \\ 
& & \qquad \quad \ \ + \theta(y^2-x^2) \, I_{n+1}(m|x|) \,
K_{n+1}(m|y|)\Big), \label{eq:Gm_gamma_n} \\  
e^{\hat\epsilon \cdot x} & \equiv & e^{x \hat\epsilon \cdot \hat x} =
2 \sum_{n=0}^{\infty} (n+1) \frac{I_{n+1}(|x|)}{|x|} \,
C_n(\hat\epsilon\cdot \hat x), \qquad x\in\mathbb{R}^4,
\la{eq:exp_expansion} \\ 
J(\hat \epsilon, u) &=& \sum_{n=0}^{\infty} z_n(u^2) \, 
C_n(\hat\epsilon \cdot \hat u), \label{eq:J_expansion} \\ 
z_n(u^2) &=& \frac{1}{4 \pi^2}  \Big[I_{n+2}(m |u|) \frac{K_0(m |u|)}{n+1}  
 + I_{n+1}(m |u|) \Big( \frac{K_1(m |u|)}{n+1}+\frac{ K_0(m |u|)}{m
   |u|}  \Big) \Big], \label{eq:z_n}
\eea
where $I_n$ is another modified Bessel function, see
Ref.~\cite{Abramowitz_Stegun}. For integrals involving modified Bessel functions,
we have also found the further references~\cite{DLMF Website,Gradshteyn_Ryzhik,Watson,Prudnikov} useful.

The expansion of the massless propagator $G_0(x-y)$ in
position space in Eqs.~(\ref{eq:G0_expansion}) and (\ref{eq:G0_d_n})
is formally equal to the corresponding expansion in momentum space and
follows immediately from the use of the generating function of the
Gegenbauer polynomials in Eq.~(\ref{eq:C_n_generating_function}) to
write down the expansion of the massive propagator in Euclidean
momentum space and then performing the limit $m \to 0$~\cite{KN_02}.

The expansion into Gegenbauer polynomials for the massive propagator
$G_m(x-y)$ can be derived as follows. It satisfies the differential
equation from Eq.~(\ref{eq:Klein_Gordon}).  Since $G_m(x-y)$ is a
scalar function, we can choose the coordinate system such that $y$
lies along the positive $\hat e_0$ axis, so that $x$ has coordinates
$(|x|\cos\phi_1, |x|\sin\phi_1, 0, 0)$ and $\cos\phi_1$ coincides with $\hat x\cdot\hat y$. 
The four-dimensional Laplacian operator for a function of $|x|$ and
$\cos\phi_1$ reads 
\be \label{eq:dAlembert}
\triangle^{(x)} = \frac{\partial^2}{\partial |x|^2} + \frac{3}{|x|}
\frac{\partial}{\partial |x|} + \frac{1}{|x|^2\sin^2\phi_1}
\frac{\partial}{\partial \phi_1} \Big(\sin^2\phi_1\;
\frac{\partial}{\partial \phi_1}\Big). 
\ee
The important fact is that the Gegenbauer polynomials (Chebyshev
polynomials of the second kind) are eigenfunctions of the angular part
of the Laplacian  operator
\be
\frac{1}{\sin^2\phi_1}  \frac{\partial}{\partial \phi_1} 
\Big(\sin^2\phi_1\; \frac{\partial}{\partial \phi_1} C_n(\cos\phi_1)\Big)
= -n(n+2) C_n(\cos\phi_1).
\ee
Thus inserting the expansion of $G_m(x-y)$ in
Eq.~(\ref{eq:Gm_expansion}) into the Klein-Gordon equation in
Eq.~(\ref{eq:Klein_Gordon}) and applying the differential operator
term by term yields the condition
\be\la{eq:cndg}
\sum_{n = 0}^{\infty} \Big\{-\frac{\partial^2}{\partial |x|^2} -
\frac{3}{|x|} \frac{\partial}{\partial |x|} + m^2 +
\frac{n(n+2)}{|x|^2} \Big\} \gamma_n(x^2,y^2) \, C_n(\hat x\cdot\hat y) =
\delta^{(4)}(x-y).  
\ee
Using the general expression
\be
\delta^{(4)}(x-y) = \frac{1}{|x|^3} \delta(|x|-|y|)
\frac{1}{\sin^2\phi_1}\delta(\phi_1-\phi_1^y) 
 \frac{1}{\sin\phi_2} \delta(\phi_2-\phi_2^y) \delta(\phi_3-\phi_3^y),
\ee
we integrate both sides of Eq.\ (\ref{eq:cndg}) over $\int_0^\pi
d\phi_2\sin\phi_2 \int_0^{2\pi} d\phi_3$ and use the completeness
relation of the Gegenbauer polynomials, 
\be
\frac{2}{\pi} \sum_{n = 0}^{\infty} C_n(\cos\phi_1) C_n(\cos\phi_1^y) =
\frac{1}{\sin^2\phi_1}\delta(\phi_1-\phi_1^y) 
\ee
to obtain 
\ba
&& 4\pi \sum_{n = 0}^{\infty} \Big\{-\frac{\partial^2}{\partial |x|^2} -
\frac{3}{|x|} \frac{\partial}{\partial |x|} 
+m^2 + \frac{n(n+2)}{|x|^2} \Big\} \gamma_n(x^2,y^2) \, C_n(\hat x \cdot
\hat y) \nonumber \\ 
&& = \frac{1}{|x|^3} \delta(|x|-|y|) \frac{2}{\pi} \sum_{n =
  0}^{\infty} C_n(\cos\phi_1^y) \, C_n(\cos\phi_1). 
\ea
Choosing again $y$ to lie along the $\hat e_0$ direction, we have $\phi_1^y=0$,
$\hat x \cdot \hat y= \cos\phi_1$ and $C_n(\cos\phi_1^y)=(n+1)$;
 comparing the series term by term we obtain the differential equation
\be
\Big\{-\frac{\partial^2}{\partial |x|^2} - \frac{3}{|x|}
\frac{\partial}{\partial |x|} 
+m^2 + \frac{n(n+2)}{|x|^2} \Big\} \gamma_n(x^2,y^2) =
\frac{n+1}{2\pi^2 |x|^3} \delta(|x|-|y|). 
\ee
Two solutions of the homogeneous equation are $K_{n+1}(m|x|)/|x|$ and
$I_{n+1}(m|x|)/|x|$. This then leads to the expression for $\gamma_n(x^2,y^2)$ given
in Eq.~(\ref{eq:Gm_gamma_n}).

We are not aware of any paper, where the expansion into Gegenbauer
polynomials of the massive propagator in position space in
Eqs.~(\ref{eq:Gm_expansion}) and (\ref{eq:Gm_gamma_n}) has been
given, even though it is a special case of Gegenbauer's Addition Theorem
(see p.\ 365 of Ref.~\cite{Watson}).
Note that at $|x|=|y|$, the function $\gamma_n(x^2,y^2)$ is
continuous, but not differentiable - there is a cusp. For $m \to 0$ we
recover from the expansion of the massive propagator the expansion for
the massless propagator.

The expansion into Gegenbauer polynomials for the exponential function
in Eq.~(\ref{eq:exp_expansion}) follows from the generating function
of the Bessel functions $I_n$ and the associated series given in
Ref.~\cite{Abramowitz_Stegun}
\be \label{eq:exp_series_I_n} 
e^{z \cos\theta} = I_0(z) + 2 \sum_{n=1}^{\infty} I_n(z) \cos(n
\theta), \qquad z \in \mathbb{C},  
\ee
by taking the derivative with respect to $\theta$ of both sides of
Eq.~(\ref{eq:exp_series_I_n}) and then using
Eq.~(\ref{eq:C_n_cos_theta}), $z = |x|$ and $\cos\theta =
\hat \epsilon \cdot \hat x$ for $x\in\mathbb{R}^4$. Similar
expressions have already been given in
Refs.~\cite{Position_space_methods_old,
  Gegenbauer_momentum_space_dim_reg, Gegenbauer_position_space}.

For the derivation of the expansion into Gegenbauer polynomials of
$J(\hat\epsilon,u)$ in Eqs.~(\ref{eq:J_expansion}) and (\ref{eq:z_n})
we start from the definition of the function in 
Eq.~(\ref{eq:J_function})
\ba\la{eq:Jini}
J(\hat \epsilon, u) &=& \int_v G_0(u-v) \, e^{m \hat\epsilon \cdot v}
G_m(v) = \sum_{n=0}^{\infty} z_n(u^2) \, C_n(\hat\epsilon\cdot \hat u). 
\ea
Inserting the expansions of the massless propagator
$G_0(u-v)$ from Eqs.~(\ref{eq:G0_expansion}) and (\ref{eq:G0_d_n}) and
of the exponential $e^{m \hat\epsilon \cdot v}$ from 
Eq.~(\ref{eq:exp_expansion}), we can use the orthogonality properties
of the Gegenbauer polynomials from 
Eq.~(\ref{eq:orthogonality_relations}) to project on the coefficients
$z_n(u^2)$ as in Eq.~(\ref{eq:zeta_n}) to obtain the intermediate
result 
\be \label{eq:z_n_prelim}
z_n(u^2) = \frac{1}{4\pi^2}\left[ \frac{1}{(m|u|)^{n+2}} \int_0^{m|u|}
\!dt \, t^{n+1} I_{n+1}(t) K_1(t) + (m |u|)^n \int_{m|u|}^\infty \!dt 
\frac{I_{n+1}(t)}{t^{n+1}} K_1(t)\right]. 
\ee
For the first integral, we use integration by parts, starting from the
observation $K_1(t) = -K_0'(t)$, and then using the well-known
identity $\frac{d}{dt}(t^{n+1}I_{n+1}(t)) = t^{n+1}I_n(t)$. We then
use the following primitives: 
\ba
\frac{d}{dt}\Big\{ t^{n+2} \Big(K_0(t) I_n(t) + K_1(t) I_{n+1}(t)\Big) \Big\}
& = & 2(n+1) t^{n+1} I_n(t) K_0(t), \label{eq:primitives_1} \\
\frac{d}{dt}\Big\{ \frac{1}{t^{n}} \Big(K_0(t) I_n(t) + K_1(t)
I_{n+1}(t)\Big) \Big\} 
& = & -\frac{2(n+1)}{t^{n+1}} \; I_{n+1}(t)
K_1(t), \label{eq:primitives_2} 
\ea
to obtain the result in Eq.~(\ref{eq:z_n}) after some slight
rearrangements.  

It is worth recording the asymptotics of the coefficients $z_n(u^2)$ for fixed $n$.
Using the known asymptotics of the modified Bessel functions, one finds
\ba\la{eq:znSmallArg}
z_n(u^2) &\stackrel{u^2\to0}{=} &  - \frac{\left( m |u| \right)^{n}}{2^{n+3} \pi^2 
  \Gamma(n+2)} \left[ \log \left( \frac{m |u|}{2} \right) +
  \gamma_{\rm E} - \frac{1}{n+1} \right] + {\rm O}\left(|u|^{n+2}
\right)   
\\
z_n(u^2) & \stackrel{u^2\to\infty}{=}& \frac{n+1}{4\pi^2 m|u|}\Big(\frac{1}{(n+1)^2} -
\frac{1}{2m | u|} +{\rm O}(u^{-2})\Big).
\la{eq:znExpNaive}
\ea
Furthermore, it is worth noting that 
\be\la{eq:znODE}
\frac{d}{d|u|}(|u| z_n(u^2)) = \frac{(n+1) I_{n+1}(m|u|)
  K_{0}(m|u|)}{4 \pi^2 m|u|}, 
\ee
and therefore, resumming the expansion in the angular variable using
Eq.~(\ref{eq:exp_expansion}), 
\be
\frac{d}{d|u|}(|u| J(\hat\epsilon,u)) = \frac{1}{8\pi^2}
e^{m\hat\epsilon\cdot u} K_0(m|u|). 
\ee
Since $J(\hat\epsilon,u)$ is logarithmically divergent for $|u| \to
0$, as can be seen from the definition in Eq.~(\ref{eq:J_function}),
we have $\lim_{|u| \to 0} (|u|J(\hat\epsilon,u))=0$. Therefore we can
solve this differential equation by simple integration to obtain the
integral representation\footnote{A very similar function appears in Ref. \cite{Blum:2017cer}.}
\be \label{eq:J_int_t}
J(\hat\epsilon,u) = \frac{1}{8\pi^2 m |u|} \int_0^{m|u|} dt\;
e^{t\hat\epsilon\cdot\hat u}\;K_0(t). 
\ee
From here it is straightforward to obtain the asymptotics of
$J(\hat\epsilon,u)$ at small and large $|u|$. 

As already mentioned, around the origin $|u| \to 0$, the function
$J(\hat\epsilon,u)$ is logarithmically divergent. The precise behavior
reads: 
\be
J(\hat\epsilon,u) \stackrel{|u|\to 0}{\sim } - \frac{1}{16\pi^2}
\log(m^2 u^2) + \frac{1}{8\pi^2}(1-\gamma_{\rm E} + \log (2)) + {\rm
  O}(|u| \log |u|). 
\ee

We note the exactly computable special case if the vector $u$ is
collinear with $\hat\epsilon$ 
\be
J(\hat\epsilon,\pm|u|\hat\epsilon) = \frac{1}{8\pi^2 }\Big( e^{\pm
  m|u|} (K_0(m|u|)\pm K_1(m|u|))\mp \frac{1}{m|u|}\Big). 
\ee
In this case, the behavior for large $|u|$ reads 
\be
J(\hat\epsilon,+|u|\hat\epsilon) \stackrel{|u|\to\infty}{\sim}
\frac{1}{4\pi\sqrt{2 \pi m|u|}}, 
\qquad\qquad
J(\hat\epsilon,-|u|\hat\epsilon) \stackrel{|u|\to\infty}{\sim}
\frac{1}{8\pi^2 m|u|}. 
\ee
For a generic direction of $u$, we obtain for large $|u|$:  
\be
J(\hat\epsilon,u) 
\stackrel{|u|\to\infty}{\sim} \frac{1}{8\pi^2 m|u|}
\frac{\pi-\theta}{\sin\theta}, \qquad  
\cos\theta = \hat\epsilon\cdot \hat u, \quad 0<\theta<\pi.
\la{eq:Jasymptgen}
\ee
This behavior at large $|u|$ is obtained from Eq.\ (\ref{eq:J_int_t})
by writing the integral $\int_0^{m|u|} = \int_0^\infty - \int_{m|u|}^\infty$.
The first term then yields Eq.\ (\ref{eq:Jasymptgen}). The second,
using the large-argument expansion of the Bessel function,
yields a correction suppressed by $\exp(-2\sin^2(\theta/2)m|u|)$.
In particular, the special `collinear' sector around $\theta=0$, in which
the function $J(\hat\epsilon,u)$ only falls off like $(m|u|)^{-1/2}$, has an angular
size  $\Delta \theta \sim (m|u|)^{-1/2}$.

\subsection{Average over the direction of the muon momentum \la{sec:muonmom_average}}

The product $J(\hat\epsilon,u) J(\hat\epsilon,x-u)$ can be viewed as a
function of $\hat\epsilon$ on the unit three-sphere, and can therefore
be expanded in scalar, vector, rank-two traceless tensor,
etc. components.  Computationally, in order to compute $S$, $V_\delta$
and $T_{\beta\delta}$ in
Eqs.~(\ref{eq:scalar_average})-(\ref{eq:tensor_average}), it is
simpler to first extract these components before performing the
convolution with the massless propagator. We therefore introduce the
notation
\ba s(x,u) &=& \Big\<
J(\hat\epsilon,u)
J(\hat\epsilon,x-u)\Big\>_{\hat\epsilon}, \label{eq:s_x_u} 
\\
v_\delta(x,u) &=& \Big\< \hat\epsilon_\delta \; J(\hat\epsilon,u)
J(\hat\epsilon,x-u)\Big\>_{\hat\epsilon}, \label{eq:v_x_u}
\\
t_{\beta\delta}(x,u) &=& \Big\<
\Big(\hat\epsilon_\beta\hat\epsilon_\delta-\frac{1}{4}\delta_{\beta\delta}\Big)
J(\hat\epsilon,u)
J(\hat\epsilon,x-u)\Big\>_{\hat\epsilon}, \label{eq:t_x_u} 
\ea 
so that
\ba S(x,y)_{\rm IR~reg.} &=& \int_u G_0(u-y) \,
s(x,u), \label{eq:S_x_y} \\  
V_\delta(x,y) &=& \int_u G_0(u-y) \,  v_\delta(x,u), \label{eq:V_x_y} 
\\
T_{\beta\delta}(x,y) &=& \int_u G_0(u-y) \,
t_{\beta\delta}(x,u). \label{eq:T_x_y}    
\ea

\subsection{Calculation of $s(x,u)$}

Inserting into Eq.~(\ref{eq:s_x_u}) the expansions of $J(\hat\epsilon,
u)$ and $J(\hat\epsilon, x-u)$ in Gegenbauer polynomials from
Eq.~(\ref{eq:J_expansion}) and using the orthogonality relations in
Eq.~(\ref{eq:orthogonality_relations}), the angular average yields
\be\la{eq:sxu_sum}
s(x,u) 
= \sum_{n=0}^\infty z_n(u^2) z_n((x-u)^2) \, \frac{C_n(\hat u\cdot
  \widehat{x-u})}{n+1}. 
\ee
We remark that the symmetry $s(x,u)=s(x,x-u)$, obvious in Eq.\ (\ref{eq:s_x_u}), 
remains manifest in Eq.\ (\ref{eq:sxu_sum}).
At $x=0$, we have
\be
s(0,u) = \sum_{n=0}^{\infty} (-1)^n \; z_n(u^2)^2 
\stackrel{|u|\to\infty}{\sim }
\frac{1}{(4\pi^2 m)^2 u^2} \sum_{n=0}^{\infty} \frac{(-1)^n}{(n+1)^2}
= \frac{1}{192\pi^2 m^2 u^2}, \label{eq:s_0_u_large_u}
\ee
where we used the first term in the large-argument expansion (\ref{eq:znExpNaive}) of $z_n(u^2)$.
Eq.~(\ref{eq:s_0_u_large_u}) provides the leading behavior
of $s(x,u)$ at large $|u|$ that will be important to deal with the IR
divergence in $S(x,y)$.
We remark that the asymptotic prediction of Eq.\ (\ref{eq:s_0_u_large_u}) for $s(x,u)$
can also be obtained directly from its definition (\ref{eq:s_x_u}), employing the asymptotic form
(\ref{eq:Jasymptgen}) of the function $J(\hat\epsilon,u)$.

\subsection{Calculation of $v_\delta(x,u)$}

We parametrize the vector components in Eq.~(\ref{eq:v_x_u}) as
follows:   
\be
v_\delta(x,u)
= x_\delta\; \mathfrak{f}^{(1)}(x^2,x\cdot u,u^2) + u_\delta \;
\mathfrak{f}^{(2)}(x^2,x\cdot u,u^2). 
\ee
Multiplying with $x_\delta$ and $u_\delta$ and solving the system of
two equations, we get 
\ba
&& \left(\begin{array}{c}  \mathfrak{f}^{(1)} \\
    \mathfrak{f}^{(2)} \end{array}\right)   = \frac{1}{x^2 u^2 -
  (x\cdot u )^2}  
\left(\begin{array}{c@{~~}c}  u^2 & -u\cdot x  \\  -u\cdot x  &
    x^2 \end{array} \right)  
\left(\begin{array}{c}   \< (x\cdot\hat\epsilon) J (\hat\epsilon,u) 
J (\hat\epsilon,x-u)\>_{\hat\epsilon} \\ 
\< (u\cdot\hat\epsilon) J (\hat\epsilon,u) 
J (\hat\epsilon,x-u)\>_{\hat\epsilon} \end{array}\right). 
\ea
Using Eq.~(\ref{eq:xiCn}) and the orthogonality
relations~(\ref{eq:orthogonality_relations}) we obtain the results  
\ba
&& \< (x\cdot\hat\epsilon) J (\hat\epsilon,u) J
(\hat\epsilon,x-u)\>_{\hat\epsilon}  
= \frac{|u|}{2} \sum_{n=0}^\infty (z_{n-1}(u^2) + z_{n+1}(u^2))
\,\frac{z_n((x-u)^2)}{n+1}\,C_n(\hat u\cdot\widehat{x-u}) 
\nonumber \\ 
&&  \quad +\frac{|x-u|}{2} \sum_{n=0}^\infty (z_{n-1}((x-u)^2) +
z_{n+1}((x-u)^2)) \,\frac{z_n(u^2)}{n+1}\, 
C_n(\hat u\cdot\widehat{x-u}), \\ 
&& \< (u\cdot\hat\epsilon) J (\hat\epsilon,u) J
(\hat\epsilon,x-u)\>_{\hat\epsilon}  
= \frac{|u|}{2} \sum_{n=0}^\infty (z_{n-1}(u^2) + z_{n+1}(u^2))
\,\frac{z_n((x-u)^2)}{n+1}\,C_n(\hat u\cdot\widehat{x-u}), 
\nonumber \\
&&  
\ea
with the convention that $z_{n}(u^2) = z_{n}((x-u)^2) = 0$ for $n \leq
-1$. In the end we get 
\ba
&& \mathfrak{f}^{(1)}(x^2,x\cdot u,u^2) = \frac{1}{2(x^2 u^2 - (x\cdot
  u )^2)} \nonumber \\ 
&&
\quad \times \sum_{n= 0}^\infty \Big(u^2 |x-u|(z_{n-1}((x-u)^2) +
z_{n+1}((x-u)^2)) z_n(u^2)  
\nonumber \\  
&& 
\quad \quad + |u| (u^2-u\cdot x) (z_{n-1}(u^2) + z_{n+1}(u^2))
\,{z_n((x-u)^2)} \Big) \frac{C_n(\hat
  u\cdot\widehat{x-u})}{n+1}, \label{eq:f1_result} 
\\
&& \mathfrak{f}^{(2)}(x^2,x\cdot u,u^2) = \frac{1}{2(x^2 u^2 - (x\cdot
  u )^2)}  \nonumber \\ 
&& 
\quad \times \sum_{n= 0}^\infty \Big( -|x-u|\,(u\cdot x)
(z_{n-1}((x-u)^2) + z_{n+1}((x-u)^2)) z_n(u^2) \nonumber \\ 
&& 
\quad \quad + |u| (x^2-u\cdot x) (z_{n-1}(u^2) + z_{n+1}(u^2))
\,{z_n((x-u)^2)} \Big) \frac{C_n(\hat
  u\cdot\widehat{x-u})}{n+1}. \label{eq:f2_result} 
\ea

We now want to study the large-$|u|$ behavior of
$\mathfrak{f}^{(i)}(x^2,x\cdot u,u^2)$. Expanding $z_n(u^2)$ and
$z_n((x-u)^2)$ for fixed $n$ and large $|u|$, one finds the result in 
Eq.~(\ref{eq:znExpNaive}) and    
\be 
z_n((x-u)^2) - z_n(u^2) =   \frac{\hat u\cdot x}{4\pi^2 m\,
  (n+1)u^2} + {\rm O}(u^{-3}). 
\ee
We note that 
\be\la{eq:CnExp}
\frac{C_n(\hat u\cdot \widehat{x-u})}{n+1} = (-1)^n \Big(1 +n(n+2)
((\hat x\cdot \hat u)^2-1) \frac{x^2}{6u^2}\Big)+ 
{\rm O}(u^{-3}).
\ee
One then sees that $\<(u\cdot\hat\epsilon) J(\hat \epsilon,u)  J(\hat
\epsilon,x-u)\>_{\hat\epsilon}$ vanishes for $x=0$.
For this reason,  we may rewrite
\ba\la{eq:f1study}
&& \<(u\cdot\hat\epsilon) J(\hat \epsilon,u)  J(\hat
\epsilon,x-u)\>_{\hat\epsilon} 
\\ && 
= \frac{|u|}{2}\sum_{n=0}^\infty 
\Big( z_{n-1}(u^2) + z_{n+1}(u^2)\Big) (-1)^n 
\Big(z_n((x-u)^2) (-1)^n \frac{C_n(\hat u\cdot\widehat{x- u})}{n+1} -
z_n(u^2)  \Big). 
\nonumber
\ea
The second factor in the sum is given in leading order by $\frac{\hat
  u\cdot x}{4\pi^2 m\, (n+1)u^2}$ for large $u$, and since the series
is then still absolutely convergent, one finds that
$\<(u\cdot\hat\epsilon) J(\hat \epsilon,u) J(\hat
\epsilon,x-u)\>_{\hat\epsilon}$ falls off at least as fast as
$1/|u|^3$. In fact, at little more work reveals that the coefficient
of the $1/|u|^3$ term vanishes.  The same argument shows that
$\<(x\cdot\hat\epsilon) J(\hat \epsilon,u) J(\hat\epsilon,x-u)\>_{\hat\epsilon}$
goes like $|u|^{-3}$ for large $|u|$,
thus showing that $\big\<\hat\epsilon_\delta J (\hat\epsilon,u) J
(\hat\epsilon,x-u)\big\>_{\hat\epsilon}$ falls off at least as fast as
$|u|^{-3}$ for large $|u|$.

\subsection{Calculation of $t_{\alpha\beta}(x,u)$}

First, calculate at $x=0$:
\ba\la{eq:bu2}
&& t_{\alpha\beta}(0,u) = \<(\hat\epsilon_\alpha\hat\epsilon_\beta -
\frac{1}{4}\delta_{\alpha\beta})  
  J(\hat\epsilon, u)  J(\hat\epsilon, -u)\>_{\hat\epsilon}
 = (u_\alpha u_\beta - \frac{u^2}{4} \delta_{\alpha\beta})\; b(u^2),
\ea
Contracting with $u_\alpha u_\beta$ and using the identity from
Eq.~(\ref{eq:xisqCn}) and the orthogonality
relations~(\ref{eq:orthogonality_relations}) within the angular
average $\< \ldots \>_{\hat\epsilon}$, we get
\ba\la{eq:bu2sum}
b(u^2) &=& \frac{1}{3u^2}  \sum_{n=0}^\infty (-1)^n\,
(z_{n-2}(u^2) + z_n(u^2) (1- \delta_{n0}) + z_{n+2}(u^2)) \,z_n(u^2)
\\ &=& \frac{1}{3u^2}   \sum_{n=0}^\infty
(-1)^n\,z_n(u^2) (z_n(u^2) (1-\delta_{n0}) + 2z_{n+2}(u^2)) 
\nonumber
\\ & \stackrel{|u|\to\infty}{\sim}& 
 \frac{1}{96\pi^2 m^2 |u|^4} \Big(-\frac{1}{\pi^2} + \frac{1}{6}\Big).
\ea
For the general case we have with $v=x-u$ the decomposition: 
\ba\la{eq:TENSOR.decomp}
t_{\alpha\beta}(x,u)
&=& (u_\alpha u_\beta - \frac{u^2}{4} \delta_{\alpha\beta})
\hat{\mathfrak{h}}^{(1)}(u^2,u\cdot v,v^2) 
\\ && + (v_\alpha v_\beta - \frac{v^2}{4} \delta_{\alpha\beta})
\hat{\mathfrak{h}}^{(2)}(u^2,u\cdot v,v^2) 
\nonumber\\ && + (u_\alpha v_\beta + v_\alpha u_\beta - \frac{1}{2}
\delta_{\alpha\beta} (u\cdot v))  
\hat{\mathfrak{h}}^{(3)}(u^2,u\cdot v,v^2).
\nonumber
\ea
Multiplying with $u_\alpha u_\beta, v_\alpha v_\beta$ and $u_\alpha
v_\beta + v_\alpha u_\beta$ and solving the system of three equations,
we get as intermediate result for the scalar functions $\hat{\mathfrak{h}}^{(k)}$ 
\ba \la{eq:Muv}
&&  \left(\begin{array}{c} \hat{\mathfrak{h}}^{(1)} \\
    \hat{\mathfrak{h}}^{(2)} \\ \hat{\mathfrak{h}}^{(3)}  
\end{array}\right) = \frac{1}{2D_{u,v}}
 \left(\begin{array}{c@{~~}c@{~~}c}
3(v^2)^2 &  u^2 v^2 +2(u\cdot v)^2 &  - 3 v^2 (u\cdot v) \\
u^2 v^2 +2(u\cdot v)^2 &  3 (u^2)^2 &  - 3 u^2(u\cdot v) \\
- 3 v^2 (u\cdot v) &  - 3 u^2(u\cdot v) &  u^2 v^2 + 2(u\cdot v)^2 
\end{array}\right)
\left( \begin{array}{c} 
s_1 \\
s_2 \\ 
s_3 
\end{array}
\right) \qquad
\ea
with $D_{u,v}= (u^2v^2-(u\cdot v)^2)^2$ and
\ba
&&  \!\!\!\!\!  s_1=\<((u\cdot\hat\epsilon)^2 - u^2 /4)
J(\hat\epsilon,u)J(\hat\epsilon,v)\>_{\hat\epsilon} 
\nonumber \\ 
&& = \frac{u^2}{4} \sum_{n=0}^\infty
\Big(z_{n-2}(u^2)+z_n(u^2) (1-\delta_{n0})+ z_{n+2}(u^2)\Big) z_n(v^2) \cdot 
 \frac{C_n(\hat u\cdot \hat v)}{n+1} , \la{eq:s1}
\\ 
&&  \!\!\!\!\!  s_2=\<((v\cdot\hat\epsilon)^2 - v^2 /4)
J(\hat\epsilon,u)J(\hat\epsilon,v)\>_{\hat\epsilon} 
\nonumber \\ 
&& = \frac{v^2}{4}\sum_{n=0}^\infty
z_n(u^2) \Big(z_{n-2}(v^2)+z_n(v^2) (1-\delta_{n0}) + z_{n+2}(v^2)\Big) \cdot 
 \frac{C_n(\hat u\cdot \hat v)}{n+1}, \la{eq:s2}
\\ 
&& \!\!\!\!\! s_3= \< ( 2 (\hat\epsilon\cdot u)(\hat\epsilon\cdot v) -
\frac{1}{2} (u\cdot v))   
J(\hat\epsilon,u)J(\hat\epsilon,v)\>_{\hat\epsilon} 
\nonumber \\ 
&& = \frac{1}{2}|u||v|  \sum_{n=0}^\infty \Big[ (z_{n-1}(u^2)+z_{n+1}(u^2)) 
(z_{n-1}(v^2)+z_{n+1}(v^2)) - (\hat u\cdot\hat v)  z_n(u^2)
z_n(v^2)\Big]  \frac{C_n(\hat u\cdot \hat v)}{n+1}. \nonumber \\
&& \la{eq:s3}
\ea
The point now is that $ \<(\hat\epsilon_\alpha\hat\epsilon_\beta -
\frac{1}{4}\delta_{\alpha\beta}) J(\hat\epsilon, u) J(\hat\epsilon,
x-u)\>_{\hat\epsilon}$ is of order ($1/u^2$) at large $u$, but with a tensor structure 
proportional to $( \hat u_\beta \hat u_\delta - \frac{1}{4}
\delta_{\beta\delta}) $.  The average over $\hat u$ in the next step
will cancel this leading contribution, so that $T_{\alpha\beta}(x,y)$
is finite.  
Anticipating the numerical implementation, we remark that it can
be advantageous to subtract a term which
vanishes upon the  $\hat u$  integration and makes the integrand
fall off faster at large $u$.

The leading asymptotic behavior of the scalar functions
$s_i(u^2,u\cdot x,x^2)$ for $|u|\to\infty$ is  
\be
s_1 \sim s_2 \sim -\frac{1}{2} s_3 \stackrel{|u|\to\infty}{\sim}
\frac{1}{128\pi^2 m^2} \,\Big( -\frac{1}{\pi^2} + \frac{1}{6}\Big).  
\ee
One also finds that, expanding in $x$ around $x=0$, and for arbitrary
$|u|$, 
\be\la{eq:s1x.eq.0}
s_1 \Big|_{x=0} =  s_2\Big|_{x=0} =  -\frac{1}{2}s_3\Big|_{x=0} =
\frac{3}{4} u^4 b(u^2). 
\ee
Furthermore, one finds 
\be\la{eq:s1primex.eq.0}
s_1 + s_2 + s_3 = {\rm O}(x^2).
\ee
Equations (\ref{eq:s1x.eq.0}) and (\ref{eq:s1primex.eq.0}) can be
shown without using the explicit expression of $z_n(u^2)$.

\section{Direct evaluation of the final convolution integral\la{sec:direct}}


In this section, we treat the final convolution integral yielding the
tensors $S$, $V_\delta$ and $T_{\beta\delta}$ (see Eqs.~(\ref{eq:S_x_y})-(\ref{eq:T_x_y}))
by performing two
angular integrations analytically, while the third angular integral as
well as the integral over the modulus are left to be done numerically.
Since this straightforward method leads to some numerical difficulties
pointed out below,
our final weight functions have been computed with the alternative method
presented in section~\ref{sec:alternative_evaluation}.
Nevertheless, this method, which was implemented as part of Ref.~\cite{NilsThesis},
provided important cross-checks
(discussed in section \ref{sec:discussion_master_formula}) for the final QED kernel.
The expressions obtained for the weight functions are also used as the starting point
for the multipole-expansion method of section~\ref{sec:alternative_evaluation};
see the text around Eqs.\ (\ref{eq:S0main}--\ref{eq:S2main}).

In order to perform the angular integrations in $\int_u$ in
Eqs.~(\ref{eq:S_x_y})-(\ref{eq:T_x_y}), we choose a coordinate system
where $x$ is pointing in the direction $\hat e_0$ and $y$ is in the
$(\hat e_0,\hat e_1)$ plane and we introduce the angle $\beta$ between
those two vectors 
\bea 
x & = & |x| (1, 0, 0, 0), \\ 
y & = & (y_0, y_1, 0, 0), \qquad y_1 \geq 0, \\ 
\hat x \cdot \hat y & = & \cos\beta, \qquad \qquad 0 \leq \beta \leq
\pi, \\  
y_0 & = & |y| \cos\beta,  \\
y_1 & = & |y| \sin\beta.  
\eea 

The vector $u$ is parametrized as follows 
\bea
u_0 & = & |u| \cos\phi_1, \nonumber \\ 
u_1 & = & |u| \sin\phi_1 \cos\phi_2, \nonumber \\
u_2 & = & |u| \sin\phi_1 \sin\phi_2 \cos\phi_3, \nonumber \\ 
u_3 & = & |u| \sin\phi_1 \sin\phi_2 \sin\phi_3, 
\eea 
with $\phi_1, \phi_2 \in [0,\pi]$, $\phi_3 \in [0,2\pi]$ and the
angular integration measure is given by 
\be 
\int d\Omega_{\hat u} = \int_{0}^{\pi} d\phi_1 \sin^2\phi_1 \int_{-1}^{1}
d\hat c_2 \int_0^{2\pi} d\phi_3 
\ee 
where $\hat c_2 \equiv \cos\phi_2$. 

From the above definitions of the vectors we get 
\bea
x \cdot u & = & |x| |u| \cos\phi_1, \\ 
u \cdot y  & = & |u| (y_0 \cos\phi_1 + y_1 \sin\phi_1 \,\hat c_2),  \\ 
(u \cdot y)^2 & = & u^2 (y_0^2 \cos^2\phi_1 + 2 y_0 y_1 \cos\phi_1
\sin\phi_1 \,\hat c_2 + y_1^2 \sin^2\phi_1 \,\hat c_2^2), \\ 
(u-y)^2 & = & u^2 - 2 u \cdot y + y^2 = y^2 + u^2 - 2 |u| y_0
\cos\phi_1 - 2 |u| y_1 \sin\phi_1 \,\hat c_2.  
\eea

We will need the following angular integrals, where $f(x,u) \equiv f(x^2,
x \cdot u, u^2) = f(x^2, |x| |u| \cos\phi_1, u^2)$ is a generic scalar
function of the vectors $x$ and $u$: 
\ba
\Big\< \frac{1}{(u-y)^2} f(x,u) \Big\>_{\hat u} & \equiv & 
\frac{1}{2\pi^2} \int d\Omega_{\hat u} \, \frac{1}{(u-y)^2} \, f(x,u)
\nonumber \\
& = &\frac{1}{\pi}\int_0^\pi d\phi_1 \sin^2\phi_1
f(x^2,|x| |u|\cos\phi_1,u^2) \nonumber \\ 
&& \quad \times {\int_{-1}^1
  \frac{d\hat c_2}{y^2+u^2-2|u| y_0\cos\phi_1 -2|u|y_1 \sin\phi_1    \,\hat c_2}} \nonumber \\
& = & \frac{-1}{2\pi |u| |y| \sin\beta} \int_{0}^{\pi} d\phi_1\sin\phi_1
f(x^2,|x| |u|\cos\phi_1,u^2) \, {\rm Log}, \la{eq:G0f}
\ea
where we introduced the abbreviation 
\be
{\rm Log} \equiv \ln \left[ \frac{y^2+u^2-2 |u| |y|\cos(\beta-\phi_1)}{
  y^2+u^2-2 |u| |y|\cos(\beta+\phi_1)} \right] . 
\ee

In a similar way one obtains 
\ba
\Big\< \frac{1}{(u-y)^2}\; (x\cdot u) f(x,u) \Big\>_{\hat u}
& = & -\frac{|x|}{2\pi |y| \sin\beta} \int_0^\pi d\phi_1 \sin\phi_1
\cos\phi_1 f(x^2,|x| |u| \cos\phi_1,u^2) \, {\rm Log}, \nonumber \\
& & \la{eq:G0xdotuf} \\   
\Big\< \frac{1}{(u-y)^2}\; (y\cdot u) f(x,u) \Big\>_{\hat u}
& = & \frac{-1}{\pi}  \int_0^\pi d\phi_1 \sin^2\phi_1
f(x^2,|x| |u| \cos\phi_1,u^2) \nonumber \\ 
&& \qquad \times\left(1 + \frac{y^2+u^2}{4|u| |y| \sin\beta
    \sin\phi_1} \, {\rm Log} \right), \la{eq:G0ydotuf} \\ 
\Big\<\frac{(y\cdot u)^2}{(u-y)^2} f(x,u))\Big\>_{\hat u} & = &
\frac{-1}{8\pi |u| |y| \sin\beta} \int_0^\pi 
d\phi_1 \sin\phi_1 f(x^2, |x| |u| \cos\phi_1,u^2) \nonumber \\ 
&& 
\qquad \times \Big[ 4 |u| |y|\sin\beta\sin\phi_1 \left(u^2+y^2 + 2 |u|
  |y| \cos\beta\cos\phi_1 \right) \nonumber \\
& & \qquad \quad + (u^2+y^2)^2 \, {\rm Log} \Big].  \la{eq:G0ydotu2f}
\ea
The integration over $\phi_3$ is always trivial and the integrations 
over $\hat c_2$ lead to simple elementary integrals.

\subsection{Calculation of the weight function $\bar{\mathfrak{g}}^{(0)}$}

Exploiting the behavior of $s(x,u)$ for large $|u|$ from
Eq.~(\ref{eq:s_0_u_large_u}), we can introduce a fixed vector $w$ and modify the integrand of $S(x,y)$
in Eq.~(\ref{eq:S_x_y}) to get an IR-regulated
function as follows 
\ba
\bar{\mathfrak{g}}^{(0)}(|x|, x \cdot y, |y|) & = & S(x,y)_{\rm IR~reg.}  
\equiv \int_{u} \, \left(G_0(u-y) \, s(x,u) 
- \frac{G_0(u)\theta(u^2-w^2)}{192\pi^2 m^2 u^2}\right) \label{eq:S_x_y_IR}
\\ 
& = & -\frac{1}{4\pi  |y|\sin\beta}\int_0^\infty du\; u^2
\int_{0}^{\pi} d\phi_1\sin\phi_1 
\la{eq:g0direct} \\ 
&& \times 
 \Big[ \frac{|y|\sin\beta\;\theta(u^2-w^2)}{192\pi m^2 u^3} + {\rm \,Log\,}
\cdot   \sum_{n=0}^\infty z_n(u^2) z_n((x-u)^2) \, \frac{C_n(\hat
  u\cdot \widehat{x-u})}{n+1}\Big], \nonumber 
\ea
where we used the result from Eq.~(\ref{eq:G0f}) for the angular
integration.  The term with $\theta(u^2 - w^2)$ in
Eq.~(\ref{eq:S_x_y_IR}) is independent of $x$ and $y$,
therefore it will not affect the final result for the QED kernel,
in which only the derivatives of the weight function
$\bar{\mathfrak{g}}^{(0)}(|x|, x\cdot y, |y|)$
with respect to $x_\alpha$ and $y_\alpha$ enter; see
Eq.~(\ref{eq:Lbar}), (\ref{eq:T_II}--\ref{eq:T_III}) and (\ref{eq:define_g_0}).

\subsection{Calculation of the weight functions $\bar{\mathfrak{g}}^{(1,2)}$}

Starting from the definition of $V_\delta(x,y)$ in
Eq.~(\ref{eq:vector_average}), we split off the integration over the
length of the vector $u$ and parametrize the angular average as
follows: 
\be \label{eq:define_g_u}
 \Big\< \frac{1}{(u-y)^2} \;\big\<\hat\epsilon_\delta J
 (\hat\epsilon,u) J
 (\hat\epsilon,x-u)\big\>_{\hat\epsilon}\Big\>_{\hat u} 
= 
x_\delta\; \mathfrak{g}_u^{(1)}(x^2,x\cdot y,y^2) + y_\delta
\;\mathfrak{g}_u^{(2)}(x^2,x\cdot y,y^2).  
\ee 
From this we get the vector weight functions in
Eq.~(\ref{eq:define_g_1_2}) via  
\be
\bar{\mathfrak{g}}^{(i)}(x^2,x\cdot y,y^2) = 
\frac{1}{2}\int_0^\infty du\,u^3 \mathfrak{g}^{(i)}_u(x^2,x\cdot
y,y^2), \quad i=1,2.
\ee

Multiplying Eq.~(\ref{eq:define_g_u}) by $x_\delta$ and $y_\delta$ we
then obtain in a similar way as before 
\ba
\left(\begin{array}{c}  \mathfrak{g}_u^{(1)} \\ 
 \mathfrak{g}_u^{(2)} \end{array}\right)   
& = & \frac{1}{x^2 y^2 -
(x\cdot y )^2}  
\left(\begin{array}{c@{~~}c}  y^2 & -x\cdot y  \\  -x\cdot y  &
    x^2 \end{array} \right)  
\nonumber \\ 
&& \qquad \times 
\left(\begin{array}{c}  
\Big\<\frac{1}{(u-y)^2}  (x^2\; \mathfrak{f}^{(1)}(x,u) + (x\cdot u)\;
\mathfrak{f}^{(2)}(x,u)) \Big\>_{\hat u} \\ 
\Big\<\frac{1}{(u-y)^2} ((x\cdot y)\; \mathfrak{f}^{(1)}(x,u) +
(y\cdot u)\; \mathfrak{f}^{(2)}(x,u)) \Big\>_{\hat u}  
\end{array}\right).
\label{eq:g1u_g2u_result}
\ea

Using the results of the angular integrals from
Eqs.~(\ref{eq:G0f}), (\ref{eq:G0xdotuf}) and (\ref{eq:G0ydotuf}), we
get the following results for the weight functions
\ba 
 && \bar{\mathfrak{g}}^{(2)}(x^2,x\cdot y,y^2) =
\frac{1}{8\pi y^2|x|\sin^3\beta}\int_0^\infty du\; u^2 \int_0^\pi{d\phi_1}
\nonumber \\ 
&& \quad \times \Big\{2 \sin\beta + \Big(\frac{y^2+u^2}{2|u||y|} -
\cos\beta\cos\phi_1\Big)\frac{\rm Log}{\sin\phi_1 }\Big\}  
\nonumber  \\ 
&& 
\quad \times \sum_{n=0}^\infty \Big( z_n(u^2) z_{n+1}((x-u)^2)
\Big[|x-u| \cos\phi_1 \frac{C_n}{n+1} + (|u|\cos\phi_1 - |x|)
\frac{C_{n+1}}{n+2} \Big] \nonumber \\ 
&& 
\quad \quad 
+ z_{n+1}(u^2) z_n((x-u)^2) \Big[(|u|\cos\phi_1-|x|)
\frac{C_n}{n+1}+ |x-u| \cos\phi_1\frac{C_{n+1}}{n+2}\Big]\Big),
\la{eq:g2Final}  
\\
&& 
\bar {\mathfrak{g}}^{(3)}(x^2,x\cdot y,y^2) 
= \frac{-1}{8\pi x^2|y|\sin\beta} \int_0^\infty du\; u^2 \int_0^\pi
d\phi_1 \sin\phi_1 \, {\rm Log} \nonumber \\ 
&& \quad
\times \sum_{n=0}^\infty \Big( z_n(u^2) z_{n+1}((x-u)^2) \Big( |x-u|
\frac{C_n}{n+1} + |u|\frac{C_{n+1}}{n+2}\Big) 
\nonumber \\ 
&& \quad \quad 
  +  z_{n+1}(u^2) z_{n}((x-u)^2) \Big(|u|\frac{C_n}{n+1} +  |x-u|
  \frac{C_{n+1}}{n+2} \Big)\Big), \la{eq:g1lincomb}
\ea
and 
\be \la{eq:g1giveng3}
\bar {\mathfrak{g}}^{(1)}(x^2,x\cdot y,y^2)
= \bar {\mathfrak{g}}^{(3)}(x^2,x\cdot y,y^2)  - \frac{|y|}{|x|}\cos\beta \, \bar {\mathfrak{g}}^{(2)}(x^2,x\cdot y,y^2) .
\ee
We choose to first compute  the scalar function $\bar {\mathfrak{g}}^{(3)}$ because
the integrand simplifies and is collinear safe for $\sin\phi_1 \to 0$.
The argument of the Gegenbauer polynomials $C_{n}$ and $C_{n+1}$ is given
by
\be\la{eq:argG}
\hat u\cdot \widehat{x-u} = -\frac{|u|-|x|\cos\phi_1}{|u-x|} .
\ee

We note that there is a large cancellation inside the curly bracket of
Eq.~(\ref{eq:g2Final}). If we call $A= 2|u| |y|/(u^2+y^2)$, such that
$0\leq A\leq 1$, then Taylor-expanding the logarithm in $A$ shows that
the curly bracket is of order $A^2$ at small $A$ (small $A$
corresponds to both $|u| \ll |y|$ and $|u| \gg |y|$).

\subsection{Calculation of the weight functions $\bar{\mathfrak{l}}^{(1,2,3)}$}

Upon integrating over the angular variables of $u$ in
Eq.~(\ref{eq:tensor_average}), the tensor decomposition reads
\ba
&& 
\Big\< \frac{1}{(u-y)^2}\<(\hat\epsilon_\alpha \hat\epsilon_\beta -
\frac{1}{4}\delta_{\alpha\beta}) \, J(\hat\epsilon, u) \, J(\hat\epsilon,
x-u)\>_{\hat\epsilon}\Big\>_{\hat u}  \nonumber \\ 
&& = (x_\alpha x_\beta -
\frac{x^2}{4}\delta_{\alpha\beta})\; 
\mathfrak{l}^{(1)}_u 
+ (y_\alpha y_\beta - \frac{y^2}{4}\delta_{\alpha\beta})\;
\mathfrak{l}^{(2)}_u 
+ (x_\alpha y_\beta + y_\alpha x_\beta  - \frac{x\cdot
  y}{2}\delta_{\alpha\beta})\; \mathfrak{l}^{(3)}_u. 
\label{eq:define_l_u}
\ea
where the scalar functions $\mathfrak{l}^{(1)}_u$ depend on $x^2, x \cdot y, y^2$, as well as
on  $|u|$. From this we get the tensor weight functions in
Eq.~(\ref{eq:define_l_1_2_3}) via    
\be \label{eq:l_u_integrated}
\bar{\mathfrak{l}}^{(i)}(x^2, x \cdot y, y^2) =
\frac{1}{2}\int_0^\infty du\; u^3\; \mathfrak{l}^{(i)}_u(x^2, x \cdot
y, y^2), \qquad i=1,2,3. 
\ee
 
Again, multiplying Eq.~(\ref{eq:define_l_u}) with $x_\alpha x_\beta$,
$y_\alpha y_\beta$ and $x_\alpha y_\beta + y_\alpha x_\beta$ and
solving as before, the weight functions $\mathfrak{l}^{(i)}_u$ are
given by
\ba
 \left(\begin{array}{c} \mathfrak{l}^{(1)}_u \\ \mathfrak{l}^{(2)}_u
     \\ \mathfrak{l}^{(3)}_u  \end{array}\right) 
= \frac{1}{2D_{x,y}} \left( \begin{array}{c@{~~}c@{~~}c}   
3(y^2)^2   & x^2 y^2 + 2(x\cdot y)^2   & -3 y^2(x\cdot y) \\
x^2 y^2 + 2(x\cdot y)^2   &   3(x^2)^2  &  -3 x^2(x\cdot y) \\
-3 y^2(x\cdot y)   & -3 x^2(x\cdot y)  & x^2 y^2 + 2(x\cdot y)^2 
\end{array}\right) \left(\begin{array}{c} v_{1,u} \\ v_{2,u}\\
  v_{3,u} \end{array}\right), \quad 
\la{eq:Mxy}
\ea
with $D_{x,y}=(x^2 y^2 - (x\cdot y)^2)^2$ and 
\ba
&& v_{1,u} = \Big\< \frac{1}{(u-y)^2} \Big\{ 
 [(u\cdot x)^2 - u^2 x^2/4]  \hat{\mathfrak{h}}^{(1)} 
 + [((x-u)\cdot x)^2 - (x-u)^2x^2/4] \hat{\mathfrak{h}}^{(2)}
\nonumber \\ 
&& \qquad 
 + [2(u\cdot x)((x-u)\cdot x) - x^2(u\cdot (x-u))/2] \hat{\mathfrak{h}}^{(3)}
\Big\} \Big\>_{\hat u},   \phantom{\underbrace{A}_{A}}
\\ 
&& v_{2,u} = 
\Big\< \frac{1}{(u-y)^2} \Big\{[(u\cdot y)^2 - u^2 y^2/4]
\hat{\mathfrak{h}}^{(1)}  
+[((x-u)\cdot y)^2 - y^2(x-u)^2/4] \hat{\mathfrak{h}}^{(2)}
\nonumber \\ 
&& \qquad + [2(u\cdot y ) ((x-u)\cdot y) - y^2(u\cdot(x-u))/2 ]
\hat{\mathfrak{h}}^{(3)}\Big\} \Big\>_{\hat u},  
\phantom{\underbrace{A}_{A}}
\\ 
&& v_{3,u} = 
\Big\< \frac{1}{(u-y)^2} \Big\{[2(x\cdot u)(y\cdot u) - u^2(x\cdot
y)/2] \hat{\mathfrak{h}}^{(1)}  
\nonumber \\ 
&& \qquad +[2(x\cdot (x-u))(y\cdot(x-u)) - (x\cdot y)(x-u)^2/2]
\hat{\mathfrak{h}}^{(2)} 
\nonumber\\ 
&&\qquad + [2(u\cdot x)(y\cdot(x-u)) + 2(u\cdot y)(x\cdot
(x-u)) - (u\cdot (x-u))(x\cdot y) ]  
\hat{\mathfrak{h}}^{(3)}\Big\} \Big\>_{\hat u}, 
\ea
where the $\hat{\mathfrak{h}}^{(i)}(u^2, u \cdot v, v^2)$ are given in
Eq.~(\ref{eq:Muv}).

Using the results of the angular integrations in Eqs.~(\ref{eq:G0f}),
(\ref{eq:G0xdotuf})-(\ref{eq:G0ydotu2f}), we get more explicitly
the following intermediate result 
\ba
 v_{1,u} &=& \frac{-x^2}{2\pi u y \sin\beta}\int_0^\pi
 d\phi_1\,\sin\phi_1\,{\rm Log} 
 \Big\{u^2[ \cos^2\phi_1 - \frac{1}{4}] \hat{\mathfrak{h}}^{(1)}
\nonumber \\ 
&& \qquad \qquad \qquad +[\frac{3}{4}x^2 - \frac{3}{2}u x \cos\phi_1 +
u^2(\cos^2\phi_1 - \frac{1}{4}) ]  \hat{\mathfrak{h}}^{(2)} \nonumber
\\  
&& \qquad \qquad \qquad + [\frac{3}{2} u x \cos\phi_1 - 2u^2
(\cos^2\phi_1 - \frac{1}{4}) ] \hat{\mathfrak{h}}^{(3)} \Big\},
\la{eq:v1} \\  
v_{2,u} &=& \frac{-1}{8\pi u y \sin\beta}\int_0^\pi
d\phi_1\,\sin\phi_1\; 
\Big[4uy\sin\beta \sin\phi_1 (u^2+y^2+2uy\cos\beta\cos\phi_1)
\nonumber \\ 
&& \qquad \qquad \qquad + (u^2+y^2)^2\,{\rm Log} \Big] 
\Big\{ \hat{\mathfrak{h}}^{(1)} +
\hat{\mathfrak{h}}^{(2)} - 2\hat{\mathfrak{h}}^{(3)} \Big\} 
\nonumber \\ 
&& - \frac{x}{2\pi u \tan\beta}\int_0^\pi d\phi_1 \,\sin\phi_1\;
 \Big[4uy\sin\beta\sin\phi_1 + (u^2+y^2)\,{\rm Log}\Big] \Big\{-
 \hat{\mathfrak{h}}^{(2)}+\hat{\mathfrak{h}}^{(3)}\Big\} 
\nonumber\\ 
&& - \frac{y}{2\pi u \sin\beta}\int_0^\pi
d\phi_1\,\sin\phi_1\;{\rm Log} \ \Big\{
[-u^2/4]\,\hat{\mathfrak{h}}^{(1)} \nonumber \\
&& \qquad \qquad 
+ [x^2\cos^2\beta - \frac{1}{4}(x^2-2ux\cos\phi_1 +
u^2)]\hat{\mathfrak{h}}^{(2)} - \frac{u}{2}[x\cos\phi_1 -
u]\hat{\mathfrak{h}}^{(3)}\Big\}, 
\la{eq:v2} \\
 v_{3,u} &=& \frac{-x }{2\pi u \tan\beta}\int_0^\pi
 d\phi_1\,\sin\phi_1\;{\rm Log} \ \Big\{[-u^2/2]
 \hat{\mathfrak{h}}^{(1)} \nonumber \\
&& \qquad \qquad +[3x^2/2-xu\cos\phi_1- u^2/2] \hat{\mathfrak{h}}^{(2)}
  +[u^2+ux\cos\phi_1] \hat{\mathfrak{h}}^{(3)} \Big\}
\nonumber \\ 
&& - \frac{1}{4\pi u y \sin\beta} \int_0^\pi d\phi_1\;\sin\phi_1
\Big[4uy\sin\beta\sin\phi_1 + (y^2+u^2)\,{\rm Log}\Big] 
\nonumber \\ 
&& \qquad 
\times \Big\{ 2xu\cos\phi_1 \hat{\mathfrak{h}}^{(1)} - 2 (x^2 - xu\cos\phi_1)
\hat{\mathfrak{h}}^{(2)} 
  + 2(x^2 -2ux\cos\phi_1)\hat{\mathfrak{h}}^{(3)}\Big\}. 
\la{eq:v3}
\ea

Similar to Eq.~(\ref{eq:l_u_integrated}), we define weight functions
$\bar{v}^{(i)}$ after the integration of $v_{i,u}$ over the length of
the vector $u$:
\be
  \bar{v}^{(i)}(x^2, x\cdot y, y^2) = \frac{1}{2}\int_0^\infty du\;
  u^3 \; v_{i,u}(x^2, x\cdot y, y^2), \qquad 
  i=1,2,3.  
\ee

Schematically, we have the following structure,
\be
\overrightarrow{\mathfrak{l}}_u = M(x,y)\;
N\;M(u,x-u)\;\overrightarrow{s}, 
\ee
where $\overrightarrow{\mathfrak{l}}_u$,
$\overrightarrow{s}\in \mathbb{R}^3$, the components of
$\overrightarrow{s}$ are given by Eqs.~(\ref{eq:s1})-(\ref{eq:s3}),
and $M$ is the symmetric $3\times 3$ matrix given in
Eqs.~(\ref{eq:Muv}) and (\ref{eq:Mxy}), which is a function of two
vectors. In particular, $\overrightarrow{\hat{\mathfrak{h}}} =
M(u,x-u)\, \overrightarrow{s}$. The linear operator $N$ corresponds to
the relations (\ref{eq:v1})-(\ref{eq:v3}), $\overrightarrow{v}_u = N\,
\overrightarrow{\hat{\mathfrak{h}}}$.

The linear operator $M(x,y)\; N\;M(u,x-u)$ leads to fairly long
algebraic expressions. We note that the final step, going from the
$\overrightarrow{v}_u$ to the $\overrightarrow{\mathfrak{l}}_u$, is a
purely algebraic one. If there are linear combinations of the
$\overrightarrow{v}_u$ that lead to simpler integrands, those can be
used and the linear combinations can be resolved in terms of the
$\overrightarrow{\mathfrak{l}}_u$ at the end. We find that the
following linear combinations have manageable expressions\footnote{The
  idea is that instead of directly applying the matrix $M(x,y)$ on
  $\overrightarrow{v}_u$, we first triangularize the linear system:
  compute $\bar{\mathfrak{l}}^{(2)}$ and $\bar{\mathfrak{l}}^{(4)} =
  2xy^2 (y\,\cos\beta\,\bar{\mathfrak{l}}^{(2)} + x\;
  \bar{\mathfrak{l}}^{(3)})$ and $\bar v^{(1)} =\frac{3}{4} x^4
  \;\bar{\mathfrak{l}}^{(1)} + x^2 y^2 (\cos^2\beta-\frac{1}{4})
  \;\bar{\mathfrak{l}}^{(2)} + \frac{3}{2} x^3 y \cos\beta\;
  \bar{\mathfrak{l}}^{(3)}$. },
\ba\la{eq:v1bar}
&& \!\! \bar v^{(1)} = \frac{-1}{4\pi y\sin\beta} 
\int_0^\infty du \,u^2\int_0^\pi d\phi_1 \,\sin\phi_1{\,\rm
  Log\,}\cdot \Big(s_1 + s_2 + s_3 \Big), 
\\
&& 
\bar{\mathfrak{l}}^{(4)} \equiv \frac{1}{\sin^2\beta} \left( 
\bar v^{(3)} -\frac{2|y|}{|x|} \cos\beta\; \bar v^{(1)} \right) =
\frac{1}{4\pi|x|\sin^3\beta} 
\int_0^\infty du \, u \int_0^\pi {d\phi_1}
\nonumber \\ 
&& \qquad \qquad \qquad \qquad \qquad \times \Big\{2\sin\beta + \Big(
\frac{u^2+y^2}{2uy}- \cos\beta \cos\phi_1\Big) 
\frac{{\rm \,Log\,}}{\sin\phi_1}\Big\} 
\nonumber\\ 
&& \qquad \qquad \qquad \qquad \qquad \times \Big(2(u\cos\phi_1-x)\;
s_1 + 2u\cos\phi_1\; s_2 + (2u\cos\phi_1-x)\; s_3\Big), 
\la{eq:l4bar} \\ 
&& \bar{\mathfrak{l}}^{(2)} = -\frac{1}{64\pi  x^2 y^5
  \sin^5\beta}\int_0^\infty \frac{du}{u^2}\int_0^\pi  
\frac{d\phi_1}{\sin^3\phi_1}
\nonumber \\ 
&& \qquad \times \Big\{\Big( 3u^4 + 8u^2y^2 + 3y^4 + 4u^2y^2\cos(2\beta) -
6uy(u^2+y^2) \cos(\beta-\phi_1) \nonumber \\
&& \qquad \quad + u^2 y^2 \cos(2(\beta-\phi_1)) + 4u^2y^2\cos(2\phi_1) -
6u^3 y\cos(\beta+\phi_1) - 6uy^3\cos(\beta+\phi_1) \nonumber \\
&& \qquad \quad + u^2y^2
\cos(2(\beta+\phi_1))\Big) \,{\rm Log}\,  
+ 12uy\sin\beta \sin\phi_1(u^2+y^2-2uy\cos\beta
\cos\phi_1) \Big\}
\nonumber\\ 
&& \qquad \times \Big([2u^2 + 3x^2 -6ux\cos\phi_1 + u^2 \cos(2\phi_1)]\;
s_1 + u^2[2+\cos(2\phi_1)]\; s_2 
\nonumber\\ 
&& \qquad \quad + u [-3x\cos\phi_1 + u (2+\cos(2\phi_1))]\; s_3 \Big). 
\la{eq:l2bar} 
\ea
Once $\bar v^{(1)}$,
$\bar{\mathfrak{l}}^{(4)}$ and $\bar{\mathfrak{l}}^{(2)}$ have been
computed, $\bar{\mathfrak{l}}^{(1)}$ and $\bar{\mathfrak{l}}^{(3)}$ are
recovered by taking successively the linear combinations
\ba\la{eq:ell3lc}
\bar{\mathfrak{l}}^{(3)} &=& \frac{1}{2x^2 y^2} 
  \bar{\mathfrak{l}}^{(4)}  
 - \frac{y}{x}\,\cos\beta\;\bar{\mathfrak{l}}^{(2)}
\ea
and 
\ba\la{eq:ell1lc}
\bar{\mathfrak{l}}^{(1)} &=& \frac{4}{3x^4}\Big(\bar v^{(1)} - x^2 y^2
(\cos^2\beta - \frac{1}{4})\; \bar{\mathfrak{l}}^{(2)} 
- \frac{3}{2} x^3 y \cos\beta \;\bar{\mathfrak{l}}^{(3)} \Big).
\ea

Next, we study the behavior of the integrands at large $u$.  The curly
bracket in Eq.~(\ref{eq:l4bar}) is of order $1/u^2$, and the bracket
containing the $s_i$ is at most of order $1/u$, based on the
properties~(\ref{eq:s1x.eq.0})-(\ref{eq:s1primex.eq.0}).  Therefore the
integrand in Eq.~(\ref{eq:l4bar}) is at most of order $1/u^2$ and the
integral is absolutely convergent (numerically, it appears to fall off
even faster, perhaps as $1/u^3$).  Similarly, in Eq.~(\ref{eq:l2bar})
the curly bracket is of order $1/u$, and the bracket containing the
$s_i$ is at most of order unity.  Therefore the integrand of
Eq.~(\ref{eq:l2bar}) is at most of order $1/u^3$ and the integral is
absolutely convergent.

The case of the integrand in Eq.~(\ref{eq:v1bar}) is a bit more subtle.
It is helpful to consider what happens at either $x=0$ or $y=0$ from
the beginning. At $y=0$, starting from Eq.\ (\ref{eq:TENSOR.decomp})
one finds that  
\be\la{eq:TENSORy.eq.0}
 T_{\alpha\beta}(x,0)
 = \frac{4}{3\pi x^2} ~(\hat x_\alpha \hat x_\beta - \frac{1}{4}
 \delta_{\alpha\beta})  \int_0^\infty du\; u \int_0^\pi d\phi_1\,
 \sin^2\phi_1\; (s_1+s_2+s_3). 
\ee
Now, it is obvious that when $x=y=0$, $\int d^4u \; G_0(u)\;
\<(\epsilon_\alpha \epsilon_\beta - \frac{1}{4}\delta_{\alpha\beta})
J(\hat\epsilon, u) J(\hat\epsilon, -u)\>_{\hat\epsilon}$ vanishes,
because $\< u_\alpha u_\beta - \frac{1}{4} u^2
\delta_{\alpha\beta}\>_{\hat u} = 0$.  How does this result emerge
from Eq.~(\ref{eq:TENSORy.eq.0}) ?  We have already seen that
$(s_1+s_2+s_3)$ vanishes to linear order (included) in $|x|$.  One
then finds that the quadratic order does not vanish, however it
vanishes upon performing the $\phi_1$ integral in
Eq.~(\ref{eq:TENSORy.eq.0}), \be \lim_{x\to 0 } \frac{1}{\pi x^2}
\int_0^\pi d\phi_1\, \sin^2\phi_1\; (s_1+s_2+s_3) = 0.  \ee It turns
out that expanding the summand of the $s_i$ in a Taylor series for
small $x$, the Taylor coefficients fall off with increasing powers of
$1/u$. This would be obvious on dimensional grounds if the muon mass
did not enter the expression. On close inspection, the only factor
that could spoil this property are the factors
$z_n(u^2-2|u||x|\cos\phi_1 + x^2)$, which are dimensionless functions
of $m\sqrt{u^2-2|u||x|\cos\phi_1 + x^2}$. However, for large argument
these functions go like $z_n(u^2) \sim \frac{1}{4\pi^2 (n+1) m |u|}$,
so that the dependence on the mass factors out and the dimensional
argument applies; note that for this leading behavior, the sum over
$n$ is still absolutely convergent.  The fact that the Taylor series
of $\int_0^\pi d\phi_1\, \sin^2\phi_1\; (s_1+s_2+s_3)$ at small $|x|$
starts at order $|x|^3$ (at the earliest) thus implies that the
expression is at most of order $|x|^3/(m^2 |u|^3)$ at large $|u|$; the
integrals over $|u|$ in Eq.~(\ref{eq:v1bar}) and in
Eq.~(\ref{eq:TENSORy.eq.0}) is then absolutely convergent.
For the numerical implementation, one option is then to subtract the
$O(x^2)$ term from $(s_1+s_2+s_3)$; this has the advantage of making
the $u$-integrand absolutely convergent in the infrared prior to the
$\phi_1$ integral.

For completeness, let us also treat the case where first $x$ is set
to zero. For $x=0$, we can use Eq.~(\ref{eq:bu2}), the expansion of
the massless propagator in Gegenbauer polynomials in
Eqs.~(\ref{eq:G0_expansion}) and (\ref{eq:G0_d_n}) and the property
(\ref{eq:xisqCn}) to find
\be
T_{\alpha\beta}(0,y)
 = \frac{1}{6} (\hat y_\alpha \hat y_\beta - \frac{1}{4} \delta_{\alpha\beta})
\int_0^\infty du\;u^5\; b(u^2)  \, d_2(u^2,y^2).
\ee

\section{Final convolution integral  via the multipole  expansion of the massless propagator}
\label{sec:alternative_evaluation}

While the method presented in the previous section can be used to numerically calculate the QED weight functions,
certain difficulties arise in special kinematic configurations of the vectors $x$ and $y$.
Especially the regime where $x$ and $y$ are near-collinear can be challenging, in view of the inverse powers of $\sin\beta$
present for instance in Eqs.\ (\ref{eq:g2Final}) or
(\ref{eq:l2bar}). Recall the definition $\cos\beta = \hat x \cdot \hat y$. 

We therefore explore a different method to obtain the QED weight functions numerically.
The idea is to use the multipole expansion of the massless propagator to obtain the 
weight functions in the form of a series of polynomials in $\cos\beta$, times a function 
of $(|x|,|y|)$. The relevant polynomials, as it turns out, are either the Gegenbauer polynomials 
themselves, or their derivatives.

Let $f(\hat x \cdot\hat u)$ be a smooth test function.
Directly integrating the expression $\frac{1}{(u-y)^2}f(\hat x \cdot
\hat u)$ over two of the three angles parametrizing $u$ yields a logarithm
(see Eq.~(\ref{eq:G0f})), which is found in the expressions for
$\bar{\mathfrak{g}}^{(0)}$ (Eq.\ \ref{eq:S_x_y_IR}), $\bar {\mathfrak{g}}^{(3)}$ (Eq.\ \ref{eq:g1lincomb})
and $\bar v^{(1)}$ (Eq.\ \ref{eq:v1bar}). As before we will introduce $\cos\phi_1 = \hat x
\cdot \hat u$. If instead one makes use of the multipole expansion of the propagator, 
as well as of the completeness and orthogonality of the Gegenbauer polynomials, 
one obtains for the same integral a sum over these
polynomials. Matching the two expressions leads to the result
\be\la{eq:S0main}
 -\frac{1}{4|u||y|\sin\beta\sin\phi_1} \,{\rm Log} = \sum_{n=0}^{\infty}  \frac{d_n(u^2,y^2)}{n+1} C_n(\cos\beta) C_n(\cos\phi_1),
\ee
which we will use in the sense of distributions, i.e.\ inserted in an
integral over $\phi_1$. The function $d_n(u^2, y^2)$ has been defined
in Eq.~(\ref{eq:G0_d_n}).  
Further useful results emerge from integrating the expressions $(\hat
u\cdot \hat y)\frac{ f(\hat x \cdot \hat u)}{(u-y)^2}$ 
and $(\hat u\cdot \hat y)^2 \frac{ f(\hat x \cdot \hat u)}{(u-y)^2}$ 
in the two different ways described above, thus leading to the
equalities 
\ba
\la{eq:S1main}
&& S_1 \equiv  -\frac{1}{4|u||y|\sin\beta} \Big(2\sin\beta +
\Big(\frac{u^2+y^2}{2|u||y|} - \cos\beta \cos\phi_1\Big)\,\frac{\rm Log}{\sin\phi_1} \Big)
\nonumber \\ && = 
\sin^2\beta\; \sin^2\phi_1
\sum_{n=1}^{\infty} 
\frac{d_n(u^2,y^2)\,C_n^{\,\prime}(\cos\beta)\;C_n^{\,\prime}(\cos\phi_1)}{n(n+1)(n+2)},
\\
&& \la{eq:S2main}
S_2 \equiv
 -\frac{1}{8u^3y^3\sin\beta\sin\phi_1} \cdot
\nonumber\\ && 
\Big\{\Big( 3u^4 + 8u^2y^2 + 3y^4 + 4u^2y^2\cos(2\beta) - 6uy(u^2+y^2) \cos(\beta-\phi_1) + u^2 y^2 \cos(2(\beta-\phi_1))
\nonumber
\\ && + 4u^2y^2\cos(2\phi_1) - 6u^3 y\cos(\beta+\phi_1) - 6uy^3\cos(\beta+\phi_1) + u^2y^2 \cos(2(\beta+\phi_1))\Big) \,{\rm Log}\, 
\nonumber\\ && + 12uy\sin\beta \sin\phi_1(u^2+y^2-2uy\cos\beta \cos\phi_1) \Big\}
\nonumber\\ &&  = 4 \sin^4\beta\; \sin^4\phi_1
\sum_{n=2}^{\infty} \frac{d_n(u^2,y^2)\,C_n^{\,\prime\prime}(\cos\beta)C_n^{\,\prime\prime}(\cos\phi_1)}
{(n-1)n(n+1)(n+2)(n+3)}.
\ea
Equality (\ref{eq:S1main}) show that two powers of $\sin^2\beta$ can be extracted
explicitly from the angular integrals
for $\bar{\mathfrak{g}}^{(2)}$ (Eq.\ \ref{eq:g2Final}) and $\bar{\mathfrak{l}}^{(4)}$ (Eq.\ \ref{eq:l4bar}).
Similarly, Eq.\ (\ref{eq:S2main}) shows that four powers of $\sin^2\beta$ can be extracted
from the angular integral for $\bar{\mathfrak{l}}^{(2)}$ (Eq.\ \ref{eq:l2bar}), 
allowing one to cancel analytically otherwise numerically problematic inverse powers of $\sin\beta$. In
this way, the case where $x$ and $y$ are exactly collinear can be
calculated directly, without the use of an extrapolation to
$\sin\beta=0$. The price one pays for this cancellation is that
the sum over the derivatives of the Gegenbauer polynomials converges somewhat
less rapidly: for instance, $C_n(1) = n+1$, while $C_n^{\,\prime}(1)= \frac{n}{3}(n+1)(n+2)$.

\subsection{Derivation of Eqs.\ (\ref{eq:S0main})-(\ref{eq:S2main})} 

Let $f(\hat x \cdot \hat u)$ be a smooth test function. As given in Eq.~(\ref{eq:G0f}), explicit integration over
the spherical-coordinate angles $\phi_2$ and $\phi_3$ yields
\ba
\Big\<\frac{1}{(u-y)^2}\; f(\hat x \cdot \hat u)\Big\>_{\hat u}
= -\frac{1}{4 |u||y|\sin\beta} \Big\< \frac{{\rm Log}}{\sin\phi_1} \cdot f(\cos\phi_1)\Big\>_{\hat u}.
\ea
On the other hand, using the expansion of the massless propagator in
Eq.\ (\ref{eq:G0_expansion}) and of the function $f$ in Gegenbauer
polynomials in Eq.~(\ref{eq:f_expansion}), 
as well as the orthogonality property
(\ref{eq:orthogonality_relations}), one finds 
\ba
\Big\<\frac{1}{(u-y)^2}\; f(\hat x \cdot \hat u)\Big\>_{\hat u}
= \sum_{n=0}^{\infty} d_n(u^2,y^2)\; \frac{C_n(\cos\beta)}{n+1}\; \<C_n(\cos\phi_1) f(\cos\phi_1)\>_{\hat u}.
\ea
Comparing the expressions yields Eq.\ (\ref{eq:S0main}).

In the same way, consider the two treatments of the following angular average,
\ba\nonumber
&& \Big\<\frac{\hat u\cdot \hat y - (\hat x\cdot \hat y)(\hat x\cdot \hat u)}{(u-y)^2}\; f(\hat x\cdot \hat u)\Big\>_{\hat u}
\\ && = 
-\frac{1}{4|u||y|\sin\beta} \Big\<\Big(2\sin\beta +
(\frac{u^2+y^2}{2|u||y|} - \cos\beta \cos\phi_1)\,\frac{\rm Log}{\sin\phi_1} \Big) f(\cos\phi_1)\Big\>_{\hat u}
\nonumber
\\ && = \sum_{n=0}^{\infty}\frac{C_n(\cos\beta)}{n+1}\,\Big\<C_n(\cos\phi_1)\Big(\frac{1}{2}(d_{n+1}+d_{n-1})
- \cos\beta \cos\phi_1 d_n\Big) f(\cos\phi_1)\Big\>_{\hat u}.
\ea
We are using the convention $d_n\equiv 0$ for $n<0$ and the argument
of the $d_n$ coefficients is $(u^2,y^2)$ throughout this subsection. 
Comparing the two equations yields an expression for $S_1$, 
\be
 S_1= \sum_{n=0}^{\infty}\frac{C_n(\cos\beta)}{n+1}C_n(\cos\phi_1)
\Big(\frac{1}{2}(d_{n+1}+d_{n-1}) - \cos\beta \cos\phi_1 d_n\Big).
\ee
We can now manipulate the sum in the following way. First apply the
cosine factors on the Gegenbauer polynomials, using
Eq.~(\ref{eq:xiCn}),
\ba
S_1 &=& \sum_{n=0}^{\infty}\frac{1}{n+1}\Big({C_n(\cos\beta)} C_n(\cos\phi_1)
\frac{1}{2}(d_{n+1}+d_{n-1}) 
\\ && -  \frac{d_n}{4} (C_{n+1}(\cos\beta)+C_{n-1}(\cos\beta)) (C_{n+1}(\cos\phi_1)+C_{n-1}(\cos\phi_1)) \Big)
\nonumber
\ea
Next, shift the summation index in the second term so as to factorize $C_n(\cos\phi_1)$, and then collect the 
terms multiplying $d_{n+1}$ and those multiplying $d_{n-1}$, 
\ba
 && S_1= \frac{1}{4}\sum_{n=0}^{\infty} \frac{C_n(\cos\phi_1)}{n+1} 
\Big[ \frac{d_{n+1}}{n+2} \Big((n+3)C_n(\cos\beta) - (n+1)C_{n+2}(\cos\beta)\Big)
\\ && \qquad \qquad \qquad\qquad  +\frac{d_{n-1}}{n}\Big((n-1)C_n(\cos\beta)-(n+1)C_{n-2}(\cos\beta)\Big)\Big].
\nonumber
\ea
Note that  $\frac{d_{n-1}}{n}$ should be interpreted as zero for $n=0$.
Now we notice that the combination appearing in the brackets can be expressed through the derivative of 
the Gegenbauer polynomials, 
\be
(n+2) C_{n-1}(z) - nC_{n+1}(z) = 2(1-z^2) C_n^{\,\prime}(z), \qquad \qquad n\geq 0.
\ee 
Then shifting again the summation index so as to factorize $C_n^{\,\prime}(\cos\beta)$,
we arrive at 
\be
 S_1 = \frac{1-\cos^2\beta}{2} \sum_{n=1}^{\infty} d_n\, \frac{C_n^{\,\prime}(\cos\beta)}{n+1}
\Big(\frac{C_{n-1}(\cos\phi_1)}{n} - \frac{C_{n+1}(\cos\phi_1)}{n+2}\Big)
\ee
(we have used the fact that $C_{n=0}^{\,\prime}(\cos\beta)=0$ to drop the $n=0$ term);
and finally, identifying again the expression for the derivative of $C_n$ in the bracket, we obtain Eq.\ (\ref{eq:S1main}).

Thirdly, one finds by the same method as above
\ba \nonumber
&& \Big\< \frac{(\hat u\cdot \hat y)^2}{(u-y)^2}\;f(\hat x\cdot \hat u)\Big\>_{\hat u}
\\ && 
= -\frac{1}{16 u^3 y^3\sin\beta} \Big\< \Big(4uy\sin\beta\sin\phi_1(u^2+y^2+{2uy}\cos\beta\cos\phi_1)+(u^2+y^2)^2{\rm Log}\Big)
      \frac{f(\cos\phi_1)}{\sin\phi_1}\Big\>_{\hat u}
\nonumber\\ && \la{eq:S2prelim}
= \frac{1}{4}\sum_{n=0}^{\infty} \frac{C_n(\cos\beta)}{n+1} (d_{n-2}+(2-\delta_{n0})d_n+d_{n+2})
\Big\< C_n(\cos\phi_1)\;f(\cos\phi_1)\Big\>_{\hat u},
\ea
which provides us with a new identity. 
The specific linear combination appearing in the QED weight function
$\bar{\mathfrak{l}}^{(2)}$ in Eq.~(\ref{eq:l2bar}) is 
\ba \nonumber
&& S_2\equiv  -\frac{1}{8u^3y^3\sin\beta\sin\phi_1} \cdot
\\ && 
\Big\{\Big( 3u^4 + 8u^2y^2 + 3y^4 + 4u^2y^2\cos(2\beta) - 6uy(u^2+y^2) \cos(\beta-\phi_1) + u^2 y^2 \cos(2(\beta-\phi_1))
\nonumber
\\ && + 4u^2y^2\cos(2\phi_1) - 6u^3 y\cos(\beta+\phi_1) - 6uy^3\cos(\beta+\phi_1) + u^2y^2 \cos(2(\beta+\phi_1))\Big) \,{\rm Log}\, 
\nonumber\\ && + 12uy\sin\beta \sin\phi_1(u^2+y^2-2uy\cos\beta \cos\phi_1) \Big\}
\nonumber\\ && = \sum_{n=0}^{\infty} \frac{C_n(\cos\beta)}{n+1}C_n(\cos\phi_1)
\Big[2d_n \Big(\frac{1}{2}+\cos(2\beta)+\cos(2\phi_1)+\frac{1}{2}\cos(2\beta)\cos(2\phi_1)\Big)
\nonumber\\ && -6\cos\beta\cos\phi_1 (d_{n+1}+d_{n-1}) + \frac{3}{2}(d_{n+2}+(2-\delta_{n0})d_n + d_{n-2}) \Big],
\la{eq:S2prelimB}
\ea
where we have used Eqs.\ (\ref{eq:S0main}), (\ref{eq:S1main}) and (\ref{eq:S2prelim}) in obtaining the second equality.
Manipulations on expression (\ref{eq:S2prelimB}) similar to those yielding Eq.\ (\ref{eq:S1main}), including the use of 
\ba
&& -2\Big((n+3)(n-1)+3\delta_{n0}\Big) C_n(z)
+(n+2)(n+3)C_{n-2}(z)
+n(n-1) C_{n+2}(z)
\nonumber \\ && = 4(1-z^2)^2C_n^{\,\prime\prime}(z), \qquad n\geq 0,
\ea
lead to Eq.\ (\ref{eq:S2main}).

\subsection{Gegenbauer expansion of the QED weight functions}
\la{sec:GegenExp}

Using the relations above,
we provide alternative expressions to compute the six required weight
functions  and their derivatives.
Using Eq.\ (\ref{eq:S0main}), one derives the following representation
for three of the weight functions with argument $(|x|,\hat
c_\beta,|y|)$, where $\hat c_\beta = \cos\beta$, 
\be\la{eq:g0g3v1}
\left(\begin{array}{c} \bar{\mathfrak{g}}^{(0)} \\ \bar{\mathfrak{g}}^{(3)} \\ \bar v^{(1)} \end{array}  \right)
=  \sum_{n=0}^{\infty} C_n(\hat c_\beta) \left[ \frac{1}{|y|^{n+2}}
\left(\begin{array}{c} \alpha^{(0)}_{n-} \\ \alpha^{(3)}_{n-}  \\  \alpha^{(1)}_{n-} \end{array}  \right)
+ |y|^n 
\left(\begin{array}{c} \alpha^{(0)}_{n+}\\ \alpha^{(3)}_{n+} \\ \alpha^{(1)}_{n+}  \end{array}  \right)\right].
\ee
Here the coefficients $\alpha^{(k)}_{n\pm}$, for $k=0,1,3$,  have argument $(|x|,|y|)$, and their functional form is
\ba
 \alpha^{(k)}_{n-}(|x|,|y|) &=& \frac{1-\frac{1}{2}\delta_{k3}}{\pi(n+1)} \int_0^{|y|}\!\!\! u^{n+3} du
\int_0^\pi\! d{\phi_1}\,\hat s_1^2
\;C_n(\hat c_1)\; \sigma_k(|x|,\hat c_1,|u|), \qquad
\\
 \alpha^{(k)}_{n+}(|x|,|y|) &=& \frac{1-\frac{1}{2}\delta_{k3}}{\pi(n+1)} \int_{|y|}^\infty  \frac{du}{u^{n-1}}
\int_0^\pi d{\phi_1}\,  \hat s_1^2 \;C_n(\hat c_1)\; \sigma_k(|x|,\hat
c_1,|u|), 
\ea
where $\hat c_1 = \cos\phi_1$ and $\hat s_1 = \sin\phi_1$. 
The functions $\sigma_k(|x|,\hat c_1,|u|)$ appearing in the coefficients $\alpha^{(k=0,1,3)}_{n\pm}$, 
as well as those appearing in $\beta^{(k=2,4)}_{n\pm}$ and
$\gamma^{(2)}_{n\pm}$ below, 
are given explicitly at the end of this subsection, Eqs.\
(\ref{eq:sigma0final})-(\ref{eq:sigma5final}). 
As a remark, we have already noted that $\bar{\mathfrak{g}}^{(0)}$ contains a logarithmic infrared divergence.
In the present representation, that divergence is entirely contained in the coefficient $\alpha^{(0)}_{0+}$,
which makes a constant contribution to $\bar{\mathfrak{g}}^{(0)}$, independent of $x$ and $y$. Since only derivatives
of $\bar{\mathfrak{g}}^{(0)}$ with respect to $x$ or $y$ appear in the QED kernel, 
the coefficient $\alpha^{(0)}_{0+}$ is never actually needed.

Next, starting from Eqs.\ (\ref{eq:g2Final}) and (\ref{eq:l4bar}), 
and using Eq.\ (\ref{eq:S1main}), one obtains the representation
\be\la{eq:g2l4}
 \left(\begin{array}{c} \bar{\mathfrak{g}}^{(2)} \\ \bar{\mathfrak{l}}^{(4)}/y^2  \end{array}  \right)
= \sum_{n=1}^{\infty} C_n^{\,\prime}(\hat c_\beta) \left[
\frac{1}{|y|^{n+3}} \left(\begin{array}{c} \beta^{(2)}_{n-} \\ \beta^{(4)}_{n-}  \end{array}  \right)
+ |y|^{n-1}  \left(\begin{array}{c} \beta^{(2)}_{n+} \\ \beta^{(4)}_{n+} \end{array}  \right)\right].
\ee
with (for $k=2,4$)
\ba
\beta^{(k)}_{n-}(|x|,|y|) &=&  \frac{-(1-\frac{1}{2}\delta_{k2})}{\pi\, n(n+1)(n+2) } \int_0^{|y|} du\, u^{n+3}  
 \int_0^\pi d\phi_1 \;\hat s^2_1\; C_n^{\,\prime}(\hat c_1) \;\sigma_k(|x|,\hat c_1,|u|), \qquad \quad
\\
\beta^{(k)}_{n+}(|x|,|y|) &=&  \frac{-(1-\frac{1}{2}\delta_{k2})}{\pi \, n(n+1)(n+2)} \int_{|y|}^\infty \frac{du}{u^{n-1}} 
 \int_0^\pi d\phi_1 \;\hat s^2_1\; C_n^{\,\prime}(\hat c_1) \;\sigma_k(|x|,\hat c_1,|u|).
\ea
Finally, using Eqs.\ (\ref{eq:l2bar}) and (\ref{eq:S2main}), we obtain the form
\ba\la{eq:l2multip}
\bar{\mathfrak{l}}^{(2)}(|x|,\hat c_\beta,|y|) &=& \sum_{n=2}^{\infty} C_n^{\,\prime\prime}(\hat c_\beta) 
\Big(\frac{1}{|y|^{n+4}} \gamma^{(2)}_{n-}(|x|,|y|) +|y|^{n-2}\gamma^{(2)}_{n+}(|x|,|y|) \Big),
\\
\gamma^{(2)}_{n-}(|x|,|y|) &=& \frac{1}{2\pi \,(n-1)\,n\,(n+1)(n+2)(n+3)} \cdot
\\ && \cdot 
\int_0^{|y|} du\; u^{n+3} \int_0^\pi d\phi_1\; \hat s^2_1\; C_n^{\,\prime\prime}(\hat c_1)\;\sigma_5(|x|,\hat c_1,|u|)
\nonumber
\\
\gamma^{(2)}_{n+}(|x|,|y|) &=& \frac{1}{2\pi \,(n-1)\,n\,(n+1)(n+2)(n+3)} \cdot
\\ && \cdot 
\int_{|y|}^\infty \frac{du}{u^{n-1}}\int_0^\pi d\phi_1\; \hat s^2_1\; C_n^{\,\prime\prime}(\hat c_1)\;\sigma_5(|x|,\hat c_1,|u|).
\nonumber
\ea

We now give the explicit expressions for the sums $\sigma_k(|x|,\hat c_1,|u|)$, $k=0,\dots,5$.
They involve the modified Bessel functions and the Gegenbauer polynomials. The argument of the $C_n$ and $C_n^{\,\prime}$
polynomials is always $(\hat u\cdot\widehat{x-u})$; this is only indicated explicitly in the relatively compact expression for $\sigma_0$,
which in fact coincides with $s(x,u)$.
For the weight functions expanded in $C_n(\hat c_\beta)$ in
Eq.~(\ref{eq:g0g3v1}), the sums are 
\ba\la{eq:sigma0final}
\sigma_0 &=& \sum_{n=0}^\infty z_n(u^2)\, z_n((x-u)^2)\, \frac{C_n(\hat u\cdot\widehat{x-u})}{n+1},
\\  \la{eq:sigma3final}
\sigma_3 &=& \frac{1}{|x|^2}\sum_{n=0}^\infty 
\Big\{ z_n(u^2) z_{n+1}((x-u)^2) \Big[ |x-u| \frac{C_n}{n+1} + |u|\frac{C_{n+1}}{n+2}\Big]
 \\ && \qquad 
  +  z_{n+1}(u^2) z_{n}((x-u)^2) \Big[|u|\frac{C_n}{n+1} +  |x-u| \frac{C_{n+1}}{n+2} \Big]\Big\}
\nonumber
\\  &=& \frac{1}{|x|}\sum_{n=0}^\infty
\Big\{ \frac{z_n(u^2) z_{n+1}((x-u)^2)}{n+2} \Big[\frac{|x|\hat s^2_1}{|x-u|}\,
\frac{C_{n+1}^{\,\prime}}{n+1} + \hat c_1 C_{n+1} \Big]
\\ && 
+  \frac{z_{n+1}(u^2) z_{n}((x-u)^2)}{n+1} \Big[- \frac{|x|\hat s^2_1}{|x-u|} \frac{C_n^{\,\prime}}{n+2}+\hat c_1 C_n\Big] \Big\},
\nonumber
\ea
and 
\ba\la{eq:sigma1final}
&& \sigma_1 \equiv s_1+s_2+s_3 =
 -\frac{x^2}{4} \sum_{n=0}^{\infty} z_n(u^2) z_n((x-u)^2) \frac{C_n}{n+1}
\\ && + \frac{u^2}{4} \sum_{n=0}^{\infty} z_n((x-u)^2)\Big(z_{n-2}(u^2) + (2-\delta_{n0})z_n(u^2) + z_{n+2}(u^2)\Big)
\frac{C_n}{n+1}
\nonumber\\ 
&& + \frac{(x-u)^2}{4} \sum_{n \geq 0} z_n(u^2)\Big( z_{n-2}((x-u)^2) + (2-\delta_{n0})z_n((x-u)^2) + z_{n+2}((x-u)^2)\Big)
\frac{C_n}{n+1}
\nonumber\\ && 
+ \frac{|u||x-u|}{4}\sum_{n=0}^{\infty} \Big\{
 \Big(z_n(u^2) z_{n+2}((x-u)^2)+2 z_n(u^2)z_n((x-u)^2)  + z_{n+2}(u^2)z_n((x-u)^2)\Big) \frac{C_{n+1}}{n+2}
\nonumber\\ &&  + \Big(z_{n-2}(u^2) z_{n}((x-u)^2)+2 z_n(u^2)z_n((x-u)^2) + z_{n}(u^2)z_{n-2}((x-u)^2)\Big) 
\frac{C_{n-1}}{n} \Big\}.
\nonumber
\ea
For the weight functions expanded in $C_n^{\,\prime}(\hat c_\beta)$ in
Eq.~(\ref{eq:g2l4}), the sums are
\ba\la{eq:sigma2final}
\sigma_2 &=& \frac{1}{|x|}\sum_{n=0}^\infty 
\Big\{ z_n(u^2) z_{n+1}((x-u)^2) \Big[|x-u| \hat c_1 \frac{C_n}{n+1}
  + (|u|\hat c_1 - |x|) 
\frac{C_{n+1}}{n+2} \Big] \qquad 
\\ && 
+ z_{n+1}(u^2) z_n((x-u)^2) \Big[(|u|\hat c_1-|x|) \frac{C_n}{n+1}+
  |x-u| \hat c_1\frac{C_{n+1}}{n+2}\Big]\Big\}
\nonumber
\\ &=&  \hat s^2_1 \sum_{n=0}^\infty
\Big\{ \frac{z_n(u^2) z_{n+1}((x-u)^2)}{n+2} \Big[\frac{|x| \hat c_1}{|x-u|} \frac{C_{n+1}^{\,\prime}}{n+1}-C_{n+1} \Big]
\\ && 
- \frac{z_{n+1}(u^2) z_n((x-u)^2)}{n+1} \Big[ \frac{|x|\hat c_1}{|x-u|}\, \frac{C_{n}^{\,\prime}}{n+2} +C_n\Big]\Big\}.
\nonumber
\ea
and 
\ba
\sigma_4 &=&  
\frac{1}{|u|}\Big((2\frac{|u|}{|x|}\hat c_1 - 1)\sigma_1  + s_2-s_1\Big).
\la{eq:sigma4final}
\ea
Finally, the sum appearing in $\bar{\mathfrak{l}}^{(2)}$  is 
\ba
\sigma_5 &=&  
\frac{1}{u^2}\Big(\frac{|u|}{x^2}(|u|(1+2\hat c_1^2) - 3|x|\hat c_1) \sigma_1  + 3\frac{|u|}{|x|}\hat c_1(s_2-s_1) + 3 s_1\Big).
\la{eq:sigma5final}
\ea
The sums $s_1$ and $s_2$ are evaluated as indicated in Eqs.\
(\ref{eq:s1}) and (\ref{eq:s2}).

For numerical purposes, it is usually preferable to evaluate the derivative of an
integrand with respect to a parameter before the integral is performed numerically.
Therefore, for completeness, we provide in appendix \ref{sec:xderivs} the expressions
of the $|x|$-derivative of the sums defining the weight functions, namely
the $\{\sigma_k\}_{k=0}^5$ as well as $s_1$ and $s_2$.
In the expressions provided in appendix, the first two derivatives of the functions $z_n(u^2)$ appear.
Thanks to Eq.\ (\ref{eq:znODE}),  they can be computed practically in an iterative fashion as follows,
\ba
\frac{\partial z_n}{\partial |u|} &=& \frac{n+1}{4\pi^2m u^2}\,{K_0(mu)I_{n+1}(m|u|)} - \frac{z_n(u)}{|u|},
\\
\frac{\partial^2 z_n}{\partial |u|^2} 
 &=& -\frac{3}{u} \Big(\frac{\partial z_n}{\partial |u|}\Big) + \frac{n(n+2)}{u^2}\,z_n(u)
- \frac{n+1}{2\pi^2u^2}\, K_1(m|u|) I_{n+1}(m|u|) .
\ea

Appendix \ref{sec:smallxy} provides  the relevant expressions to obtain the QED weight functions and the full kernel
at $x=0$ or at $y=0$. The motivation for investigating these special cases is twofold.
First, the expressions simplify compared to the general case in that they have one fewer integral or infinite sum.
Thus, their evaluation is significantly faster and provides a cross-check for the numerics of the general case,
which should approach the special cases in the appropriate limits.
Second, when we consider modifications of the QED kernel via subtractions in section \ref{sec:KernelSubtractions}, 
the kernel at $x=0$ or $y=0$ will be needed explicitly.
The QED kernel for $y=x$ can be obtained from the case $y=0$ using the property (\ref{eq:Lx,x-y}).

\section{Numerical evaluation of the QED kernel}
\label{sec:discussion_master_formula}

The basic idea of our approach is to precompute and store the weight
functions, since by O(4) symmetry they are functions of three
variables. This stands in stark contrast with the QED kernel itself,
which is a function of eight real variables and has 384 independent
components. Since up to two derivatives with respect to components of $x$ and $y$ act
on the tensors $S(x,y)$, $V_\delta(x,y)$ and $T_{\beta\delta}(x,y)$, chain rules are used
to convert these derivatives, when they act on the weight functions,
into derivatives with respect to the variables $(|x|,\,\hat c_\beta\equiv \hat x\cdot \hat y,\,|y|)$,
for instance
\be
\partial_\alpha^{(x)} = \hat x_\alpha \frac{\partial}{\partial |x|} 
+ \frac{1}{|x|} (\hat y_\alpha - \hat c_\beta \hat x_\alpha) \frac{\partial}{\partial \hat c_\beta}.
\ee
The chain-rule based expressions for the tensors $T^{\rm A}_{\alpha\beta\delta}(x,y)$ in terms of
the weight functions are given in appendix~\ref{sec:chainrT} ($A=\;$I, II, III).
With these rank-three tensors at hand, and with the Dirac traces ${\cal G}^{A}$ computed upon initialization,
the QED kernel is obtained via a simple linear combination, Eq.\ (\ref{eq:Lbar}).

We have pursued two strategies to numerically compute the
QED weight function.  The first is based on Eqs.\ (\ref {eq:g0direct},
\ref{eq:g2Final}, \ref{eq:g1lincomb}, \ref{eq:v1bar}, \ref{eq:l4bar},
\ref{eq:l2bar}), followed by taking the appropriate linear
combinations. In this strategy, the weight functions are computed on a
three-dimensional grid, each direction representing one of the
variables $|x|$, $\hat c_\beta$ and $|y|$.
While we will not describe this implementation in detail (see~\cite{NilsThesis} for more information),
it is worth mentioning that the logarithm appearing in each of the six equations
referenced above required a dedicated treatment in the regions where its argument vanishes.
The second strategy,
which is the one we opted for in our subsequent tests and lattice QCD
calculations, consists in calculating the coefficients of the
weight-function expansion in Gegenbauer polynomials according to
Eqs.\ (\ref{eq:g0g3v1}, \ref{eq:g2l4}, \ref{eq:l2multip}). The
coefficients are functions of $|x|$ and $|y|$ and carry an index
corresponding to the order of the polynomial in $\hat c_\beta$
which they multiply.  Implementing both strategies with two
independent codes allowed us to have a valuable cross-check of our
results. 
In the following, we describe a sample of the results
obtained with the second strategy and the most important technical
aspects involved in the numerical calculation.
It is worth mentioning at this point the order-of-magnitude computational cost
of precomputing the required weight functions provided at~\cite{KQEDcode}:
it amounted in total to about three weeks on a dual-core laptop.

The three tensor weight functions as well as the derivatives of the
scalar weight function appearing in the kernel $\bar{\cal
  L}_{[\rho\sigma];\mu\nu\lambda}(x,y)$ for given values of $|x|$ and
$\hat c_\beta$ are displayed in
Fig.~\ref{fig:dg0l123}. The result of the numerical integration is
shown as a curve. At $y=0$, we confront the numerical results with
the Taylor expansion of the weight functions obtained in appendix
\ref{sec:smallxy} and observe good agreement~\footnote{For
  $\bar{\mathfrak{l}}^{(2)}$, we did not derive a prediction at $y=0$
because it is not needed for the QED kernel.}.
In addition, the large-$|y|$ asymptotics for the
derivatives of the scalar function $\bar{\mathfrak{g}}^{(0)}$ are determined
in appendix~\ref{sec:largeYasympt} and displayed for $m |y|>5$ in
the three left panels of Fig.~\ref{fig:dg0l123}.
Similarly, Fig.~\ref{fig:g1g2} shows the two vector weight functions.  
The scalar and tensor weight functions have unit of ${\rm GeV}^{-2}$, while 
the vector ones have unit of ${\rm GeV}^{-1}$. It is natural to use the muon mass
to build dimensionless combinations. We note that
all weight functions are smooth functions of $|y|$, and that they have
rather different magnitudes in units of the muon mass.  The scalar and the first vector weight
function are largest, the other weight functions being at least an
order of magnitude smaller. We have found this hierarchy to be fairly
generic. For the reader's convenience, we have collected a few numerical values
of the weight functions in Table~\ref{tab:tabWF}.
Quantitative checks have been performed against the weight functions resulting from the first computational
strategy described in the previous paragraph.
For instance, at the reference point $m_\mu|x|=0.436$, $\hat c_\beta=-0.59375$ and $m_\mu|y|=0.654$,
all derivatives required for the QED kernel have been compared; with $m_\mu$ set to unity,
the largest absolute difference was found in
$\frac{\partial^2 g_1}{\partial |x|\partial \hat c_\beta}$, and amounted to $7\times 10^{-9}$.

We next describe some of the numerical techniques we have used to
arrive at the results presented in Fig.~\ref{fig:dg0l123} and
\ref{fig:g1g2}. We first note that we have worked in double precision throughout,
and have not found it necessary to employ further enhanced arithmetic precision.
In the representation of Eqs.\ (\ref{eq:g0g3v1}, \ref{eq:g2l4}, \ref{eq:l2multip}) of the weight functions, one
has to carry out a two-dimensional integral of an integrand which is
represented as an infinite sum over products of modified Bessel
functions and a Gegenbauer polynomial or its derivatives.  One
integration variable represents the angle between the position vectors
$x$ and $u$, the other is the norm of $u$.  The most important
numerical task is thus to evaluate efficiently a sum involving the
modified Bessel functions and the Gegenbauer polynomial or its derivatives. 

We have evaluated strings of modified Bessel functions (e.g. $K_0$,
$K_1$, \dots, $K_{n_{\rm max}}$) using routines inspired by those given
  in~\cite{NR}. The most important aspect is that the modified Bessel
  functions of the second kind ($K_n$) can be evaluated using the
  recursion relation among them in the direction of increasing index
  $n$, while those of the first kind ($I_n$) must be evaluated in a
  downward recursion, starting from a sufficiently large $n$. Since we
  need these functions for a wide range of $n$, we store them on the
  fly during the recursion.

Next, we use the Clenshaw algorithm (see for instance~\cite{NR}) to perform the sum,
exploiting the recursion relations 
\ba \label{eq:clen1}
C_{n+1}(z) &=& 2z C_n(z) - C_{n-1}(z) \qquad\qquad\qquad (n\geq 1),
\\
C_{n+1}^{\,\prime}(z) &=& 2z\, \frac{n+1}{n} C_n^{\,\prime}(z) - \frac{n+2}{n} C_{n-1}^{\,\prime}(z)
\qquad (n\geq 1), \label{eq:clen2}
\\
C_{n+1}^{\,\prime\prime}(z) &=& 2z\,\frac{n+1}{n-1}  C_n^{\,\prime\prime}(z) - \frac{n+3}{n-1} C_{n-1}^{\,\prime\prime}(z)
\qquad  (n\geq 2)\label{eq:clen3}
\ea
among the polynomials. Thus none of the $C_n$, $C_n^{\,\prime}$ and
$C_n^{\,\prime\prime}$ are evaluated explicitly in the calculation of the sums.

We have performed the integration using the integrator {\tt
  cubature}~\cite{cubature}.  This integrator is able to perform numerical
integrals on a multi-dimensional rectangular region.  For calculating
the weight functions, we have mostly used the $p$-adaptive cubature
routine, which uses a tensor product of Clenshaw-Curtis quadrature
rules; the degree of the rules is doubled along each dimension until
convergence is achieved. An advantage of the {\tt cubature} package is
that it allows for a vector of integrands. Since it is the different
coefficients $\alpha_m^{(k)}(|x|,|y|)$ ($k=0,1,3$),
$\beta_m^{(k)}(|x|,|y|)$ ($j=2,4$) and $\gamma^{(2)}_m(|x|,|y|)$ that
are being calculated, and that they all involve the same sums (one of
the $\sigma_i$ ($i=0,\dots,5$)), only the Gegenbauer polynomial
$C_m(\hat u\cdot \hat x)$ (or its derivative) must be reevaluated for
different values of the index $m$. This saves a significant amount of
computations.  Because the Gegenbauer polynomial is strongly
oscillating for large $m$, the calculation of the large-order
coefficients to a given relative precision dominates the computing
time.

We have turned the integral over $|u|$ from $|y|$ to $\infty$ into an
integral from $0$ to $1/|y|$ by making the change of variables $u_1 =
1/|u|$. Given that we want to compute the coefficients of the Gegenbauer-polynomial expansion for all $|y|$,
a considerable amount of computing time is saved by simply observing that 
\ba
 \alpha^{(k)}_{m-}(|x|,|y|+\Delta|y|)  &=&  \alpha^{(k)}_{m-}(|x|,|y|) 
\\ && + \frac{1-\frac{1}{2}\delta_{k3}}{\pi(m+1)} \int_{|y|}^{|y|+\Delta |y|}\!\!\! u^{m+3} du
\int_0^\pi\! d{\phi_1}\,\hat s_1^2
\;C_m(\hat c_1)\; \sigma_k(|x|,\hat c_1,|u|),
\nonumber\\
 \alpha^{(k)}_{m+}(|x|,|y|-\Delta|y|) &=&  \alpha^{(k)}_{m+}(|x|,|y|) 
\\ && + \frac{1-\frac{1}{2}\delta_{k3}}{\pi(m+1)} \int_{|y|-\Delta|y|}^{|y|}  \frac{du}{u^{m-1}}
\int_0^\pi d{\phi_1}\,  \hat s_1^2 \;C_m(\hat c_1)\; \sigma_k(|x|,\hat c_1,|u|),
\nonumber
\ea
and similarly for the other coefficients. In this way, the cost of
increasing the resolution $1/\Delta|y|$ with which the coefficients
are computed is very low. We also recall that the derivative of the coefficients
with respect to $|y|$ are needed to obtain the QED kernel.
We have used the fact that one can easily obtain these $|y|$-derivatives analytically at practically zero
computational cost. Indeed,  since 
\be
\frac{1}{|y|^{m+2}} \frac{\partial}{\partial|y|}\alpha^{(k)}_{m-}(|x|,|y|) + |y|^m \frac{\partial}{\partial|y|} \alpha^{(k)}_{m+}(|x|,|y|) = 0,
\ee
we have for instance
\be
\frac{\partial}{\partial|y|}\bar{\mathfrak{g}}^{(0)} = \sum_{m\geq 0} C_m(\hat c_\beta) 
\left[-(m+2) \frac{1}{|y|^{m+3}} \alpha^{(0)}_{m-}  + m|y|^{m-1}  \alpha^{(0)}_{m+} \right],
\ee
and similar expressions apply to all six weight functions. The derivatives with respect to $\hat c_\beta$
are of course simply obtained by analytically deriving the Gegenbauer polynomial. 
Finally, the $|x|$ dependence of the weight functions appears only in the sums $\sigma_k(|x|,\hat c_1,|u|)$.
Their first $|x|$-derivative is given analytically in Appendix~\ref{sec:xderivs}.
Only the second $|x|$-derivative of the sums $\sigma_2$ and $\sigma_3$
was computed numerically by taking a finite difference (with step size $m\delta|x|= 10^{-3}$) of the analytically obtained first derivative.

\begin{figure}[t]
\centerline{\includegraphics[width=0.5\textwidth]{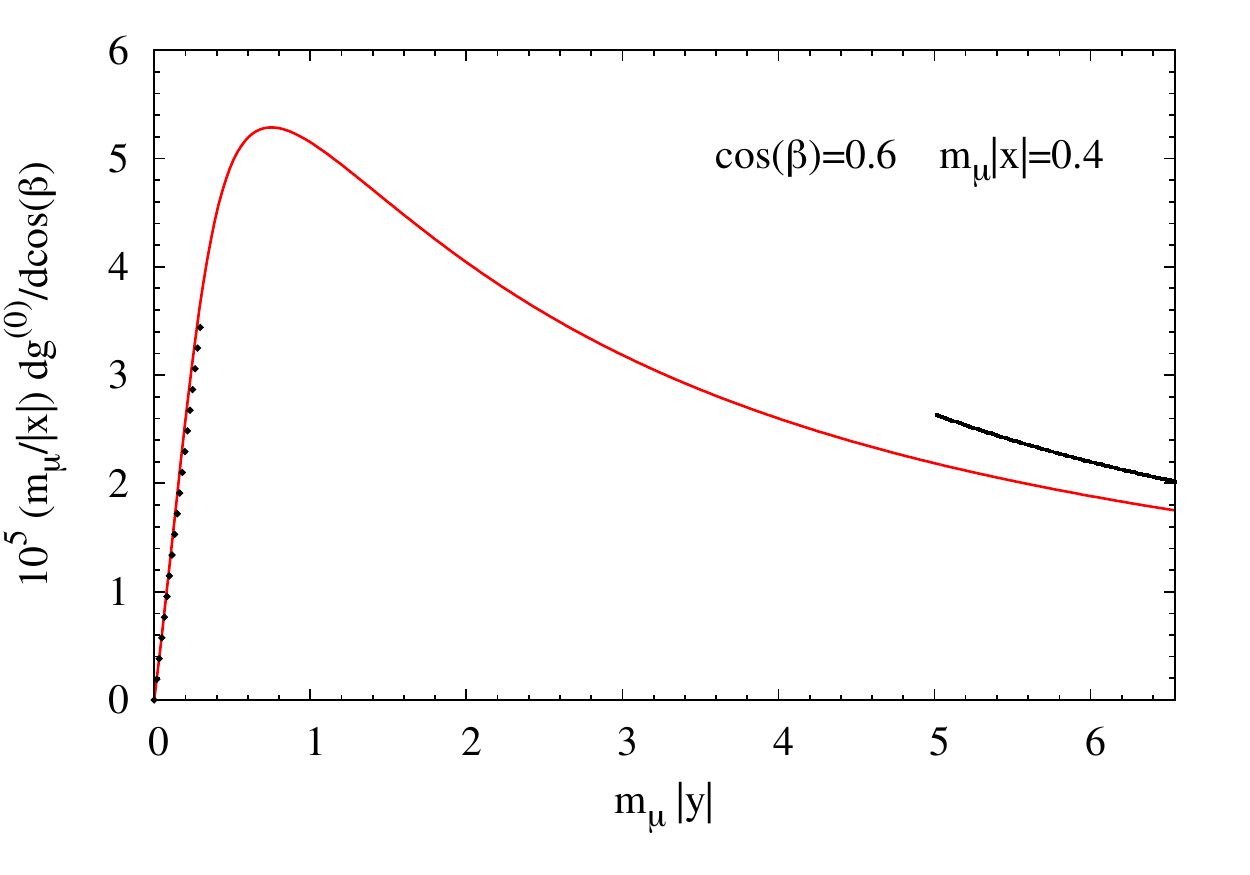}\includegraphics[width=0.5\textwidth]{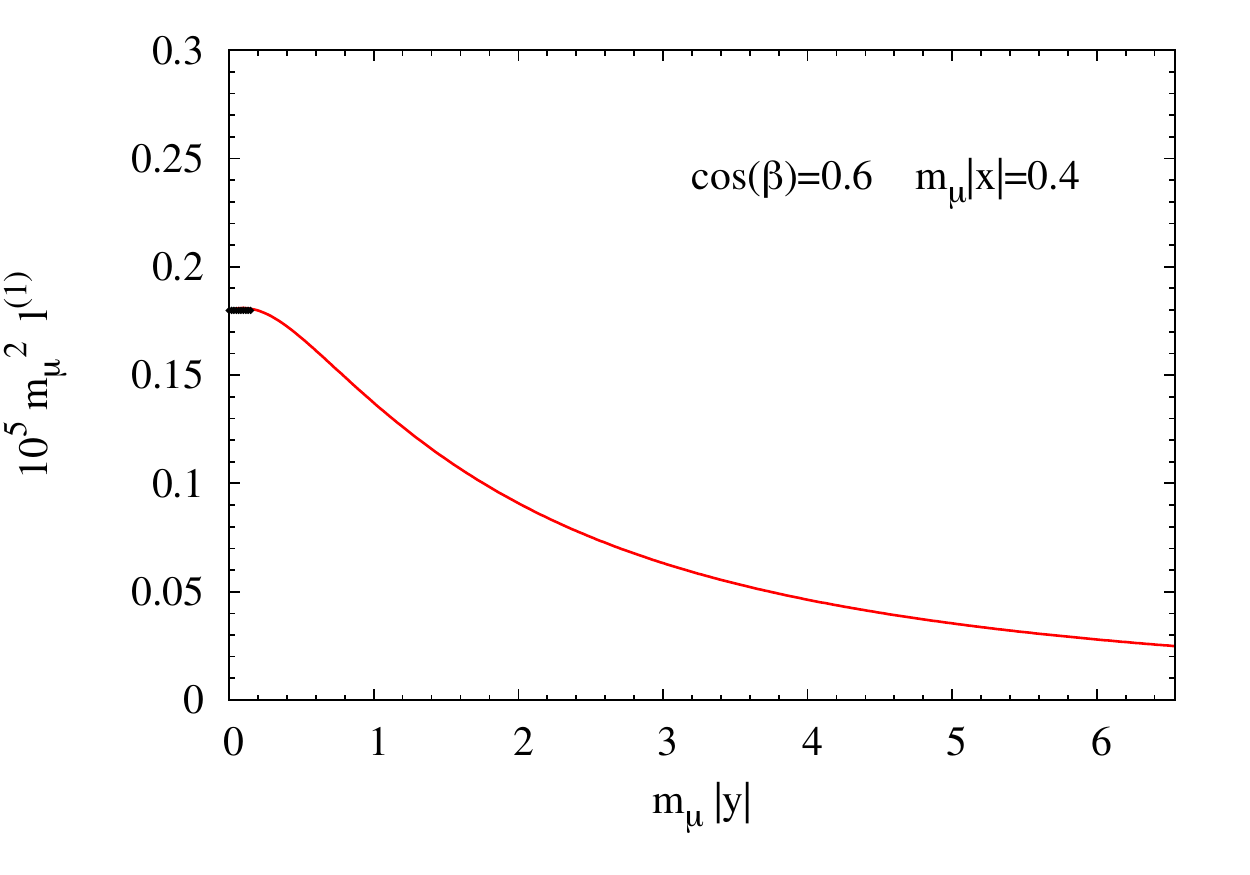}}
 \centerline{\includegraphics[width=0.5\textwidth]{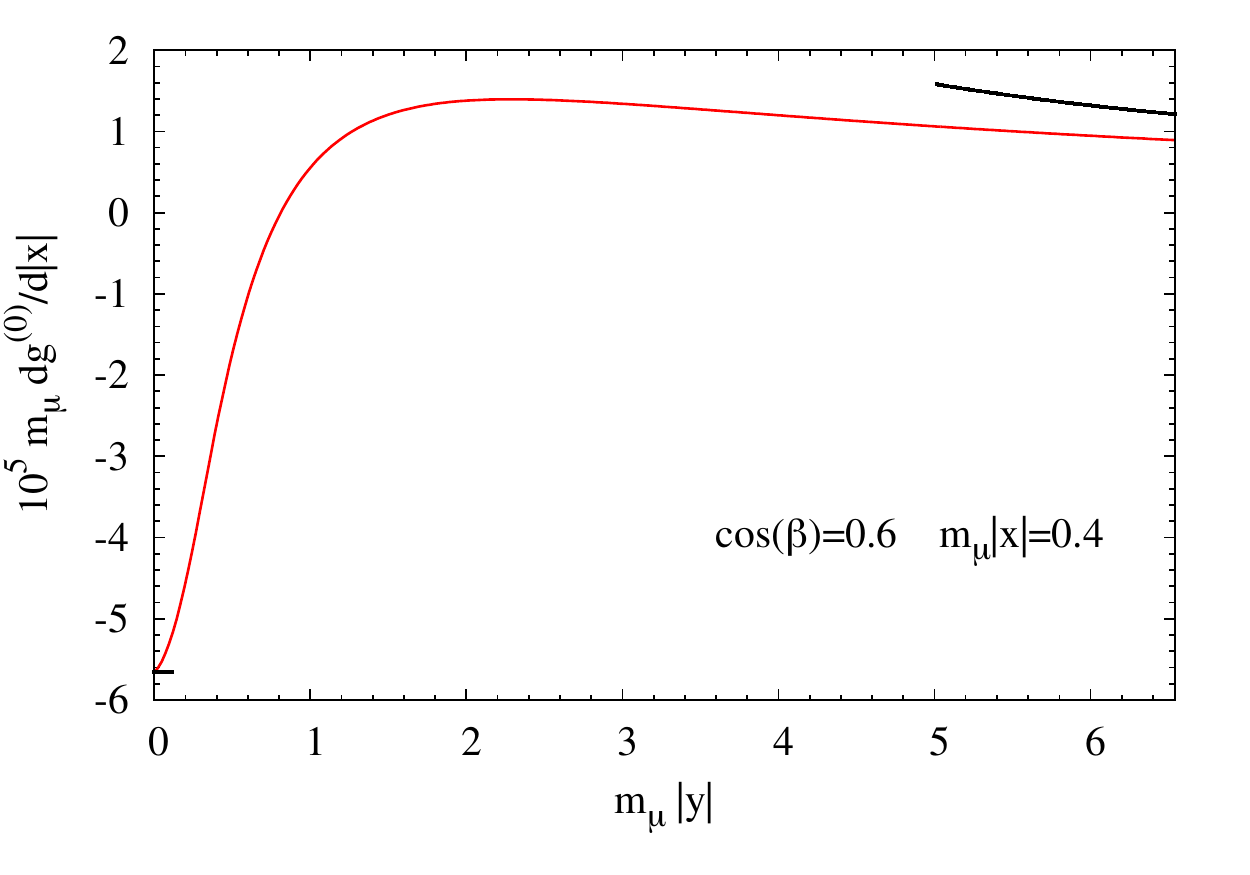}\includegraphics[width=0.5\textwidth]{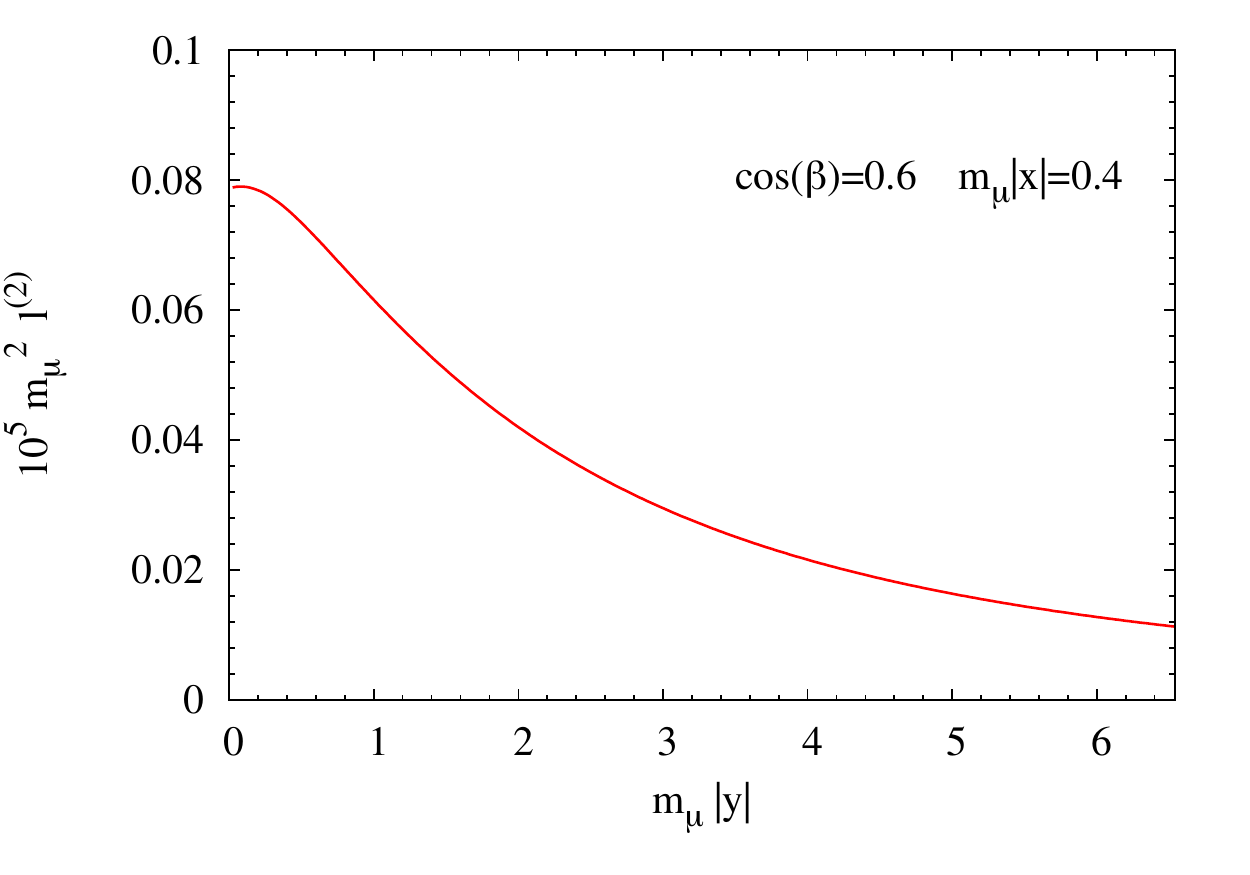}}
 \centerline{\includegraphics[width=0.5\textwidth]{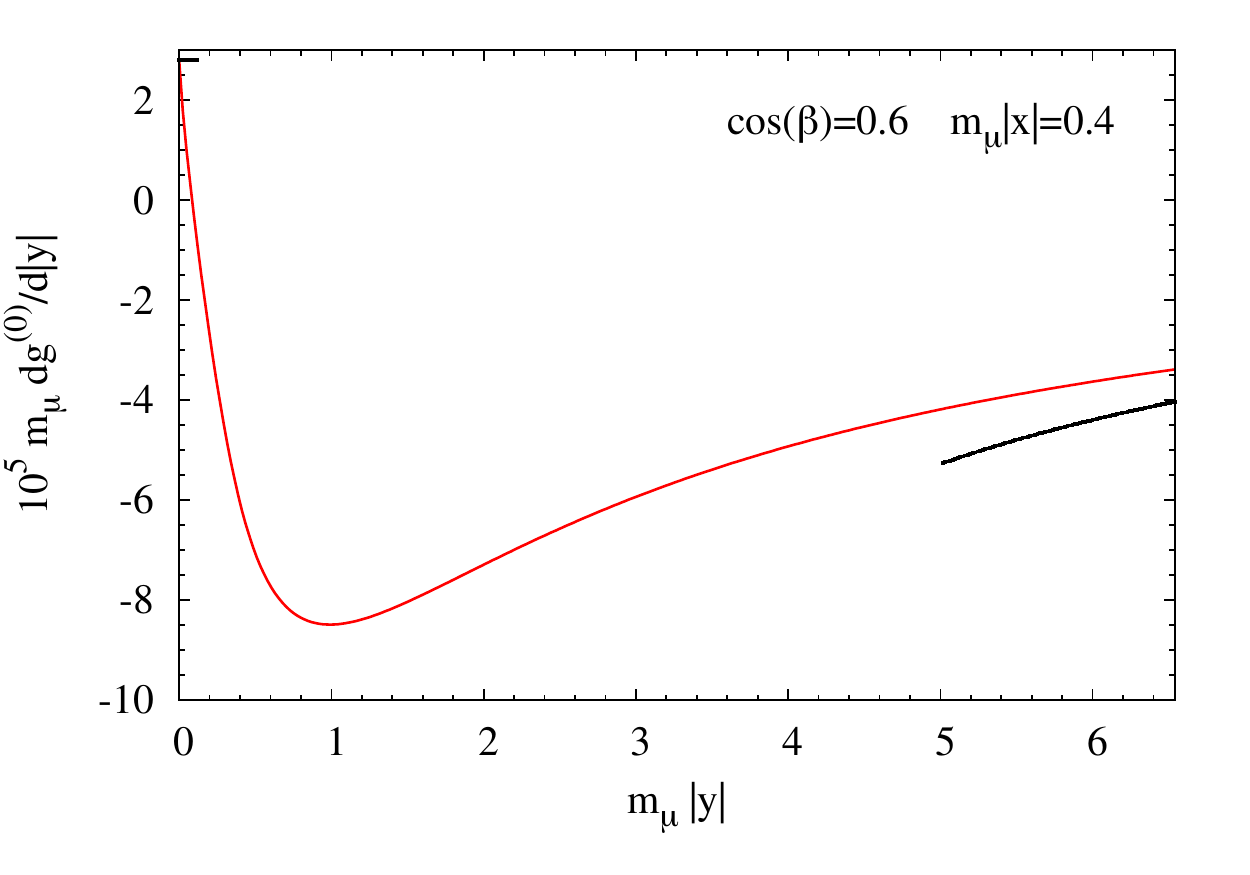}\includegraphics[width=0.5\textwidth]{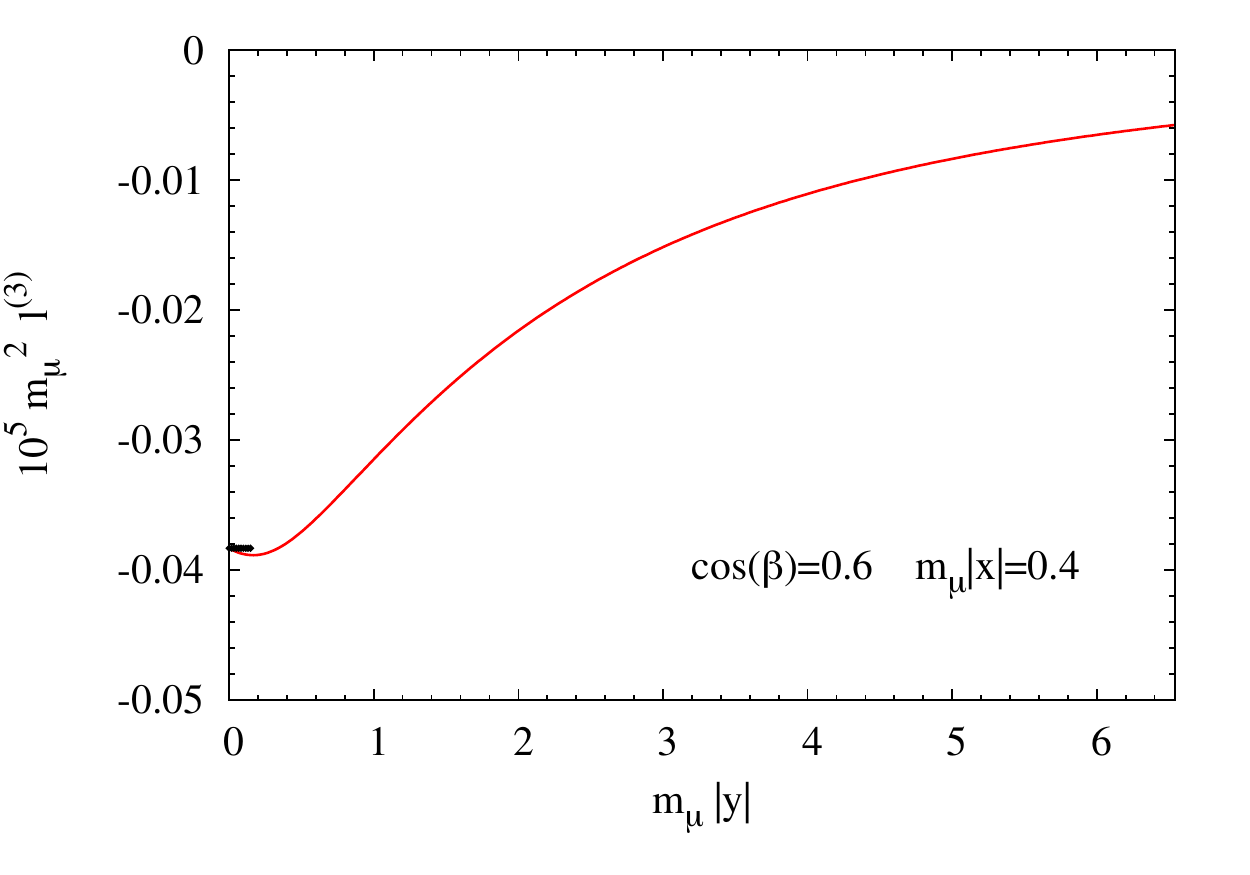}}
 \vspace{-0.2cm}
\caption{Left panels: the derivatives of the scalar weight function $\bar{\mathfrak{g}}^{(0)}$ appearing 
  in the QED kernel for given values of $|x|$ and $\cos \beta\equiv \hat x\cdot \hat y$.
  The analytic predictions for the large-$y$ asymptotics are displayed as thick solid lines.
  Right panels: the three tensor weight functions $\bar{\mathfrak{l}}^{(1)}$, $\bar{\mathfrak{l}}^{(2)}$, $\bar{\mathfrak{l}}^{(3)}$.
  The leading term of the Taylor expansion around the origin, which is O$(|y|^0)$ except for the top left panel,
  is indicated for all cases but $\bar{\mathfrak{l}}^{(2)}$; note that the latter weight function is multiplied by a tensor of order $|y|^2$
  in the QED kernel.
\la{fig:dg0l123}}
\end{figure}

\begin{figure}[t]
\centerline{\includegraphics[width=0.5\textwidth]{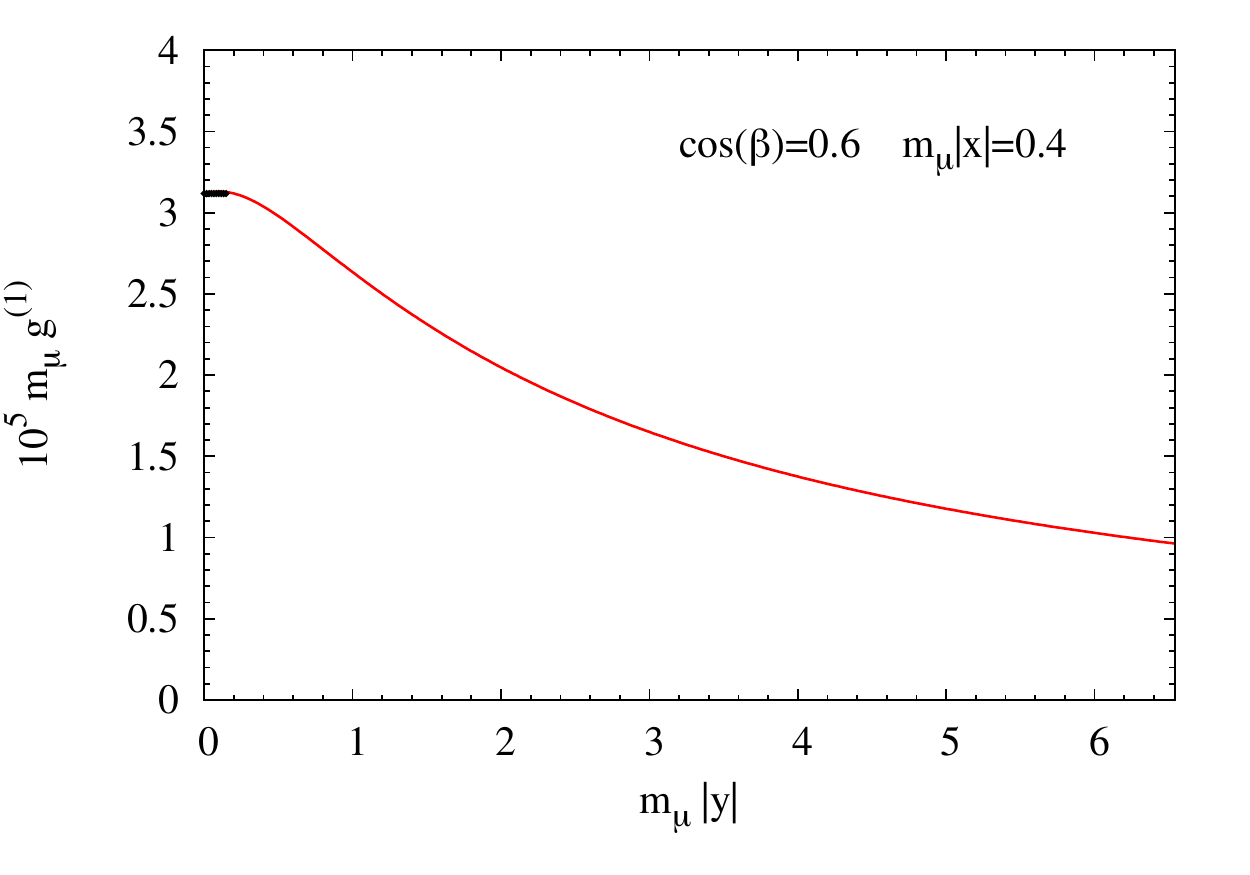}
  \includegraphics[width=0.5\textwidth]{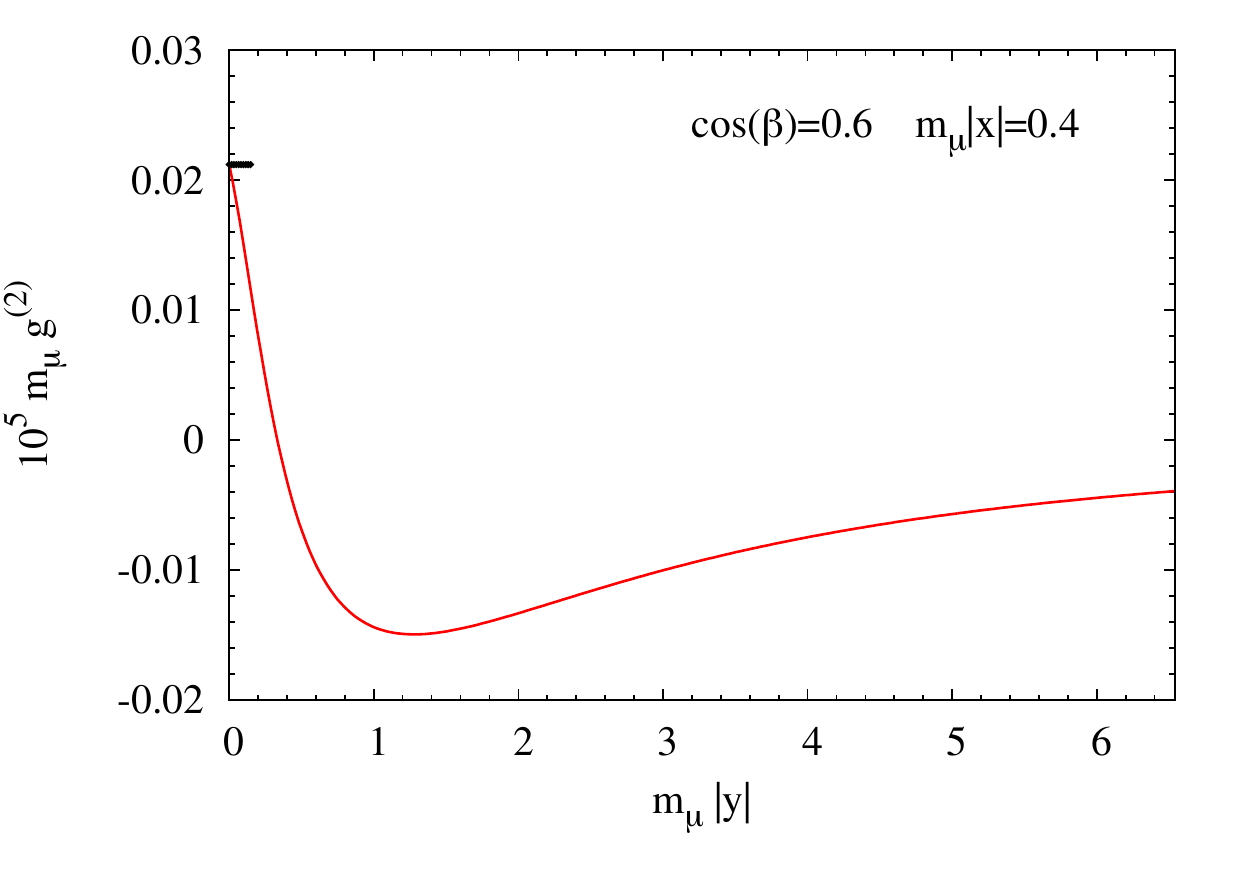}}
\caption{The vector weight functions $\bar{\mathfrak{g}}^{(1)}$ and $\bar{\mathfrak{g}}^{(2)}$  appearing 
  in the QED kernel for given values of $|x|$ and $\hat c_\beta\equiv \hat x\cdot \hat y$, the muon mass
  being denoted by $m_\mu$.
  The leading term of the Taylor expansion around the origin (O$(|y|^0)$) is indicated as a short horizontal line.
\la{fig:g1g2}}
\end{figure}






\begin{table}
\begin{tabular}{cr@{~~~~~}r@{~~~~~}r}
  \hline
$10^5\times$ weight fct.   &    $|y|= 0.532889$  &   $|y|=0.872000$ & $|y|=2.39800$ \\ 
  \hline
  $d\bar{\mathfrak{g}}^{(0)}/d\cos\beta$ & 2.023 &   2.099  & 1.465  \\
    $d\bar{\mathfrak{g}}^{(0)}/d|x|$ &   -1.489 &   0.2071  &  1.393 \\
  $ d\bar{\mathfrak{g}}^{(0)}/d|y|$ & -7.317 &  -8.443 &  -6.713 \\
  $\bar{\mathfrak{g}}^{(1)}$ &   2.958  &    2.723 &  1.870 \\
  $\bar{\mathfrak{g}}^{(2)}$ &   -0.007883 & -0.01357 &  -0.01197 \\
  $ \bar{\mathfrak{l}}^{(1)}$ &  0.1653 &  0.1448 &  0.07812 \\
    $ \bar{\mathfrak{l}}^{(2)}$ & 0.07271 & 0.06462 & 0.03633 \\
  $ \bar{\mathfrak{l}}^{(3)}$ & -0.03674 &  -0.03297 &  -0.01867 \\
  \hline
  \end{tabular}
\caption{\la{tab:tabWF} The weight functions (multiplied by $10^{5}$) for $|x|=0.4$, $\cos\beta=0.6$ and
three different values of $|y|$. The muon mass is set to unity throughout this table.}
\end{table}


An important question in the method based on the series in Gegenbauer polynomials is, how
many terms are needed to reach a good approximation to the weight function.
The answer obviously depends on  $|x|$ and $|y|$. Consider the case of the scalar weight function,
which is given by the integral of the massless propagator $G_0(y-u)$ multiplied with $s(x,u)$, for fixed vectors $x$ and $y$.
If one used the expansion of $s(x,u)$ in $C_n(\hat x\cdot \hat u)$, together with the 
expansion of $G_0(y-u)$ in $C_n(\hat y\cdot \hat u)$, the result of the angular integration (see Eq.\ (\ref{eq:orthogonality_relations}))
would be to give the expansion of the scalar weight function in  $C_n(\hat x\cdot \hat y)$, with coefficients 
proportional to the product of the coefficients in the two series appearing in the integrand.
The expansion of the scalar weight function in $C_n(\hat x\cdot \hat y)$ is precisely the series whose coefficients we compute numerically.
Thus, for that series to converge rapidly, it is sufficient that for all $u$, of the multipole expansion of the massless propagator
and the expansion of $s(x,u)$ in $C_n(\hat x\cdot \hat u)$, at least one converges rapidly.
The multipole expansion of $G_0(y-u)$ converges poorly when $|u|=|y|$; 
the expansion of $s(x,u)$ in $C_n(\hat x\cdot \hat u)$ converges poorly when $|u|=|x|$.
Thus to guarantee that the series in $C_n(\hat x\cdot \hat y)$  converges rapidly, one must avoid the case $|x|\approx|y|$.
The latter condition however defines a surface of codimension one in the space of $(x,y)$, and the poor convergence thus affects 
a substantial fraction of the sampled $(x,y)$ points. However, we can use the symmetry (\ref{eq:g0symmetry}) to compute the weight function with 
the argument $y$ exchanged for $(x-y)$, which in general will improve the convergence of the series in $C_n(\hat x\cdot \hat y)$,
since the value of $|x|$ is then substantially different from $|x-y|$. The only case where the convergence cannot be improved by using 
the symmetry property of the weight function is when the points $(0,x,y)$ form an equilateral triangle. The condition for this to happen
however is a subset of codimension two. Thus if the convergence of the series cannot be controlled in a region $||x|-|y||<\Delta$ and 
$||x|-|x-y||<\Delta$, an integral over the QED kernel with a function which is smooth for an equilateral-triangle constellation of $(0,x,y)$
will suffer an error of order $\Delta^2$. To guarantee an accurate computation of the QED kernel in the equilateral-triangle constellation,
additional computations would be required. In the practical applications of the QED kernel, we have not found it necessary to 
improve further the quality of its evaluation for an overall accuracy on $a_\mu^{\rm HLbL}$ on the order of one percent.
The considerations above apply to all weight functions, and the symmetry relations (\ref{eq:g0symmetry}-\ref{eq:l3symmetry}) can be exploited
to always compute the weight functions with the most favorable arguments for the purpose of the expansion in Gegenbauer polynomials.

Some trial-and-error was necessary to find out an appropriate extent
and step size\footnote{The required extent of the grid is dictated by the physics entering the correlation
function $i\widehat \Pi$; it was chosen large enough for the $\pi^0$ exchange
to be reproduced with subpercent precision at $m_\pi=135\,$MeV. The step size should be small enough
that the achieved precision on the grid points does not get `spoiled' entirely by the interpolation. Since
the weight functions vary more rapidly at small arguments $|x|$ and $|y|$, a relatively small step size
was chosen in this region.} for the grid in $|x|$ and $|y|$. 
Both variables were sampled up to $|x|_{\rm max}=|y|_{\rm max}=6.54 m^{-1}$.  For the
variable $m|y|$, which is cheap to sample finely, we use a step size
of $\delta_{\rm g} = 0.0242222$.  For the variable $m|x|$, we use the same
fine step size for small $m|x|$ up to $m|x|=0.363333$, and a step size
three times bigger for the larger $m|x|$. The number of coefficients
of the Gegenbauer-polynomial expansion computed for a given pair
$(|x|,|y|)$ was chosen\footnote{The stability of the resulting weight functions was tested
by varying the number of terms in the Gegenbauer-polynomial series.} to be $(8+ {\rm floor}(5m|x|))$.
For $m|y|<\delta_{\rm g}$, we use the Taylor-expansion given in
appendix~\ref{sec:smallxy} in order to interpolate the QED kernel
between the point at $m|y|=0$ and the first point of the grid at $m|y|
= \delta_{\rm g}$. We proceed similarly in the case $m|x|<\delta_{\rm g}$.  The
large-$|y|$ asymptotics derived in appendix~\ref{sec:largeYasympt}
were not used in the numerical implementation, they only served
to acquire a qualitative understanding of the large-distance behavior
of the QED kernel.

As sketched at the beginning of this section,
given precomputed coefficients of the Gegenbauer polynomial expansion of the weight functions on a grid in $|x|$ and $|y|$,
the remaining computational tasks to obtain the QED kernel at a given $(x,y)$ point are: to evaluate the sums yielding
the weight functions (again using the Clenshaw algorithm), interpolate
the weight functions (and their derivatives), apply the chain rules to obtain the $T^{({\rm A})}_{\alpha\beta\delta}$ tensors and
finally to perform the contraction of these with the $ {\cal G}^{\rm A}_{\delta[\rho,\sigma]\mu\alpha\nu\beta\lambda}$ tensors.
We have written a code in the \verb|C| programming language to perform these tasks.
Some implementation details and a link to the code can be found in appendix~\ref{app:code}.

\section{Example calculations of the four-point amplitude $i\widehat\Pi$}
\la{sec:ipihats}

In this section we derive explicit expressions for
$i\widehat\Pi_{\rho;\mu\nu\lambda\sigma}(x,y)$ in several models that
are relevant for understanding the corresponding tensor in QCD. Our
primary goal is to use these four-point functions to test (in
section~\ref{sec:numtests}) the validity of the coordinate-space
approach to $a_\mu^{\rm HLbL}$ developed here. In addition, we will
gain insight into the shape and range of the integrand, information
which is valuable in preparing the lattice-QCD calculation.

But first, we review the most important properties of the tensor 
$i\widehat\Pi_{\rho;\mu\nu\lambda\sigma}(x,y)$.

\subsection{General properties of $i\widehat\Pi$}
\la{sec:iPihatgenprop}
The rank-five tensor has the Bose symmetries
\be\la{eq:Bose}
\widehat \Pi_{\rho;\mu\nu\lambda\sigma}(x,y) = \widehat \Pi_{\rho;\nu\mu\lambda\sigma}(y,x) = 
\widehat \Pi_{\rho;\lambda\nu\mu\sigma}(-x,y-x).
\ee
Combining the two generators  of Bose symmetries from Eq.\ (\ref{eq:Bose}), we obtain a finite symmetry group with $3!$ elements.
Note that while the first equality in Eq.\ (\ref{eq:Bose}) follows immediately from the definition of $\widehat \Pi$,
the second one requires translation symmetry and the property that 
\be\la{eq:JJJJarea}
\int d^4z \,\<j_\mu(x) j_\nu(y)j_\sigma(z)j_\lambda(0)\> = 0,
\ee
which holds in infinite volume as a consequence of the observation
that a conserved current can be written as a total divergence,
\be
j_\sigma(z) = \partial^{(z)}_\alpha (z_\sigma j_\alpha(z)),
\ee
and Gauss's theorem.

The tensor $\widehat \Pi$ has the further discrete symmetry 
\be
\widehat \Pi_{\rho;\mu\nu\lambda\sigma}(-x,-y) = -\widehat \Pi_{\rho;\mu\nu\lambda\sigma}(x,y)
\ee
as a consequence of the space and Euclidean-time reversal symmetries of QCD.
Current conservation implies
\ba
\partial_\mu^{(x)} \widehat\Pi_{\rho;\mu\nu\lambda\sigma}(x,y) =0 ,
\\
\partial_\nu^{(y)} \widehat\Pi_{\rho;\mu\nu\lambda\sigma}(x,y) =0 ,
\\
(\partial_\lambda^{(x)}+\partial_\lambda^{(y)}) \widehat\Pi_{\rho;\mu\nu\lambda\sigma}(x,y) = 0.
\ea
For the last result, we have again assumed infinite volume and translation symmetry, and made use of Eq.\ (\ref{eq:JJJJarea}).
For the same reasons, we have the properties
\be\la{eq:iPihatarea}
\int_x \widehat\Pi_{\rho;\mu\nu\lambda\sigma}(x,y) = \int_y \widehat\Pi_{\rho;\mu\nu\lambda\sigma}(x,y) = 0,
\ee
which will be exploited in the numerical tests of section~\ref{sec:KernelSubtractions}.
These properties hold in the continuum formulation of QED or QCD. In a
theory like scalar QED (see subsection~\ref{sec:scalarQED} below), the
appropriate contact terms, as predicted by the Ward identities of
current conservation, must be included into the definition of
$i\widehat\Pi$ in order for Eq.\ (\ref{eq:iPihatarea}) to be
satisfied.

We finally give the representation of $i\widehat\Pi$ in terms of the Euclidean momentum-space HLbL amplitude.
The latter is related to the position-space four-point function of the 
electromagnetic current via a triple Fourier transform,
\be
\Big\<j_\mu(x) j_\nu(y)j_\sigma(z) j_\lambda(0)\Big\>
 = \int_{q_1,q_2,q_3} e^{i(q_1\cdot x+q_2\cdot y+q_3\cdot z)}\Pi_{\mu\nu\sigma\lambda}(q_1,q_2,q_3).
\ee
The quantity of interest, $i\widehat\Pi_{\rho;\mu\nu\lambda\sigma}(x,y)$, can then be calculated using the equation
\be\la{eq:iPihatFromPiQ}
i\widehat\Pi_{\rho;\mu\nu\lambda\sigma}(x,y) = 
-i \int_{q_1,q_2} e^{i(q_1\cdot x+q_2\cdot y)} \frac{\partial}{\partial q_{3\rho}} \Pi_{\mu\nu\sigma\lambda}(q_1,q_2,q_3)\Big|_{q_3=0}.
\ee
This relation is particularly useful when form factors are introduced to describe the coupling of mesons to photons,
as in the case of the pion-pole contribution.

\subsection{Pion-pole contribution to hadronic light-by-light
  scattering in the VMD model \label{sec:Pihat_vmd}}

As a starting point for calculating the contribution of the pion pole, we take the Minkowski-space 
expression for the HLbL amplitude given in~\cite{KN_02}. To convert it to Euclidean space, we use the 
prescription given in~\cite{Gerardin:2017ryf}, whereby we arrive at 
\ba
&& \Pi_{\mu\nu\sigma\lambda}(q_1,q_2,q_3)\Big|_{\pi_0} = 
\frac{{\cal F}(-q_1^2,-q_2^2) {\cal F}(-q_3^2,-(q_1+q_2+q_3)^2)}{(q_1+q_2)^2+m_\pi^2}
\epsilon_{\mu\nu\alpha\beta} \,q_{1\alpha}\, q_{2\beta}
\;\epsilon_{\sigma\lambda\gamma\delta} \,q_{3\gamma} \,(q_1+q_2)_{\delta}
\nonumber\\ && 
+ \frac{{\cal F}(-q_1^2,-(q_1+q_2+q_3)^2){\cal F}(-q_2^2,-q_3^2)}{(q_2+q_3)^2+m_\pi^2}
\epsilon_{\mu\lambda\alpha\beta}\,q_{1\alpha}\,(q_2+q_3)_\beta 
\;\epsilon_{\nu\sigma\gamma\delta} \,q_{2\gamma}\,q_{3\delta}
\nonumber\\ &&
+ \frac{{\cal F}(-q_1^2,-q_3^2){\cal F}(-q_2^2,-(q_1+q_2+q_3)^2)}{(q_1+q_3)^2+m_\pi^2}
\epsilon_{\mu\sigma\alpha\beta} \,q_{1\alpha}\,q_{3\beta}\;\epsilon_{\nu\lambda\gamma\delta}\, q_{2\gamma}\,(q_1+q_3)_\delta.
\la{eq:Pipi0}
\ea
From here the required derivative $\frac{\partial}{\partial q_{3\rho}} \Pi_{\mu\nu\sigma\lambda}(q_1,q_2,q_3)|_{q_3=0}$
is obtained straightforwardly. Before Fourier transforming it with respect to $q_1$ and $q_2$, 
we choose a specific parametrization of the  form factor.

\subsubsection{Vector-meson dominance parametrization of the form factor} 

In the following, we use the vector-meson dominance (VMD) model for the
transition form factor, 
\be
{\cal F}(-q_1^2,-q_2^2) 
= \frac{c_\pi}{(q_1^2+m_V^2)(q_2^2+m_V^2)}, \qquad 
\ee
its normalization $c_\pi = -\frac{N_c m_V^4}{12\pi^2 F_\pi}$ being determined by the axial anomaly.
Inserting this form into the expression for $\frac{\partial}{\partial q_{3\rho}} \Pi_{\mu\nu\sigma\lambda}(q_1,q_2,q_3)|_{q_3=0}$,
and the latter into (\ref{eq:iPihatFromPiQ}), and 
rewriting the expression using coordinate-space propagators, one finds the master expression 
\ba
&& i \widehat\Pi_{\rho;\mu\nu\lambda\sigma}(x,y)
= \frac{c_\pi^2}{m_V^2(m_V^2-m_\pi^2)} \frac{\partial}{\partial x_\alpha}\frac{\partial}{\partial y_\beta}
\Big\{ 
 \epsilon_{\mu\nu\alpha\beta}\epsilon_{\sigma\lambda\rho\gamma}
\Big(\frac{\partial}{\partial x_\gamma}+\frac{\partial}{\partial y_\gamma}\Big) 
K_\pi(x,y)
\nonumber\\ && + \epsilon_{\mu\lambda\alpha\beta} \epsilon_{\nu\sigma\gamma\rho}
 \frac{\partial}{\partial y_\gamma} 
K_\pi(y-x,y)
+ \epsilon_{\mu\sigma\alpha\rho}\epsilon_{\nu\lambda\beta\gamma} 
\frac{\partial}{\partial x_\gamma}
K_\pi(x,x-y)
 \Big\},
\la{eq:PihatPi0Master}
\ea
where 
\be\la{eq:KpiDef}
K_\pi(x,y) \equiv \int_u \Big(G_{m_\pi}(u) - G_{m_V}(u)\Big)G_{m_V}(x-u) G_{m_V}(y-u) = K_\pi(y,x).
\ee
We remind the reader that we denote by $G_M$ the scalar propagator with mass $M$; see Eq.\ (\ref{eq:massive_scalar_prop}).

Thus the main task is to compute the scalar function $K_\pi(x,y)$, which depends on three scalar quantities, 
$|x|$, $\cos\beta\equiv\hat x\cdot \hat y$ and $|y|$.
The three derivatives that must be applied onto the function $ K_\pi(x,y)$ are computed using a lengthy chain rule.

The goal is therefore to compute $K_\pi(x,y)$ and the derivatives with respect to the scalar variables
analytically as far as possible.
Using the expansion of the scalar propagator in Gegenbauer polynomials and exploiting their orthogonality property, 
one obtains immediately an expansion of $K_\pi(x,y)$  in $C_n(\hat x\cdot \hat y)$.
Via the change of integration variable $u\to x-u$ in Eq.\ (\ref{eq:KpiDef}) and the subsequent
Gegenbauer expansion of the propagators, one arrives at the alternative expansion
\ba\la{eq:Kpibis}
K_\pi(x,y) &=& \sum_{n\geq0} C_n(\widehat{x-y}\cdot\hat x)\;b_n(|x|,|x-y|),
\\
b_n(|x|,|x-y|) &=&  \frac{2\pi^2}{n+1} \int_0^\infty d|u|\;|u|^3 G_{m_V}(u)
\;\delta\gamma_n(x^2,u^2) \;\gamma_n((x-y)^2,u^2),
\ea
 with $  \delta\gamma_n(x^2,u^2) = \gamma_n(x^2,u^2)|_{m_\pi} -
  \gamma_n(x^2,u^2)|_{m_V}$ the difference of the expansion coefficients of two
  massive propagators with different masses in Gegenbauer
  polynomials; see Eq.~(\ref{eq:Gm_gamma_n}) for the explicit
  expression for $\gamma_n(x^2, u^2)$. Because the difference of two
propagators is only logarithmically divergent at the origin, the
convergence of the multipole expansion is improved. A further,
analogous expression for $K_\pi(x,y)$ expanded in
$C_n(\widehat{x-y}\cdot\hat y)$ is obtained by interchanging $x$ and
$y$ on the right-hand side of Eq.\ (\ref{eq:Kpibis}).

An important consideration in the evaluation of $K_\pi(x,y)$, very
similar to the discussion in the penultimate paragraph of section \ref{sec:discussion_master_formula}, is the
following.  For the sum in Eq.\ (\ref{eq:Kpibis}) to be rapidly
convergent, at least one of the $\delta\gamma_n(x^2,u^2)$ and
$\gamma_n((x-y)^2,u^2)$ should decrease rapidly with $n$, for all $u$.
This works when the ratio of $|x|$ and $|x-y|$ is not too close to
unity.  If this is not the case, the expansion in
$C_n(\widehat{x-y}\cdot\hat y)$ can be used.  The only problematic
constellation is when the points $(0,x,y)$ form an equilateral
triangle; in this case, a more sophisticated technique would be
required. However, in an integral over $x$ at fixed $y$ for instance,
this case represents a set of codimension~2.  In practice, this means
that if the sum, truncated at $n_{\rm max}$, does not provide an
accurate estimate of $K_\pi(x,y)$ in a range $|x|=|y|(1\pm\epsilon)$
and $\cos\beta=\frac{\pi}{3}(1\pm\epsilon)$, where $\epsilon$ shrinks when $n_{\rm max}$ increases, 
the error on the resulting integral is of order $\epsilon^2$, since the integrand is
regular in the equilateral constellation. In our numerical implementation, 
we chose $n_{\rm max}=64$.

Explicitly, the form of the coefficients is
\ba
b_n(|x|,\Delta) &\stackrel{\Delta\leq |x|}{=}& \frac{(n+1)m_V}{8\pi^4 |x| \Delta} 
\\ && \Big\{K_{n+1}(m_V \Delta) \Big(K_{n+1}(m_\pi |x|) G^1_n(m_\pi,m_V,\Delta)  - K_{n+1}(m_V |x|) G^1_n(m_V,m_V,\Delta)\Big)
\nonumber\\&& + I_{n+1}(m_V\Delta)  \Big(K_{n+1}(m_\pi|x|) (G^{2}_n(m_\pi,m_V,\Delta)-G^{2}_n(m_\pi,m_V,|x|))
\nonumber\\ &&          \qquad\qquad \quad        -K_{n+1}(m_V|x|) (G^{2}_n(m_V,m_V,\Delta)-G^{2}_n(m_V,m_V,|x|))\Big)
\nonumber\\ && +  I_{n+1}(m_V\Delta)\Big(I_{n+1}(m_\pi|x|) G^3_n(m_\pi,m_V,|x|)- I_{n+1}(m_V|x|) G^3_n(m_V,m_V,|x|)\Big)\Big\},
\nonumber\\ b_n(|x|,\Delta)  &\stackrel{|x|\leq\Delta}{=} &  \frac{(n+1)m_V}{8\pi^4 |x| \Delta} 
\\ && \Big\{K_{n+1}(m_V \Delta) \Big(K_{n+1}(m_\pi |x|) G^1_n(m_\pi,m_V,|x|)  - K_{n+1}(m_V |x|) G^1_n(m_V,m_V,|x|)\Big)
\nonumber\\&& + K_{n+1}(m_V\Delta)  \Big(I_{n+1}(m_\pi|x|) (G^{2}_n(m_V,m_\pi,|x|)-G^{2}_n(m_V,m_\pi,\Delta))
\nonumber\\ &&   \qquad\qquad \quad     -I_{n+1}(m_V|x|) (G^{2}_n(m_V,m_V,|x|)-G^{2}_n(m_V,m_V,\Delta))\Big)
\nonumber\\ && + I_{n+1}(m_V\Delta)\Big(I_{n+1}(m_\pi|x|) G^3_n(m_\pi,m_V,\Delta)- I_{n+1}(m_V|x|) G^3_n(m_V,m_V,\Delta)\Big)\Big\}
\nonumber
\ea
with
\ba
G^1_n(m_1,m_2,|x|) &=& \int_0^{|x|} du\; K_1(m_Vu) I_{n+1}(m_1 u)  I_{n+1}(m_2u),
\\
G^{2}_n(m_1,m_2,|x|) &=& \int_{|x|}^s  du\;K_1(m_Vu) I_{n+1}(m_1 u) K_{n+1}(m_2 u),
\\
G^3_n(m_1,m_2,|x|) &=&  \int_{|x|}^\infty du\;K_1(m_Vu) K_{n+1}(m_1 u) K_{n+1}(m_2u).
\ea
As a reminder, $I_n$ and $K_n$ denote the modified Bessel functions.
Note that the dependence of $G^{2}_n$ on the upper limit $s$ of integration drops out in the functions $b_n$.
In our approach to the numerical implementation, 
the values of the functions $G^{1,2,3}_n$ are computed and stored on a regular one-dimensional grid.
Then, when the function $K_\pi(x,y)$ and its derivatives are needed at a certain target point, 
the more favorable expansion in $C_n(\widehat{x-y}\cdot\hat x)$ and $C_n(\widehat{x-y}\cdot\hat y)$ is chosen,
and an interpolation is performed in the variables $|x|$ and $|x-y|$, at the target value of 
 $\cos\alpha = \widehat{x-y}\cdot \hat x$. A chain rule relates the derivatives with respect to 
one of the variables $(|x|,\cos\beta,|y|)$ to those with respect to one of the variables $(|x|,\cos\alpha,|x-y|)$.
With up to three derivatives involved, the chain rule is best generated with a symbolic manipulation
program.
A further element used in the evaluation of the derivatives is based on the observation
\be
(-\triangle^{(y)}+M_V^2) K_\pi(x,y)  =  G_{m_V}(x-y) \Big( G_{m_\pi}(y) - G_{m_V}(y)\Big).
\ee
This equation is used to express the second and higher derivatives with respect to $|y|$ in terms 
of the derivatives with respect to $\cos\beta$ of the same order and in terms of lower derivatives.

\subsubsection{Tests performed }

From the momentum-space expression 
\be
K_\pi(x,y) = (m_V^2-m_\pi^2)
\int_{q_1,q_2} \frac{e^{i(q_1\cdot x+q_2\cdot y)} }{[(q_1+q_2)^2+m_\pi^2](q_1^2+m_V^2)(q_2^2+m_V^2)[(q_1+q_2)^2+m_V^2]},
\ee
 one easily obtains 
\ba\la{eq:MasterCheck}
\int_{x,y} K_\pi(x,y) = \frac{m_V^2-m_\pi^2}{m_\pi^2 m_V^6},
\ea
which provides a test of the numerical implementation of
$K_\pi(x,y)$. 
More differential information can also be obtained,
\ba
\int_{y} K_\pi(x,y) &=& 
\frac{1}{m_V^2}\Big(\frac{1}{m_V^2-m_\pi^2}(G_{m_\pi}(x)-G_{m_V}(x))-\frac{K_0(m_V|x|)}{8\pi^2}\Big).
\ea
Also, using integration by parts and the result (\ref{eq:MasterCheck}), one shows that 
\ba
&& \int_{x,y} \delta_{\mu\nu}x_\lambda(x_\rho y_\sigma - y_\rho x_\sigma) i\widehat\Pi_{\rho;\mu\nu\lambda\sigma}(x,y)
 = -\int_{x,y} \delta_{\mu\nu}y_\lambda(x_\rho y_\sigma - y_\rho x_\sigma) i\widehat\Pi_{\rho;\mu\nu\lambda\sigma}(x,y)
\nonumber\\ && = \frac{1}{2}\int_{x,y} (\delta_{\nu\lambda}x_\mu-\delta_{\mu\lambda} y_\nu)
(x_\rho y_\sigma - y_\rho x_\sigma) i\widehat\Pi_{\rho;\mu\nu\lambda\sigma}(x,y)
 = \frac{3}{\pi^4 m_\pi^2 F_\pi^2}.
\ea
The three integrals test respectively the third, second and first term of Eq.\ (\ref{eq:PihatPi0Master}),
which contains three terms in total.

\subsection{Lepton-loop contribution to light-by-light scattering in QED
\label{sec:Pihat_lepton}}

In this subsection, we present the perturbative calculation of the
fermion loop contribution to $i\widehat\Pi_{\rho;\mu\nu\lambda\sigma}(x,y)$.

\subsubsection{The coordinate-space four-point function of the electromagnetic current}

We start out by writing out the expression for the quark-connected contribution of the unintegrated coordinate-space four-point function
of the electromagnetic current, valid in an arbitrary U($N_c$) gauge field background.
We note the important property 
\be\la{eq:g5herm}
S(x,y) = \gamma_5 S(y,x)^\dagger \gamma_5
\ee
of the fermion propagator, where the dagger acts on the Dirac indices. This property holds in an arbitrary background gauge field.

Performing the six fully-connected Wick contractions, we note that
due to the property (\ref{eq:g5herm}) they pair up, corresponding to
the fermion number flowing in opposite directions. One obtains
\be\la{eq:master4QED}
\<j_\mu(x) \;j_\nu(y)\; j_\sigma(z)\; j_\lambda(0)\>
= - 2{\rm Re}\{ I^{(1)}_{\mu\nu\sigma\lambda}(x,y,z,0)  + I^{(2)}_{\mu\nu\sigma\lambda}(x,y,z,0) + I^{(3)}_{\mu\nu\sigma\lambda}(x,y,z,0)\},
\ee
with
\ba\la{eq:wickI1}
 I^{(1)}_{\mu\nu\sigma\lambda}(x,y,z,w) &=& {\rm Tr}\{ \gamma_\mu S(x,y)\gamma_\nu S(y,z) \gamma_\sigma S(z,w) \gamma_\lambda S(w,x)  \},
\\  \la{eq:wickI2}
 I^{(2)}_{\mu\nu\sigma\lambda}(x,y,z,w) &=& {\rm Tr}\{\gamma_\mu S(x,y)\gamma_\nu S(y,w) \gamma_\lambda S(w,z) \gamma_\sigma S(z,x)  \}
=  I^{(1)}_{\mu\nu\lambda\sigma}(x,y,w,z) ,
\\ \la{eq:wickI3}
 I^{(3)}_{\mu\nu\sigma\lambda}(x,y,z,w) &=& {\rm Tr}\{ \gamma_\lambda S(w,y)\gamma_\nu S(y,z) \gamma_\sigma S(z,x) \gamma_\mu S(x,w) \}
= I^{(1)}_{\lambda\nu\sigma\mu}(w,y,z,x).
\ea

\subsubsection{Vanishing background gauge field: the QED case}

Recall the free Dirac fermion propagator $S(x,y)$ in Euclidean position space,
\ba
S(w+x,w) \equiv \int \frac{d^4p}{(2\pi)^4}\; \frac{-ip_\mu\gamma_\mu+m}{p^2+m^2}\; e^{ip\cdot x}
 &=& \frac{m^2}{4\pi^2 |x|} \Big[ \gamma_\mu x_\mu \,\frac{K_2(m|x|)}{|x|} + K_1(m|x|)\Big],  ~~~~
\ea
with
\ba
S(w+x,w) &=& \frac{x_\mu \,\gamma_\mu}{2\pi^2 (x^2)^2} \qquad \qquad (m=0)
\ea
in the massless case. In the free theory, the propagator is actually Hermitian 
with respect to the Dirac indices, so that the general property (\ref{eq:g5herm})  holds even without the dagger.
From here on, the expressions in this section assume a single fermion flavor with unit electric charge.
Thus the calculation can be interpreted as a treatment of the contribution of a lepton of mass $m$ 
to light-by-light scattering in $a_\mu$. 
Also, in a vanishing background field, we note the translation-invariance property
\be
I^{(j)}_{\mu\nu\sigma\lambda}(x,y,z,w) = I^{(j)}_{\mu\nu\sigma\lambda}(x-w,y-w,z-w,0), \qquad j=1,2,3,
\ee
which can be combined with the Bose symmetry of the photons to write
\ba
&& I^{(2)}_{\mu\nu\sigma\lambda}(x,y,z,0) = I^{(1)}_{\mu\nu\lambda\sigma}(x,y,0,z) = I^{(1)}_{\mu\nu\lambda\sigma}(x-z,y-z,-z,0),
\\ && I^{(3)}_{\mu\nu\sigma\lambda}(x,y,z,0)  = I^{(1)}_{\lambda\nu\sigma\mu}(0,y,z,x) = I^{(1)}_{\lambda\nu\sigma\mu}(-x,y-x,z-x,0).
\ea
Furthermore, the expression inside the curly bracket in Eq.\ (\ref{eq:master4QED}) is already real.
In the free massless case, one thus obtains
\ba
I^{(1)}_{\mu\nu\sigma\lambda}(x,y,z,0) &=& \frac{(-x_\alpha) (x-y)_\beta (y-z)_\gamma z_\delta}{(2\pi^2)^4 |x|^4 |x-y|^4 |y-z|^4 |z|^4} 
{\rm Tr}\{ \gamma_\alpha \gamma_\mu \gamma_\beta \gamma_\nu \gamma_\gamma \gamma_\sigma \gamma_\delta \gamma_\lambda\},
\\
I^{(2)}_{\mu\nu\sigma\lambda}(x,y,z,0) &=& \frac{y_\alpha (-z_\beta) (z-x)_\gamma (x-y)_\delta}{(2\pi^2)^4 |y|^4 |z|^4 |x-z|^4 |x-y|^4} 
{\rm Tr}\{ \gamma_\nu \gamma_\alpha \gamma_\lambda \gamma_\beta \gamma_\sigma \gamma_\gamma \gamma_\mu \gamma_\delta\}, \phantom{\Bigg()}
\\
I^{(3)}_{\mu\nu\sigma\lambda}(x,y,z,0) &=& \frac{(y-z)_\alpha (z-x)_\beta \; x_\gamma \;(-y_\delta)}{(2\pi^2)^4 |y-z|^4 |z-x|^4 |x|^4 |y|^4} 
{\rm Tr}\{ \gamma_\nu \gamma_\alpha \gamma_\sigma \gamma_\beta \gamma_\mu \gamma_\gamma \gamma_\lambda \gamma_\delta\}.
\ea
In the free massive case, the result is 
\ba\la{eq:I1res}
&& I^{(1)}_{\mu\nu\sigma\lambda}(x,y,z,0) = \Big(\frac{m}{2\pi}\Big)^8 \Big[
\nonumber \\ &&
 \frac{(-x_\alpha) (x-y)_\beta (y-z)_\gamma z_\delta\; K_2(m|x|) K_2(m|x-y|)K_2(m|y-z|)K_2(m|z|)}{|x|^2 |x-y|^2 |y-z|^2 |z|^2}\cdot 
\nonumber \\ && \qquad \quad \cdot {\rm Tr}\{ \gamma_\alpha \gamma_\mu \gamma_\beta \gamma_\nu \gamma_\gamma \gamma_\sigma \gamma_\delta \gamma_\lambda\} 
\phantom{\bigg(\bigg)}
 \\ && + \frac{ K_1(m|x|) K_1(m|x-y|)K_1(m|y-z|)K_1(m|z|)}{ |x| |x-y| |y-z| |z|}
 {\rm Tr}\{  \gamma_\mu  \gamma_\nu  \gamma_\sigma  \gamma_\lambda\}
\nonumber \\ && +  \frac{(-x_\alpha) (x-y)_\beta \; K_2(m|x|) K_2(m|x-y|)K_1(m|y-z|)K_1(m|z|)}{ |x|^2 |x-y|^2 |y-z| |z|}
 {\rm Tr}\{ \gamma_\alpha \gamma_\mu \gamma_\beta \gamma_\nu  \gamma_\sigma \gamma_\lambda\}
\nonumber  \\ && +  \frac{(-x_\alpha) (y-z)_\gamma \; K_2(m|x|) K_1(m|x-y|)K_2(m|y-z|)K_1(m|z|)}{|x|^2 |x-y| |y-z|^2 |z|}
   {\rm Tr}\{ \gamma_\alpha \gamma_\mu  \gamma_\nu \gamma_\gamma \gamma_\sigma \gamma_\lambda\}
\nonumber  \\ && +  \frac{(-x_\alpha) z_\delta\; K_2(m|x|) K_1(m|x-y|)K_1(m|y-z|)K_2(m|z|)}{|x|^2 |x-y| |y-z| |z|^2}
   {\rm Tr}\{ \gamma_\alpha \gamma_\mu  \gamma_\nu  \gamma_\sigma \gamma_\delta \gamma_\lambda\}
\nonumber  \\ && +  \frac{ (x-y)_\beta (y-z)_\gamma \; K_1(m|x|) K_2(m|x-y|)K_2(m|y-z|)K_1(m|z|)}{|x| |x-y|^2 |y-z|^2 |z|}
   {\rm Tr}\{  \gamma_\mu \gamma_\beta \gamma_\nu \gamma_\gamma \gamma_\sigma \gamma_\lambda\}
\nonumber  \\ && +  \frac{(x-y)_\beta  z_\delta\; K_1(m|x|) K_2(m|x-y|)K_1(m|y-z|)K_2(m|z|)}{|x| |x-y|^2 |y-z| |z|^2}
   {\rm Tr}\{  \gamma_\mu \gamma_\beta \gamma_\nu  \gamma_\sigma \gamma_\delta \gamma_\lambda\}
\nonumber  \\ && +  \frac{ (y-z)_\gamma z_\delta\; K_1(m|x|) K_1(m|x-y|)K_2(m|y-z|)K_2(m|z|)}{|x| |x-y| |y-z|^2 |z|^2}
   {\rm Tr}\{  \gamma_\mu  \gamma_\nu \gamma_\gamma \gamma_\sigma \gamma_\delta \gamma_\lambda\}
 \Big].
\nonumber
\ea
To evaluate the coordinate-space four-point function, 
it is thus sufficient to program the function $I^{(1)}_{\mu\nu\sigma\lambda}(x,y,z,0)$, 
and call it three times to com\-pute the four-point function $\<j_\mu(x) \;j_\nu(y)\; j_\sigma(z)\; j_\lambda(0)\>$.
As for the Dirac traces, it is straightforward to compute and store the 65536 components of 
${\rm Tr}\{ \gamma_\alpha \gamma_\mu \gamma_\beta \gamma_\nu \gamma_\gamma \gamma_\sigma \gamma_\delta \gamma_\lambda\}$
once for all times.

\subsubsection{Calculation of $i\widehat\Pi_{\rho;\mu\nu\lambda\sigma}(x,y) $}

From Eq.\ (\ref{eq:master4QED}), in order to compute 
\be
i\widehat\Pi_{\rho;\mu\nu\lambda\sigma}(x,y) 
=  2 \int_z z_\rho\;
 {\rm Re}\{ I^{(1)}_{\mu\nu\sigma\lambda}(x,y,z,0)  + I^{(2)}_{\mu\nu\sigma\lambda}(x,y,z,0) + I^{(3)}_{\mu\nu\sigma\lambda}(x,y,z,0)\},
\ee
it is sufficient to compute the two integrals
\ba 
\widehat\Pi^{(1)}_{\rho;\mu\nu\lambda\sigma}(x,y) &\equiv&  2\,{\rm Re}\int d^4z \; z_\rho\;I^{(1)}_{\mu\nu\sigma\lambda}(x,y,z,0),
\\ 
\Pi^{(r,1)}_{\mu\nu\lambda\sigma}(x,y)  &\equiv&  2\,{\rm Re} \int d^4z\; I_{\mu\nu\sigma\lambda}^{(1)}(x,y,z,0).
\ea
Indeed, for the second term, we make use of the property
\be
I^{(2)}_{\mu\nu\sigma\lambda}(x,y,z,0) = I^{(1)}_{\mu\nu\lambda\sigma}(x,y,0,z)
= I_{\nu\lambda\sigma\mu}^{(1)}(y,0,z,x) = I^{(1)}_{\nu\lambda\sigma\mu}(y-x,-x,z-x,0)
\ee
(where we have performed a cyclic permutation of the arguments in the second equality),
from which there follows 
\ba
&& 2\int d^4z\; z_\rho\, I_{\mu\nu\sigma\lambda}^{(2)}(x,y,z,0) 
= 2\int d^4z\; (z_\rho+x_\rho) I^{(1)}_{\nu\lambda\sigma\mu}(y-x,-x,z,0)
\\ && = \widehat\Pi^{(1)}_{\rho;\nu\lambda\mu\sigma}(y-x,-x) + x_\rho\, \Pi^{(r,1)}_{\nu\lambda\mu\sigma}(y-x,-x).
\nonumber 
\ea
Similarly, the third term can be expressed as 
\be
I^{(3)}_{\mu\nu\sigma\lambda}(x,y,z,0) = I^{(1)}_{\lambda\nu\sigma\mu}(0,y,z,x) 
= I^{(1)}_{\lambda\nu\sigma\mu}(-x,y-x,z-x,0),
\ee
so that 
\ba
&& 2\int d^4z\; z_\rho\, I^{(3)}_{\mu\nu\sigma\lambda}(x,y,z,0)
= 2\int d^4z\; (z_\rho+x_\rho) I^{(1)}_{\lambda\nu\sigma\mu}(-x,y-x,z,0)
\\ && = \widehat\Pi^{(1)}_{\rho;\lambda\nu\mu\sigma}(-x,y-x) + x_\rho\,\Pi^{(r,1)}_{\lambda\nu\mu\sigma}(-x,y-x).
\nonumber
\ea
Thus $i\widehat\Pi_{\rho;\mu\nu\lambda\sigma}(x,y)$ can be expressed through the functions 
$\widehat\Pi^{(1)}_{\rho;\mu\nu\lambda\sigma}(x,y)$ and $\Pi^{(r,1)}_{\nu\lambda\mu\sigma}(x,y)$ via
\ba
i\widehat\Pi_{\rho;\mu\nu\lambda\sigma}(x,y) &=& \widehat\Pi^{(1)}_{\rho;\mu\nu\lambda\sigma}(x,y)
\\ && +\widehat\Pi^{(1)}_{\rho;\nu\lambda\mu\sigma}(y-x,-x) + x_\rho\, \Pi^{(r,1)}_{\nu\lambda\mu\sigma}(y-x,-x)
\nonumber  \\ && +  \widehat\Pi^{(1)}_{\rho;\lambda\nu\mu\sigma}(-x,y-x) + x_\rho\,\Pi^{(r,1)}_{\lambda\nu\mu\sigma}(-x,y-x).
\nonumber 
\ea

It remains to perform the required integrals. The result is
\ba\la{eq:d4zI1}
&&  \widehat\Pi^{(1)}_{\rho;\mu\nu\lambda\sigma}(x,y) 
\\ \nonumber && = 2\Big(\frac{m}{2\pi}\Big)^8 \Big[
 \frac{(-x_\alpha) (x-y)_\beta  K_2(m|x|) K_2(m|x-y|)}{|x|^2 |x-y|^2}\cdot   {f}_{\rho\delta\gamma}(y)
 \cdot {\rm Tr}\{ \gamma_\alpha \gamma_\mu \gamma_\beta \gamma_\nu \gamma_\gamma \gamma_\sigma \gamma_\delta \gamma_\lambda\} 
\\ \nonumber && 
 +  \frac{  K_1(m|x|) K_1(m|x-y|)}{|x| |x-y| }\cdot {f}_{\rho\delta\gamma}(y)\cdot 
   {\rm Tr}\{  \gamma_\mu  \gamma_\nu \gamma_\gamma \gamma_\sigma \gamma_\delta \gamma_\lambda\}
\\ \nonumber &&  + \frac{ K_1(m|x|) K_1(m|x-y|)}{ |x| |x-y| } \,g_\rho(y)\cdot
 {\rm Tr}\{  \gamma_\mu  \gamma_\nu  \gamma_\sigma  \gamma_\lambda\}
\nonumber \\ && +  \frac{(-x_\alpha) (x-y)_\beta \; K_2(m|x|) K_2(m|x-y|)}{ |x|^2 |x-y|^2}
g_\rho(y)\cdot 
 {\rm Tr}\{ \gamma_\alpha \gamma_\mu \gamma_\beta \gamma_\nu  \gamma_\sigma \gamma_\lambda\}
\nonumber  \\ && 
 +  \frac{(-x_\alpha) \; K_2(m|x|) K_1(m|x-y|)}{|x|^2 |x-y| }
h_{\rho\gamma}(y)\cdot 
   {\rm Tr}\{ \gamma_\alpha \gamma_\mu  \gamma_\nu \gamma_\gamma \gamma_\sigma \gamma_\lambda\}
\nonumber \\ && 
+  \frac{ (x-y)_\beta  \; K_1(m|x|) K_2(m|x-y|)}{|x| |x-y|^2 }
h_{\rho\gamma}(y)\cdot 
   {\rm Tr}\{  \gamma_\mu \gamma_\beta \gamma_\nu \gamma_\gamma \gamma_\sigma \gamma_\lambda\}
\nonumber \\ && 
 +  \frac{(-x_\alpha) \; K_2(m|x|) K_1(m|x-y|)}{|x|^2 |x-y| }
\hat f_{\rho\delta}(y)\cdot 
   {\rm Tr}\{ \gamma_\alpha \gamma_\mu  \gamma_\nu  \gamma_\sigma \gamma_\delta \gamma_\lambda\}
\nonumber \\ && 
+  \frac{(x-y)_\beta \; K_1(m|x|) K_2(m|x-y|)}{|x| |x-y|^2}
\hat f_{\rho\delta}(y)\cdot 
   {\rm Tr}\{  \gamma_\mu \gamma_\beta \gamma_\nu  \gamma_\sigma \gamma_\delta \gamma_\lambda\}
\Big]
\ea
and 
\ba
&& \Pi^{(r,1)}_{\mu\nu\lambda\sigma}(x,y)  
\\ \nonumber && = 2\Big(\frac{m}{2\pi}\Big)^8 \Big[
 \frac{(-x_\alpha) (x-y)_\beta \; K_2(m|x|) K_2(m|x-y|)}{|x|^2 |x-y|^2 }\cdot 
l_{\gamma\delta}(y) \cdot  {\rm Tr}\{ \gamma_\alpha \gamma_\mu \gamma_\beta \gamma_\nu \gamma_\gamma \gamma_\sigma \gamma_\delta \gamma_\lambda\} 
\\ \nonumber &&
+ \frac{ K_1(m|x|) K_1(m|x-y|)}{ |x| |x-y| } \cdot 
p(|y|) \cdot  {\rm Tr}\{  \gamma_\mu  \gamma_\nu  \gamma_\sigma  \gamma_\lambda\}
\nonumber \\ && +  \frac{(-x_\alpha) (x-y)_\beta \; K_2(m|x|) K_2(m|x-y|) }{ |x|^2 |x-y|^2 }\cdot
p(|y|) \cdot {\rm Tr}\{ \gamma_\alpha \gamma_\mu \gamma_\beta \gamma_\nu  \gamma_\sigma \gamma_\lambda\}
\nonumber  \\ &&
+  \frac{(-x_\alpha)  \; K_2(m|x|) K_1(m|x-y|)}{|x|^2 |x-y| }\cdot
q_\gamma(y) \cdot   {\rm Tr}\{ \gamma_\alpha \gamma_\mu  \gamma_\nu \gamma_\gamma \gamma_\sigma \gamma_\lambda\}
\nonumber  \\ &&
 +  \frac{ (x-y)_\beta  \; K_1(m|x|) K_2(m|x-y|)}{|x| |x-y|^2 }\cdot
q_\gamma(y) \cdot    {\rm Tr}\{  \gamma_\mu \gamma_\beta \gamma_\nu \gamma_\gamma \gamma_\sigma \gamma_\lambda\}
\nonumber  \\ &&
 +  \frac{(-x_\alpha) \; K_2(m|x|) K_1(m|x-y|)}{|x|^2 |x-y|}\cdot
q_\delta(y) \cdot   {\rm Tr}\{ \gamma_\alpha \gamma_\mu  \gamma_\nu  \gamma_\sigma \gamma_\delta \gamma_\lambda\}
\nonumber \\ && 
+  \frac{(x-y)_\beta \; K_1(m|x|) K_2(m|x-y|)}{|x| |x-y|^2}\cdot 
q_\delta(y) \cdot    {\rm Tr}\{  \gamma_\mu \gamma_\beta \gamma_\nu  \gamma_\sigma \gamma_\delta \gamma_\lambda\}
\nonumber \\ && 
+ \frac{ K_1(m|x|) K_1(m|x-y|)}{|x| |x-y| }\cdot
l_{\gamma\delta}(y) 
\cdot   {\rm Tr}\{  \gamma_\mu  \gamma_\nu \gamma_\gamma \gamma_\sigma \gamma_\delta \gamma_\lambda\} \Big].
\nonumber
\ea
The functions appearing in the expressions above are
\ba
&& \hat f_{\rho\delta}(y) \equiv  \int_z z_\rho z_\delta \frac{K_1(m|y-z|)}{|y-z|} \, \frac{K_2(m|z|)}{|z|^2}
= \frac{\pi^2}{m^3}\Big\{ \hat y_\rho \hat y_\delta \; m|y| K_1(m|y|)  + \delta_{\rho\delta} K_0(m|y|) \Big\}, \qquad 
\\
&& {f}_{\rho\delta\gamma}(y) 
 \equiv \int_z z_\rho\, z_\delta\,(y-z)_\gamma \frac{K_2(m|y-z|)}{|y-z|^2}\, \frac{K_2(m|z|)}{|z|^2}
 = - \frac{1}{m}\frac{\partial}{\partial y_\gamma} \hat f_{\rho\delta}(y) 
\\ && = \frac{\pi^2}{m^3}\Big\{ \hat y_\gamma \hat y_\delta \hat y_\rho\, m|y|K_2(m|y|)
      + ( \delta_{\rho\delta} \hat y_\gamma - \delta_{\gamma\rho} \hat y_\delta - \delta_{\gamma\delta}\hat y_\rho) \,K_1(m|y|)\Big\},
\\
&& g_\rho(y) \equiv \int d^4z\; z_\rho \frac{K_1(m|y-z|)}{|y-z|}\; \frac{K_1(m|z|)}{|z|} = \frac{\pi^2}{m^2} y_\rho\, K_0(m|y|),
\\
&& h_{\rho\gamma}(y) \equiv \int d^4z \; z_\rho\, (y-z)_\gamma\, \frac{K_2(m|y-z|)}{|y-z|^2}\, \frac{K_1(m|z|)}{|z|}
= - \frac{1}{m} \frac{\partial}{\partial y_\gamma} g_\rho(y) 
\\ && = \frac{\pi^2}{m^3} \Big(\hat y_\gamma \hat y_\rho\, m|y|\,K_1(m|y|) - \delta_{\gamma\rho} K_0(m|y|)\Big),
\\ 
&& l_{\gamma\delta}(y) \equiv \int d^4z\; z_\delta(y-z)_\gamma \frac{K_2(m|y-z|)K_2(m|z|)}{|y-z|^2 |z|^2}
\\ && \qquad \quad 
=  \frac{2\pi^2}{m^2}\Big( \hat y_\gamma \hat y_\delta\, K_2(m|y|) - \delta_{\gamma\delta}\, \frac{K_1(m|y|)}{m|y|}\Big),
\\
&& p(|y|) \equiv \int d^4z\; \frac{K_1(m|y-z|)}{|y-z|} \frac{K_1(m|z|)}{|z|}  = \frac{2\pi^2}{m^2}\, K_0(m|y|),
\\
&& q_\gamma(y) \equiv  \int d^4z\; (y-z)_\gamma\frac{K_2(m|y-z|)}{|y-z|^2}\, \frac{K_1(m|z|)}{|z|}  = -\frac{y_\gamma}{m|y|} p'(|y|) 
= \frac{2\pi^2}{m^2}\,\hat y_\gamma\, K_1(m|y|). \qquad 
\ea
The integrals are performed by using the Gegenbauer expansion of the massive scalar propagator, Eq.\ (\ref{eq:Gm_expansion}).
Then, in the case of $p(|y|)$, which is proportional to the convolution of two scalar propagators,
one makes use of the integrals~\cite{Prudnikov}\footnote{In Eq.\ (\ref{eq:i1}), ${\rm Re}(\lambda)>-1$ is assumed
and in Eq.\ (\ref{eq:i2}), ${\rm Re}(m)>0$ is assumed. The prime denotes the derivatives of the Bessel function
with respect to their argument, e.g.\ $K_0'(mr)=-K_1(mr)$.}
\ba\la{eq:i1}
\int_0^r dz\,z K_\lambda(mz)\, I_\lambda(mz)  &=&  \frac{r^2}{2}
\Big[ \Big(1+\frac{\lambda^2}{m^2r^2}\Big)I_\lambda(mr) K_\lambda(mr) - I_\lambda'(mr) K_\lambda'(mr) \Big] - \frac{\lambda}{2m^2},
\qquad 
\\ \la{eq:i2}
\int_r^\infty dz\,z K_\lambda(mz)^2  &=& \frac{r^2}{2} (K_\lambda'(mr))^2 - \frac{1}{2} \Big(r^2 + \frac{\lambda^2}{m^2}\Big) K_\lambda(mr)^2.
\ea
In other radial integrals, one can first reduce the order of the Bessel functions using integration by parts.

\subsection{Pion-loop contribution to light-by-light scattering in scalar QED}
\la{sec:scalarQED}

In this subsection we present in some detail the calculation of the charged-pion-loop contribution
to $i\widehat\Pi$ in the framework of scalar QED.
In this framework, the pions are approximated as point particles; it should be noted that
the absence of form factors associated with the $\gamma \pi\pi$ vertex leads to an
$a_\mu^{\rm HLbL}$ contribution almost three times larger than the dispersively evaluated ``pion box''
(see Ref.~\cite{Colangelo:2017fiz}).

Since some expressions are quite long and only one mass appears in the entire calculation,
we denote the pion propagator simply by $G$ rather than $G_{m_\pi}$.
Also, the position-space vectors of the four vertices of the light-by-light amplitude
will generally be denoted by $(X^1,X^2,X^3,X^4)$, rather than $(x,y,z,0)$,
in order to notationally exploit the high degree of permutation symmetry of the amplitude.

The Euclidean Lagrangian for a massive complex scalar field minimally coupled to an external gauge field is 
\be
{\cal L} = (\partial_\mu +ieA_\mu) \phi^* (\partial_\mu-ieA_\mu) \phi + m_{\pi}^2 \phi^* \phi.
\ee
It is convenient to introduce the generating functional 
\be
Z[A_\mu] = \int D\phi D\phi^*\;e^{-S[\phi,\phi^*,A_\mu]}.
\ee
We want to compute the connected four-point function of the gauge field $A_\mu(x)$,
\be
\tilde\Pi_{\mu_1\mu_2\mu_3\mu_4}(X^1,X^2,X^3,X^4) \equiv 
\left.\frac{\delta^4\log Z}{\delta A_{\mu_1}(X^1)\delta A_{\mu_2}(X^2)\delta A_{\mu_3}(X^3)\delta A_{\mu_4}(X^4)}\right|_{A_\mu=0}.
\ee
Its relation to the desired function $i\widehat\Pi$ is given below in Eq.\ (\ref{eq:iPihatfromPitilde}).

Let 
\be
j_\mu(x) = i(\phi^* \partial_\mu \phi - \phi \partial_\mu \phi^*)
\ee
be the electromagnetic current (in units of $e$) associated with the complex scalar field $\phi$.
We split up the calculation of $\tilde \Pi$ into three contributions,
\be
\tilde \Pi_{\mu_1\mu_2\sigma\mu_3}(X^1,X^2,z,X^3) =
\sum_{n=0,1,2} \tilde \Pi^{(n)}_{\mu_1\mu_2\sigma\mu_3}(X^1,X^2,z,X^3),
\ee
where $\tilde \Pi^{(n)}$ is the contribution to $\tilde \Pi$ resulting from $n$ insertions of the
Lagrangian term $\Delta{\cal L} = e^2 \phi^*\phi A_\mu A_\mu$ and $(4-2n)$  insertions
of the electromagnetic current.

\subsubsection{Four-point function of the current}
As the main contribution to the four-point function of the gauge field $A_\mu(x)$, 
we compute the four-point function of the electromagnetic current
\ba
\tilde \Pi^{(0)}_{\mu_1 \mu_2 \mu_3 \mu_4}(X^1, X^2, X^3, X^4) 
&\equiv& \Big\< j_{\mu_1}(X^1) j_{\mu_2}(X^2)  j_{\mu_3}(X^3) j_{\mu_4}(X^4)\Big\>
\\ &=& \sum_{{\rm A}={\rm I,II,III}}\tilde \Pi_{\mu_1 \mu_2 \mu_3 \mu_4}^{(0),{\rm A}}(X^1, X^2, X^3, X^4).
\la{eq:PiAs}
\ea
There are 16 individual four-point functions generated by the two terms of each current. Each one gives rise to 6 Wick contractions.
Thus there are 96 Wick contractions in total.

The three types of terms that we distinguish in Eq.\ (\ref{eq:PiAs}) 
differ by the number of derivatives acting on $\phi$ and on the $\phi^*$ respectively,
\ba
\tilde \Pi_{\mu_1 \mu_2 \mu_3 \mu_4}^{(0),{\rm I}}(X^1, X^2, X^3, X^4) &=& 
 \Big\< \prod_{k=1}^4 \phi(X^k)^* \partial_{\mu_k} \phi(X^k) \Big\> + 
 \Big\< \textrm{herm. conjug.} \Big\> ,
\\
\tilde \Pi_{\mu_1 \mu_2\mu_3 \mu_4}^{(0),{\rm II}}(X^1, X^2, X^3, X^4) &=&
- \sum_{l=1}^4 \Big\{\Big\<  \phi(X^l)\partial_{\mu_l} \phi(X^l)^*\; \prod_{k\neq l}^4 \phi(X^k)^* \partial_{\mu_k} \phi(X^k)\Big\>
\\ && \qquad \quad 
 + \Big\<  \textrm{herm. conjug.} \Big\>\Big\},
\nonumber
\\
\tilde \Pi_{\mu_1 \mu_2\mu_3 \mu_4}^{(0),{\rm III}}(X^1, X^2, X^3, X^4) &=&
\sum_{k<l}^4\Big\< \phi(X^k)\partial_{\mu_k} \phi(X^k)^*  \cdot \phi(X^l)\partial_{\mu_l} \phi(X^l)^*\cdot
\\ &&  \qquad \cdot \prod_{j\neq k,l}^4 \phi(X^j)^* \partial_{\mu_j} \phi(X^j)\Big\>.
\nonumber
\ea
We note that $\tilde \Pi^{(0),{\rm I,II,III}}$ contain respectively 12, 48 and 36 Wick contractions.

We set $(X^3,\mu_3):=(z,\sigma)$ and then $(X^4,\mu_4):= (X^3,\mu_3)$, and write out the $z$-dependence explicitly,
since we want to integrate over $z$ at a later stage. We use the group of permutations ${\cal S}_n$, which 
contains $n!$ elements. The result of the Wick contractions is
\ba
\tilde \Pi_{\mu_1 \mu_2\sigma \mu_3}^{(0),{\rm I}}(X^1, X^2, z, X^3) &=&
2 \sum_{\pi\in{\cal S}_3} \partial_\sigma^{z} G(z-X^{\pi(3)})~
\partial_{\mu_{\pi(3)}}^{X^{\pi(3)}} G(X^{\pi(3)}-X^{\pi(2)})~
\\ && \qquad \partial_{\mu_{\pi(2)}}^{X^{\pi(2)}} G(X^{\pi(2)}-X^{\pi(1)})~
\partial_{\mu_{\pi(1)}}^{X^{\pi(1)}} G(X^{\pi(1)}-z),
\nonumber
\ea
and 
\ba
&& \tilde \Pi_{\mu_1 \mu_2\sigma \mu_3}^{(0),{\rm II}}(X^1, X^2, z, X^3) =
 2 \sum_{\pi\in{\cal S}_3} \Big\{
\\ && 
~~ \partial_\sigma^z \partial_{\mu_{\pi(3)}}^z G(z-X^{\pi(3)})~ G(X^{\pi(3)}-X^{\pi(2)})~
  \partial_{\mu_{\pi(2)}}^{X^{\pi(2)}} G(X^{\pi(2)}-X^{\pi(1)})~
 \partial_{\mu_{\pi(1)}}^{X^{\pi(1)}} G(X^{\pi(1)}-z)
\nonumber \\ && 
+ \partial_{\mu_{\pi(3)}}^{X^{\pi(3)}} \partial_{\sigma}^{X^{\pi(3)}} G(X^{\pi(3)}-z)~ G(z-X^{\pi(2)})~
 \partial_{\mu_{\pi(2)}}^{X^{\pi(2)}} G(X^{\pi(2)}-X^{\pi(1)}) ~\partial_{\mu_{\pi(1)}}^{X^{\pi(1)}} G(X^{\pi(1)}-X^{\pi(3)})
\nonumber\\ && 
+ \partial_{\mu_{\pi(3)}}^{X^{\pi(3)}} \partial_{\mu_{\pi(2)}}^{X^{\pi(3)}}  G(X^{\pi(3)}-X^{\pi(2)}) ~ G(X^{(2)}-z)~
 \partial_\sigma^z G(z-X^{\pi(1)})\, \partial_{\mu_{\pi(1)}}^{X^{\pi(1)}} G(X^{\pi(1)}-X^{\pi(3)})
\nonumber\\ &&
+ \partial_{\mu_{\pi(3)}}^{X^{\pi(3)}} \partial_{\mu_{\pi(2)}}^{X^{\pi(3)}}  G(X^{\pi(3)}-X^{\pi(2)}) ~ G(X^{\pi(2)}-X^{\pi(1)})~
\partial_{\mu_{\pi(1)}}^{X^{\pi(1)}} G(X^{\pi(1)}-z)~ \partial_\sigma^z G(z-X^{\pi(3)}) \Big\},
\nonumber
\ea
as well as 
\ba
&& \tilde \Pi_{\mu_1 \mu_2\sigma \mu_3}^{(0),{\rm III}}(X^1, X^2, z, X^3) =
2 \sum_{\pi\in{\cal S}_3} \Big\{
\\ &&
~~\partial_\sigma^z \partial_{\mu_{\pi(3)}}^{z} G(z-X^{\pi(3)})~ 
\partial_{\mu_{\pi(1)}}^{X^{\pi(1)}}\partial_{\mu_{\pi(2)}}^{X^{\pi(1)}} G(X^{\pi(1)}-X^{\pi(2)})~
G(X^{\pi(3)}-X^{\pi(1)}) ~G(X^{\pi(2)}-z) 
\nonumber\\ && +
\partial_{\mu_{\pi(3)}}^z G(z-X^{\pi(3)})~ \partial_\sigma^{X^{\pi(2)}}\partial_{\mu_{\pi(2)}}^{X^{\pi(2)}} G(X^{\pi(2)}-z)~
\partial_{\mu_{\pi(1)}}^{X^{\pi(1)}} G(X^{\pi(1)}-X^{\pi(2)})~ G(X^{\pi(3)}-X^{\pi(1)})
\nonumber\\ && + 
\partial_{\sigma}^{X^{\pi(1)}} G(X^{\pi(1)}-z)~ 
\partial_{\mu_{\pi(1)}}^{X^{\pi(3)}}\partial_{\mu_{\pi(3)}}^{X^{\pi(3)}} G(X^{\pi(3)}-X^{\pi(1)})~ 
\partial_{\mu_{\pi(2)}}^{X^{\pi(2)}} G(X^{\pi(2)}-X^{\pi(3)})~ G(z-X^{\pi(2)})
\Big\}.
\nonumber
\ea

\subsubsection{One-tadpole contributions}
Now to the  contributions of the four-point function of $A_\mu(x)$ involving exactly one tadpole,
coming from the term $\Delta{\cal L} = e^2 \phi^*\phi A_\mu A_\mu$ in the Lagrangian. Note that a factor 
two appears because the term is quadratic in $A_\mu$.
\ba
&& \tilde \Pi^{(1)}_{\mu_1 \mu_2 \mu_3 \mu_4}(X^1,X^2,X^3,X^4)
= - 2\sum_{k<l} \delta_{\mu_k \mu_l} \delta(X^k-X^l)\;\Big\<\phi^*(X^k) \phi(X^k)\; \prod_{j\neq k,l} j_{\mu_j}(X^j) \Big\>
\nonumber\\ && = 2
\sum_{k<l} \delta_{\mu_k \mu_l} \delta(X^k-X^l)\; \Big\{ 
\Big\< \phi^*(X^k) \phi(X^k)\; \prod_{j\neq k,l} \phi(X^j)^*\partial_{\mu_j}\phi(X^j) \Big\> + \Big\<\textrm{herm. conjug.}\Big\>
\nonumber\\ && -  \Big\<\phi^*(X^k) \phi(X^k)\; \sum_{j\neq k,l} 
\phi(X^j)^*\partial_{\mu_j}\phi(X^j) \Big(\phi(X^m)\partial_{\mu_m}\phi(X^m)^*\Big)_{m=10-(j+k+l)}\Big\> \Big\}.
\ea
Disregarding the overall factor of two, 
there are six permutations $(k<l)$ and each gives rise to eight Wick contractions, yielding a total of 48 such contractions.
Now set $(X^3,\mu_3):=(z,\sigma)$ and $(X^4,\mu_4):= (X^3,\mu_3)$.
There are three permutations in which $z$ appears in the delta function, and three where it does not.

\ba
&& \tilde \Pi^{(1)}_{\mu_1 \mu_2 \sigma \mu_3 }(X^1,X^2,z,X^3)
= 2\sum_{\pi\in{\cal S}_3} \Big\{
 \delta_{\sigma \mu_{\pi(1)}}\delta(z-X^{\pi(1)}) 
\\ && \cdot \Big(\partial_{\mu_{\pi(3)}}^{X^{\pi(3)}} G(X^{\pi(3)}-X^{\pi(2)})
~ G(X^{\pi(1)}-X^{\pi(3)}) ~ \partial_{\mu_{\pi(2)}}^{X^{\pi(2)}} G(X^{\pi(2)}-X^{\pi(1)}) 
\nonumber\\ && + G(X^{\pi(1)}-X^{\pi(2)}) ~\partial_{\mu_{\pi(3)}}^{X^{\pi(3)}} G(X^{\pi(3)}-X^{\pi(1)})
~ \partial_{\mu_{\pi(2)}}^{X^{\pi(2)}}G(X^{\pi(2)} -X^{\pi(3)}) 
\nonumber\\ && + \partial_{\mu_{\pi(3)}}^{X^{\pi(3)}}\partial_{\mu_{\pi(2)}}^{X^{\pi(3)}} G(X^{\pi(3)}-X^{\pi(2)})
~  G(X^{\pi(1)}-X^{\pi(3)}) ~ G(X^{\pi(2)}-X^{\pi(1)})
\nonumber\\ && + \partial_{\mu_{\pi(2)}}^{X^{\pi(1)}} G(X^{\pi(1)}-X^{\pi(2)}) ~
\partial_{\mu_{\pi(3)}}^{X^{\pi(3)}} G(X^{\pi(3)}-X^{\pi(1)}) ~ G(X^{\pi(2)}-X^{\pi(3)}) \Big)
\nonumber\\ && +  \delta_{\mu_{\pi(1)}\mu_{\pi(2)}} \delta(X^{\pi(1)}-X^{\pi(2)})
\nonumber\\ && \cdot 
 \Big(\partial_{\mu_{\pi(3)}}^{X^{\pi(3)}}  G(X^{\pi(3)}-z) ~ G(X^{\pi(1)}-X^{\pi(3)})~ \partial_\sigma^z G(z-X^{\pi(1)})
\nonumber\\ && +  G(X^{\pi(1)}-z)~ \partial_{\mu_{\pi(3)}}^{X^{\pi(3)}} G(X^{\pi(3)} - X^{\pi(1)})~ \partial_\sigma^z G(z-X^{\pi(3)})
\nonumber\\ && + \partial_{\mu_{\pi(3)}}^{X^{\pi(3)}}\partial_{\sigma}^{X^{\pi(3)}} G(X^{\pi(3)}-z) ~ G(X^{\pi(1)}-X^{\pi(3)})~ G(z-X^{\pi(1)})
\nonumber\\ && + \partial_{\sigma}^{X^{\pi(1)}} G(X^{\pi(1)}-z) ~\partial_{\mu_{\pi(3)}}^{X^{\pi(3)}}G(X^{\pi(3)}-X^{\pi(1)}) ~
 G(z-X^{\pi(3)})\Big)\Big\}.
\nonumber
\ea

\subsubsection{Two-tadpole contributions}
Finally, the contribution containing two tadpoles has the form 
\ba
&& \tilde \Pi^{(2)}_{\mu_1 \mu_2 \mu_3 \mu_4}(X^1,X^2,X^3,X^4)
\\ && = 4 \sum_{l=1}^3  \delta_{\mu_4\mu_l}\, \delta(X^4-X^l) \,\delta_{\mu_j\mu_k}\, \delta(X^j-X^k)  \Big\<(\phi^*\phi)(X^4)\, (\phi^*\phi)(X^l) \Big\>,
\nonumber
\ea
where it is understood that $(j,k,l)$ form a permutation of $(1,2,3)$. The expression can also be written as 
\ba
&& \tilde \Pi^{(2)}_{\mu_1 \mu_2 \mu_3 \mu_4}(X^1,X^2,X^3,X^4)
\\ && = 2 \sum_{\pi\in{\cal S}_3} 
 \delta_{\mu_4\mu_{\pi(1)}}\, \delta(X^4-X^{\pi(1)}) \,\delta_{\mu_{\pi(2)}\mu_{\pi(3)}}\, 
     \delta(X^{\pi(2)}-X^{\pi(3)})  \Big\<(\phi^*\phi)(X^4)\, (\phi^*\phi)(X^{\pi(2)}) \Big\>,
\nonumber
\ea
Performing the contractions, one obtains
\ba
&& \tilde \Pi^{(2)}_{\mu_1 \mu_2 \sigma \mu_3 }(X^1,X^2,z,X^3)
\\ && = 2 \sum_{\pi\in{\cal S}_3} 
 \delta_{\sigma\mu_{\pi(1)}}\, \delta(z-X^{\pi(1)}) \,\delta_{\mu_{\pi(2)}\mu_{\pi(3)}}\, 
     \delta(X^{\pi(2)}-X^{\pi(3)})  G(X^{\pi(1)}-X^{\pi(2)})^2.
\nonumber
\ea

\subsubsection{Test of the Ward identity }

The Ward identity for current conservation reads
\be\la{eq:WI}
\partial_\sigma^z (\tilde
 \Pi^{(0)}_{\mu_1\mu_2\sigma\mu_3}
+ \tilde \Pi^{(1)}_{\mu_1\mu_2\sigma\mu_3} 
+ \tilde \Pi^{(2)}_{\mu_1\mu_2\sigma\mu_3})(X^1,X^2,z,X^3)=0.
\ee
Taking into account the Green's function property (\ref{eq:Klein_Gordon}) of the scalar propagator
as well as the identity
$\delta'(z-x_1) f(z) = \delta'(z-x_1) f(x_1) - \delta(z-x_1) f'(x_1)$,
a straightforward but tedious calculation yields
\ba
&& \partial_\sigma^z \tilde \Pi^{(0)}_{\mu_1\mu_2\sigma\mu_3}(X^1,X^2,z,X^3)
= -2 \sum_{\pi\in{\cal S}_3} \partial_{\mu_{\pi(1)}}^z \delta(z-X^{\pi(1)}) \Big\{
\\ && ~~ G(X^{\pi(1)}-X^{\pi(2)})~ \partial_{\mu_2}^{X^2}G(X^2-X^3)~ \partial_{\mu_3}^{X^3} G(X^{\pi(3)}-X^{\pi(1)})
\nonumber\\ && + G(X^{\pi(1)}-X^{\pi(2)})~ \partial_{\mu_{\pi(2)}}^{X^2} ~G(X^{\pi(2)}-X^{\pi(3)})~ 
\partial_{\mu_{\pi(3)}}^{X^3} G(X^{\pi(3)}-X^{\pi(1)})
\nonumber \\ && 
+ \partial_{\mu_{\pi(3)}}^{X^{\pi(1)}}G(X^{\pi(1)}-X^{\pi(3)})~ \partial_{\mu_{\pi(2)}}^{X^2} ~G(X^{\pi(2)}-X^{\pi(1)})~
G(X^{\pi(3)}-X^{\pi(2)})
\nonumber \\ && 
+ \partial_{\mu_{\pi(3)}}^{X^{\pi(3)}}\partial_{\mu_{\pi(2)}}^{X^{\pi(3)}} G(X^{\pi(3)}-X^{\pi(2)})~
G(X^{\pi(1)}-X^{\pi(3)})~ G(X^{\pi(2)}-X^{\pi(1)}) 
\Big\}.
\nonumber 
\ea
Similarly, using again Eq.\ (\ref{eq:Klein_Gordon}), one finds
\ba
&& \partial_\sigma^z \tilde \Pi^{(1)}_{\mu_1\mu_2\sigma\mu_3}(X^1,X^2,z,X^3) = 
2\sum_{\pi\in{\cal S}_3} \Big\{ \partial_{\mu_{\pi(1)}}^z \delta(z-X^{\pi(1)}) 
\\ && \cdot \Big(\partial_{\mu_{\pi(3)}}^{X^{\pi(3)}} G(X^{\pi(3)}-X^{\pi(2)})
~ G(X^{\pi(1)}-X^{\pi(3)}) ~ \partial_{\mu_{\pi(2)}}^{X^{\pi(2)}} G(X^{\pi(2)}-X^{\pi(1)}) 
\nonumber\\ && + G(X^{\pi(1)}-X^{\pi(2)}) ~\partial_{\mu_{\pi(3)}}^{X^{\pi(3)}} G(X^{\pi(3)}-X^{\pi(1)})
~ \partial_{\mu_{\pi(2)}}^{X^{\pi(2)}}G(X^{\pi(2)} -X^{\pi(3)}) 
\nonumber\\ && + \partial_{\mu_{\pi(3)}}^{X^{\pi(3)}}\partial_{\mu_{\pi(2)}}^{X^{\pi(3)}} G(X^{\pi(3)}-X^{\pi(2)})
~  G(X^{\pi(1)}-X^{\pi(3)}) ~ G(X^{\pi(2)}-X^{\pi(1)})
\nonumber\\ && + \partial_{\mu_{\pi(2)}}^{X^{\pi(1)}} G(X^{\pi(1)}-X^{\pi(2)}) ~
\partial_{\mu_{\pi(3)}}^{X^{\pi(3)}} G(X^{\pi(3)}-X^{\pi(1)}) ~ G(X^{\pi(2)}-X^{\pi(3)}) \Big)
\nonumber\\ && - \delta_{\mu_{\pi(1)}\mu_{\pi(2)}} \delta(X^{\pi(1)}-X^{\pi(2)}) 
\; \partial_{\mu_{\pi(3)}}^{z} \delta(X^{\pi(3)}-z)~ G(X^{\pi(3)}-X^{\pi(1)})^2
\Big\}.
\nonumber
\ea
Finally, 
\ba
&& \partial_\sigma^z \tilde \Pi^{(2)}_{\mu_1\mu_2\sigma\mu_3}(X^1,X^2,z,X^3) =
\\ && = 2 \sum_{\pi\in{\cal S}_3} 
 \partial_{\mu_{\pi(1)}}^z\, \delta(z-X^{\pi(1)}) \,\delta_{\mu_{\pi(2)}\mu_{\pi(3)}}\, 
     \delta(X^{\pi(2)}-X^{\pi(3)})  G(X^{\pi(1)}-X^{\pi(3)})^2.
\nonumber
\ea
Thus one verifies that the Ward identity Eq.\ (\ref{eq:WI}) is satisfied: the terms with  two delta functions
cancel between $\tilde\Pi^{(1)}$ and $\tilde\Pi^{(2)}$, while the terms with a single delta function
cancel between $\tilde\Pi^{(0)}$ and $\tilde\Pi^{(1)}$.

\subsubsection{The expression for $i\widehat \Pi_{\rho;\mu_1\mu_2\mu_3\sigma}(X^1,X^2)$}

We recall the relation
\be\la{eq:iPihatfromPitilde}
i\widehat \Pi_{\rho;\mu_1\mu_2\mu_3\sigma}(X^1,X^2) 
= -\int d^4z\; z_\rho \; \tilde \Pi_{\mu_1\mu_2\sigma\mu_3}(X^1,X^2,z,X^3)\Big|_{X^3=0},
\ee
and decompose the rank-five tensor according to
\ba\la{eq:iPihatPionDecomp}
i\widehat \Pi_{\rho;\mu_1\mu_2\mu_3\sigma}(X^1,X^2) &=& \sum_{n=0,1,2} i\widehat \Pi^{(n)}_{\rho;\mu_1\mu_2\mu_3\sigma}(X^1,X^2),
\\
i\widehat \Pi^{(n)}_{\rho;\mu_1\mu_2\mu_3\sigma}(X^1,X^2) &\equiv &
 -\int d^4z\; z_\rho \; \tilde \Pi^{(n)}_{\mu_1\mu_2\sigma\mu_3}(X^1,X^2,z,X^3)\Big|_{X^3=0}.
\ea

We will make use of the integral
\ba
H_\rho(X^1,X^3) &=&\int d^4z \, z_\rho \,G(z-X^3) \,G(z-X^1)
\\  &=& \left(\frac{m_{\pi}}{4\pi^2}\right)^2 \Big[g_\rho(X^3-X^1) + X^1_\rho\, p(|X^3-X^1|)  \Big],
\ea
where
\ba
g_\rho(y) &\equiv& \int d^4z\; z_\rho\; \frac{K_1(m_{\pi}|y-z|)}{|y-z|}\; \frac{K_1(m_{\pi}|z|)}{|z|} = \frac{\pi^2}{m_{\pi}^2} y_\rho\, K_0(m_{\pi}|y|),
\\
p(|y|) &\equiv& \int d^4z\; \frac{K_1(m_{\pi}|y-z|)}{|y-z|} \frac{K_1(m_{\pi}|z|)}{|z|}  = \frac{2\pi^2}{m_{\pi}^2}\, K_0(m_{\pi}|y|).
\ea
Simplifying, one finds
 \be
 H_\rho(X^1,X^3) = \frac{1}{16\pi^2} K_0(m_{\pi}|X^3-X^1|)\; (X^1_\rho+X^3_\rho).
 \ee
What is needed in the following is the
antisymmetrized derivative 
\be
\partial_{[\sigma}^{X^3} H_{\rho]}(X^1,X^3) = - \frac{1}{4} G(X^3-X^1)\, 
[X^1_\rho,X^3_\sigma],
\ee
where we have introduced the notation
$ [X^1_\rho,X^3_\sigma] \equiv X^1_\rho X^3_\sigma - X^3_\rho X^1_\sigma $.

A straightforward but lengthy calculation then leads to the expressions 
\ba\la{eq:PihatPionL0}
&& i\widehat \Pi^{(0)}_{[\rho;\mu_1\mu_2\mu_3\sigma]}(X^1,X^2)
\\ &&  =  \sum_{\pi\in{\cal S}_3} \Big\{
  G(X^{\pi(3)}-X^{\pi(2)})~  \partial_{\mu_{\pi(2)}}^{X^{\pi(2)}} G(X^{\pi(2)}-X^{\pi(1)})~
\partial_{\mu_{\pi(1)}}^{X^{\pi(1)}}\partial_{\mu_{\pi(3)}}^{X^{\pi(3)}} G(X^{\pi(1)}-X^{\pi(3)})~ [X^{\pi(1)}_\rho,X^{\pi(3)}_\sigma]
\nonumber \\ && +
  G(X^{\pi(3)}-X^{\pi(2)})~  \partial_{\mu_{\pi(2)}}^{X^{\pi(2)}} G(X^{\pi(2)}-X^{\pi(1)})~
 G(X^{\pi(1)}-X^{\pi(3)})
(\delta_{\mu_{\pi(1)}\rho} \delta_{\sigma\mu_{\pi(3)}}  - \delta_{\rho\mu_{\pi(3)}} \delta_{\sigma\mu_{\pi(1)}} )
\nonumber \\ && +
  G(X^{\pi(3)}-X^{\pi(2)})~  \partial_{\mu_{\pi(2)}}^{X^{\pi(2)}} G(X^{\pi(2)}-X^{\pi(1)})~
\partial_{\mu_{\pi(1)}}^{X^{\pi(1)}}G(X^{\pi(1)}-X^{\pi(3)}) ~ 
(\delta_{\sigma\mu_{\pi(3)}} X^{\pi(1)}_\rho - \delta_{\rho\mu_{\pi(3)}} X^{\pi(1)}_\sigma ) 
\nonumber \\ && +
  G(X^{\pi(3)}-X^{\pi(2)})~  \partial_{\mu_{\pi(2)}}^{X^{\pi(2)}} G(X^{\pi(2)}-X^{\pi(1)})~
\partial_{\mu_{\pi(3)}}^{X^{\pi(3)}}G(X^{\pi(1)}-X^{\pi(3)})
~(\delta_{\rho\mu_{\pi(1)}} X^{\pi(3)}_\sigma - \delta_{\sigma\mu_{\pi(1)}} X^{\pi(3)}_\rho )
\nonumber \\ && 
+ \frac{1}{2} \partial_{\mu_{\pi(2)}}^{X^{\pi(2)}} G(X^{\pi(2)}-X^{\pi(1)}) ~\partial_{\mu_{\pi(1)}}^{X^{\pi(1)}} G(X^{\pi(1)}-X^{\pi(3)})
\partial_{\mu_{\pi(3)}}^{X^{\pi(3)}}  G(X^{\pi(2)}-X^{\pi(3)})~ [X^{\pi(3)}_\rho,X^{\pi(1)}_\sigma-X^{\pi(2)}_\sigma]
\nonumber \\ && + \frac{1}{2}  \partial_{\mu_{\pi(3)}}^{X^{\pi(3)}} G(X^{\pi(3)}-X^{\pi(2)}) 
~\partial_{\mu_{\pi(2)}}^{X^{\pi(2)}} G(X^{\pi(2)}-X^{\pi(1)}) G(X^{\pi(3)}-X^{\pi(1)}) 
 (\delta_{\sigma\mu_{\pi(1)}} X^{\pi(3)}_\rho - \delta_{\rho\mu_{\pi(1)}} X^{\pi(3)}_\sigma) 
\nonumber \\ && + \frac{1}{2}   G(X^{\pi(3)}-X^{\pi(2)}) 
~\partial_{\mu_{\pi(2)}}^{X^{\pi(2)}} G(X^{\pi(2)}-X^{\pi(1)}) \partial_{\mu_{\pi(1)}}^{X^{\pi(1)}} G(X^{\pi(3)}-X^{\pi(1)}) 
 (\delta_{\sigma\mu_{\pi(3)}} X^{\pi(2)}_\rho - \delta_{\rho\mu_{\pi(3)}} X^{\pi(2)}_\sigma) 
\nonumber\\ && 
+ \partial_{\mu_{\pi(3)}}^{X^{\pi(3)}} \partial_{\mu_{\pi(2)}}^{X^{\pi(3)}}  G(X^{\pi(3)}-X^{\pi(2)}) ~ G(X^{\pi(2)}-X^{\pi(1)})~
\partial_{\mu_{\pi(1)}}^{X^{\pi(1)}}  G(X^{\pi(1)}-X^{\pi(3)})~ [X^{\pi(1)}_\rho,X^{\pi(2)}_\sigma-X^{\pi(3)}_\sigma] 
\nonumber\\ && + 
\partial_{\mu_{\pi(3)}}^{X^{\pi(3)}} \partial_{\mu_{\pi(2)}}^{X^{\pi(3)}}  G(X^{\pi(3)}-X^{\pi(2)}) ~ G(X^{\pi(2)}-X^{\pi(1)})~
G(X^{\pi(1)}-X^{\pi(3)}) 
(\delta_{\sigma\mu_{\pi(1)}} X^{\pi(3)}_\rho - \delta_{\rho\mu_{\pi(1)}} X^{\pi(3)}_\sigma ) 
\Big\},
\nonumber
\ea
\ba\la{eq:PihatPionL1and2}
&&  i\widehat \Pi^{(1)}_{[\rho;\mu_1\mu_2\mu_3\sigma]}(X^1,X^2)
+  i\widehat \Pi^{(2)}_{[\rho;\mu_1\mu_2\mu_3\sigma]}(X^1,X^2)
= \sum_{\pi\in{\cal S}_3} \Big\{
 ( \delta_{\rho \mu_{\pi(1)}} X^{\pi(1)}_\sigma - \delta_{\sigma \mu_{\pi(1)}} X^{\pi(1)}_\rho) 
\qquad 
\\ && \cdot \Big(
 2 G(X^{\pi(1)}-X^{\pi(2)}) ~\partial_{\mu_{\pi(3)}}^{X^{\pi(3)}} G(X^{\pi(3)}-X^{\pi(1)})
~ \partial_{\mu_{\pi(2)}}^{X^{\pi(2)}}G(X^{\pi(2)} -X^{\pi(3)}) 
\nonumber\\ && + \partial_{\mu_{\pi(2)}}^{X^{\pi(1)}} G(X^{\pi(1)}-X^{\pi(2)}) ~
\partial_{\mu_{\pi(3)}}^{X^{\pi(3)}} G(X^{\pi(3)}-X^{\pi(1)}) ~ G(X^{\pi(2)}-X^{\pi(3)}) 
\nonumber\\ && +
\partial_{\mu_{\pi(3)}}^{X^{\pi(3)}}\partial_{\mu_{\pi(2)}}^{X^{\pi(3)}} G(X^{\pi(3)}-X^{\pi(2)})
~ G(X^{\pi(1)}-X^{\pi(3)}) ~ G(X^{\pi(2)}-X^{\pi(1)})
\nonumber\\ && + 
 \delta_{\mu_{\pi(2)}\mu_{\pi(3)}}
\delta(X^{\pi(2)}-X^{\pi(3)})  ~ G(X^{\pi(2)}-X^{\pi(1)})^2 \Big) 
\nonumber\\ && +  \delta_{\mu_{\pi(2)}\mu_{\pi(3)}} \delta(X^{\pi(2)}-X^{\pi(3)})  ~ G(X^{\pi(2)}-X^{\pi(1)})^2 
 ( \delta_{\sigma \mu_{\pi(1)}} X^{\pi(2)}_\rho - \delta_{\rho \mu_{\pi(1)}} X^{\pi(2)}_\sigma)
\Big\}.
\nonumber
\ea
The setting  of $X^3$ to zero is implied in the two equations above (see Eq.\ (\ref{eq:iPihatfromPitilde})).
In summary, the charged-pion-loop contribution to the function
$i\widehat\Pi$ is given by 
 Eqs.\ (\ref{eq:iPihatPionDecomp}), (\ref{eq:PihatPionL0}) and (\ref{eq:PihatPionL1and2}).

A final step is required if one wants to perform numerical integrations over $i\widehat\Pi$ in order
to obtain $a_\mu^{\rm HLbL}$, namely to isolate the delta-function-like contributions, which we also call
contact contributions. The last two terms of Eq.\ (\ref{eq:PihatPionL1and2}) are explicitly contact contributions.
Eq.~(\ref{eq:PihatPionL0}), however, also contributes contact terms, because second derivatives of the scalar propagator appear.
Using the defining property (\ref{eq:Klein_Gordon}) of the propagator, these second-derivative terms can be written as 
\be
\partial_\mu \partial_\nu G(x) = \left(\delta_{\mu\nu} - 4\frac{x_\mu x_\nu}{x^2}\right) \frac{G'(x)}{|x|}
 + \frac{x_\mu x_\nu}{x^2}\,m_{\pi}^2\,G(x) - \frac{1}{4} \delta_{\mu\nu}\,\delta(x),
\ee
where $G'(x) = \frac{d}{d|x|} G(x) = \frac{-m_{\pi}^2}{4\pi^2}\frac{K_2(m_{\pi}|x|)}{|x|}$.
We have taken into account the fact that, applied to smooth test functions,
$\left(\delta_{\mu\nu} - 4\frac{x_\mu x_\nu}{x^2}\right) \delta(x) = 0$,
so that $\frac{x_\mu x_\nu}{x^2}\delta(x)$ can be substituted by $\frac{1}{4}\delta_{\mu\nu} \delta(x)$.
Collecting all the contributions proportional to a delta function
in $i\widehat\Pi_{[\rho;\mu_1\mu_2\mu_3\sigma]}(X^1,X^2)$, we find in total
\ba\la{eq:Pihat_contact}
 i\widehat \Pi_{[\rho;\mu_1\mu_2\mu_3\sigma]}(X^1,X^2)_{\rm contact} 
 &=& \frac{3}{2} \Big[ 
\delta_{\mu_{2}\mu_{3}}\,\delta(X^2)\,G(X^1)^2\, (\delta_{\rho\mu_1}X^1_\sigma - \delta_{\sigma\mu_1} X^1_\rho)
 \\ && 
- \delta_{\mu_{1}\mu_{2}}\,\delta(X^1-X^2)\,G(X^2)^2\, (\delta_{\rho\mu_3}X^2_\sigma - \delta_{\sigma\mu_3}X^2_\rho)
\nonumber \\ &&
+ \delta_{\mu_{1}\mu_{3}}\,\delta(X^1)\,G(X^2)^2\, (\delta_{\rho\mu_2}X^2_\sigma - \delta_{\sigma\mu_2} X^2_\rho)\Big].
\nonumber
\ea
We note that these terms are integrable; for instance, the first term goes like $1/|X^1|^3$ for small $|X^1|$.

\section{Applications and tests of the QED kernel \la{sec:numtests}}

In this section, we combine the QED kernel as computed in
section~\ref{sec:alternative_evaluation} with the four-point functions
$i\widehat\Pi$ given explicitly in section~\ref{sec:ipihats} in order
to check whether the coordinate-space method developed here reproduces
known results. Furthermore, the method is tested for the case that the
QED fermion loop is computed on the lattice, reproducing the known
result after taking the continuum limit. Finally, in subsection
\ref{sec:LQCD} an overview of results obtained for the
fully connected part of the lattice QCD four-point function is presented.

\subsection{Improved kernels \la{sec:KernelSubtractions}}

The master formula, given by Eq.~(\ref{eq:master_formula}), can be written in a slightly different way to optimize the lattice QCD calculation. First, as already noted in Ref.~\cite{Blum:2017cer}, the QED weight function is not uniquely defined. This freedom can be used to obtained a better behaved integrand with smaller statistical and systematic uncertainties. As shown below, it turns out to be a crucial ingredient for practical lattice QCD calculations. The idea is to remove large fluctuations or large cancellations in the integrand that do not affect the central value in the continuum and infinite volume but increase the statistical error and/or systematic effects in the estimator.
Second, a naive implementation of the master formula in lattice QCD calculations is rather expensive. 
But the numerical cost can be considerably reduced by using a different formula, equivalent in the infinite volume limit. These two improvements are  discussed next. 

In the continuum and in infinite volume, the conservation of the electromagnetic current implies Eq.\ (\ref{eq:iPihatarea}),
namely that the integral of the four-point function $i\widehat\Pi_{\rho;\mu\nu\lambda\sigma}(x,y)$ over $x$ without any
$x$-dependent weight-factor vanishes. The same observation applies to the integral over the coordinate-vector $y$.
Therefore, in infinite volume, any function which depends only on $x$ or $y$ can be added to the QED kernel without affecting the final result. On the infinite lattice, and with Wilson fermions, this property still holds at finite lattice spacing if one uses the conserved vector current, see Eq.~(\ref{eq:consvec}). When using local vector currents, the result holds once the continuum limit has been taken.
We will consider four kernels $\Lb^{(n)}$ which differ only by such subtractions,  
\begin{subequations}
\label{eq:kernels}
\begin{align}
\Lb^{(0)}_{[\rho,\sigma],\mu\nu\lambda}(x,y) &= \Lb_{[\rho,\sigma],\mu\nu\lambda}(x,y) \,, \\
\Lb^{(1)}_{[\rho,\sigma],\mu\nu\lambda}(x,y) &= \Lb_{[\rho,\sigma],\mu\nu\lambda}(x,y) - \frac{1}{2}  \Lb_{[\rho,\sigma],\mu\nu\lambda}(x,x) - \frac{1}{2}  \Lb_{[\rho,\sigma],\mu\nu\lambda}(y,y) \,, \\
\Lb^{(2)}_{[\rho,\sigma],\mu\nu\lambda}(x,y) &= \Lb_{[\rho,\sigma],\mu\nu\lambda}(x,y) - \Lb_{[\rho,\sigma],\mu\nu\lambda}(0,y) - \Lb_{[\rho,\sigma],\mu\nu\lambda}(x,0)\,, \\
\Lb^{(3)}_{[\rho,\sigma],\mu\nu\lambda}(x,y) &= \Lb_{[\rho,\sigma],\mu\nu\lambda}(x,y) - \Lb_{[\rho,\sigma],\mu\nu\lambda}(x,x) + \Lb_{[\rho,\sigma],\mu\nu\lambda}(0,x) - \Lb_{[\rho,\sigma],\mu\nu\lambda}(0,y) \,.
\end{align}
\end{subequations}
In addition to vanishing when both arguments vanish,
\be
\Lb^{(n)}_{[\rho,\sigma],\mu\nu\lambda}(0,0) = 0,\qquad (n=0,1,2,3),
\ee
(see Eq.\ (\ref{eq:LbarEq0at0}) for the non-trivial cases $n=0$ and $n=2$),
the subtracted kernels vanish in various special configurations,
\begin{align}
\Lb^{(1)}(x,x) = 0 \,, \quad \Lb^{(2)}(x,0) = \Lb^{(2)}(0,y) = 0 \,, \quad \Lb^{(3)}(x,0) = \Lb^{(3)}(x,x) = 0 \,.
\end{align}

In \cite{Asmussen:2019act,Chao:2020kwq} we introduced a kernel tuneable by an arbitrary parameter $\Lambda$ that approaches $\Lb^{(0)}$ when $\Lambda \rightarrow \infty$ and $\Lb^{(2)}$ when $\Lambda\rightarrow 0$,
\begin{align}\label{eq:lamsub}
\Lb^{(\Lambda)}_{[\rho,\sigma];\mu\nu\lambda}(x,y) &= \Lb_{[\rho,\sigma];\mu\nu\lambda}(x,y)\nonumber \\ 
&-\partial_\mu^{(x)} (x_\alpha e^{-\Lambda m^2 x^2/2}) \Lb_{[\rho,\sigma];\alpha\nu\lambda}(0,y) - \partial_\nu^{(y)} (y_\alpha e^{-\Lambda m^2 y^2/2})\Lb_{[\rho,\sigma];\mu\alpha\lambda}(x,0),
\end{align}
as we empirically found that with $\Lb^{(2)}$ and $\Lb^{(3)}$, the integrand was too long-ranged, while with $\Lb^{(0)}$ and $\Lb^{(1)}$ it was too peaked at short distances. In our most recent works~\cite{Chao:2021tvp,Chao:2022xzg}, we presented results exclusively with $\Lambda=0.4$.

\begin{figure}[t]
        \includegraphics*[width=0.18\linewidth]{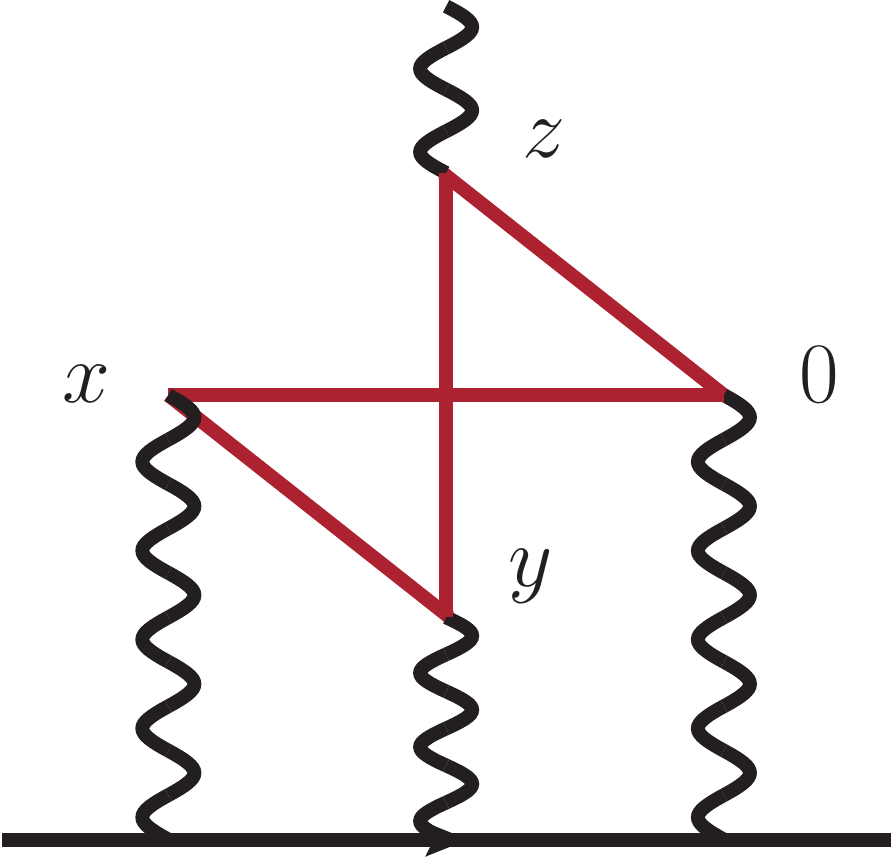} \hspace{1cm}
        \includegraphics*[width=0.18\linewidth]{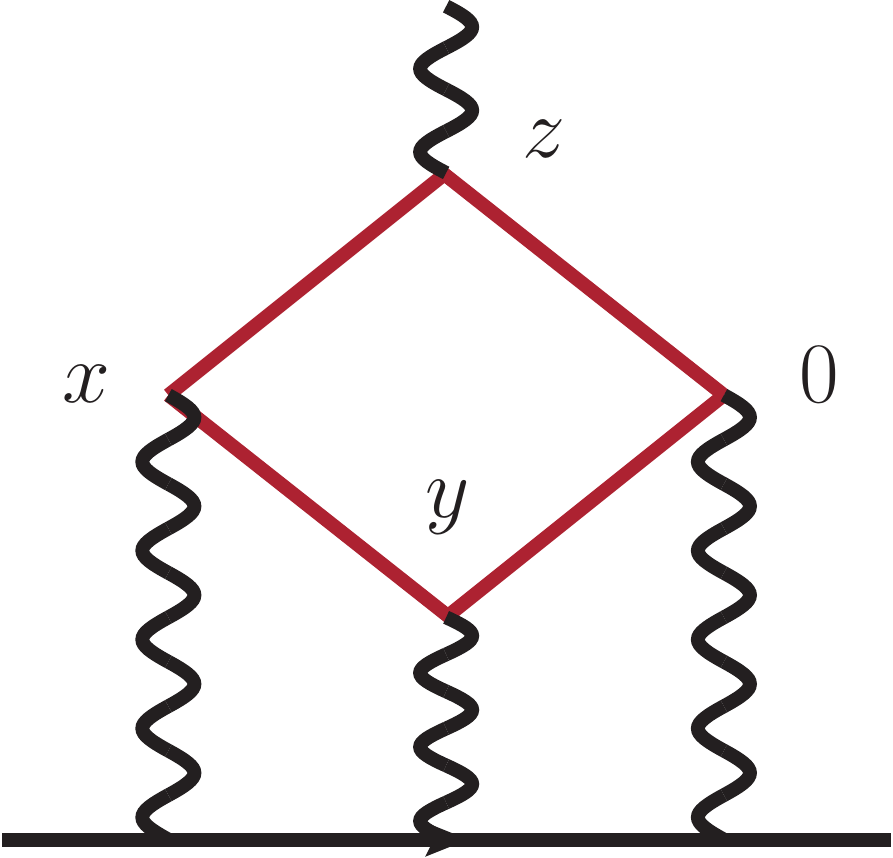} \hspace{1cm}
        \includegraphics*[width=0.18\linewidth]{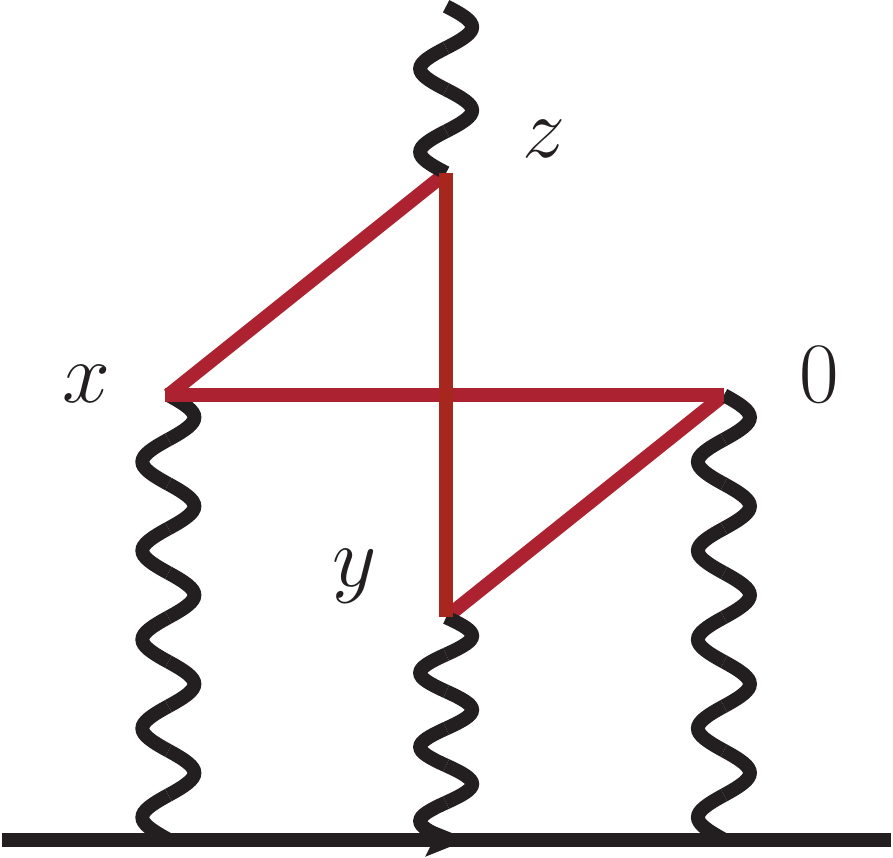}
                        
        \caption{Wick contractions for the connected contribution in Method 1.
        }
        \label{fig:wickM1}
\end{figure}

For each quark flavour $f$,
the quark-connected part of the hadronic four-point function $\Pi$ involves three different Wick contractions.
Each of those appears twice, with opposite fermion-number flow, resulting in a purely real contribution; see Eq.\ (\ref{eq:master4QED})
and Fig.~\!\ref{fig:wickM1}.
Computing explicitly all three contractions, and using the master formula given by Eq.~(\ref{eq:master_formula}), amounts to what we call Method~1~\cite{Chao:2020kwq}. To estimate all three contractions, we first compute point-to-all propagators with sources located at the origin and at the site $y$. Then, we perform sequential inversions using the propagators, summed over $z$ and with the weight factor $z_{\rho}$, as sequential sources to finally contract both results and summing over $x$. Since one needs to anti-symmetrize between $\rho$ and $\sigma$, it amounts to 6 sequential inversions for each primary inversion.

Alternatively, we could choose the first Wick contraction, $I^{(1)}_{\mu\nu\sigma\lambda}(x,y,z,w)$ of Eq.\ (\ref{eq:wickI1}) corresponding
to the leftmost diagram in Fig.~\ref{fig:wickM1},
as a reference and swap the vertices at the level of the muon line -- an idea that was already exploited in~\cite{Blum:2015gfa}.
Defining the gauge-field average of the three Wick contractions as follows,
\be
\widetilde\Pi^{(j)}_{\mu\nu\sigma\lambda}(x,y,z) \equiv \<-2\,{\rm Re}\;I^{(j)}_{\mu\nu\sigma\lambda}(x,y,z,0)\>,
\qquad (j=1,2,3),
\ee
we obtain the  Method~2~\cite{Chao:2020kwq} estimator ($A=0,1,2,3$ or $\Lambda$)
\begin{multline}\label{eq:master2}
  \amu^{\rm HLbL} = -\frac{me^6}{3} \int {d}^4y \int {d}^4x  \, \\  \Big\{ \, \Big(  \Lb^{(A)}_{[\rho,\sigma],\mu\nu\lambda}(x,y) + \Lb^{(A)}_{[\rho,\sigma],\nu\mu\lambda}(y,x) - \Lb^{(A)}_{[\rho,\sigma],\lambda\nu\mu}(x,x-y)\Big)  \,\int d^4z\, z_\rho\; \widetilde\Pi^{(1)}_{\mu\nu\sigma\lambda}(x,y,z)
  \\
+ \Lb^{(A)}_{[\rho,\sigma],\lambda\nu\mu}(x,x-y)  \, x_{\rho} \,\int {d}^4z \,  \widetilde\Pi^{(1)}_{\mu\nu\sigma\lambda}(x,y,z) \Big\} \,.
\end{multline}
For the quark-connected contribution, and with $A\doteq \Lambda$, this equation is the starting point for all our lattice-QCD results 
in Refs.~\cite{Chao:2021tvp,Chao:2022xzg}, as well as the final results in Ref.~\cite{Chao:2020kwq}.
The advantage of this representation is that
all propagators can be expressed in terms of the two point-to-all propagators with sources located at the origin and on site $y$
by exploiting the $\gamma_5$-hermiticity relation (\ref{eq:g5herm}).
Eq.\ (\ref{eq:master2}) can be proven starting from the master formula (\ref{eq:master_formula}) and using the identities 
\begin{subequations}
\label{eq:Pi_sym}
\begin{align}
  \widetilde\Pi^{(2)}_{\mu\nu\sigma\lambda}(x,y,z) &= \widetilde\Pi^{(1)}_{\nu\mu\sigma\lambda}(y,x,z) \,, \\
\widetilde\Pi^{(3)}_{\mu\nu\sigma\lambda}(x,y,z) &= \widetilde\Pi^{(1)}_{\lambda\nu\sigma\mu}(-x,y-x,z-x) \,
\end{align}
\end{subequations}
as well as Eq.\ (\ref{eq:LbarIsOdd}).
In practice, we reduce the master formulae to a one-dimensional
integral over the variable $|y|$; the integrand then differs between Method~1 and Method~2, even in the continuum limit. The advantage of Method~2 is that only one additional propagator needs to be computed for each value of $y$: for $N$ values of $|y|$, the number of quark propagators that need to be computed is $N+1$ compared to $7(N+1)$ for Method~1, where sequential inversions are used.
In addition, combining all possible pairs of quark propagators allows one to compute O$(N^2)$ independent data, which may include multiple statistical samples of the same $|y|$. 

In general, the last term in Eq.~(\ref{eq:master2}) does not vanish, in spite of Eq.\ (\ref{eq:JJJJarea}) holding,
because $\Pi^{(1)}_{\mu\nu\sigma\lambda}(x,y,z)$ is only one of three contributing Wick contractions
to the four-point function. In fact, as a Ward identity following from current conservation, one can show that
\be
\int d^4z\; \widetilde\Pi^{(1)}_{\mu\nu\sigma\lambda}(x,y,z) = (-2y_\sigma)\;
\Big\<{\rm Re}\,{\rm Tr}\{\gamma_\mu S(x,y) \gamma_\nu S(y,0) \gamma_\lambda S(0,x)  \}\Big\>.
\ee
However, for a pseudoscalar-pole contribution, this term vanishes,
\begin{align}\la{eq:tildePi1area}
  \int {d}^4z \,  \Pi^{(1);\pi^0}_{\mu\nu\sigma\lambda}(x,y,z) = 0 \,,
 \end{align}
because\footnote{As noted in~\cite{Asmussen:2019act,Chao:2021tvp}, the quark-level Wick contraction
 $\Pi^{(1)}_{\mu\nu\sigma\lambda}(x,y,z)$ does not contain
 the diagram in which the $\pi^0$ propagates between the pair $(0, y)$ and the pair
 $(x,z)$ of vertices, which corresponds to the third (and last) term in Eq.\ (\ref{eq:Pipi0}).
 The first two terms, which do contribute, vanish at $q_3=0$, leading to the conclusion (\ref{eq:tildePi1area}).} the Fourier-transform of
$\Pi^{(1);\pi^0}_{\mu\nu\sigma\lambda}(x,y,z)$ contains an explicit
factor of $q_3$, the momentum dual to $z$. As a consequence, we expect
smaller finite-size effect on this term.

In summary, Method~2 is numerically far cheaper to apply. It leads to
a mild broadening of the overall integrand in $|y|$ (see Fig.~\ref{fig:pi0m1m2},
as well as Figs.~18 and~19 of Ref.~\cite{Chao:2020kwq}), which a suitable tuning of the parameter $\Lambda$
can counteract. Thus the combination of Method 2 and the kernel $\Lb^{(\Lambda)}$
was invaluable to our lattice QCD results in Refs.~\cite{Chao:2020kwq,Chao:2021tvp,Chao:2022xzg}.

\subsection{Tests in the continuum}

As a first check of our master formula, and to gain some insight on the shape of the integrand, we compute several contributions to $a_\mu^{\rm LbL}$ using the expressions for $\widehat{\Pi}$ derived in the previous section, and compare them with the known results obtained in momentum space. 

\subsubsection{The pion-pole contribution to $a_\mu^{\rm HLbL}$}

\begin{figure}[t]
\includegraphics[width=0.49\textwidth]{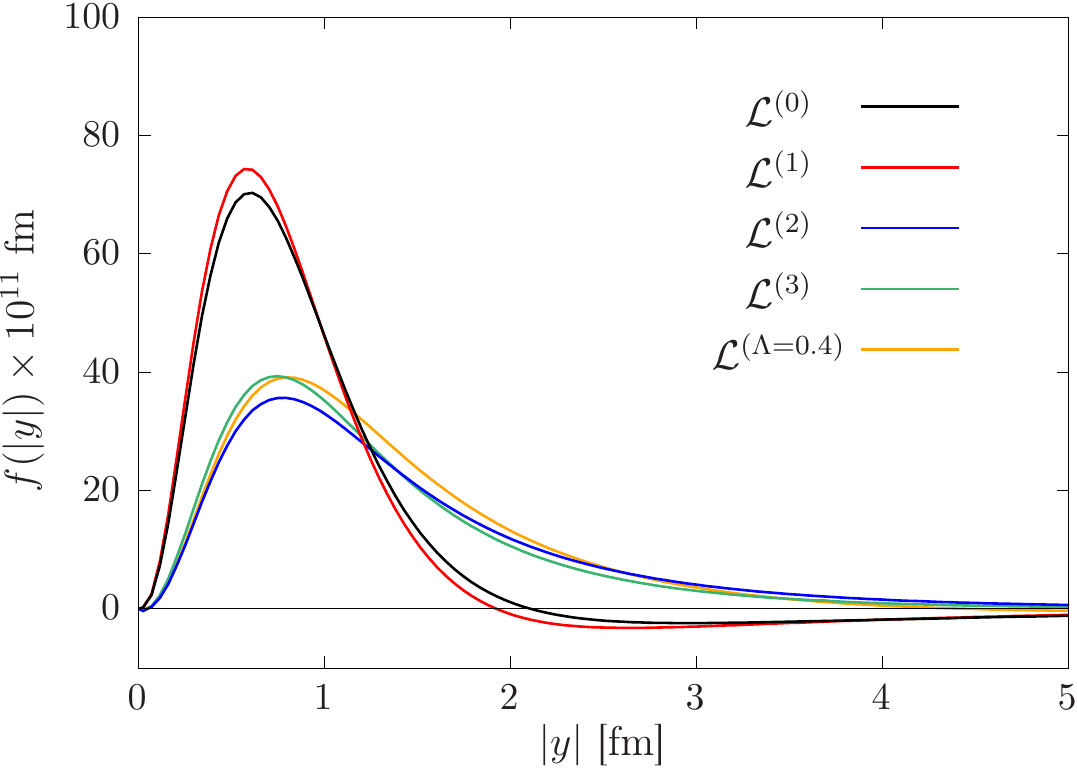}
\includegraphics[width=0.49\textwidth]{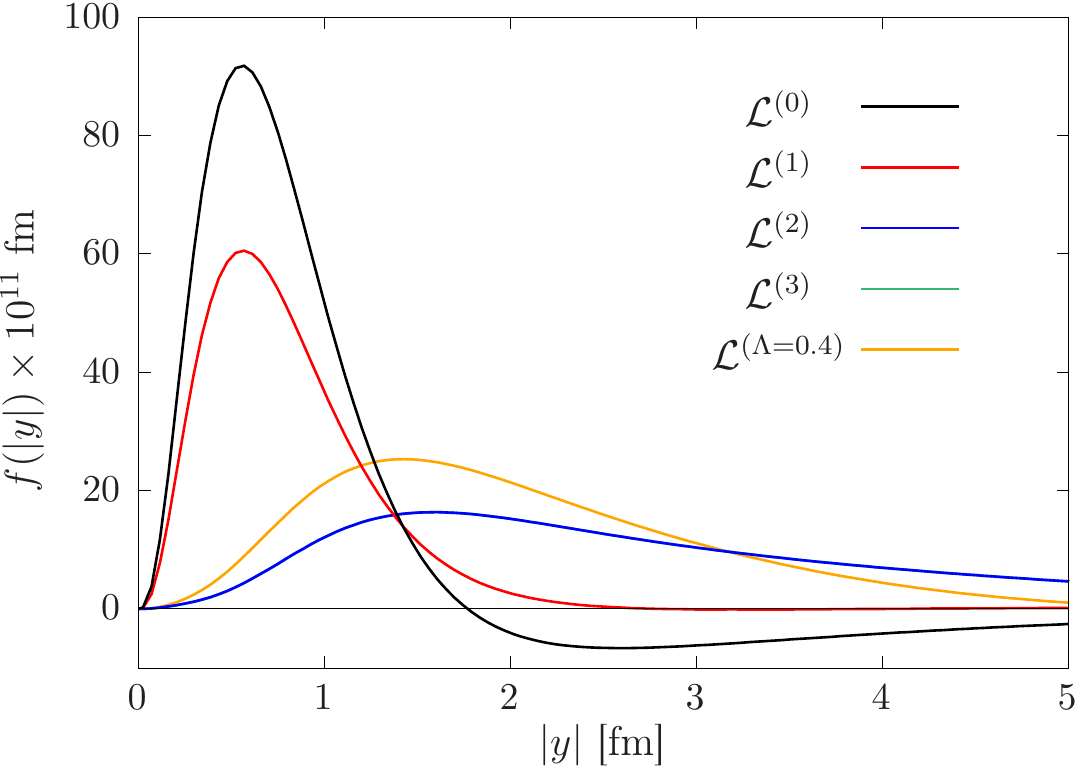}
\caption{\la{fig:pi0m1m2} The integrand with respect to the variable $|y|$
leading to $a_\mu^{\rm HLbL}$ for the neutral pion-pole with a VMD
form factor and a pion mass of 135 MeV. 
The integrals over $|x|$ and $\cos\beta$ have already been performed at this stage.
The different curves correspond to the kernels introduced in Eq.~(\ref{eq:kernels}) and Eq.~(\ref{eq:lamsub}). Left: Method 1. Right: Method 2. Note that for method~2, the kernel $\Lb^{(2)}$ and $\Lb^{(3)}$ yield identical curves. }
\end{figure}

The pion-pole contribution is estimated using the expression of $\widehat{\Pi}$ derived in Sec.~\ref{sec:Pihat_vmd}, assuming a VMD transition form factor. We note that, when using Method 2, one needs to find the correct mapping between the three contractions in momentum space (see Eq.~(\ref{eq:Pipi0})) and the reference contraction in position space. This can be done using partially-quenched chiral perturbation theory and we refer the reader to Ref~\cite{Chao:2021tvp} for more details. 
After integration over $|x|$ and $\cos\beta$, the integrand, as a function of $|y|$, is displayed in Fig.~\ref{fig:pi0m1m2} for both Methods 1 and 2 and all four kernels $\Lb^{(n)}$ with a pion mass of $m_{\pi} = 135$~MeV. 
For pion masses in the range [135 - 600]~MeV, we reproduce the results, obtained from the three-dimensional integral representation in momentum space in Ref.~\cite{Jegerlehner:2009ry}, at the percent level. The results are summarized in Table~\ref{tab:pi0VMD}.
Using the standard kernel, one observes that this contribution is remarkably long-range with a negative tail at large $|y|$. In particular, one needs a very large lattice $L\gg 5~$fm to capture the negative tail at the physical pion mass. When using the method~1, the integrands corresponding to the kernel 2 and 3 are less peaked at short distances, approach zero faster at long distances and remain positive. When using Method 2, the integrand for both kernels 2 and 3 are identical 
 but also more long range. Since this setup is considerably cheaper for practical lattice QCD calculations, one can attempt to correct for finite-size effects on this contribution by computing the pion-transition form factor on the same set of ensembles, as done in Ref.~\cite{Gerardin:2019vio}. See Ref.~\cite{Chao:2021tvp} for a practical implementation.

\begin{table}[h!]
\caption{Results for the pion-pole contribution to $a_\mu^{\rm HLbL}$
  in units of $10^{-11}$ assuming a VMD transition form factor as
  discussed in the text. We use $m_V = 775.49~$MeV and $F_{\pi} =
  92.4~$MeV. The results are obtained using the standard kernel
  $\Lb^{(0)}$ and for both methods~1 and 2. We also provide the
  deviation to the results obtained from the expressions in
  Ref.~\cite{Jegerlehner:2009ry}.  } 
\vskip 0.1in
\begin{tabular}{c@{\hskip 01em}c@{\hskip 01em}c@{\hskip 01em}c@{\hskip 01em}c@{\hskip 01em}c}
	\hline
$m_{\pi}$~[MeV]	&	 $a_\mu^{\rm HLbL}$~\cite{Jegerlehner:2009ry} 	&	 Method 1  	&	deviation &	Method 2 	&	deviation	\\
\hline
135 &   	57.00	&	57.21	&	$+0.37$\%& 	57.33	&	$+0.57$\%	\\
200 &	42.84	&	42.91	&	$+0.16$\%&	42.83	&	$-0.02$\%		\\
300 &	29.64	&	29.63	&	$-0.03$\%	&	29.64	&	$+0.00$\%	\\
400 &	21.75	&	21.71	&	$-0.18$\%	&	21.71	&	$-0.18$\%	\\
600 &	13.10	&	13.07	&	$-0.22$\%	&	13.07	&	$-0.22$\% 	\\
\hline
 \end{tabular} 
\label{tab:pi0VMD}
\end{table}

\subsubsection{The lepton-loop contribution to $a_\mu^{\rm LbL}$}

\begin{figure}[t]
\includegraphics[width=0.49\textwidth]{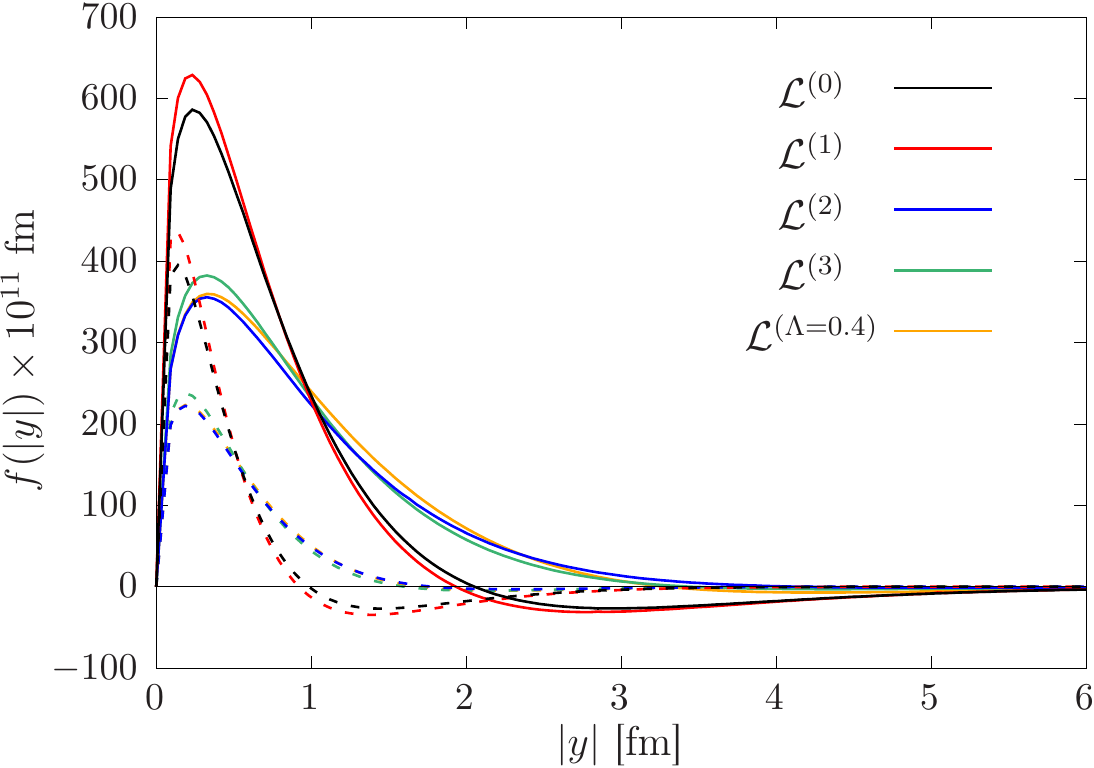}
\includegraphics[width=0.49\textwidth]{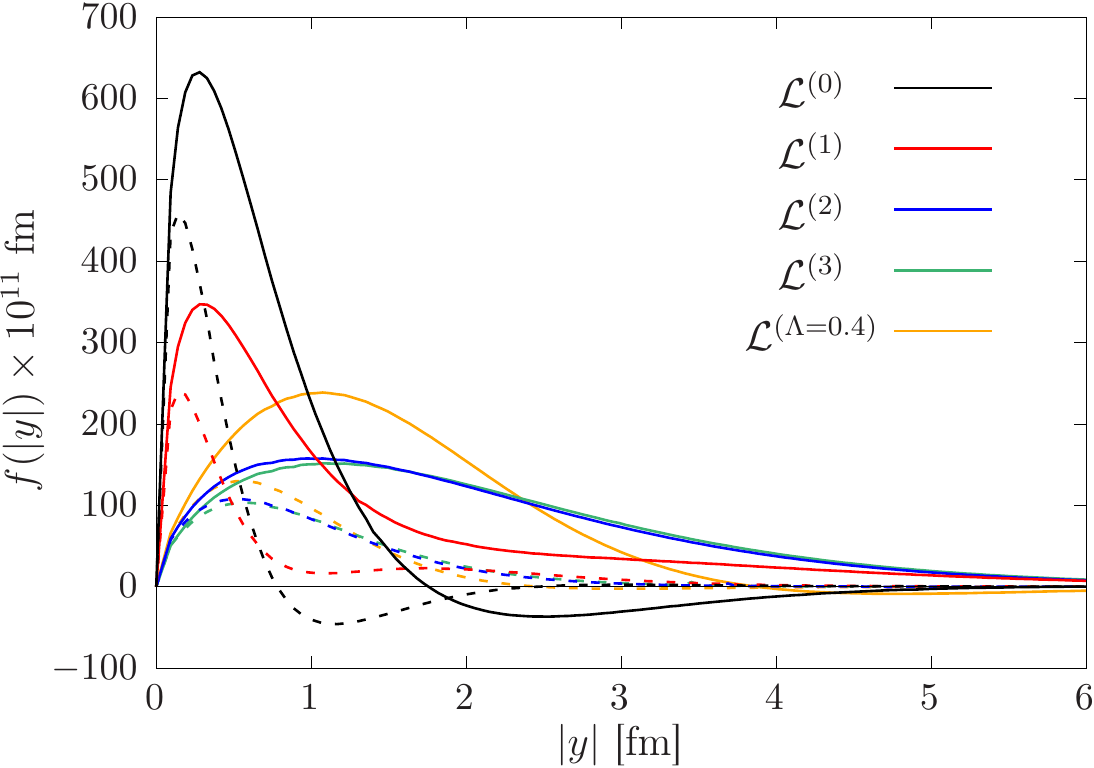}

\caption{\la{fig:leptonloop} The integrand with respect to the variable $|y|$
leading to $a_\mu^{\rm LbL}$ for the lepton-loop contribution with $m_l/m=1$ (full lines) and $m_l/m = 2$ (dashed lines).
The integrals over $|x|$ and $\cos\beta$ have already been performed at this stage.
The different curves correspond to the kernels introduced in Eq.\ (\ref{eq:kernels}) and Eq.\ (\ref{eq:lamsub}).
Left: Method 1. Right: Method 2.}
\end{figure}

The lepton-loop contribution to $a_\mu^{\rm LbL}$ is estimated using
the expression of $\widehat{\Pi}$ derived in
Sec.~\ref{sec:Pihat_lepton} with $m_{l}/m_{\mu} = 1/2, 1, 2$ where
$m_{l}$ is the mass of the lepton in the loop. The results are
summarized in Table~\ref{tab:lepton} and the shape of the integrand,
as a function of $|y|$, is shown in Fig.~\ref{fig:leptonloop}. For
$m_{l} = m, 2m$ we reproduce the analytically known results
for $a_\mu^{\rm LbL}$ in QED~\cite{Laporta:1992pa,
  Passera_private_communication} with a precision of about 1\%. As can
be seen from the plots, the integrand is quite steep close to the
origin and we probe the QED kernel at short distances. We observe that
the height of the peak grows for smaller masses of the lepton in the
loop. For $m_{l} = m/2$ this rise to the peak is very steep,
and we observe a 2.3\% deviation from the exact result.
It thus appears difficult to obtain $a_\mu^{\rm LbL}$ at the percent level
with our implementation of the QED kernel for such a long-range contribution.
As for the pion, the integrand resulting from the standard kernel also exhibits a long
negative tail. Again, the kernel 2 and 3 are peaked at short ranges when
using Method 1, a feature that does not hold when switching to Method 2.

\begin{table}[h!]
\caption{Results, precision and deviation of the lepton-loop
  contribution to $a_\mu^{\rm LbL}$ (in units of $10^{-11}$)
  computed in the continuum with kernel $\overline{\cal L}^{(0)}$
   compared to the known results~\cite{Laporta:1992pa,
    Passera_private_communication}. The first uncertainty originates
  from the three-dimensional numerical integration, the second from
  the extrapolation of the integrand to small $|y|$.} 
\vskip 0.1in
\begin{tabular}{l@{\hskip 01em}c@{\hskip 01em}c@{\hskip 01em}c@{\hskip 01em}c}
	\hline
$m_{l}/m_{\mu}$	&	 $a_\mu^{\rm LbL}$~(exact) 	&	 $a_\mu^{\rm LbL}$  	&	precision & deviation	\\
\hline
1/2 &  1229.07	&  1257.5(6.2)(2.4) &  0.5\% &	2.3\%	\\
1   &   464.97	&   470.6(2.3)(2.1) &  0.7\% &	1.2\%	\\
2   &   150.31	&   150.4(0.7)(1.7) &  1.2\% &	0.06\%	\\
\hline
 \end{tabular} 
\label{tab:lepton}
\end{table}

\subsubsection{The charged-pion loop contribution to $a_\mu^{\rm HLbL}$}

Using the master formula (\ref{eq:master_formula}) with kernel ${\cal
  L}^{(2)}$ defined in Eq.\ (\ref{eq:kernels}) and $i\widehat\Pi$
given by Eqs.\ (\ref{eq:iPihatPionDecomp}), (\ref{eq:PihatPionL0}) and
(\ref{eq:PihatPionL1and2}), we compute the contribution of a
physical-mass charged pion to $a_\mu^{\rm HLbL}$ in the scalar QED
framework.  After the $x$ integral has been performed, no contributing
delta-function contributions are left, so that the integrand can be
displayed straightforwardly; see figure~\ref{fig:piloopm1L2}.
With $a_\mu^{\rm HLbL} = -43.9 \times 10^{-11}$ we reproduce the
result $-43.86(5) \times 10^{-11}$ obtained analytically in
Ref.~\cite{Kuhn:2003pu} with a rapidly converging series expansion in
$(m/m_{\pi^+})^2 = 0.573092$ at the per-mille level.

\begin{figure}[h]
\centerline{\includegraphics[width=0.6\textwidth]{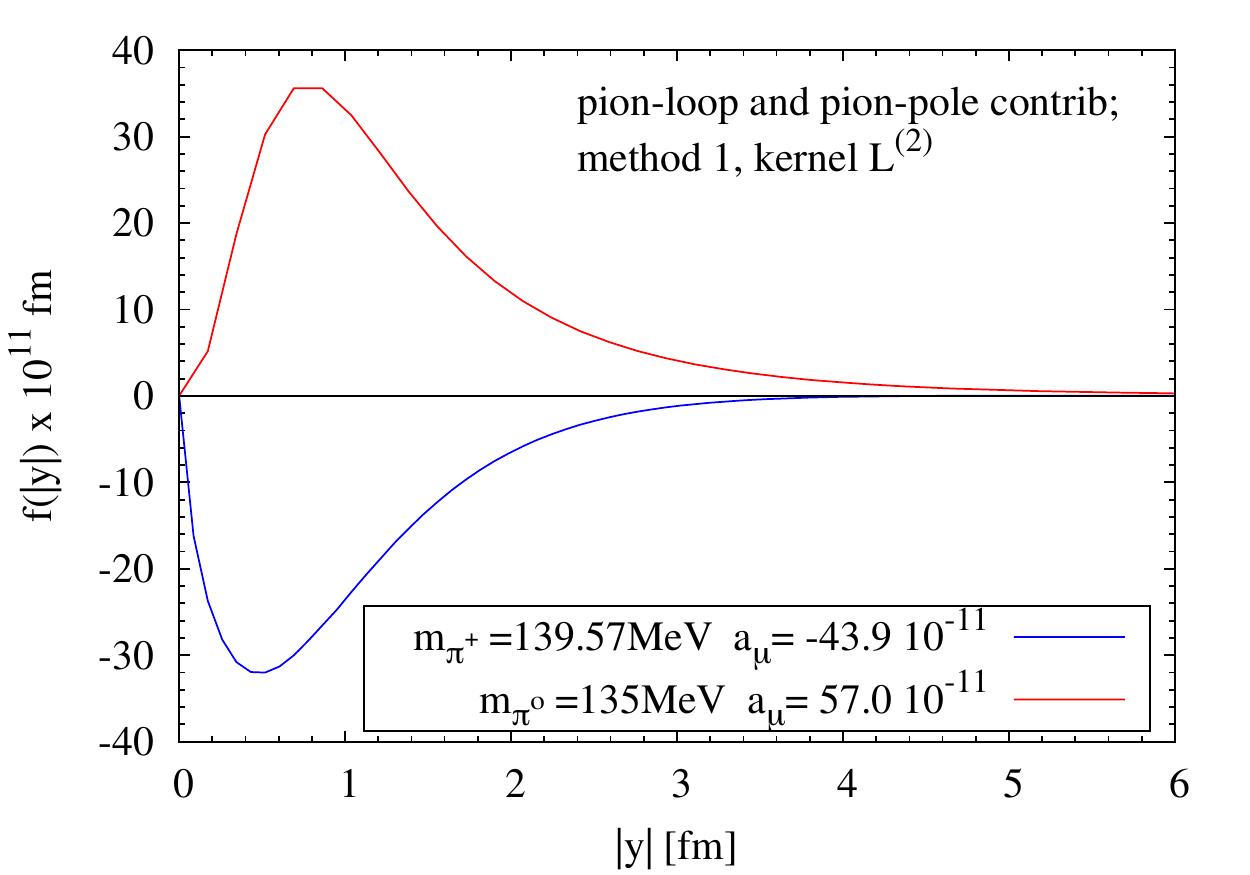}}
\caption{\la{fig:piloopm1L2} The integrand with respect to the variable $|y|$
leading to $a_\mu^{\rm HLbL}$ for the charged-pion loop computed in scalar QED,
together (for comparison) with the integrand for the neutral pion-pole computed with a VMD form factor.
The integrals over $|x|$ and $\cos\beta$ have already been performed at this stage.
Here Method 1 is used with kernel ${\cal L}^{(2)}$. }
\end{figure}

\subsection{The lepton-loop on the lattice}

We will now show the results of performing a full lattice calculation of the lepton loop contribution using both Methods 1 and 2. This was a first step towards the full lattice QCD calculations, and an important benchmark of our implementation of the position-space approach. 

The correlation function is computed on a $L^4$ lattice using unit gauge links, (anti-)periodic boundary conditions in (time-)space and Wilson fermions.
For a fixed vertex position $y$, the sums over the sites $x$ and $z$ in Eq.~(\ref{eq:master_formula}) 
are performed explicitly. 
After which we have a one-dimensional integral that can be sampled sufficiently finely using the variable $|y|$, for all values of $y=(n,n,n,n)$ with $0<n<L/2$. We also consider both the local ($l$) and the conserved $(c)$ vector currents, which have different discretization effects,
\begin{subequations}
\begin{align}
j_{\mu}^{(l)}(x) &= \psib(x) \gamma_{\mu} \mathcal{Q} \psi_j(x) \,,\\
j_{\mu}^{(c)}(x) &= \frac{1}{2} \left( \psib(x+a\hat{\mu})(1+\gamma_{\mu}) U^{\dag}_{\mu}(x)  \mathcal{Q} \psi(x) - \psib(x) (1-\gamma_{\mu} ) U_{\mu}(x)  \mathcal{Q} \psi(x+a\hat{\mu}) \right)  \,.
\label{eq:consvec}
\end{align}  
\end{subequations}
Here, $\mathcal{Q}$ is the lepton charge in units of $e$.
Let  $m_{0}$ be the bare subtracted lepton mass, related to the hopping parameter $\kappa_\ell$ via
$a m_{0}=(1/\kappa_{\ell} - 1/\kappa_{\rm cr})/2$ with $\kappa_{\rm cr}=1/8$.
In the free theory, the local vector current has the advantage of being automatically O$(a)$-improved, if one uses as multiplicative renormalization factor $(1 + b_{\rm V} a m_{0})$ with $b_{\rm V} = 1$~\cite{Sint:1997jx}. While no multiplicative renormalization of the conserved vector current is needed, an additive improvement term with a coefficient $c_{\rm V} = 1/2$ is required to remove `on-shell' O($a$) lattice artifacts, but not included here (see e.g.~\!\cite{Gerardin:2018kpy}). Indeed, for both current discretizations, additional lattice artifacts scaling linearly with the lattice spacing are expected to arise from the region where two or more currents are separated by a distance on the order of the lattice spacing. Our line of constant physics is defined by a constant renormalized mass of the lepton in the loop, $m_{\ell}=\;$cst. Including O$(a)$ effects, one has $m_{\ell} = Z_{\rm m} \, m_{0} (1 + b_{\rm m} a m_{0} )$, with $Z_{\rm m}=1$ and $b_{\rm m} = -1/2$~\cite{Sint:1997jx}, so that the bare subtracted quark mass has to be adjusted in the simulation. Finally, we are working with the QCD code and the result must be divided by $N_c=3$, the number of colors.

\begin{figure}[t!]
	\centering
	\includegraphics*[width=0.44\linewidth]{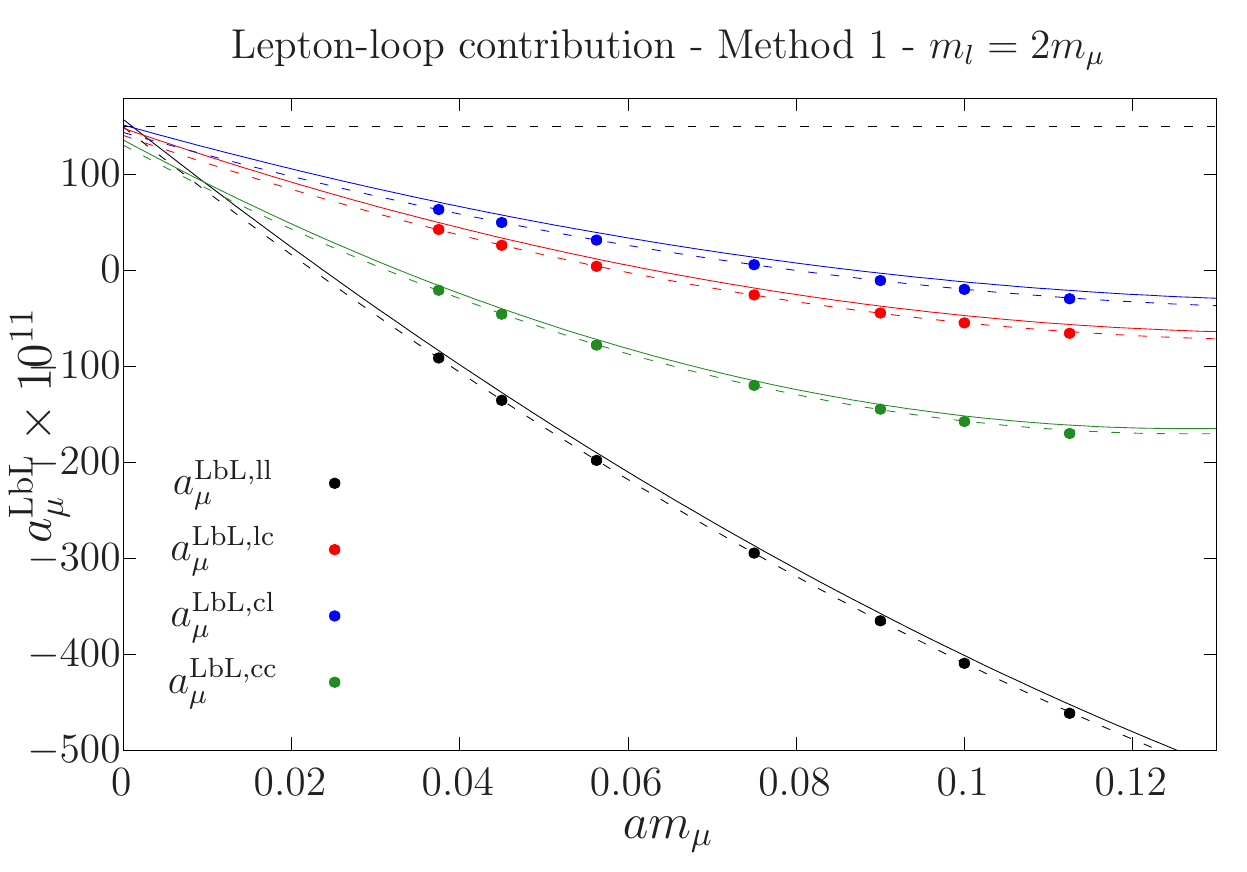}
	\includegraphics*[width=0.44\linewidth]{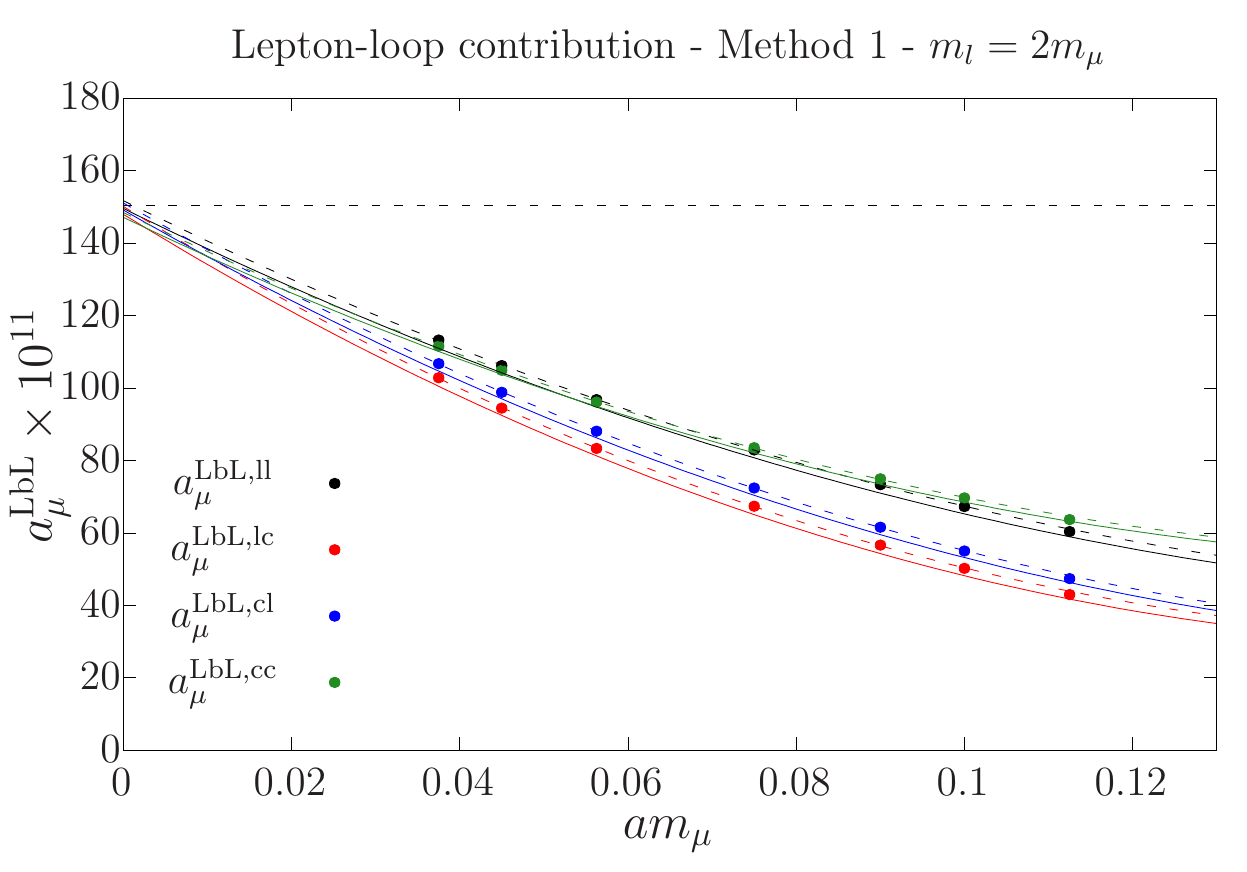}
	
	\caption{Continuum extrapolation of the lepton-loop contribution using the kernel $\Lb^{(0)}$ (left panel) and $\Lb^{(2)}$ (right panel) using Method 1. We use $m_l = 2 m_{\mu}$.  The colors correspond to different discretizations of the correlation function.
        The horizontal dashed line represents the exact result.} 	
	\label{fig:lepton_loop_lat_M1}
\end{figure}

We have used seven lattices, with the same physical volume $L\,m_{l}=7.2$, and the continuum extrapolation is performed assuming the simple functional form 
\begin{equation}
a_{\mu}^{\rm LbL}(a) = a_{\mu}^{\rm LbL}(0) + \alpha \, a + \beta \, a^2 \,,
\end{equation}
where only the four lattices with the smallest lattice spacings are
included in the fit. The result using the first strategy with $m_l = 2
m_{\mu}$ is shown in Fig.~\ref{fig:lepton_loop_lat_M1} for the kernel
$\Lb^{(0)}$ (left panel) and for the subtracted kernel $\Lb^{(2)}$
(right panel);

 Note the very different ranges on the $y$-axis. In both
cases, we use four different discretizations of the correlation
function (all combinations of local and conserved vector currents at
sites $x$ and $z$). The continuum extrapolation, at fixed volume, is
given by the dashed lines. To estimate the 
correction due to finite-size effects, a new set of two lattices, with larger volumes, are
used. The finite-size effect correction is assumed to be independent
of the lattice spacing and is estimated as the difference between the
small and the large volumes at a given lattice spacing. The corrected
results are finally given by the plain lines. One observes much
smaller discretization effects for the kernel $\Lb^{(2)}$ in the right
panel where one only has to extrapolate $a_\mu^{\rm LbL} \times
10^{11}$ from about 100 to 140, compared to rather large
extrapolations with kernel $\Lb^{(0)}$ in the left panel (even from
negative values). The results of the continuum extrapolations are collected in table~\ref{tab:M1leptlooplat}.
The same observation applies to the following two
figures. 
We also note that very similar results are obtained with the
kernel $\Lb^{(3)}$.

\begin{table}[b!]
  \caption{Results for $a_\mu^{\rm LbL}$ in units of $10^{-11}$
    from the lepton loop of mass $m_l=2m_\mu$ obtained with Method~1 by extrapolating the lattice results displayed
    in Fig.\ \ref{fig:lepton_loop_lat_M1} to the continuum.
    Two different kernels and four different discretizations of $i\widehat\Pi$ are used. The exact result is
    $a_\mu^{\rm LbL} = 150.31$~\cite{Laporta:1992pa,Passera_private_communication}.}
\vskip 0.1in
  \centering
  \begin{tabular}{l@{\hskip 01em}c@{\hskip 01em}c@{\hskip 01em}c@{\hskip 01em}c}
    \hline
 $m_l=2m_\mu$  &  ll &  lc &  cl & cc \\
    \hline
    Kernel $\overline{\cal L}^{(0)}$ & 157.8  & 148.5    & 152.0   & 136.3  \\
    Kernel $\overline{\cal L}^{(2)}$ & 149.5 &  147.9 & 149.1 & 147.1 \\
    \hline
  \end{tabular}
  \label{tab:M1leptlooplat}
  \end{table}

\begin{figure}[t!]
	\centering
	\includegraphics*[width=0.44\linewidth]{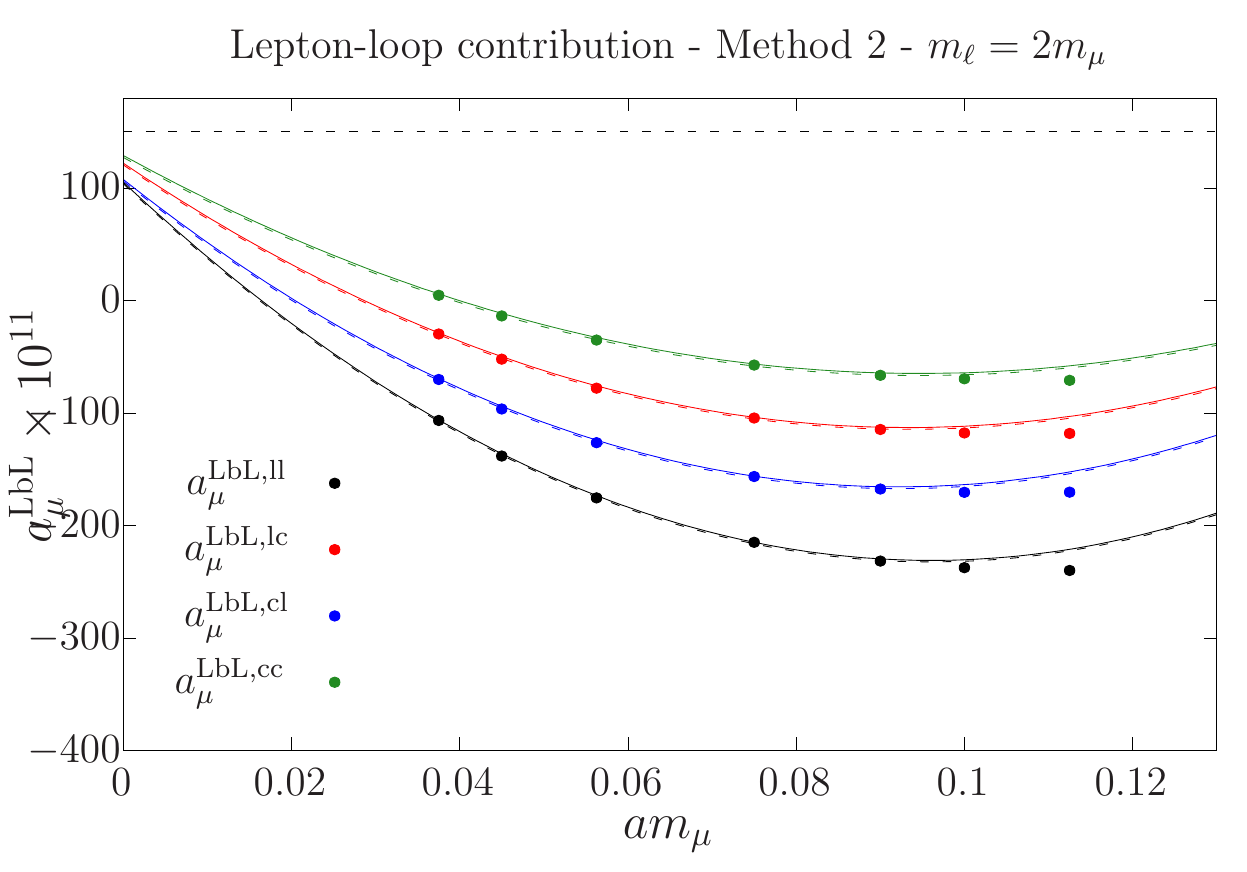}
	\includegraphics*[width=0.44\linewidth]{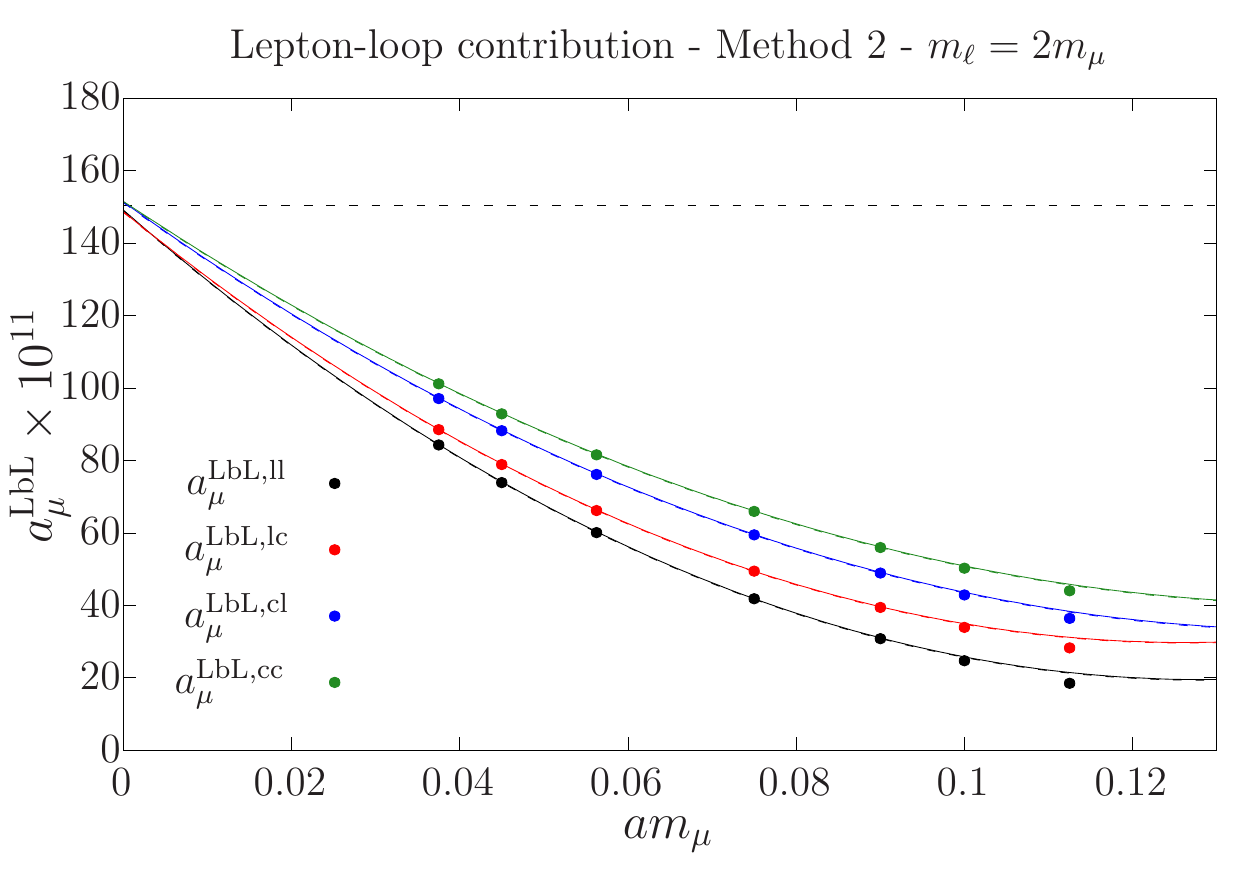}
	
	\caption{Continuum extrapolation of the lepton loop contribution for the two kernels $\Lb^{(0)}$ (left panel) and $\Lb^{(2)}$ (right panel) using Method 2. We use $m_l = 2 m_{\mu}$. The colors correspond to different discretizations of the correlation function.
        The horizontal dashed line represents the exact result.} 		
	\label{fig:lepton_loop_lat_M2_eq}
\end{figure}

\begin{figure}[t!]
	\centering
	\includegraphics*[width=0.44\linewidth]{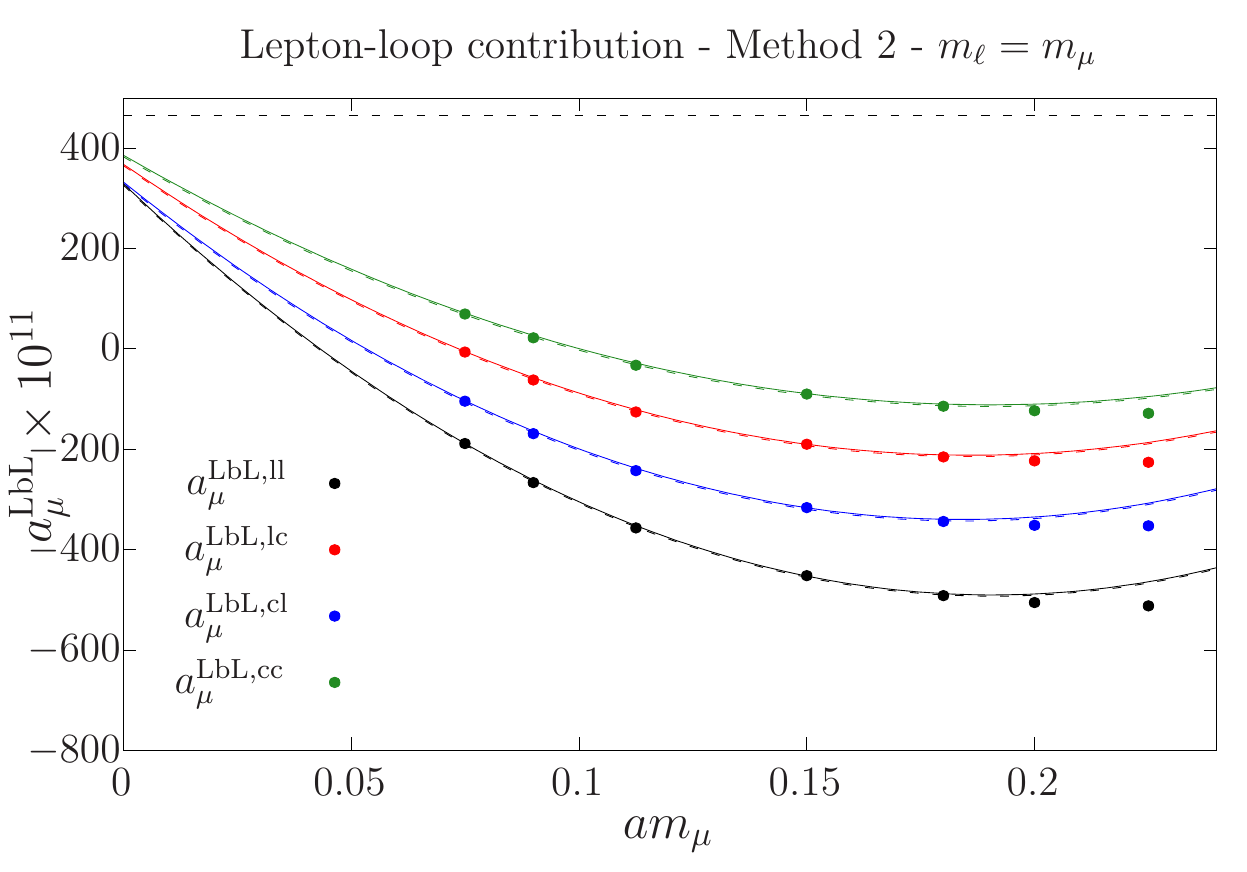}
	\includegraphics*[width=0.44\linewidth]{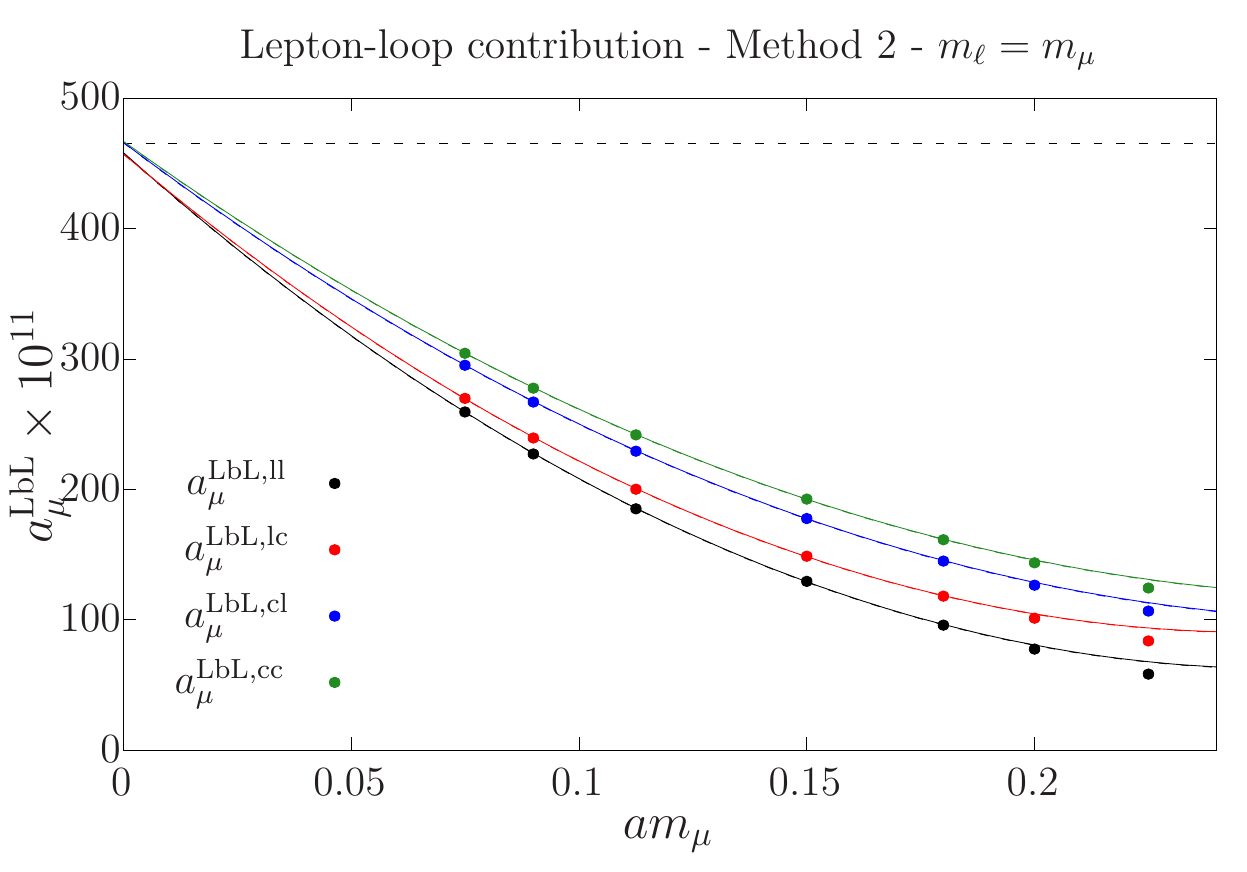}
	
	\caption{Same as Fig.~\ref{fig:lepton_loop_lat_M2_eq} but with $m_l = m_{\mu}$.}	
	\label{fig:lepton_loop_lat_M2_dif}
\end{figure}

The results using Method 2 are shown in Figs.~\ref{fig:lepton_loop_lat_M2_eq} and \ref{fig:lepton_loop_lat_M2_dif} for $m_{l} = 2 m_{\mu}$  and $m_{l} = m_{\mu}$ respectively. As before, the left panel corresponds to the standard kernel $\Lb^{(0)}$ and the right panel to the subtraction $\Lb^{(2)}$.  Again, the continuum extrapolation is much easier using the subtracted kernel. The resuls of the continuum extrapolations are collected in table~\ref{tab:M2leptlooplat}.
Further results for the lepton loop, computed on the lattice with the kernel $\bar{\cal L}^{(\Lambda)}$, can be found
in appendix~B of Ref.\ \cite{Chao:2020kwq}.

\begin{table}[h!]
  \caption{Results for $a_\mu^{\rm LbL}$ in units of $10^{-11}$
    from the lepton loop of mass $m_l=2m_\mu$ or $m_l=m_\mu$
    obtained with Method~2 by extrapolating the lattice results displayed in
    Figs.\ \ref{fig:lepton_loop_lat_M2_eq} and \ref{fig:lepton_loop_lat_M2_dif} to the continuum.
    Two different kernels and four discretizations of $i\widehat\Pi$ employing different
    combinations of local (l) and conserved (c) vector currents are used. The exact results are
    $a_\mu^{\rm LbL} = 150.31$ and $464.97$ respectively~\cite{Laporta:1992pa,Passera_private_communication}.}
\vskip 0.1in
  \centering
  \begin{tabular}{l@{\hskip 01em}c@{\hskip 01em}c@{\hskip 01em}c@{\hskip 01em}c}
    \hline
$m_l=2m_\mu$ &  ll &  lc &  cl & cc \\
    \hline
    Kernel $\overline{\cal L}^{(0)}$ & 105.2  & 122.1    & 107.7   & 129.0  \\
    Kernel $\overline{\cal L}^{(2)}$ & 149.1  & 148.6  & 151.3  & 151.5 \\
    \hline    \hline
$m_l=m_\mu$ &  ll &  lc &  cl & cc \\
    \hline
    Kernel $\overline{\cal L}^{(0)}$ & 329.3  & 368.3    & 333.0   & 386.2  \\
    Kernel $\overline{\cal L}^{(2)}$ & 458.2 & 457.3  & 466.1 & 466.9 \\
    \hline
  \end{tabular}
    \label{tab:M2leptlooplat}
  \end{table}

\subsection{Overview of lattice QCD results for the quark-connected contribution}\label{sec:LQCD}

In this section, we briefly review a subset of the results obtained to date for
the hadronic-light-by-light contribution $a_\mu^{\rm HLbL}$ using the
QED kernel derived above, restricting ourselves to the quark-connected contribution.
The relevant publications are~\cite{Chao:2020kwq,Chao:2021tvp,Chao:2022xzg},
and some preliminary results can also be found in the earlier
proceedings contributions~\cite{Asmussen:2018oip,Asmussen:2019act}.
All these calculations have been performed on gauge ensembles provided by the
Coordinated Lattice Simulations (CLS) initiative~\cite{Bruno:2014jqa}. The ensembles were
generated using three flavours of non-perturbatively O($a$)-improved
Wilson fermions and with the tree-level O($a^2$)-improved Symanzik gauge action.

Reference~\cite{Chao:2020kwq} focuses on QCD with degenerate $u,d,s$ quark,
corresponding to $m_\pi= m_K \simeq 420$\,MeV.
It contains results for the connected contribution
obtained either with Method~1 or Method~2.
Focussing first on Method~1 with the kernel choice $\overline{\cal L}^{(\Lambda=0.16)}$,
the difference in the integrand between choosing
a conserved or a local lattice vector current at vertex $x$ was tested and found
to be modest\footnote{See Fig.\ 8 of Ref.\ \cite{Chao:2020kwq}.
The difference is about 10\% around the peak of the integrand, at $|y|\simeq 0.4$\,fm,
and not statistically significant for $|y|>0.6$\,fm.}. The cutoff effect on $a_\mu^{\rm HLbL}$ for an ensemble with
lattice spacing of 0.076\,fm turned out to be on the order of 10\%,
with a significant uncertainty on this estimate.
Finite-size effects were probed directly by comparing
two ensembles differing only by their volume, $m_\pi L = 4.4$ and 6.4, and
their size found to be roughly consistent with the finite-size effects expected for the $\pi^0$
pole contribution. 
As for Method~2, the kernel $\overline{\cal L}^{(\Lambda=0.40)}$ was found to be a good choice,
and the use of local and conserved currents was
investigated as well. The final choice fell on four local currents,
although the final integrand with respect to the variable $|y|$ was found to be similar if the
current at vertex $z$ was replaced by a conserved current.
The size of both cutoff effects and finite-size effects was similar to Method~1.
Hence, given the lower computational cost of Method~2, the latter method was selected
for all subsequent calculations.

\begin{figure}[t!]
        \centering
        \includegraphics*[width=0.64\linewidth]{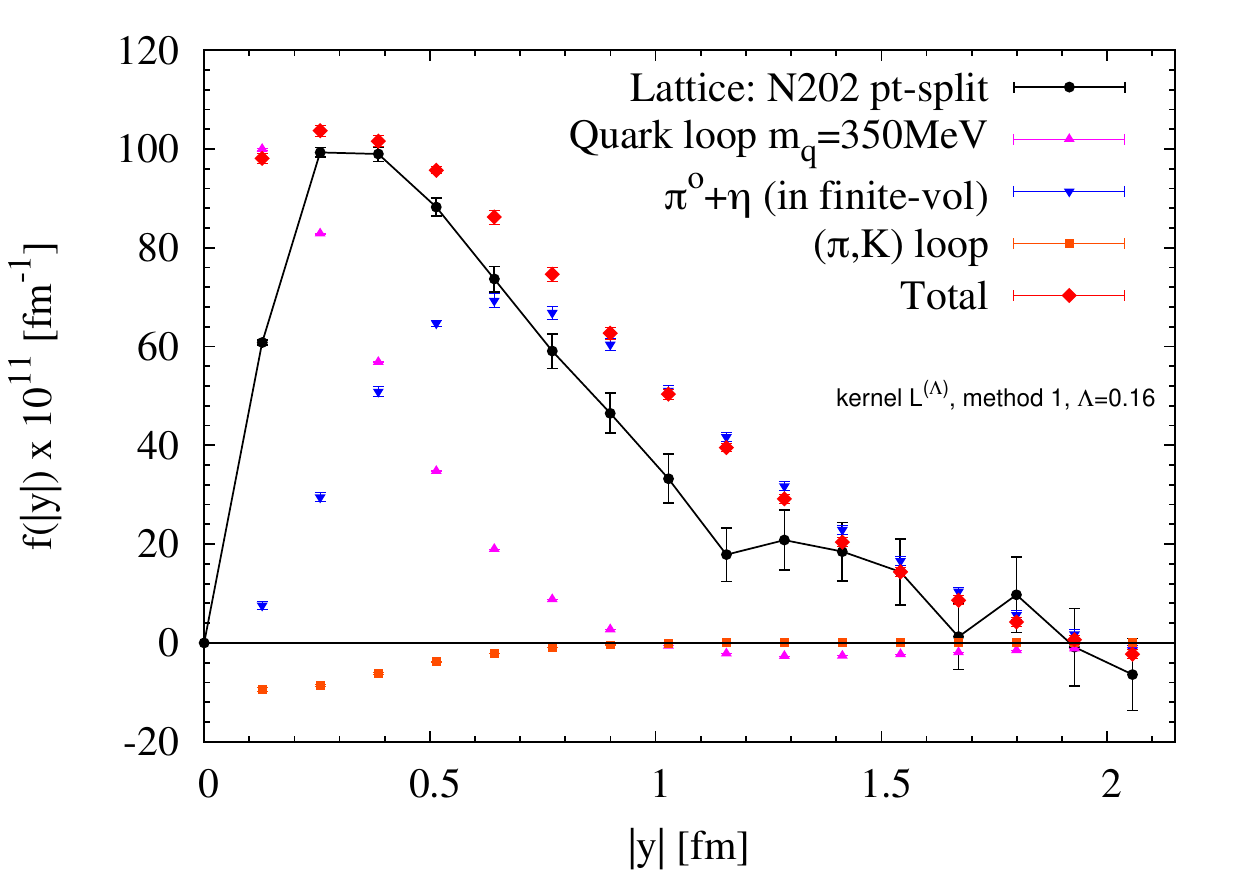} 
        \caption{Integrand for the connected contribution using Method~1 with $\overline{\cal L}^{(\Lambda= 0.16)} $ on ensemble N202 of size $48^3\times 128$, lattice spacing 0.064\,fm~\cite{Bruno:2016plf} with $m_\pi=m_K=421$\,MeV. The lattice data use a point-split current at $x$. The integrand is compared to the prediction for the pole contributions of the $\pi^0$ and $\eta$
mesons with a VMD transition form factor, which is expected to provide a good approximation to the tail. In addition, an attempt to describe the
 short-distance contribution with  a constituent-quark loop with a quark mass of 350\,MeV is made. Figure reproduced from~\cite{Chao:2020kwq}.}         \label{fig:N202_CQM}
\end{figure}

Figure~\ref{fig:N202_CQM} shows the integrand for the
quark-connected contribution within Method~1 obtained in lattice QCD
with degenerate $u,d,s$ quarks. It is compared to predictions based
on the calculations of section~\ref{sec:ipihats}: the quark loop with
a `constituent mass' of 350\,MeV, the $\pi^0$ and $\eta$ pole
contributions, as well as the charged pion loop.  For the latter two
contributions, a modification of the prediction applicable to the full
HLbL amplitude must be applied in order to account for the fact
that only the quark-connected contribution is considered. This leads
to an enhancement of the combined $\pi^0$ and $\eta$ contribution by a
factor of three~\cite{Bijnens:2016hgx,Gerardin:2017ryf}.  The combined
contribution of pion and kaon loops, computed in the framework of
scalar QED, must instead be divided by a factor of
three~\cite{Chao:2020kwq} to match the contribution of quark-connected
diagrams. Figure~\ref{fig:N202_CQM} illustrates that a
semi-quantitative understanding of the integrand can be gained via
these fairly simple calculations.

Analogous plots (based on Method~2 with $\Lambda=0.40$) at lighter
pion masses can be found in~\cite{Chao:2021tvp}, for which similar
qualitative observations can be made.  The $\pi^0$ pole contribution becomes
increasingly dominant as the pion mass is lowered towards its physical
value.  In the case of a heavy quark propagating in the
loop~\cite{Chao:2022xzg}, the upward trend as a function of the
lattice spacing is seen most clearly. The latter calculation allowed
for a state-of-the-art estimate of the charm-quark contribution to
$a_\mu^{\rm HLbL}$.

\section{Conclusions\la{sec:concl}}

In this paper we have presented extensive details of our calculation
of the `QED kernel' needed in the Lorentz-covariant coordinate-space
method for computing the hadronic light-by-light contribution to the
muon $(g-2)$ in lattice QCD.  At the core of this QED kernel is the
amplitude represented by the graph of
Fig.~\!\ref{fig:diagram_I_position_space}.  It is an unusual amplitude
in particle physics in that it involves both the plane-wave
propagation of the muon in the initial and final state, and the
emission by the muon of massless particles (photons) propagating to
definite coordinate-space positions. It is thereby a mixed
momentum-space and coordinate-space amplitude.  We remark that such
mixed amplitudes, although quite complex, may have further interesting
applications in quantum-field theoretic calculations~\cite{Position_space_methods_new,Parrino_master_thesis}.

We were able to carry out the calculation analytically up to and
including the averaging over the direction of the muon
momentum. Numerical methods were used only in the final convolution
integral, which can be interpreted as yielding the static potential
generated by a certain (analytically known) electric charge
distribution in four space dimensions.  While the angular integral of
this final convolution can perhaps still be handled analytically, and
we have made some progress in this direction~\cite{SchroederThesis},
one angular integral still had to be performed numerically in our two
implementations of sections \ref{sec:direct} and
\ref{sec:alternative_evaluation}.

For the practical purpose of computing $a_\mu^{\rm HLbL}$ on the
lattice, we have tested the robustness of the numerics by reproducing
a number of known light-by-light contributions, typically at the
one-percent level.  Such a precision is sufficient for the foreseeable
future, given that the current precision goal for $a_\mu^{\rm HLbL}$
is to reach the 10\% level. As noted in the introduction, our
implementation of the QED kernel has already been applied in lattice
QCD calculations of $a_\mu^{\rm
  HLbL}$~\cite{Chao:2020kwq,Chao:2021tvp,Chao:2022xzg}. Given the phenomenological
importance of the muon $(g-2)$, the present paper serves to document
and underpin a central aspect of these results.

The idea of treating photon propagators in the continuum,
infinite-volume theory has been applied in other contexts as well,
most recently in Ref.\!~\cite{Biloshytskyi:2022ets}. One simpler
application is the fully covariant coordinate-space method for the HVP
contribution to the muon $(g-2)$~\cite{Meyer:2017hjv}.  Another
interesting application concerns the calculation of the QED
self-energy of stable hadrons without power-law finite-size
effects~\cite{Feng:2018qpx}. Further applications will probably follow
in the future.

\section*{Acknowledgements}

We thank Henryk Czyz for discussions and providing references about
position-space methods, Hartmut~Wittig and Georg~von~Hippel for many
discussions about the status of the $(g-2)_\mu$ puzzle.
This work has been supported by the European Research Council (ERC) under
the European Union's Horizon 2020 research and innovation programme
through grant agreement 771971-SIMDAMA, as well as by the Deutsche
Forschungsgemeinschaft (DFG) through the Collaborative Research Centre
1044, the research unit FOR 5327
“Photon-photon interactions in the Standard Model and beyond -
exploiting the discovery potential from MESA to the LHC” (grant
458854507)
and through the Cluster of Excellence \emph{Precision Physics,
Fundamental Interactions, and Structure of Matter} (PRISMA+ EXC
2118/1) within the German Excellence Strategy (Project ID
39083149). The project leading to this publication has also received
funding from the Excellence Initiative of Aix-Marseille University -
A*MIDEX, a French “Investissements d’Avenir” programme,
AMX-18-ACE-005.  Calculations for this project were partly performed
on the HPC clusters ``Clover'' and ``HIMster II'' at the
Helmholtz-Institut Mainz and ``Mogon II'' at JGU Mainz.
Our lattice-QCD programs use the deflated SAP+GCR solver from the openQCD
package~\cite{Luscher:2012av}.  We are grateful to our colleagues in the CLS
initiative for sharing ensembles.

\appendix

\section{The tensors $T_{\alpha\beta\delta}^{\rm A}(x,y)$ in terms of the weight functions \label{sec:chainrT}}

In this appendix, we consider the weight functions $\bar{\mathfrak{g}}^{(0)}$, $\bar{\mathfrak{g}}^{(1)}$, $\bar{\mathfrak{g}}^{(2)}$,
 $\bar{\mathfrak{l}}^{(1)}$, $\bar{\mathfrak{l}}^{(2)}$, $\bar{\mathfrak{l}}^{(3)}$
as being functions of $(|x|,\hat c_\beta \equiv \hat x\cdot \hat y, |y|)$
and use the notation $\partial^{[j]}$ to denote the derivative of a weight function
with respect to its $j$'th argument.
The chain rule for $T^{({\rm I})}_{\alpha\beta\delta}(x,y)$ reads
\ba
&& T^{{\rm I}}_{\alpha\beta\delta}(x,y) =
\delta_{\beta\delta} \frac{x_\alpha}{|x|} 
\Big\{ \partial^{[1]}\bar{\mathfrak{g}}^{(1)} + \partial^{[1]}\bar{\mathfrak{g}}^{(2)} - \frac{\hat c_\beta}{|x|}\Big(\partial^{[2]}\bar{\mathfrak{g}}^{(1)}+\partial^{[2]}\bar{\mathfrak{g}}^{(2)}\Big)\Big\}
\nonumber\\ && + \delta_{\beta\delta}\frac{y_\alpha}{|x||y|} \Big\{ \partial^{[2]}\bar{\mathfrak{g}}^{(1)} + \partial^{[2]}\bar{\mathfrak{g}}^{(2)}\Big\}
\nonumber \\ && + (\delta_{\alpha\delta} x_\beta + \delta_{\alpha\beta}x_\delta)\frac{1}{|x|}
\Big\{\partial^{[1]}\bar{\mathfrak{g}}^{(1)}+ \Big(\frac{1}{|y|} - \frac{\hat c_\beta}{|x|} \Big) \partial^{[2]}\bar{\mathfrak{g}}^{(1)} \Big\}
\nonumber \\ && + \delta_{\alpha\delta} \frac{y_\beta}{|y|} \Big\{ \partial^{[3]}\bar{\mathfrak{g}}^{(1)} + \Big(\frac{1}{|x|} - \frac{\hat c_\beta}{|y|} \Big)\partial^{[2]}\bar{\mathfrak{g}}^{(1)}\Big\}
\nonumber \\ && + \delta_{\alpha\beta} \frac{y_\delta}{|x|} \Big\{ \partial^{[1]}\bar{\mathfrak{g}}^{(2)} + \Big( \frac{1}{|y|}- \frac{\hat c_\beta}{|x|}\Big) \partial^{[2]}\bar{\mathfrak{g}}^{(2)} \Big\}
\nonumber \\ && + \frac{x_\alpha x_\beta x_\delta}{x^2} \Big\{
\partial^{[1]}\partial^{[1]}\bar{\mathfrak{g}}^{(1)} +  \Big(\frac{1}{|y|} - 2\frac{\hat c_\beta}{|x|}\Big) \partial^{[1]}\partial^{[2]}\bar{\mathfrak{g}}^{(1)}
- \frac{\hat c_\beta}{|x|} \Big(\frac{1}{|y|}-\frac{\hat c_\beta}{|x|}\Big) \partial^{[2]}\partial^{[2]}\bar{\mathfrak{g}}^{(1)}
\nonumber \\ && \qquad\qquad - \frac{1}{|x|} \partial^{[1]}\bar{\mathfrak{g}}^{(1)} + \frac{1}{|x|} \Big(\frac{3\hat c_\beta}{|x|} - \frac{1}{|y|} \Big) \partial^{[2]} \bar{\mathfrak{g}}^{(1)} \Big\}
\nonumber \\ && + \frac{y_\alpha y_\beta y_\delta}{|x| \,y^2} 
\Big\{ \partial^{[2]}\partial^{[3]} \bar{\mathfrak{g}}^{(2)} + \Big(\frac{1}{|x|} - \frac{\hat c_\beta}{|y|}\Big)\partial^{[2]}\partial^{[2]}\bar{\mathfrak{g}}^{(2)} 
- \frac{1}{|y|}  \partial^{[2]} \bar{\mathfrak{g}}^{(2)} \Big\}
\nonumber \\ && + \frac{x_\alpha x_\beta y_\delta}{|x|^2} \Big\{
\partial^{[1]}\partial^{[1]} \bar{\mathfrak{g}}^{(2)} + \Big( \frac{1}{|y|} - 2\frac{\hat c_\beta}{|x|}\Big) \partial^{[1]}\partial^{[2]} \bar{\mathfrak{g}}^{(2)}
- \frac{\hat c_\beta}{|x|} \Big(\frac{1}{|y|} - \frac{\hat c_\beta}{|x|}\Big) \partial^{[2]}\partial^{[2]}\bar{\mathfrak{g}}^{(2)} 
\nonumber \\ && \qquad\qquad - \frac{1}{|x|} \partial^{[1]} \bar{\mathfrak{g}}^{(2)} 
 + \frac{1}{|x|} \Big( \frac{3\hat c_\beta}{|x|} - \frac{1}{|y|}\Big)\partial^{[2]}\bar{\mathfrak{g}}^{(2)} \Big\} 
\nonumber \\ && + \frac{y_\alpha y_\beta x_\delta}{|x|\,|y|^2} \Big\{
\partial^{[2]}\partial^{[3]}\bar{\mathfrak{g}}^{(1)} + \Big( \frac{1}{|x|} - \frac{\hat c_\beta}{|y|}\Big)\partial^{[2]}\partial^{[2]}\bar{\mathfrak{g}}^{(1)}
- \frac{1}{|y|}  \partial^{[2]} \bar{\mathfrak{g}}^{(1)} \Big\}
\nonumber \\ && + \frac{y_\alpha x_\beta x_\delta}{|x|^2|y|} \Big\{
\partial^{[1]}\partial^{[2]}\bar{\mathfrak{g}}^{(1)} + \Big(\frac{1}{|y|} - \frac{\hat c_\beta}{|x|}\Big)\partial^{[2]}\partial^{[2]}\bar{\mathfrak{g}}^{(1)} 
- \frac{1}{|x|} \partial^{[2]}\bar{\mathfrak{g}}^{(1)} \Big\}
\nonumber \\ && + \frac{x_\alpha y_\beta y_\delta}{|x||y|} \Big\{
\partial^{[1]}\partial^{[3]} \bar{\mathfrak{g}}^{(2)} - \frac{\hat c_\beta}{|x|} \partial^{[2]}\partial^{[3]} \bar{\mathfrak{g}}^{(2)}
+ \Big(\frac{1}{|x|} - \frac{\hat c_\beta}{|y|} \Big) \partial^{[1]}\partial^{[2]} \bar{\mathfrak{g}}^{(2)} 
\nonumber \\ && \qquad\qquad  - \frac{\hat c_\beta}{|x|} \Big(\frac{1}{|x|} - \frac{\hat c_\beta}{|y|}\Big)  \partial^{[2]}\partial^{[2]} \bar{\mathfrak{g}}^{(2)}
 + \frac{1}{|x|}  \Big(\frac{\hat c_\beta}{|y|} -\frac{1}{|x|}  \Big) \partial^{[2]}\bar{\mathfrak{g}}^{(2)} \Big\}
\nonumber \\ && + \frac{x_\alpha y_\beta x_\delta}{|x||y|} \Big\{ \partial^{[1]}\partial^{[3]} \bar{\mathfrak{g}}^{(1)} 
- \frac{\hat c_\beta }{|x|} \partial^{[2]} \partial^{[3]} \bar{\mathfrak{g}}^{(1)}
+ \Big(\frac{1}{|x|} - \frac{\hat c_\beta}{|y|}\Big) \Big(\partial^{[1]}\partial^{[2]}\bar{\mathfrak{g}}^{(1)} - \frac{\hat c_\beta}{|x|}\partial^{[2]}\partial^{[2]}\bar{\mathfrak{g}}^{(1)}\Big)
\nonumber \\ && \qquad\qquad + \frac{1}{|x|} \Big(\frac{\hat c_\beta}{|y|}-\frac{1}{|x|}\Big) \partial^{[2]} \bar{\mathfrak{g}}^{(1)}\Big\}
\nonumber \\ && + \frac{y_\alpha x_\beta y_\delta}{|x|^2|y|} \Big\{\partial^{[1]}\partial^{[2]}\bar{\mathfrak{g}}^{(2)} 
+ \Big(\frac{1}{|y|} - \frac{\hat c_\beta}{|x|}\Big) \partial^{[2]}\partial^{[2]} \bar{\mathfrak{g}}^{(2)}
- \frac{1}{|x|} \partial^{[2]} \bar{\mathfrak{g}}^{(2)}\Big\}.
\ea

Similarly, the chain rule for $T^{{\rm II}}_{\alpha\beta\delta}$ and $T^{{\rm III}}_{\alpha\beta\delta}$ read
\ba
&& \frac{1}{m} T^{{\rm II}}_{\alpha\beta\delta}(x,y) = 
\delta_{\beta\delta} \frac{x_\alpha}{|x|}
\left\{
\frac{1}{4}
\left(
\partial^{[1]} \bar{\mathfrak{g}}^{(0)} - \frac{\hat c_\beta}{|x|}
\partial^{[2]} \bar{\mathfrak{g}}^{(0)} 
\right)
- \frac{x^2}{4} 
\left(
\partial^{[1]} \bar{\mathfrak{l}}^{(1)} - \frac{\hat c_\beta}{|x|}
\partial^{[2]} \bar{\mathfrak{l}}^{(1)} 
\right)
\right.    
\nonumber \\
&&
\left. 
\qquad \qquad \qquad - \frac{y^2}{4}
    \left( \partial^{[1]} \bar{\mathfrak{l}}^{(2)} - \frac{\hat
      c_\beta}{|x|} \partial^{[2]} \bar{\mathfrak{l}}^{(2)} 
    \right)
- \frac{|x| |y| \hat c_\beta}{2} 
\left(
\partial^{[1]} \bar{\mathfrak{l}}^{(3)} - \frac{\hat c_\beta}{|x|}
\partial^{[2]} \bar{\mathfrak{l}}^{(3)} 
\right)
- \frac{|x|}{2} \bar{\mathfrak{l}}^{(1)}
\right\}
\nonumber \\
&& + \delta_{\beta\delta} \frac{y_\alpha}{|x| |y|}
\left\{
\frac{1}{4}
\partial^{[2]} \bar{\mathfrak{g}}^{(0)}
- \frac{x^2}{4} \partial^{[2]} \bar{\mathfrak{l}}^{(1)}
- \frac{y^2}{4} \partial^{[2]} \bar{\mathfrak{l}}^{(2)}
- \frac{|x| |y| \hat c_\beta}{2} \partial^{[2]} \bar{\mathfrak{l}}^{(3)}
- \frac{|x| |y|}{2} \bar{\mathfrak{l}}^{(3)} 
\right\}
\nonumber \\
&& + \left( \delta_{\alpha\beta} x_\delta + \delta_{\alpha\delta}
x_\beta \right) \left\{ \bar{\mathfrak{l}}^{(1)} \right\}
\nonumber \\ 
&& + \left( \delta_{\alpha\beta} y_\delta + \delta_{\alpha\delta}
y_\beta \right) \left\{ \bar{\mathfrak{l}}^{(3)} \right\}
\nonumber \\ 
&& + \frac{x_\alpha x_\beta x_\delta}{|x|}
\left\{
\partial^{[1]} \bar{\mathfrak{l}}^{(1)} - \frac{\hat c_\beta}{|x|} \partial^{[2]} \bar{\mathfrak{l}}^{(1)}
\right\}
\nonumber \\
&& + \frac{y_\alpha y_\beta y_\delta}{|x| |y|}
\left\{ \partial^{[2]} \bar{\mathfrak{l}}^{(2)} \right\} 
\nonumber\\
&& + \frac{y_\alpha x_\beta x_\delta}{|x| |y|}
\left\{ \partial^{[2]} \bar{\mathfrak{l}}^{(1)} \right\} 
\nonumber \\ 
&& + \frac{x_\alpha y_\beta x_\delta}{|x|}
\left\{
\partial^{[1]} \bar{\mathfrak{l}}^{(3)} - \frac{\hat c_\beta}{|x|} \partial^{[2]} \bar{\mathfrak{l}}^{(3)}
\right\}
\nonumber \\
&& + \frac{x_\alpha x_\beta y_\delta}{|x|}
\left\{
\partial^{[1]} \bar{\mathfrak{l}}^{(3)} - \frac{\hat c_\beta}{|x|} \partial^{[2]} \bar{\mathfrak{l}}^{(3)}
\right\}
\nonumber \\
&& + \frac{x_\alpha y_\beta y_\delta}{|x|}
\left\{
\partial^{[1]} \bar{\mathfrak{l}}^{(2)} - \frac{\hat c_\beta}{|x|} \partial^{[2]} \bar{\mathfrak{l}}^{(2)}
\right\}
\nonumber \\
&& 
+ \frac{y_\alpha x_\beta y_\delta}{|x| |y|}
\left\{ \partial^{[2]} \bar{\mathfrak{l}}^{(3)} \right\} 
\nonumber \\
&& 
+ \frac{y_\alpha y_\beta x_\delta}{|x| |y|}
\left\{ \partial^{[2]} \bar{\mathfrak{l}}^{(3)} \right\} 
\ea
and
\ba
&& \frac{1}{m} T^{{\rm III}}_{\alpha\beta\delta}(x,y) = 
\left( \delta_{\beta\delta} x_\alpha + \delta_{\alpha\beta} x_\delta \right) 
\left\{
\bar{\mathfrak{l}}^{(1)} + \bar{\mathfrak{l}}^{(3)}
\right\}
\nonumber \\
&& + \left( \delta_{\beta\delta} y_\alpha + \delta_{\alpha\beta} y_\delta \right) 
\left\{
\bar{\mathfrak{l}}^{(2)} + \bar{\mathfrak{l}}^{(3)}
\right\}
\nonumber \\
&& + \delta_{\alpha\delta} \frac{x_\beta}{|x|}
\left\{
\frac{1}{4} \left( 
\partial^{[1]} \bar{\mathfrak{g}}^{(0)} + \left(\frac{1}{|y|} - \frac{\hat
  c_\beta}{|x|} \right) \partial^{[2]} \bar{\mathfrak{g}}^{(0)} \right) \right.
\nonumber \\
&& 
\qquad \qquad 
- \frac{x^2}{4} \left(
\partial^{[1]} \bar{\mathfrak{l}}^{(1)} + \left(\frac{1}{|y|} -
\frac{\hat c_\beta}{|x|} \right) \partial^{[2]} \bar{\mathfrak{l}}^{(1)} \right)  
\nonumber \\
&& 
\qquad \qquad
- \frac{y^2}{4} \left(
\partial^{[1]} \bar{\mathfrak{l}}^{(2)} + \left(\frac{1}{|y|} -
\frac{\hat c_\beta}{|x|} \right) \partial^{[2]} \bar{\mathfrak{l}}^{(2)} \right)
\nonumber \\
&& 
\qquad \qquad 
\left. - \frac{|x| |y| \hat c_\beta}{2} \left(
\partial^{[1]} \bar{\mathfrak{l}}^{(3)} + \left(\frac{1}{|y|} - \frac{\hat
  c_\beta}{|x|} \right) \partial^{[2]} \bar{\mathfrak{l}}^{(3)} \right) 
- \frac{|x|}{2} \left( \bar{\mathfrak{l}}^{(1)} +
\bar{\mathfrak{l}}^{(3)} \right) \right\}  
\nonumber \\
&& + \delta_{\alpha\delta} \frac{y_\beta}{|y|}
\left\{
\frac{1}{4} \left( 
\partial^{[3]} \bar{\mathfrak{g}}^{(0)} + \left(\frac{1}{|x|} - \frac{\hat
  c_\beta}{|y|} \right) \partial^{[2]} \bar{\mathfrak{g}}^{(0)} \right) \right.
\nonumber \\
&& 
\qquad \qquad 
- \frac{x^2}{4} \left(
\partial^{[3]} \bar{\mathfrak{l}}^{(1)} + \left(\frac{1}{|x|} -
\frac{\hat c_\beta}{|y|} \right) \partial^{[2]} \bar{\mathfrak{l}}^{(1)} \right)  
\nonumber \\
&& 
\qquad \qquad
- \frac{y^2}{4} \left(
\partial^{[3]} \bar{\mathfrak{l}}^{(2)} + \left(\frac{1}{|x|} -
\frac{\hat c_\beta}{|y|} \right) \partial^{[2]} \bar{\mathfrak{l}}^{(2)} \right)
\nonumber \\
&& 
\qquad \qquad 
\left. - \frac{|x| |y| \hat c_\beta}{2} \left(
\partial^{[3]} \bar{\mathfrak{l}}^{(3)} + \left(\frac{1}{|x|} - \frac{\hat
  c_\beta}{|y|} \right) \partial^{[2]} \bar{\mathfrak{l}}^{(3)} \right) 
- \frac{|y|}{2} \left( \bar{\mathfrak{l}}^{(2)} +
\bar{\mathfrak{l}}^{(3)} \right) \right\}  
\nonumber \\
&& + \frac{x_\alpha x_\beta x_\delta}{|x|}
\left\{
\partial^{[1]} \bar{\mathfrak{l}}^{(1)} + \left(\frac{1}{|y|} -
\frac{\hat c_\beta}{|x|} \right) \partial^{[2]} \bar{\mathfrak{l}}^{(1)}
\right\}
\nonumber \\
& & + \frac{y_\alpha y_\beta y_\delta}{|y|}
\left\{
\partial^{[3]} \bar{\mathfrak{l}}^{(2)} + \left(\frac{1}{|x|} -
\frac{\hat c_\beta}{|y|} \right) \partial^{[2]} \bar{\mathfrak{l}}^{(2)}
\right\}
\nonumber \\
&& + \frac{x_\alpha y_\beta x_\delta}{|y|}
\left\{
\partial^{[3]} \bar{\mathfrak{l}}^{(1)} + \left(\frac{1}{|x|} -
\frac{\hat c_\beta}{|y|} \right) \partial^{[2]} \bar{\mathfrak{l}}^{(1)}
\right\}
\nonumber \\
&& + \frac{y_\alpha x_\beta x_\delta}{|x|}
\left\{
\partial^{[1]} \bar{\mathfrak{l}}^{(3)} + \left(\frac{1}{|y|} -
\frac{\hat c_\beta}{|x|} \right) \partial^{[2]} \bar{\mathfrak{l}}^{(3)}
\right\}
\nonumber \\
&& + \frac{x_\alpha x_\beta y_\delta}{|x|}
\left\{
\partial^{[1]} \bar{\mathfrak{l}}^{(3)} + \left(\frac{1}{|y|} -
\frac{\hat c_\beta}{|x|} \right) \partial^{[2]} \bar{\mathfrak{l}}^{(3)}
\right\}
\nonumber \\
&& + \frac{y_\alpha x_\beta y_\delta}{|x|}
\left\{
\partial^{[1]} \bar{\mathfrak{l}}^{(2)} + \left(\frac{1}{|y|} -
\frac{\hat c_\beta}{|x|} \right) \partial^{[2]} \bar{\mathfrak{l}}^{(2)}
\right\}
\nonumber \\
&& + \frac{y_\alpha  y_\beta x_\delta}{|y|}
\left\{
\partial^{[3]} \bar{\mathfrak{l}}^{(3)} + \left(\frac{1}{|x|} -
\frac{\hat c_\beta}{|y|} \right) \partial^{[2]} \bar{\mathfrak{l}}^{(3)}
\right\}
\nonumber \\
&& + \frac{x_\alpha y_\beta y_\delta}{|y|}
\left\{
\partial^{[3]} \bar{\mathfrak{l}}^{(3)} + \left(\frac{1}{|x|} -
\frac{\hat c_\beta}{|y|} \right) \partial^{[2]} \bar{\mathfrak{l}}^{(3)}
\right\}.
\ea

\section{Derivatives of the integrands for the six weight functions with respect to $|x|$ \la{sec:xderivs}}

In this appendix, we provide the expressions of the $|x|$-derivatives of the relevant sums
(see Eqs.\ \ref{eq:s1}--\ref{eq:s3} as well as Eqs.\ \ref{eq:sigma0final})-(\ref{eq:sigma5final}) 
entering the expression for the six weight functions parametrizing the QED kernel.

A few notational remarks are in order. In this appendix, we write the
argument of $z_n$ as the four-vector $u$ instead of $u^2$ and use the
notation $ z_n' \equiv \frac{\partial }{\partial |u|} z_n$, not
$\frac{\partial}{\partial |u|^2}$.  Also, we set $v=x-u$. The argument
of the Gegenbauer polynomials $C_n$ and their derivatives is always $(\hat
u\cdot \widehat{x-u})$.

We obtain 
\ba
\frac{\partial}{\partial|x|}  \sigma_0 &=& 
\frac{|x|-|u|\hat c_1}{|u-x|} \sum_{n=0}^{\infty} 
z_n(u)z_n'(v)\,\frac{C_n}{n+1}
\\ && +\frac{|u||x||\hat s_1^2}{|x-u|^3}\sum_{n=0}^{\infty} z_n(u) z_n(v) \frac{C_n^{\,\prime}}{n+1}\;,
\nonumber
\ea
\ba
\frac{\partial\sigma_3}{\partial|x|}  &=&
 \frac{\hat c_1 }{|x|} \sum_{n=0}^{\infty} 
\Big\{ z_n(u) \Big[-\frac{z_{n+1}(v)}{|x|} + \frac{|x|-|u|\hat c_1}{|x-u|} z_{n+1}'(v)\Big] \frac{C_{n+1}}{n+2} 
\\ && \qquad   + z_{n+1}(u) \Big[-\frac{z_{n}(v)}{|x|} + \frac{|x|-|u|\hat c_1}{|x-u|} z_{n}'(v)\Big] \frac{C_n}{n+1}\Big\}
\nonumber\\ && + \frac{\hat s_1^2}{|x-u|^2} \sum_{n=0}^{\infty} \Big\{
\Big(\Big[|u|\hat c_1 - \frac{ (|x|-|u|\hat c_1)}{n+1} \Big] \frac{z_n(u) z_{n+1}(v)}{|x-u|}
 +  \frac{|x|-|u|\hat c_1}{n+1} z_n(u) z_{n+1}'(v)\Big) \frac{C_{n+1}^{\,\prime}}{n+2}
\nonumber\\ && +\Big(\Big[ |u|\hat c_1  +\frac{|x|-|u|\hat c_1}{n+2} \Big] \frac{z_{n+1}(u) z_n(v)}{|x-u|}
  - \frac{|x|-|u|\hat c_1}{n+2} z_{n+1}(u) z_n'(v) \Big)\frac{C_n^{\,\prime}}{n+1} 
\Big\}
\nonumber\\ && 
+ \frac{|x||u|\hat s_1^4}{|x-u|^4}\sum_{n=0}^{\infty} \frac{1}{(n+1)(n+2)}\Big\{
z_n(u)z_{n+1}(v) C^{\,\prime\prime}_{n+1} - z_{n+1}(u) z_n(v) C^{\,\prime\prime}_{n} \Big\},
\nonumber
\ea
\ba
&& \frac{1}{\hat s_1^2}\frac{\partial}{\partial|x|} \sigma_2 =
-\frac{|x|-|u|\hat c_1}{|x-u|}\sum_{n=0}^{\infty}
\left\{ \frac{z_n(u)z_{n+1}'(v)}{n+2}C_{n+1} + \frac{z_{n+1}(u)z_n'(v)}{n+1} C_n\right\}
\\ &&
 + \frac{x^2\,|u|\hat c_1\hat s_1^2}{|x-u|^4} \sum_{n=0}^{\infty} 
\frac{1}{(n+1)(n+2)}\Big\{z_n(u)z_{n+1}(v)\, C_{n+1}'' -  z_{n+1}(u)z_n(v) C_n''\Big\}
\nonumber\\ && + \frac{1}{|x-u|^3}\sum_{n=0}^{\infty}\frac{1}{(n+1)(n+2)}\, \Big\{ 
\nonumber\\ &&
\Big[ |x| |x-u| \hat c_1(|x|-|u|\hat c_1) z_{n+1}'(v) 
+ |u|\Big(|u|\hat c_1-|x|\hat c_1^2-(n+1) |x|\hat s_1^2\Big) z_{n+1}(v) \Big]
z_n(u) C_{n+1}'
\nonumber\\ &&
 - \Big[ |x||x-u|\hat c_1(|x|-|u|\hat c_1) z_n'(v) + |u|
\Big(|u|\hat c_1-|x|\hat c_1^2 + (n+2)|x|\hat s_1^2\Big) z_n(v)\Big]
z_{n+1}(u)C_n' \Big\},
\nonumber
\ea
\ba
&& \frac{\partial}{\partial |x|}\sigma_1 =
-\frac{|x|}{2}\sum_{n=0}^{\infty} z_n(u)z_n(v) \frac{C_n}{n+1}
\\ && - \frac{x^2}{4} \frac{|x|-|u|\hat c_1}{|x-u|} \sum_{n=0}^{\infty}z_n(u)z_n'(v) \frac{C_n}{n+1}
\nonumber\\ && + \frac{u^2}{4}\frac{|x|-|u|\hat c_1}{|x-u|}\sum_{n=0}^{\infty} z_n'(v)
\Big(z_{n-2}(u)+(2-\delta_{n0})z_n(u)+z_{n+2}(u)\Big) \frac{C_n}{n+1}
\nonumber\\ && + \frac{(x-u)^2}{4}\frac{|x|-|u|\hat c_1}{|x-u|}\sum_{n=0}^{\infty} z_n(u)
\Big(z_{n-2}'(v)+(2-\delta_{n0})z_n'(v) + z_{n+2}'(v)\Big)\frac{C_n}{n+1}
\nonumber\\ && + \frac{(|x|-|u|\hat c_1)}{2}
\sum_{n=0}^{\infty} z_n(u) \Big(z_{n-2}(v) + (2-\delta_{n0})z_n(v) + z_{n+2}(v)\Big)\frac{C_n}{n+1}
\nonumber\\ && + \frac{|u|(|x|-|u|\hat c_1)}{4|x-u|}\sum_{n=0}^{\infty} \Big\{
\Big(z_n(u) z_{n+2}(v) + 2z_n(u) z_n(v) + z_{n+2}(u) z_n(v) \Big) \frac{C_{n+1}}{n+2}
\nonumber\\ && + \Big(z_{n-2}(u)z_n(v) + 2z_n(u)z_n(v) + z_n(u)z_{n-2}(v)\Big) \frac{C_{n-1}}{n} \Big\}
\nonumber\\ && + \frac{|u|}{4} (|x|-|u|\hat c_1) \sum_{n=0}^{\infty} \Big\{
\Big( z_n(u) z_{n+2}'(v) + 2z_n(u)z_n'(v) + z_{n+2}(u) z_n'(v)\Big) \frac{C_{n+1}}{n+2}
\nonumber\\ && + \Big( z_{n-2}(u) z_n'(v) + 2z_n(u)z_n'(v) + z_n(u) z_{n-2}'(v)\Big) \frac{C_{n-1}}{n}
\Big\}
\nonumber\\ && + \frac{|x||u|\hat s_1^2}{|x-u|^3}
\bigg\{
 -\frac{x^2}{4} \sum_{n=0}^{\infty} z_n(u) z_n(v) \frac{C_n^{\,\prime}}{n+1}
\nonumber\\ && + \frac{u^2}{4} \sum_{n=0}^{\infty} z_n(v)\Big(z_{n-2}(u) + (2-\delta_{n0})z_n(u) + z_{n+2}(u)\Big)
\frac{C_n^{\,\prime}}{n+1}
\nonumber\\ && 
+ \frac{(x-u)^2}{4} \sum_{n \geq 0} z_n(u)\Big( z_{n-2}(v) + (2-\delta_{n0})z_n(v) + z_{n+2}(v)\Big)
\frac{C_n^{\,\prime}}{n+1}
\nonumber\\ && 
+ \frac{|u||x-u|}{4}\sum_{n=0}^{\infty} \Big\{
 \Big(z_n(u) z_{n+2}(v)+2 z_n(u)z_n(v)  + z_{n+2}(u)z_n(v)\Big) \frac{C_{n+1}^{\,\prime}}{n+2}
\nonumber\\ &&  + \Big(z_{n-2}(u) z_{n}(v)+2 z_n(u)z_n(v) + z_{n}(u)z_{n-2}(v)\Big) 
\frac{C_{n-1}^{\,\prime}}{n} \Big\}
\bigg\},
\nonumber
\ea
\ba
 \frac{\partial s_1}{\partial |x|} &=& \frac{u^2}{4} \frac{|x|-|u|\hat c_1}{|x-u|}
\sum_{n=0}^{\infty} \Big(z_{n-2}(u)+(1-\delta_{n0})z_n(u) + z_{n+2}(u)\Big) z_n'(v)\,\frac{C_n}{n+1}
 \\ && + \frac{|u|^3|x|\hat s_1^2}{4|x-u|^3}\sum_{n=0}^{\infty} 
\Big(z_{n-2}(u) + (1-\delta_{n0})z_n(u) + z_{n+2}(u)\Big)z_n(v) \frac{C_n^{\,\prime}}{n+1}\;,
\nonumber
\ea
\ba
\frac{\partial s_2}{\partial |x|} &=&
\frac{|x|-|u|\hat c_1}{2} \sum_{n=0}^{\infty} z_n(u) \Big(z_{n-2}(v) + (1-\delta_{n0})z_n(v) + z_{n+2}(v)\Big) \frac{C_n}{n+1}
\\ && +\frac{|x-u|(|x|-|u|\hat c_1)}{4} \sum_{n=0}^{\infty} z_n(u) 
\Big(z_{n-2}'(v) + (1-\delta_{n0}) z_n'(v) + z_{n+2}'(v)\Big) \frac{C_n}{n+1}
\nonumber\\ && + \frac{|x||u|\hat s_1^2}{4|x-u|} \sum_{n=0}^{\infty} z_n(u) \Big(z_{n-2}(v) + (1-\delta_{n0})z_n(v) + z_{n+2}(v)\Big)
\frac{C_n^{\,\prime}}{n+1}\;,
\nonumber
\ea
\ba
\frac{\partial\sigma_4}{\partial|x|} &=& \frac{1}{|u|}\Big(  -\frac{2|u|}{x^2}\hat c_1\sigma_1 
+ (2\frac{|u|}{|x|}\hat c_1-1)\frac{\partial\sigma_1}{\partial|x|} + \frac{\partial s_2}{\partial |x|}
- \frac{\partial s_1}{\partial |x|}\Big),
\\
\frac{\partial\sigma_5}{\partial|x|} &=& \frac{1}{u^2}\Big(\frac{|u|}{|x|^3}  (-2 |u|(1+2\hat c_1^2) + 3|x| \hat c_1) \sigma_1
+ \frac{|u|}{x^2}(|u|(1+2\hat c_1^2) - 3|x|\hat c_1)\frac{\partial \sigma_1}{\partial |x|} 
\\ && -3\frac{|u|}{x^2} \hat c_1 (s_2-s_1) 
+ 3\frac{|u|}{|x|} \hat c_1 (\frac{\partial s_2}{\partial |x|}-\frac{\partial s_1}{\partial |x|})
+ 3\frac{\partial s_1}{\partial |x|}\Big).
\nonumber
\ea


\section{Expansion of the kernel for small arguments\la{sec:smallxy}}

This appendix provides the relevant expressions to obtain the QED weight functions and the full kernel
at $x=0$ or at $y=0$.
In the following, we refer to various functions defined mainly in subsection~\ref{sec:GegenExp}.

\subsection{The regime of small $|x|$}

In this subsection, the argument of the $z_n$ is always $u^2$ if not
explicitly specified.  As in appendix \ref{sec:xderivs},
$z_n'$ means $\frac{\partial}{\partial |u|} z_n$, not
the derivative with respect to $u^2$.  As for the sums $\sigma_k$
($1\leq k\leq 5$), we recall that their generic arguments are
$(|x|,\hat c_1 \equiv \hat x\cdot\hat u, |u|)$.

We begin by giving the first terms of the Taylor expansion of the sums $\sigma_k$ that enter the calculation
of the QED weight functions $\bar{\mathfrak{g}}^{(0)}$, $\bar{\mathfrak{g}}^{(1)}$, $\bar{\mathfrak{g}}^{(2)}$, $\bar{\mathfrak{l}}^{(1)}$,
$\bar{\mathfrak{l}}^{(2)}$, $\bar{\mathfrak{l}}^{(3)}$. In several cases, we express the result in terms of auxiliary sums collected 
below in Eqs.\ (\ref{eq:sigma1-22}--\ref{eq:sigma5-13}). We obtain
\ba
\sigma_0\big|_{x=0} &=& \sum_{n=0}^{\infty}  (-1)^n\, z_n^2\;,
\\
\sigma_3\big|_{x=0}  &=& \hat s_1^2 \sum_{n=0}^\infty (-1)^n 
(1 + \frac{2n}{3}) \frac{z_n z_{n+1}}{|u|} 
~+~ \hat c_1^2 \sum_{n=0}^\infty (-1)^n \Big(z_n z_{n+1}' -
z_{n+1} z_n' \Big),
\\
\lim_{|x|\to0} \frac{\sigma_1}{x^2} & = & 
\frac{4\hat c_1^2-1}{12} \sigma_1^{(2,2)}(u),
\\
\lim_{|x|\to0} \frac{\sigma_2}{|x|}& = & \hat c_1 \hat s_1^2 
\sigma_2^{(1,2)}(u) ,
\\
\lim_{|x|\to0} \frac{\sigma_4}{|x|} & = & -\frac{2}{3}\hat c_1 \hat s_1^2\, \sigma^{(1,2)}_4(u),
\\
\sigma_5\big|_{x=0}  & = & \frac{2}{3} \hat s_1^4\sigma_5^{(0,2)}(u).
\ea
In addition, the following intermediate results are needed,
\ba
\lim_{|x|\to0} \frac{s_2-s_1}{|x|} & = & -\frac{|u|\hat c_1 }{2}
\sigma_1^{(2,2)}(u),
\\
s_1\big|_{x=0} & = & \frac{u^2}{4} \sigma_1^{(2,2)}(u).
\ea

For some of the sums, we will need the expansion to one order higher.
In particular, we record
\ba
&& \frac{1}{2} \frac{\partial^2(s_2-s_1)}{\partial|x|^2} \Big|_{x=0} = \frac{1}{4} \sigma_1^{(2,2)}(u)
 + \frac{|u|\hat c_1^2}{2} \sigma_{s_1}^{(1,1)}(u) ,
\\
&&  \frac{\partial s_1}{\partial|x|} \Big|_{x=0} =  -\frac{\hat c_1 u^2}{4} \,\sigma_{s_1}^{(1,1)}(u),
\ea
as well as the results
\ba
\frac{\partial\sigma_3}{\partial |x|} \Big|_{x=0} & = &  \hat c_1^3 \sum_{n=0}^\infty (-1)^n
\left(- \frac{z_n z_{n+1}'' - z_{n+1} z_n''}{2} \right)  \\ 
&& +  \hat c_1 \hat s_1^2 \sum_{n=0}^\infty (-1)^n 
\left( \frac{(2n+3) z_n z_{n+1}}{2 u^2} 
- \frac{(2n+9) z_n z_{n+1}'}{6|u|} 
+ \frac{(3-2n) z_{n+1} z_n'}{6|u|} \right) ,
\nonumber \\ 
 \frac{1}{6}\frac{\partial^3\sigma_1}{\partial|x|^3} \Big|_{x=0} &=&
2 \Big( C_1(\hat c_1)\, \hat t_{\sigma_1}^{(3,1)}(u) +  C_3(\hat c_1)\,\hat u_{\sigma_1}^{(3,1)}(u)\Big),
\\
\frac{1}{2} \frac{\partial^2\sigma_2}{\partial|x|^2}\Big|_{x=0} & = &  \hat s_1^4 \sum_{n=0}^\infty (-1)^n
\left(- \frac{(2n+3) z_n z_{n+1}}{6 u^2} 
+ \frac{z_n z_{n+1}' - z_{n+1} z_n'}{2|u|} \right) \nonumber \\ 
&& +  \hat s_1^2 \hat c_1^2 \sum_{n=0}^\infty (-1)^n
\left(\frac{(2n+3) z_n z_{n+1}}{3 u^2} 
- \frac{(n+3) z_n z_{n+1}'}{3|u|} 
- \frac{n z_{n+1} z_{n}'}{3|u|}
\right. \nonumber \\
& & \qquad \qquad \qquad \qquad 
\left. + \frac{z_n z_{n+1}'' - z_{n+1} z_n''}{2} \right) ,
\\
  \frac{\partial\sigma_5}{\partial|x|}\Big|_{x=0} &=&  \frac{2 \hat c_1 \hat s_1^4}{3} \;  \sigma_5^{(1,3)}(u).
\ea

The auxiliary sums appearing in the results above are defined as follows,
\ba \la{eq:sigma1-22}
 \sigma_1^{(2,2)}(u) &=& \sum_{n=0}^{\infty}(-1)^n z_n\Big((1-\delta_{n0})z_n+2z_{n+2}\Big) = 3u^2b(u^2),
\\  \la{eq:sigma2-12}
\sigma_2^{(1,2)}(u) &=& \sum_{n=0}^\infty (-1)^n \left(z_{n+1}
  z_n' - z_n z_{n+1}'  
  + (1+\frac{2n}{3})\,\frac{z_n z_{n+1}}{|u|}\right),
\\ \sigma^{(1,2)}_4(u) &=& \sigma_1^{(2,2)}(u), \phantom{\frac{1}{1}}
\\ \sigma_5^{(0,2)}(u) &=& \sigma_1^{(2,2)}(u), \phantom{\frac{1}{1}}
\\
 \sigma_{s_1}^{(1,1)}(u) &=&\sum_{n=0}^{\infty} (-1)^n \Big[ (1-\delta_{n0}) z_n z_n' + z_n z_{n+2}' + z_{n+2} z_n' \Big],
\\
 \hat t_{\sigma_1}^{(3,1)}(u)  &=&  - \frac{1}{24|u|}
\sigma_1^{(2,2)}(u) - \frac{1}{48} \sigma_{s_1}^{(1,1)}(u) ,
\label{eq:t_sigma1} 
\\
 \hat u_{\sigma_1}^{(3,1)}(u) &=& \frac{1}{48} \sigma_5^{(1,3)}(u) ,
\\ \la{eq:sigma5-13}
 \sigma_5^{(1,3)}(u) &=& \frac{\sigma_1^{(2,2)}(u)}{|u|} - \sigma_{s_1}^{(1,1)}(u).
\ea
We recall that $b(u^2)$ was first introduced in Eqs.\ (\ref{eq:bu2}--\ref{eq:bu2sum}).

\subsubsection{The scalar weight function}

The behavior of $\bar{\mathfrak{g}}^{(0)}$ around $x=0$ is determined by 
a small number of coefficients $\alpha^{(0)}_{m\pm}$; see Eq.\ (\ref{eq:g0g3v1}).
All $\alpha^{(0)}_{m\pm}$ vanish at $|x|=0$, except
\ba
\alpha_{0-}^{(0)}(0,|y|) &=& \frac{1}{2}\sum_{n=0}^{\infty}(-1)^n \int_0^{|y|} d|u|\; |u|^3\;  z_n^2
\ea
and $\alpha_{0+}^{(0)}$; the latter, whose integral is infrared divergent at large $|u|$, is however not needed.
As for the $|x|$-derivative of the coefficients at the origin, they all vanish except
\ba
\frac{\partial \alpha_{1+}^{(0)}}{\partial |x|}\Big|_{|x|=0} = 
\lim_{|x|\to0}\alpha_{1+}^{(0)}(|x|,|y|)/|x| &=& -\frac{1}{8} \sum_{n=0}^{\infty}(-1)^n  \int_{|y|}^\infty d|u| \; z_n z_n'
\qquad 
\\ &=& \frac{1}{16} \sum_{n=0}^{\infty} (-1)^n z_n(y^2)^2,
\nonumber\\ 
\frac{\partial  \alpha_{1-}^{(0)}}{\partial |x|}\Big|_{|x|=0} = 
\lim_{|x|\to0}\alpha_{1-}^{(0)}(|x|,|y|)/|x| &=& -\frac{1}{8} \sum_{n=0}^{\infty}(-1)^n \int_0^{|y|} d|u|\; |u|^4 \;  z_n z_n'
\qquad 
\\ &=& \frac{1}{2} \alpha_{0-}^{(0)}(|x|=0,|y|) - |y|^4 \lim_{|x|\to0}\alpha_{1+}^{(0)}(|x|,|y|)/|x| .
\nonumber
\ea
Combining these observations, one finds for the actually needed derivatives of the scalar weight function
\ba \la{eq:dg0dx.x.eq.0}
\left.\frac{\partial}{\partial |x|} \bar {\mathfrak{g}}^{(0)}\right|_{x=0} &=& 2\hat c_\beta \Big(\frac{1}{|y|^3}
\frac{\alpha^{(0)}_{1-}(|x|,|y|)}{|x|}  + |y| \frac{\alpha^{(0)}_{1+}(|x|,|y|)}{|x|}\Big)_{x=0},
\\ \la{eq:dg0dcb.x.eq.0}
\left.\frac{\partial}{\partial\hat c_\beta} \bar {\mathfrak{g}}^{(0)}\right|_{x=0} &= & 2|x| \Big(\frac{1}{|y|^3}
\frac{\alpha^{(0)}_{1-}(|x|,|y|)}{|x|}  + |y| \frac{\alpha^{(0)}_{1+}(|x|,|y|)}{|x|}\Big)_{x=0} +{\rm O}(x^2),
\\
\la{eq:dg0dy.x.eq.0}
\left.\frac{\partial}{\partial |y|} \bar{\mathfrak{g}}^{(0)}\right|_{x=0} &= &  -\frac{2}{|y|^3} \alpha^{(0)}_{0-}(|x|=0,|y|).
\ea
From Eqs.\ (\ref{eq:dg0dx.x.eq.0}--\ref{eq:dg0dcb.x.eq.0}) one sees that
\[
\frac{\partial}{\partial |x|} \bar{\mathfrak{g}}^0 - \frac{\hat c_\beta }{ |x|}
  \frac{\partial}{\partial \hat c_\beta} \bar{\mathfrak{g}}^0 = {\rm O}(|x|).
\]
This and further similar relations for the other weight functions will be exploited to arrive
at the final expressions of the tensors $T^{\rm A}_{\alpha\beta\delta}(x,y)$, Eqs.\ (\ref{eq:TIabd_xeq0}--\ref{eq:TIIIabd_xeq0}) below.

\subsubsection{The vector weight functions}

In the Taylor expansion of a QED weight function $\bar{\mathfrak{g}}^{(k)}$, we denote by $\bar {\mathfrak{g}}^{(k,n)}$ the term of order $|x|^n$.
For $\bar{\mathfrak{g}}^{(2)}$, one finds $\bar {\mathfrak{g}}^{(2,0)}=0$ and 
\ba\la{eq:g21}
\bar {\mathfrak{g}}^{(2,1)} &=& -\frac{|x|\hat c_\beta}{12}\Big[\frac{1}{|y|^5}\int_0^{|y|}d|u|\; |u|^5\; \sigma_2^{(1,2)}(u) 
+ |y| \int_{|y|}^\infty \frac{d|u|}{|u|}\; \sigma_2^{(1,2)}(u)\Big] ,
\ea
and 
\ba \la{eq:g22}
\bar {\mathfrak{g}}^{(2,2)} &=& - \frac{x^2}{6} \Big[\frac{1}{|y|^4}\int_0^{|y|}d|u|\,|u|^4 \sigma_2^{(2,1)}(u) + \int_{|y|}^\infty d|u| \,\sigma_2^{(2,1)}(u)\Big]
\\ && - \frac{x^2}{30}(6 \hat c_\beta^2 - 1) \Big[  \frac{1}{|y|^6}\int_0^{|y|}d|u|\,|u|^6 \sigma_2^{(2,3)}(u) 
     + |y|^2\int_{|y|}^\infty \frac{d|u|}{|u|^2} \,\sigma_2^{(2,3)}(u)\Big], \nonumber 
\\ 
\sigma_2^{(2,1)}(u) &=& \frac{-1}{48 u^2}\sum_{n=0}^{\infty}(-1)^n\Big\{ |u| z_{n+1}\Big((2n+15)z_n'+3|u|z_n''\Big) 
\\ && \qquad + z_n\Big((9+6n)z_{n+1} + |u|((2n-9)z_{n+1}'-3|u|z_{n+1}'')\Big)\Big\}, \nonumber 
\\ \la{eq:sigma2-23}
\sigma_2^{(2,3)}(u) &=& \frac{5}{96 u^2}\sum_{n=0}^{\infty}(-1)^n\Big\{|u| z_{n+1}\Big((3-2n)z_n'-3|u|z_n''\Big)
\\ && \qquad  + z_n \Big((9+6n) z_{n+1} + |u| (3|u|z_{n+1}''-(9+2n) z_{n+1}')\Big) \Big\}. \nonumber
\ea

Similarly for $\bar{\mathfrak{g}}^{(3)}$:
\ba
\bar {\mathfrak{g}}^{(3,0)} &=& \frac{1}{2} \Big[ \frac{1}{|y|^2}\int_0^{|y| }d|u|\, |u|^3 \sigma_3^{(0,0)}(u) +\int_{|y|}^\infty d|u|\, |u|\, \sigma_3^{(0,0)}(u)\Big]
\\ && + \frac{4\hat c_\beta^2-1}{6} \Big[\frac{1}{|y|^4} \int_0^{|y|} d|u|\,|u|^5\,\sigma_3^{(0,2)}(u)
  + y^2 \int_{|y|}^\infty \frac{d|u|}{|u|}\, \sigma_3^{(0,2)}(u)\Big],   \nonumber
\\
\sigma_3^{(0,0)}(u) &=& \frac{1}{8|u|}\sum_{n=0}^{\infty}(-1)^n\Big\{ -|u| z_{n+1} z_n' + z_n((2n+3)z_{n+1} + |u| z_{n+1}')\Big\},
\\
\sigma_3^{(0,2)}(u) &=& -\frac{1}{8}\sigma_2^{(1,2)}(u)
\ea
(with $\sigma_2^{(1,2)}(u)$ defined in Eq.\ (\ref{eq:sigma2-12})), as well as 
\ba
\bar {\mathfrak{g}}^{(3,1)} &=&  \frac{|x|\hat c_\beta}{2}\Big[\frac{1}{|y|^3}\int_0^{|y|}d|u| \,|u|^4\,\sigma_3^{(1,1)}(u) + |y|\int_{|y|}^\infty d|u|\,\sigma_3^{(1,1)}(u) \Big]
\\ && + \frac{|x|\hat c_\beta (2 \hat c_\beta^2 -1)}{2}\Big[\frac{1}{|y|^5}\int_0^{|y|}d|u| \,|u|^6\,\sigma_3^{(1,3)}(u) 
 + |y|^3\int_{|y|}^\infty \frac{d|u|}{|u|^2}\,\sigma_3^{(1,3)}(u) \Big], \nonumber
\\
\sigma_3^{(1,1)}(u)&=& \frac{1}{48u^2}\sum_{n=0}^{\infty} (-1)^n\Big\{ |u|z_{n+1} ((3-2n)z_n' + 3|u|z_n'') 
\\ &&       + z_n((6n+9)z_{n+1} - |u|((2n+9) z_{n+1}' + 3|u| z_{n+1}'')) \Big\},\nonumber
\\
\sigma_3^{(1,3)}(u) &=& -\frac{1}{5} \sigma_2^{(2,3)}(u)
\ea
(with $\sigma_2^{(2,3)}(u)$ given in Eq.\ (\ref{eq:sigma2-23})).
These results imply, for the actually needed weight function $\bar{\mathfrak{g}}^{(1)}$, 
\ba\la{eq:g10}
\bar {\mathfrak{g}}^{(1,0)} &=& \frac{1}{2} \Big[ \frac{1}{|y|^2}\int_0^{|y| }d|u|\, |u|^3 \sigma_3^{(0,0)}(u) +\int_{|y|}^\infty d|u|\, |u|\, \sigma_3^{(0,0)}(u)\Big]
\\ && + \frac{1}{48} \Big[\frac{1}{|y|^4} \int_0^{|y|} d|u|\,|u|^5\,\sigma_2^{(1,2)}(u)
  + y^2 \int_{|y|}^\infty \frac{d|u|}{|u|}\, \sigma_2^{(1,2)}(u)\Big],
\nonumber
\\
\bar {\mathfrak{g}}^{(1,1)} &=& |x|\hat c_\beta \Big[ 
\frac{1}{2}\Big(\frac{1}{|y|^3}\int_0^{|y|}d|u| \,|u|^4\,(\sigma_3^{(1,1)}(u)+\frac{1}{3}\sigma_2^{(2,1)}(u))
 + |y|\int_{|y|}^\infty d|u|\,(\sigma_3^{(1,1)}(u)+\frac{1}{3}\sigma_2^{(2,1)}(u)) \Big)
\nonumber\\ && + \frac{1}{15} \Big( \frac{1}{|y|^5}\int_0^{|y|}d|u| \,|u|^6\,\sigma_2^{(2,3)}(u) 
 + |y|^3\int_{|y|}^\infty \frac{d|u|}{|u|^2}\,\sigma_2^{(2,3)}(u) \Big) \Big].
\la{eq:g11}
\ea

\subsubsection{The tensor weight functions}

Similar to above, in the Laurent series of a QED weight function
$\bar{\mathfrak{l}}^{(k)}$, we denote by $\bar{\mathfrak{l}}^{(k,n)}$
the term of order $|x|^n$.  First, using the same notation for the
auxiliary weight function $\bar v^{(1)}$, we find
\ba
\bar v^{(1,2)} &=&  \frac{x^2(4\hat c_\beta^2-1)}{72} \Big(\frac{1}{|y|^4} \int_0^{|y|}d|u|\,|u|^5\,\sigma_1^{(2,2)}(u)
+ y^2 \int_{|y|}^\infty \frac{d|u|}{|u|}\, \sigma_1^{(2,2)}(u)\Big),
\ea
\ba 
\bar v^{(1,3)} &=& \frac{|x|^3C_1(\hat c_\beta)}{2}     \Big(\frac{1}{|y|^3} \int_0^{|y|}d|u|\, |u|^4\,\hat t_{\sigma_1}^{(3,1)}(u)
+ |y| \int_{|y|}^\infty d|u|\, \hat t_{\sigma_1}^{(3,1)}(u) \Big)
\\ && + \frac{|x|^3C_3(\hat c_\beta)}{4} \Big(\frac{1}{|y|^5}  \int_0^{|y|}d|u|\,|u|^6 \hat u_{\sigma_1}^{(3,1)}(u)
 + |y|^3 \int_{|y|}^\infty \frac{d|u|}{|u|^2}\, \hat u_{\sigma_1}^{(3,1)}(u) \Big).
\nonumber
\ea
We can now proceed to determining the first two non-trivial terms for weight function $\bar{\mathfrak{l}}^{(4)}$,
\ba
\bar{\mathfrak{l}}^{(4,1)} &=& \frac{|x|\,\hat c_\beta}{9}
\Big[ \frac{1}{|y|^3} \int_0^{|y|} d|u|\,|u|^5\,\sigma^{(1,2)}_4(u) + |y|^3 \int_{|y|}^\infty \frac{d|u|}{|u|}\; \sigma^{(1,2)}_4(u)\Big],
\ea
\ba
\bar{\mathfrak{l}}^{(4,2)} &=& \frac{-x^2 }{3} \Big(\frac{1}{y^2}\int_0^{|y|}d|u|\,|u|^4 \sigma_4^{(2,1)}(u) + y^2 \int_{|y|}^\infty d|u| \,\sigma_4^{(2,1)}(u) \Big)
\\ && 
- \frac{x^2 (6 \hat c_\beta^2 - 1)}{15}\Big(\frac{1}{|y|^4}\int_0^{|y|}d|u| \,|u|^6\,\sigma_4^{(2,3)}(u) + |y|^4 \int_{|y|}^\infty \frac{d|u|}{|u|^2} \sigma_4^{(2,3)}(u)\Big),
\nonumber
\\
\sigma_4^{(2,1)}(u) &=& -4 \hat t_{\sigma_1}^{(3,1)}(u),
\\ 
\sigma_4^{(2,3)}(u) &=& -\frac{5}{24} \sigma_5^{(1,3)}(u).
\ea
Next come the first two terms for weight function $\bar{\mathfrak{l}}^{(2)}$,
\ba \la{eq:l20}
\bar{\mathfrak{l}}^{(2,0)} &=& \frac{1}{18} \Big( \frac{1}{|y|^6}\int_0^{|y|}d|u|\,|u|^5 \,\sigma_5^{(0,2)}(u)
+ \int_{|y|}^\infty \frac{d|u|}{|u|}\, \sigma_5^{(0,2)}(u) \Big),
\\ \la{eq:l21}
\bar{\mathfrak{l}}^{(2,1)} &=& 
\frac{|x|\hat c_\beta}{24}
\Big(\frac{1}{|y|^7} \int_0^{|y|}d|u|\,|u|^6 \sigma_5^{(1,3)}(u) + |y| \int_{|y|}^\infty \frac{d|u|}{|u|^2} \sigma_5^{(1,3)}(u) \Big).
\ea
Now from Eq.\ (\ref{eq:ell3lc}), we obtain  $\bar{\mathfrak{l}}^{(3)}$ via
\ba
\bar{\mathfrak{l}}^{(3)} &=& \frac{1}{2x^2 y^2 } \bar{\mathfrak{l}}^{(4)}
 - \frac{|y|}{|x|}\,\hat c_\beta\;\bar{\mathfrak{l}}^{(2)},
\ea
so that the ${\rm O}(1/|x|)$ contribution cancels out,
\ba
\bar{\mathfrak{l}}^{(3,-1)} &=& \frac{1}{2x^2 y^2 } \bar{\mathfrak{l}}^{(4,1)} - \frac{|y|}{|x|}\,\hat c_\beta\;\bar{\mathfrak{l}}^{(2,0)}
= 0.
\ea
The leading contribution is finite,
\ba\la{eq:l30}
\bar{\mathfrak{l}}^{(3,0)} &=& \frac{1}{2x^2 y^2 } \bar{\mathfrak{l}}^{(4,2)}
- \frac{|y|}{|x|}\,\hat c_\beta\;\bar{\mathfrak{l}}^{(2,1)}
\\ &=& \frac{-1}{6  } \Big[\frac{1}{|y|^4}\int_0^{|y|}d|u|\,|u|^4 \sigma_4^{(2,1)}(u) +  \int_{|y|}^\infty d|u| \,\sigma_4^{(2,1)}(u)
\\ && 
+ \frac{1}{24} \Big(\frac{1}{|y|^6} \int_0^{|y|}d|u|\,|u|^6 \sigma_5^{(1,3)}(u) + y^2 \int_{|y|}^\infty \frac{d|u|}{|u|^2} \sigma_5^{(1,3)}(u) \Big)\Big].
\nonumber
\ea
Similarly, one finds that the ${\rm O}(1/x^2)$ contribution to (see Eq.\ (\ref{eq:ell1lc}))
\be\la{eq:ell1lcb}
\bar{\mathfrak{l}}^{(1)}  = \frac{4}{3x^4}\Big(\bar v_1 - x^2 y^2 (\hat c_\beta^2 - \frac{1}{4})\; \bar{\mathfrak{l}}^{(2)}
- \frac{3}{2} x^3 y \hat c_\beta \;\bar{\mathfrak{l}}^{(3)} \Big)
\ee
vanishes,
\be
\bar{\mathfrak{l}}^{(1,-2)} = 0.
\ee
For the contribution of order $1/|x|$,  it is useful to decompose the 
expression in the basis of the $C_m(\hat c_\beta)$, calculating the coefficient
$\bar{\mathfrak{l}}^{(1,-1,m)}$ of $C_m(\hat c_\beta)$ using
$\bar{\mathfrak{l}}^{(1,-1,m)}=\frac{2}{\pi}\int_0^\pi d\beta \,\sin^2\beta\, C_m(\hat c_\beta) \bar{\mathfrak{l}}^{(1,-1)}$.
Before forming the linear combination (\ref{eq:ell1lcb}),
we find that there are $m=1$ and $m=3$ components, but they cancel in the linear combination,
  $\bar{\mathfrak{l}}^{(1,-1,1)}=\bar{\mathfrak{l}}^{(1,-1,3)} =0$,
so that 
\be
\bar{\mathfrak{l}}^{(1,-1)} = 0.
\ee
Thus $\bar{\mathfrak{l}}^{(1)}$ is finite in the limit $|x|\to0$.

\subsubsection{The limit $|x|\to0$ for the tensors $T^{\rm A}_{\alpha\beta\delta}(x,y)$} 

With the help of the chain rules for the tensors $T^{\rm A}_{\alpha\beta\delta}(x,y)$ given in appendix~\ref{sec:chainrT},
we find the following, finite expressions for these three tensors at $|x|\to0$,
\ba\la{eq:TIabd_xeq0}
T^{\rm I}_{\alpha\beta\delta}(0,y) &=& (\delta_{\alpha\delta}\hat y_\beta + \delta_{\beta\delta} \hat y_\alpha) 
\left(\frac{\partial \bar{\mathfrak{g}}^{(1)}}{|x|\partial\hat c_\beta}\right) 
+ \delta_{\alpha\beta} y_\delta    
\left(\frac{\partial^2  \bar{\mathfrak{g}}^{(2)}}{\partial|x|^2}- {\hat c_\beta^2}\frac{\partial^2  \bar{\mathfrak{g}}^{(2)}}{x^2\partial\hat c_\beta^2}\right)    
\\ &&     
 + \delta_{\alpha\delta}\hat y_\beta \Big(\frac{\partial  \bar{\mathfrak{g}}^{(1)}}{\partial|y|}\Big)
+ (\delta_{\beta\delta} \hat y_\alpha + \delta_{\alpha\beta} \hat y_\delta) \Big(\frac{\partial  \bar{\mathfrak{g}}^{(2)}}{|x|\partial\hat c_\beta}\Big)
\nonumber\\ &&  + \hat y_\alpha \hat y_\beta \hat y_\delta \;\frac{1}{|x|}
\Big(\frac{|y|}{|x|}\,\frac{\partial }{\partial\hat c_\beta}
+|y|\frac{\partial }{\partial|y|} - 1 \Big) \frac{\partial  \bar{\mathfrak{g}}^{(2)}}{\partial\hat c_\beta} \,,
\nonumber\\  \la{eq:TIIabd_xeq0}
\frac{1}{m}T^{\rm II}_{\alpha\beta\delta}(0,y) &=& 
\Big(\delta_{\alpha\beta} y_\delta + \delta_{\alpha\delta} y_\beta - \frac{y_\alpha}{2}\delta_{\beta\delta} \Big)\bar{\mathfrak{l}}^{(3)}
+ \Big(y_\beta y_\delta - \frac{y^2}{4} \delta_{\beta\delta}\Big) \hat y_\alpha\left(\frac{\partial \bar{\mathfrak{l}}^{(2)}}{|x|\partial\hat c_\beta} \right)
\\ && + \frac{\delta_{\beta\delta}}{4} \hat y_\alpha
\left(\frac{\partial   \bar{\mathfrak{g}}^{(0)}}{|x|\partial\hat c_\beta}\right),
\nonumber\\  \la{eq:TIIIabd_xeq0}
\frac{1}{m}T^{\rm III}_{\alpha\beta\delta}(0,y) &=&
\Big(\delta_{\beta\alpha} y_\delta + \delta_{\beta\delta} y_\alpha - \frac{y_\beta}{2}\delta_{\alpha\delta} \Big)
\Big(\bar{\mathfrak{l}}^{(2)} +\bar{\mathfrak{l}}^{(3)} \Big)
\\ && + \Big(y_\alpha y_\delta - \frac{y^2}{4} \delta_{\alpha\delta}\Big) \hat y_\beta
\left(\frac{\partial \bar{\mathfrak{l}}^{(2)}}{|x|\partial\hat c_\beta}+\frac{\partial\bar{\mathfrak{l}}^{(2)}}{\partial |y|} \right)
 +\frac{\delta_{\alpha\delta}}{4} \hat y_\beta
\left(  \frac{\partial  \bar{\mathfrak{g}}^{(0)}}{|x|\partial\hat c_\beta} + \frac{\partial   \bar{\mathfrak{g}}^{(0)}}{\partial|y|}\right).
\nonumber
\ea
The right-hand side should be evaluated in the limit $|x|\to0$.
Using the results respectively in Eqs.~(\ref{eq:g21}, \ref{eq:g11},
\ref{eq:g22}, \ref{eq:g10}), (\ref{eq:l30}, \ref{eq:l21},
\ref{eq:dg0dcb.x.eq.0}) and (\ref{eq:l20}, \ref{eq:l30}, \ref{eq:l21},
\ref{eq:dg0dcb.x.eq.0}, \ref{eq:dg0dy.x.eq.0}) the evaluation of the three tensors is
easily performed. We remark that the $\hat c_\beta$ dependence cancels everywhere, 
in particular in the second term of Eq.\ (\ref{eq:TIabd_xeq0}), which contains an explicit factor of $\hat c_\beta^2$.

\subsection{The regime of small $|y|$}

In the following, we give without further comment the relevant expressions for the weight functions in the limit of $|y|\to0$.

\subsubsection{The scalar weight function}

\ba
\frac{\partial \bar {\mathfrak{g}}^{(0)}}{\partial |x|}\Big|_{y=0} 
&=& \frac{1}{\pi} \int_0^\infty d|u|\,|u|\int_0^\pi d\phi_1\;\hat s_1^2\;\frac{\partial \sigma_0}{\partial|x|}(|x|,\hat c_1,|u|)
= \frac{\partial\alpha_{0+}^{(0)}(|x|,0)}{\partial|x|},
\\
\frac{\partial \bar {\mathfrak{g}}^{(0)}}{\partial\hat c_\beta}\Big|_{y=0} &=&
 \frac{ |y|}{\pi}\int_0^\infty d|u| \int_0^\pi d\phi_1\;\hat s_1^2\; C_1(\hat c_1) \,\sigma_0(|x|,\hat c_1,|u|)
= 2 |y| \alpha_{1+}^{(0)}(|x|,0),
\\
\frac{\partial \bar {\mathfrak{g}}^{(0)}}{\partial |y|}\Big|_{y=0}
&=& C_1(\hat c_\beta)\, \alpha_{1+}^{(0)}(|x|,0).
\qquad 
\ea

\subsubsection{The vector weight functions}
\ba
\bar {\mathfrak{g}}^{(1)}\Big|_{y=0} &=& \frac{1}{2\pi}\int_0^\infty d|u|\,|u|\int_0^\pi d\phi_1\;\hat s_1^2\;\sigma_3(|x|,\hat c_1,|u|) = \alpha^{(3)}_{0+}(|x|,0),
\\
\bar {\mathfrak{g}}^{(2)}\Big|_{y=0} &=& \frac{-1}{3\pi}\int_0^\infty d|u|\int_0^\pi d\phi_1\;\hat s_1^2\; \sigma_2(|x|,\hat c_1,|u|) = 2\, \beta^{(2)}_{1+}(|x|,0),
\\
\frac{\partial \bar {\mathfrak{g}}^{(1)}}{\partial |y|}\Big|_{y=0} &=&
C_1(\hat c_\beta)  \Big[-\frac{1}{2|x|} \bar {\mathfrak{g}}^{(2)}\Big|_{y=0} 
+ \frac{1}{2\pi} \int_0^\infty d|u|\int_0^\pi d\phi_1\;\hat s_1^2\; \hat c_1\; \sigma_3(|x|,\hat c_1,|u|)\Big]
\\ &=& C_1(\hat c_\beta)  \Big[-\frac{\beta^{(2)}_{1+}(|x|,0)}{|x|}   + \alpha^{(3)}_{1+}(|x|,0)\Big].
\ea

\subsubsection{The tensor weight functions}
\ba
\bar{\mathfrak{l}}^{(1)}\Big|_{y=0} &=& \frac{4}{3\pi |x|^4}\int_0^\infty d|u|\;|u|\int_0^\pi d\phi_1\;\hat s_1^2\;\sigma_1(|x|,\hat c_1,|u|) 
= \frac{4}{3|x|^4} \alpha^{(1)}_{0+}(|x|,0),
\\
\bar{\mathfrak{l}}^{(3)}\Big|_{y=0} &=& \frac{-1}{3\pi\,x^2}  \int_0^\infty d|u|\int_0^\pi d\phi_1\;\hat s_1^2 \;\sigma_4(|x|,\hat c_1,|u|)
= \frac{\beta^{(4)}_{1+}(|x|,0)}{x^2},
\\
\frac{\partial \bar{\mathfrak{l}}^{(1)}}{\partial |y|}\Big|_{y=0} &=& \frac{C_1(\hat c_\beta)}{|x|}
\Big[\frac{4}{3\pi|x|^3}  \int_0^\infty d|u|\int_0^\pi d\phi_1\;\hat s_1^2\,\hat c_1\;\sigma_1(|x|,\hat c_1,|u|)
-  \; \bar{\mathfrak{l}}^{(3)}\Big|_{y=0}\Big]
\\ &=& C_1(\hat c_\beta) \Big[ \frac{4}{3|x|^4} \alpha^{(1)}_{1+}(|x|,0) - \frac{1}{|x|^3} \beta^{(4)}_{1+}(|x|,0)\Big]  .
\ea

\subsubsection{The limit $|y|\to0$ for the tensors $T^{\rm A}_{\alpha\beta\delta}(x,y)$}

We find the finite result
\ba
T_{\alpha\beta\delta}^{\rm I}(x,0) &=& \Big(\delta_{\alpha\delta} \hat x_\beta + \delta_{\beta\delta}\hat x_\alpha + \delta_{\alpha\beta} \hat x_\delta 
- \hat x_\alpha \hat x_\beta \hat x_\delta\Big) \Big(\frac{\partial \bar{\mathfrak{g}}^{(1)}}{\partial|x|} \Big)
+ \hat x_\alpha \hat x_\beta x_\delta \Big(\frac{\partial^2\bar{\mathfrak{g}}^{(1)}}{\partial|x|^2}\Big)
~~~~\\ && +\Big( \delta_{\alpha\delta} \hat x_\beta + \delta_{\alpha\beta}\hat x_\delta - \hat x_\alpha \hat x_\beta \hat x_\delta \Big) 
\Big( \frac{\partial \bar{\mathfrak{g}}^{(1)}}{|y|\partial\hat c_\beta} \Big)
+ \hat x_\alpha \hat x_\beta x_\delta \Big(\frac{\partial^2 \bar{\mathfrak{g}}^{(1)}}{|y|\partial|x|\partial\hat c_\beta}\Big)
\nonumber\\ && + \delta_{\beta\delta}\,\hat x_\alpha \Big(\frac{\partial\bar{\mathfrak{g}}^{(2)}}{\partial |x|}\Big),
\nonumber\\
\frac{1}{m}T^{\rm II}_{\alpha\beta\delta}(x,0) &=& \frac{\delta_{\beta\delta}\,\hat x_\alpha}{4} \Big(\frac{\partial \bar{\mathfrak{g}}^{(0)}}{\partial |x|}\Big)
\\ && +\Big(\delta_{\alpha\beta}x_\delta + x_\beta\delta_{\alpha\delta} - \frac{x_\alpha}{2} \delta_{\beta\delta}\Big) \,\bar{\mathfrak{l}}^{(1)}
\nonumber + \Big(x_\beta x_\delta - \frac{x^2}{4}\delta_{\beta\delta}\Big)\,\hat x_\alpha \Big(\frac{\partial \bar{\mathfrak{l}}^{(1)}}{\partial |x|}\Big),
\\ \frac{1}{m}T^{\rm III}_{\alpha\beta\delta}(x,0) &=&
\frac{\delta_{\alpha\delta}}{4}\,\hat x_\beta
\Big( \frac{\partial \bar{\mathfrak{g}}^{(0)}}{\partial|x|} + \frac{\partial \bar{\mathfrak{g}}^{(0)}}{|y|\partial\hat c_\beta}  \Big)
\\ && + \Big(\delta_{\alpha\beta}x_\delta +x_\alpha \delta_{\beta\delta} - \frac{x_\beta}{2} \delta_{\alpha\delta}\Big) 
\Big(\bar{\mathfrak{l}}^{(1)}+\bar{\mathfrak{l}}^{(3)}\Big)
\nonumber \\ && + \Big(x_\alpha x_\delta - \frac{x^2}{4}\delta_{\alpha\delta}\Big) \hat x_\beta
\Big(\frac{\partial\bar{\mathfrak{l}}^{(1)}}{\partial|x|} +  \frac{\partial\bar{\mathfrak{l}}^{(1)}}{|y|\partial\hat c_\beta} \Big),
\nonumber
\ea
where the right-hand side should be evaluated in the limit $|y|\to0$.
Using the results above in this subsection, this evaluation is easily performed.

\section{Contribution of the scalar function $S(x,y)$ to the QED kernel: large-$|y|$ asymptotics}
\la{sec:largeYasympt}

The scalar weight function is given by 
\be\la{eq:Sxy_apdx}
S(x,y) = \int_{u,{\rm IR-reg.}} G_0(y-u) s(x,u),
\ee
with $s(x,u)\sim |u|^{-2}$  at large $|u|$, see Eq.\ (\ref{eq:s_0_u_large_u}).
Thus the $y$-dependence of $S(x,y)$ corresponds to the static potential induced (in four space dimensions)
by a charge distribution given by $s(x,u)$,  $x$ playing the role of a fixed position vector.
The function $S(x,y)$ itself is logarithmically infrared-divergent, however 
$-\frac{\partial}{\partial |y|} S(x,y)$, which corresponds to the radial electric field,
is finite.
The electric field generated by a charge distribution falling like $|u|^{-2}$
is of order $|y|^{-1}$, in any number of dimensions greater than two.
This is easiest obtained by applying Gauss' law to a sphere of radius $|y|$ and one finds
\be
-\frac{\partial}{\partial |y|} S(x,y) \stackrel{|y|\to\infty}{=} \frac{1}{384\pi^2 m^2  |y|},
\ee
independent of $x$, which is kept fixed.
Although we have obtained this from the region of large $u$, it is clear that the integral over $|u|$ 
from 0 to a finite $|u|_{\rm max}$ cannot generate an electric field falling off as slowly as $|y|^{-1}$;
instead it generates  an O($|y|^{-3}$) field.

We now proceed to determining the leading behavior of $\frac{\partial}{\partial |x|} S(x,y)$
and $\frac{\partial}{\partial \hat c_\beta} S(x,y)$ at large $|y|$.
For this purpose it is useful to write (similarly to Eq.\ (\ref{eq:f1study}))
\be
s(x,u) - s(0,u) = \sum_{n\geq 0} (-1)^n z_n(u^2) \Big[z_n((x-u)^2) (-1)^n \frac{C_n(\hat u\cdot\widehat{x- u})}{n+1} -
z_n(u^2)  \Big],
\ee
and to realize that the expression inside the square bracket has the asymptotic large-$|u|$ behavior
$\frac{\hat  u\cdot x}{4\pi^2 m\, (n+1)u^2}$.
The series is then still absolutely convergent and one finds
 \be
 s(x,u) - s(0,u) \stackrel{|u|\to\infty}{=}   \frac{1}{192\pi^2 m^2|u|^3} \, {x\cdot \hat u}.
\ee
Thus we get, using Eq.\ (\ref{eq:Sxy_apdx}) and the multipole expansion (\ref{eq:G0_expansion}--\ref{eq:G0_d_n})
of the photon propagator $G_0(y-u)$,
\ba
\frac{\partial \bar{\mathfrak{g}}^{(0)}}{\partial |x|} &=&  \frac{\hat c_\beta}{768\pi^2 m^2|y|},
\\
\frac{\partial \bar{\mathfrak{g}}^{(0)}}{\partial \hat c_\beta} &=&  \frac{|x|}{768\pi^2 m^2|y|}.
\ea
Therefore the scalar contribution to the tensors $T_{\alpha\beta\delta}^{\rm II}(x,y)$ and 
$T_{\alpha\beta\delta}^{\rm III}(x,y)$ is respectively
\ba
\frac{m}{4}\delta_{\beta\delta} \partial_\alpha^{(x)} S(x,y) \stackrel{|y|\to\infty}{=}
\frac{\delta_{\beta\delta} \, \hat y_\alpha}{3072\pi^2 m|y|} (1+{\rm O}((m|y|)^{-1})),
\\
\frac{m}{4}\delta_{\alpha\delta} (\partial_\beta^{(x)}+\partial_\beta^{(y)}) S(x,y) \stackrel{|y|\to\infty}{=}
\frac{-\delta_{\alpha\delta} \,\hat y_\beta}{3072\pi^2 m|y|} (1+{\rm O}((m|y|)^{-1})).  
\ea
We note that this result is consistent with the scalar contribution to the rank-three tensors
satisfying Eq.\ (\ref{eq:TIIandTIII}).

\section{Our version of the kernel code}\label{app:code}

Our implementation of the QED kernel \verb|KQED| can be found in~\cite{KQEDcode}, it is licensed under version 3 of the GNU public license~\cite{GPLv3}. The library is built using GNU automake and is intended to be linked as a static library. An example for integrating the lepton loop using \verb|hcubature|~\cite{cubature} can be found in the companion code \verb|KAMU|~\cite{KAMUcode}, which illustrates how to link to \verb|KQED| and initialize it.

\verb|KQED| includes a look-up-table of the Chebyshev coefficients as a file in single precision, which is entirely read upon initialization (and the crc32c of it is computed for correctness), although all computations are performed in double precision. The code can be compiled with \verb|OpenMP| to use multi-threading whereby the kernel at coordinates $x$ and $y$ can be called safely within a parallel region. The code makes heavy use of AVX/FMA intrinsics to speed up the calculation of the various terms $S(x,y),V_\delta(x,y),$ and $T_{\beta\delta}(x,y)$, particularly in the Clenshaw recurrences (Eqs.~(\ref{eq:clen1}), (\ref{eq:clen2}) and (\ref{eq:clen3})). Where possible, loop-fusion is performed on all necessary weight-function derivatives (see Sec.~\ref{sec:discussion_master_formula}), as well as an internal re-mapping of neighboring elements on the grid to SIMD lanes.

The code compiles a binary which performs some simple unit tests and provides a stress-test for the time taken to compute a fixed number of expensive kernel calls. A script is also included in the package to regression-test the kernel against various $x$ and $y$, and to check the multi-threading equivalence. Heavily loop-unrolled and optimised versions of the subtracted kernels (with 4 arbitrary $\Lambda$ terms Eq.~(\ref{eq:lamsub})) are available as this was one of the most costly parts of our calculation in~\cite{Chao:2020kwq,Chao:2021tvp} and~\cite{Chao:2022xzg}.

\end{document}